\documentclass[12pt]{report}

\usepackage{amsmath} 
\usepackage{amssymb} 
\usepackage{amsfonts}

\usepackage{graphicx,color}                 
\usepackage{subfigure}

\usepackage{rotating}                         
\usepackage{multirow}
\usepackage{enumerate}

\usepackage{tikz}
\usetikzlibrary{arrows,shapes,snakes}
\tikzstyle{block}=[draw opacity=0.7,line width=1.4cm]

\usepackage{setspace}
\renewcommand{\em}[1]{{\textit{#1}}}

\newcommand{\rem}[1]{\em{\textbf{Remark}}: #1}
\newcommand{\rems}[1]{\em{\textbf{Remarks}}: #1}

\newcommand{\sindex}[1]{\em{#1}\index{#1}}

\newcommand{\tagref}[1]{\tag{$\ref{#1}$}}

\makeatletter

\newcommand{\Rmnum}[1]{\expandafter\@slowromancap\romannumeral #1@}
\makeatother 

\graphicspath{{./}{Figs/}{Plots/}}

\title{Recursive Bayesian Filters for Data Assimilation}
\author{Xiaodong Luo}

\begin{document}

\baselineskip=18pt plus1pt

\setcounter{secnumdepth}{3}
\setcounter{tocdepth}{2}

\maketitle
\thispagestyle{empty}


\begin{center}
\LARGE{ \ \\[+5cm] To my parents}
\end{center}



\thispagestyle{empty} 
\section*{\centering Recursive Bayesian Filters for Data Assimilation}
\subsection*{\centering Xiaodong Luo}
\subsection*{\centering University College, University of Oxford}
\subsection*{\centering A thesis submitted for the degree of \em{Doctor of Philosophy}}
\subsection*{\centering Trinity 2009}

Data assimilation refers to any approach designed to improve the estimation of system states or parameters by exploiting the additional information contained in the observations of a dynamical system. In practice, a mathematical model is often not the complete description of the underlying physical process for various reasons. In this case, data assimilation can be used to narrow the gap between the estimation from a mathematical model and reality.

In data assimilation applications, one is often confronted by three problems: nonlinearity, non-Gaussianity and high dimensionality. This dissertation is thus dedicated to studying some data assimilation methods that aim to address these problems.

First of all, we consider two types of nonlinear Kalman filter, the ensemble Kalman filter (EnKF) and the sigma point Kalman filter (SPKF), for data assimilation in nonlinear Gaussian systems. To reduce the computational cost of the SPKF in high dimensional systems, we introduce the reduced rank SPKF.  

Then we proceed to study the Gaussian sum filter (GSF) for data assimilation in nonlinear non-Gaussian systems. A GSF essentially consists of a set of parallel nonlinear Kalman filters. For this reason, we call a nonlinear Kalman filter a ``base filter'' of the GSF. The aforementioned EnKF and reduced rank SPKF can both be used as base filters of a GSF. To reduce the computational cost of a GSF, we also propose an auxiliary algorithm. We show that, if the reduced rank SPKF-based GSF is equipped with the auxiliary algorithm and implemented in parallel, it can achieve almost the same computational speed as the reduced rank SPKF itself. With suitable parameters in the reduced rank SPKF-based GSF or the EnKF-based GSF, the GSF normally outperforms its base filter. 

\thispagestyle{empty}
\thispagestyle{empty}
\pagenumbering{roman}
\tableofcontents
\listoffigures

\clearpage
\pagenumbering{arabic}

\pagenumbering{arabic}
\pagestyle{plain}
\normalsize
\chapter{Introduction} \label{ch0: introduction}

\section{Motivation}
In this dissertation, data assimilation is referred to as any technique that incorporates information from observations into a dynamical system in order to improve the estimation of system states or parameters \cite{Nichols2002}. Throughout this dissertation, our focus will be on the state estimation problem. In principle, the parameter estimation problem can be recast as a state estimation problem by treating the parameters as some unobserved system states \cite{Wan2002}. 

To see the demand for data assimilation in practice, we note that a practical model is usually not a complete description of the underlying physical process in the real world. For example, the limitation of our knowledge in understanding nature, and the model resolution that a modern computer can afford are two of the factors that make a practical model deviate, to some extent, from the underlying physical process. As a result, data assimilation is often employed to narrow the gap between the estimation from a practical model and reality by exploiting the information contained in observations of the underlying physical process.

An an example, we use global numerical weather prediction (NWP) to illustrate the role of data assimilation in practice.   

\begin{figure*}[!t]
\centering
\hspace*{-0.5in} \includegraphics[width=1.01\textwidth]{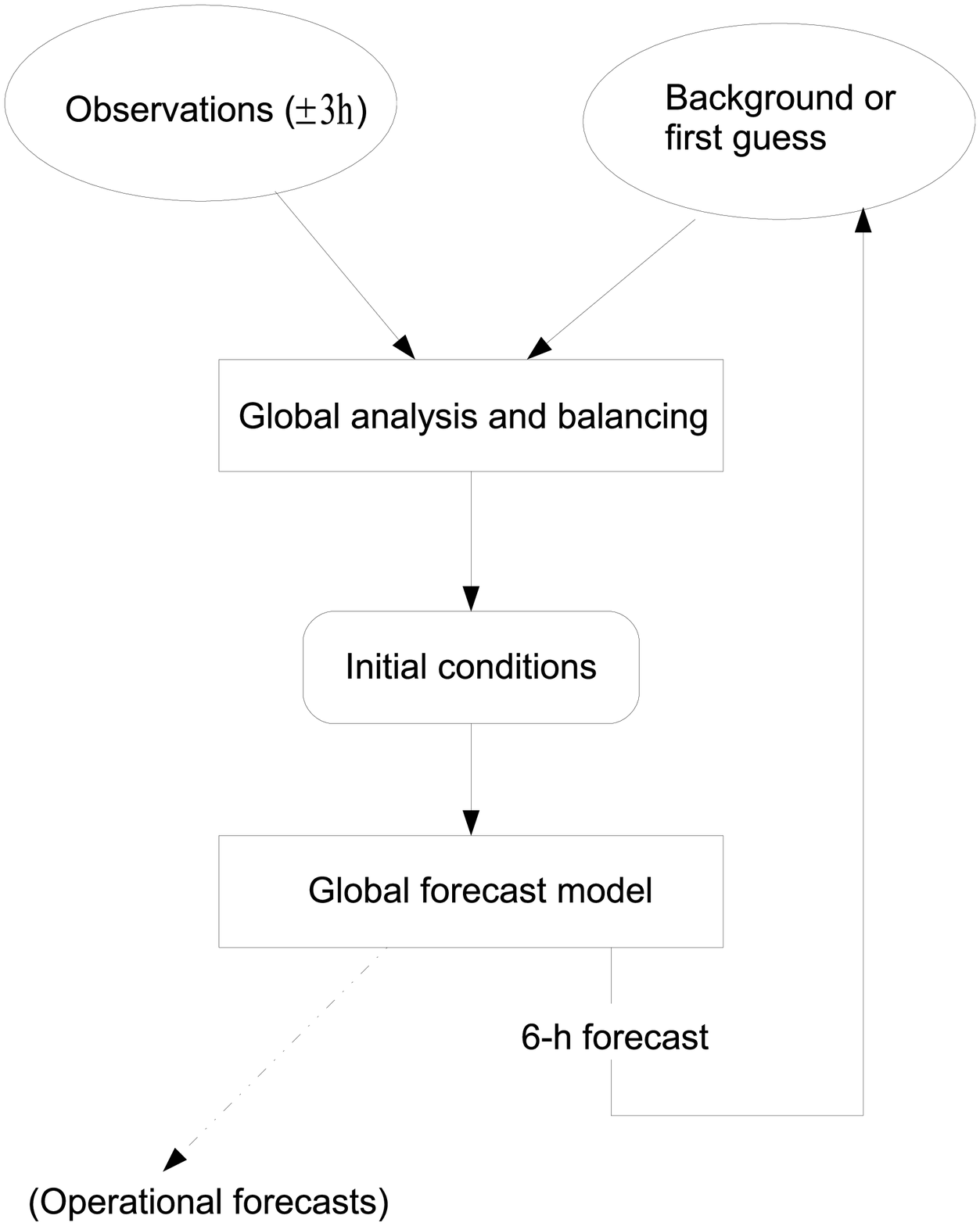} 
\caption{ \label{fig:ch0_illustration_da} Flow diagram of a typical 6-hour data assimilation cycle in global numerical weather prediction. After Fig.~1.4.2 of \cite{Kalnay-atmospheric}. }
\end{figure*} 

In global NWP, the primitive equations that govern the evolution of the atmosphere are derived from the conservation laws of momentum, energy, gas and water masses, together with the equation of state for ideal gases \cite[p. 32]{Kalnay-atmospheric}. Theoretical solution of these governing equations is intractable. Therefore, one has to discretize the governing equations to obtain a numerical solution instead. With discretization, a global prediction model typically has millions of state variables, with a resolution of $50$--$100$ kilometer \cite[pp. 13, 127]{Kalnay-atmospheric} \footnote{The most recent configuration of a global model may have a better resolution. For example, the model used by the Met Office in UK has a mid-latitude resolution of approximately $40$ km, with the number of state variables in the order of $10^7$. For details, see the Met Office website http://www.metoffice.gov.uk/science/creating/daysahead/nwp/um\_config.html.}. Because of the limitation in model resolution, there may be some subgrid-scale physical processes that cannot be resolved \cite[Ch.~4]{Kalnay-atmospheric}.
 
NWP is an initial-value problem, in the sense that we need the present states of the atmosphere, normally called ``initial conditions'', to predict its evolution in the future. For this reason, the determination of the initial conditions is one of the important practices in NWP. In Fig.~\ref{fig:ch0_illustration_da} we show a typical 6-hour data assimilation cycle of present-day operational NWP systems (after Fig.~1.4.2 of \cite{Kalnay-atmospheric}). For convenience of discussion, we suppose that an assimilation cycle starts at time $t$ and ends at time $t+6$ (in unit of hours). One evolves the estimation of the initial conditions at $t$ forward, so as to obtain a first guess at $t+6$. Then the observations between $t-3$ and $t+3$ are incorporated to calibrate the first guess through some data assimilation method, for example, three or four dimensional variational data assimilation \cite[Ch.~5]{Kalnay-atmospheric} (Note that the calibrated initial conditions have to satisfy the governing equations, which is the reason to include ``balancing'' in Fig.~\ref{fig:ch0_illustration_da}). After calibration, one obtains an improved estimation of the initial conditions at $t+6$. With this information, on one hand, one evolves the improved estimation at $t+6$ forward to obtain a first guess of the initial conditions at $t+12$, and then incorporates the observations between $t+3$ and $t+9$ to calibrate the first guess, and so on. On the other hand, one also evolves the improved estimation at $t+6$ forward without any calibration for longer times, for example, $24$-$72$ hours, for the purpose of operational weather forecasts.  
   
Data assimilation can also be adopted in many other fields, for example, climate prediction \cite{Anderson1996}, oceanography \cite{Park2009}, hydrology \cite{Park2009}, petroleum engineering \cite{Oliver2008}, bioinformatics \cite{Nagasaki2006,Yoshida2008}, finance and econometrics \cite{Wang2002,Wells1996}, to name but a few. In general, data assimilation practices in those fields (including NWP) are often confronted by the following three problems: 
\begin{itemize}
\item Nonlinearity: the systems under assimilation are nonlinear;
\item Non-Gaussianity: the probability distributions of the systems under assimilation are non-Gaussian;
\item High dimensionality: the dimensions of the systems under assimilation are very high. Therefore the computational cost is very expensive. 
\end{itemize}

The first two problems, nonlinearity and non-Gaussianity, often make the well-established methods for linear Gaussian systems fail to attain the optimal estimations of the underlying physical processes. Much research has been conducted in the field of estimation theory to address these two problems. For examples, see \cite{Anderson-optimal,Jazwinski1970,Maybeck-stochastic,Simon2006} and the references therein. The last problem of high dimensionality, affecting the computational speed of a data assimilation algorithm, is an important factor for real-time applications. This is frequently discussed in the data assimilation community from a practical point of view. For examples, see \cite{Daley-atmospheric,Evensen2006,Kalnay-atmospheric} and the references therein.

The aim of this dissertation is to study and develop some sequential data assimilation methods, in an attempt to address the above three problems. In this regard, we have three major objectives to achieve, as will be stated in \S~\ref{ch0:objectives_and_approaches}. Before that, however, we would like to introduce some concepts that will be frequently used in this dissertation.

\section{Some concepts in data assimilation}
\subsection{Classification of data assimilation methods} \label{ch0:sec_classification}
Depending on the relative positions between the most recent observations and the system states to be estimated, data assimilation methods can be classified as three categories: predictive algorithms, filtering algorithms, and smoothing algorithms \cite[p. 10]{Simandl2006}, as will be explained below. 

Let $\mathbf{x}_i$ be the system state to be estimated at time $i$, $\mathbf{Y}_k = \left \{\mathbf{y}_k, \mathbf{y}_{k-1}, \dotsb,  \right \}$ be the collection of historical observations available up to and including time $k$, with $\mathbf{y}_j$ being the observation made at instant $j$ ($j \le k$). To use the information contents of $\mathbf{Y}_k$ to improve the estimation of $\mathbf{x}_i$,
\begin{itemize}
\item the estimation method is a predictive algorithm if $k<i$;
\item the estimation method is a filtering algorithm if $k=i$;
\item the estimation method is a smoothing algorithm if $k>i$. 
\end{itemize}

In the data assimilation community, it is customary to classify data assimilation as either a \em{sequential} and or a \em{retrospective} (non-sequential) method \cite{Bouttier-data}. A \sindex{sequential data assimilation} method is an algorithm that utilizes the information contents of the observations up to and including the time when the system state is to be estimated. This is usually used for real time estimation problems. In contrast, a \sindex{retrospective data assimilation} method incorporates not only observations from the past, but also those in the future (relative to the system state to be estimated), which is often applied to the exercise of re-analysis \cite{Bouttier-data}. Thus by definition, a retrospective data assimilation method is a smoothing algorithm, while in general a sequential data assimilation method is a combination of predictive and filtering algorithms. 

\subsection{Dynamical and observation systems}

A \sindex{dynamical system} is a mathematical description of a process that ``consists of a set of possible states, together with a rule that determines the present state in terms of past states'' \cite[p. 2]{Alligood1996}. An \sindex{observation system} is a description of how the observations of a dynamical system are made. In this dissertation, we will follow the notations suggested by Ide et al. \cite{Ide-unified} as far as possible. We normally use the symbol $\mathbf{x}$ to denote a state vector and the symbol $\mathbf{y}$ to denote an observation vector. 

For illustration, let us take the following system 
\begin{subequations} \label{sec: DA systems}
\begin{align}
\label{dynamical system} & \mathbf{x}_{k+1} = \mathcal{M}_{k+1,k} ( \mathbf{x}_k ) + \mathbf{u}_{k} \, , \\
\label{observer} & \mathbf{y}_{k}=\mathcal{H}_{k} ( \mathbf{x}_{k}) + \mathbf{v}_{k} \, ,
\end{align}
\end{subequations}
as an example. Eq.~(\ref{dynamical system}) represents a dynamical system, where $\mathbf{x}_{k}$ denotes the state vector at time $k$, $\mathbf{u}_k$ is the \sindex{dynamical noise}, and $\mathcal{M}_{k,k+1}$ is the \sindex{transition operator}. We define the \em{state space} as the set of all possible system states. Thus the transition operator $\mathcal{M}_{k,k+1}$ maps a state space onto the state space itself. On the other hand, Eq.~(\ref{observer}) represents the corresponding observation system, where $\mathcal{H}_{k}$ is the \sindex{observation operator} at time $k$, $\mathbf{y}_{k}$ is the corresponding observation vector, and $\mathbf{v}_{k}$ is the \sindex{observation noise}. Similarly, we define the \em{observation space} as the set of all possible observations. Thus the observation operator $\mathcal{H}_{k}$ maps a state space onto an observation space. In this dissertation, we normally denote the dimension of the state space by $m$, and the dimension of the observation space by $m^{obv}$ for distinction.

\subsection{Truth, background and analysis}
In this dissertation, the ``true state'' of a physical process at a given time $k$ is referred to as the realization of the process at that instant\footnote{If there is any randomness in the underlying physical process, the ``true state'' at a given instant may not be unique. In this case, it appears more appropriate to call state estimation ``state tracking'', following the argument in \cite{Judd2009}. However, we will follow the convention and still use the term ``state estimation'' throughout this dissertation.}. Following the convention in the data assimilation community, we often call the true state the ``truth''. 

The \sindex{background} at a given time $k$ represents the prior information of the true state at that instant \cite{Bouttier-data}. It can be considered as an analogue to the concept of ``prior distribution'' in Bayesian statistics. Similarly, the \sindex{analysis} at a given time $k$ is the output of a data assimilation algorithm at that instant. In sequential data assimilation, the analysis is normally updated from the background by incorporating the incoming observation. An analysis can be considered as an analogue to the concept of ``posterior distribution'' in Bayesian statistics. 

In this dissertation, normally we will denote the truth, the background and the analysis at instant $k$ by $\mathbf{x}^{tr}_k$, $\mathbf{x}^b_k$, and $\mathbf{x}^a_k$, respectively.

\subsection{Error statistics}\label{ch0_error_statistics}

It is customary to use some offset quantities from the ``true states'' to describe the uncertainties in data assimilation. Following the convention in the data assimilation community \cite{Bouttier-data}, we normally call these offsets ``errors''. Let $\mathbf{x}^{tr}_k$ be the truth at instant $k$, then we can define the following two types of estimation errors \cite{Bouttier-data}:
\begin{itemize}
\item \em{Background error}\index{background error} $\mathbf{\epsilon}^b_k$ at instant $k$: $\mathbf{\epsilon}^b_k =\mathbf{x}^b_k - \mathbf{x}^{tr}_k$.
\item \em{Analysis error}\index{analysis error} $\mathbf{\epsilon}^a_k$ at instant $k$: $\mathbf{\epsilon}^a_k = \mathbf{x}^a_k - \mathbf{x}^{tr}_k$.
\end{itemize}
Correspondingly, the error covariances are defined as \cite{Bouttier-data}:

\begin{itemize}
\item \em{Background error covariance}\index{background error covariance} $\mathbf{P}_k^b$ at instant $k$: $\mathbf{P}_k^b = \mathbb{E} ( (
    \mathbf{\epsilon}^b_k - \mathbb{E} \mathbf{\epsilon}^b_k )
  (\mathbf{\epsilon}^b_k - \mathbb{E} \mathbf{\epsilon}^b_k )^{T}
)$,

\item \em{Analysis error covariance}\index{analysis error covariance} $\mathbf{P}_k^a$ at instant $k$: $\mathbf{P}_k^a = \mathbb{E} ( (
    \mathbf{\epsilon}^a_k - \mathbb{E} \mathbf{\epsilon}^a_k )
  (\mathbf{\epsilon}^a_k - \mathbb{E} \mathbf{\epsilon}^a_k )^{T}
)$,
\end{itemize}
where $\mathbb{E}$ denotes the expectation, and the superscript $T$ means transpose. For convenience, we may also call $\mathbf{P}_k^b$ and $\mathbf{P}_k^a$ \em{background covariance} and \em{analysis covariance}, respectively.  

\section{Objectives and approaches} \label{ch0:objectives_and_approaches}

\subsection{Objectives} \label{ch0:sec_objectives}
In this dissertation we focus on studying sequential data assimilation methods, more specifically, the ensemble Kalman filter (EnKF) and other types of recursive nonlinear filters for data assimilation in high dimensional systems. In this regard, we have three major objectives.  

One objective is to understand various sequential data assimilation algorithms from the point of view of recursive Bayesian estimation (RBE). RBE is a general probabilistic approach that recursively estimates the probability density function (pdf) of an underlying physical process over time. It provides a uniform framework to interpret and derive sequential data assimilation algorithms in various situations, as will be shown in subsequent chapters. 
  
Another objective is to introduce a few filters, including the sigma point Kalman filters (SPKFs) and the sigma point Gaussian sum filters (SPGSFs). The SPKFs \cite{Simandl2006,Merwe2004} were developed to assimilate nonlinear/Gaussian systems. Here by ``nonlinear/Gaussian'' we mean the scenario, where there exists nonlinearity in the dynamical and/or observation system(s), and the underlying system states, together with the dynamical and observation noise, are all assumed to follow some Gaussian distributions\footnote{Similar notations like ``linear/Gaussian'' and ``nonlinear/non-Gaussian'' will be frequently adopted in this dissertation. Their meanings shall be interpreted in a similar way.}. Like the extended Kalman filter (EKF) \cite[ch. 8]{Anderson-optimal}, the SPKFs are extensions of the original Kalman filter \cite{Kalman-new,Kalman1961} to nonlinear/Gaussian systems (all such extensions will be called nonlinear Kalman filters in this dissertation), but they are particularly designed to attack the problem of nonlinearity without the need to compute the derivatives of a nonlinear function. Instead, they all require the generation of some special system states, called sigma points, for the purpose of approximations. For this reason, they are normally known as the sigma point Kalman filters or derivative-free filters \cite{Simandl2006,Merwe2004}. On the other hand, the SPGSFs, are extensions of the SPKFs to nonlinear/non-Gaussian systems. The basic idea of a Gaussian sum filter (GSF) is to use a set of Gaussian distributions to approximate the pdfs of the underlying system states, as well as the dynamical and observation noise if necessary. It can be shown that a GSF essentially consists of a set of parallel nonlinear Kalman filters (cf. Chapter~\ref{ch5:spgsf}), while a SPGSF is just a GSF that consists of the SPKFs.

Our last objective is to increase the computational efficiency of the aforementioned filters in high dimensional systems. As the computational cost (essentially, the computational speed) is often of a concern in practice, one may not wish to directly apply the above filters to assimilate high dimensional systems. Instead, some modifications can be introduced to increase their computational efficiencies. For this purpose, we will present some strategies that aim to reduce the computational cost and/or increase the computational speed of the aforementioned filters.

\subsection{Approaches} \label{ch0: methodology}
We will mainly employ two approaches, namely least squares estimation (LSE) and recursive Bayesian estimation (RBE), to interpret and derive data assimilation algorithms in this dissertation. With the knowledge of both the dynamical and observation noise, RBE is a uniform framework that can be used to derive the EnKFs, the SPKFs and the SPGSFs, as will be shown in subsequent chapters. LSE is equivalent to RBE in linear/Gaussian scenarios, but in general it may differ from RBE in nonlinear and/or non-Gaussian cases. For this reason, in this dissertation, we will adopt RBE more frequently. Nevertheless, there are still some algorithms, for example, the Kalman filter with fading memory (cf. \S ~\ref{ch1: KF fading}), that can be better understood from the standpoint of LSE.

The idea of LSE is first to specify a cost function $J$ with respect to the system states. $J$ is often quadratic, but its concrete form might be case-dependent. 
The optimal estimation $\hat{\mathbf{x}}^{opt}$ of a system state $\mathbf{x}$ is the one that minimizes the cost function $J$ \footnote{In some situations, one may instead define a ``benefit'' function (or utility function). In this case, the optimal estimation is the one that maximizes utility.}, i.e.,
\begin{equation}
\hat{\mathbf{x}}^{opt} = \underset{\mathbf{x}}{\text{argmin}} \, J \, .
\end{equation}
As an example, in the next chapter we will apply LSE to derive the conventional Kalman filter in discrete linear/Gaussian systems.

To illustrate the idea of RBE, let us take Eq.~(\ref{sec: DA systems}) as the system under assimilation. Let $p ( \mathbf{x}_{k} | \mathbf{Y}_{k-1} )$ be the \sindex{prior pdf} of the state $\mathbf{x}_{k}$ conditioned on the observations $\mathbf{Y}_{k-1} = \left \{ \mathbf{y}_{k-1}, \mathbf{y}_{k-2}, \dotsb \right \}$. Once a new observation $ \mathbf{y}_k$ is available, one updates the prior pdf to the \sindex{posterior pdf}  $p ( \mathbf{x}_{k} | \mathbf{Y}_{k} )$ according to Bayes' rule. Based on the dynamical system Eq.~(\ref{dynamical system}), one can compute the prior pdf $p ( \mathbf{x}_{k+1} | \mathbf{Y}_{k} )$ at the next assimilation cycle. Concretely, suppose that $\mathbf{x}_k$ is an $m$-dimensional state vector in the $m$-dimensional real space $\mathbb{R}^{m}$ at time $k$. One can formulate the mathematical description of RBE as follows \cite{Arulampalam2002}:
\begin{subequations} \label{ch0:BRR}
\begin{align}
\label{BRR:update}  p ( \mathbf{x}_{k} | \mathbf{Y}_{k} ) =& \dfrac{ p ( \mathbf{y}_{k} | \mathbf{x}_{k}  ) p ( \mathbf{x}_{k} | \mathbf{Y}_{k-1} ) }{\int p ( \mathbf{y}_{k} | \mathbf{x}_{k}  ) p ( \mathbf{x}_{k} | \mathbf{Y}_{k-1} )  d\mathbf{x}_{k}} \, , \\
\label{BRR:prediction} p ( \mathbf{x}_{k+1} | \mathbf{Y}_{k} ) =& \int p ( \mathbf{x}_{k+1} | \mathbf{x}_{k} )  p ( \mathbf{x}_{k} | \mathbf{Y}_{k} ) d\mathbf{x}_{k} \, ,
\end{align}
\end{subequations}
where $p ( \mathbf{y}_{k} | \mathbf{x}_{k}  )$ is equal to the value of $p (\mathbf{v}_{k} )$ evaluated at $\mathbf{v}_{k} = \mathbf{y}_{k} - \mathcal{H}_{k} ( \mathbf{x}_{k})$ (by Eq.~(\ref{observer})) and conditioned on $\mathbf{x}_{k}$, and $p ( \mathbf{x}_{k+1} | \mathbf{x}_{k}  )$ is equal to the value of $p (\mathbf{u}_{k} )$ evaluated at $\mathbf{u}_{k} = \mathbf{x}_{k+1}-\mathcal{M}_{k,k+1} ( \mathbf{x}_{k})$ (by Eq.~(\ref{dynamical system})) and conditioned on $\mathbf{x}_{k}$. Note that in Eq.~(\ref{ch0:BRR}), we dropped the domain of definition $\mathbb{R}^{m}$ of $\mathbf{x}_k$ in the integrals with respect to $\mathbf{x}_k$ for notational convenience. This convention will be adopted throughout this dissertation when it causes no confusion.

\section{Principal new results}
This dissertation consists of some materials drawn from the following research works.
\begin{itemize}
\item[\textbf{W1.}] Xiaodong Luo and Irene Moroz. ``State Estimation in High Dimensional Systems: The Method of The Ensemble Unscented Kalman Filter'', \em{Inference and Estimation in Probabilistic Time-Series Models} (Cambridge, United Kingdom, 18-20 June 2008).

\item[\textbf{W2.}] Xiaodong Luo and Irene Moroz. ``Ensemble Kalman filter with the unscented transform.'' \em{Physica D} 238 (2009): 549-562.

\item[\textbf{W3.}] Xiaodong Luo and Irene Moroz. ``Sigma point Kalman filters for large-scale systems.'' submitted.

\item[\textbf{W4.}] Xiaodong Luo and Irene Moroz. ``Sigma Point Gaussian Sum Filters \Rmnum{1}: Theory.'' submitted.

\item[\textbf{W5.}] Xiaodong Luo and Irene Moroz. ``Sigma Point Gaussian Sum Filters \Rmnum{2}: Application to high dimensional systems.'' submitted. 
\end{itemize}

For all the works (\textbf{W1} -- \textbf{W5}), I developed the algorithms, wrote the codes, ran the numerical experiments, and wrote the manuscripts.

The principal new results in these research works are:

\begin{itemize}
\item In \textbf{W1} I proposed the reduced rank version of the scaled unscented Kalman filter (SUKF). In this way, the computational cost of the SUKF in high dimensional systems can be reduced.  

\item \textbf{W2} is an extension of the work \textbf{W1}. In \textbf{W2} I considered the implementation of the reduced rank SUKF in the form of a square root filter.   

\item In \textbf{W3} I reviewed two types derivative-free filters, including the SUKF, and the family of divided difference filters (DDFs). Apart from the reduced rank SUKF introduced in \textbf{W1} and \textbf{W2}, I also proposed the reduced rank DDFs, in the form of square root filters.   

\item In \textbf{W4} and \textbf{W5} I explored the idea of Gaussian sum filter (GSF). I used different nonlinear Kalman filters, including the ensemble Kalman filter, the SUKF, and the DDFs, as the base filters of the GSF. I also proposed an auxiliary algorithm in order to reduce the potential computational cost of the GSF and increase its stability. 

\end{itemize}

\section{Outline of this dissertation}
In what follows we provide an outline of the whole dissertation. We will point out our original works with the marker $^*$ in appropriate places, while in the unmarked places, we are following previous works in the literature by default.  

In Chapter~\ref{ch1: KF linear} we study the data assimilation problem in linear/Gaussian systems, which is the foundation of the data assimilation algorithms in subsequent chapters. We apply RBE to solve the problem, which leads to the well-known Kalman filter. We also derive the same result from the point of view of LSE. This will help us to understand one important variant of the Kalman filter in this dissertation, namely the Kalman filter with fading memory (KF-FM), designed to improve the robustness of the filter. We introduce the square root filter (SRF) as another variant of the Kalman filter in order to increase the numerical accuracy and stability of the filter. For these benefits, all the nonlinear filters to be introduced in this dissertation will be implemented in both the forms of the KF-FM and the SRF.

In Chapters~\ref{ch2: EnKF} - \ref{ch4:ddfs} we consider the data assimilation problem in nonlinear/Gaussian systems. We review some extensions of the original Kalman filter to nonlinear/Gaussian systems from the point of view of RBE. We also develop some reduced rank versions for some of these extensions. 

Chapter~\ref{ch2: EnKF} focuses on reviewing the ensemble Kalman filter (EnKF). There are two major types of the EnKFs in the literature, called the stochastic EnKF and the ensemble square root filter (EnSRF), respectively. In general, all the implementations of the EnKF in the literature can be deemed as different approximation schemes to approximate the integrals in RBE numerically. To improve the performance of the EnKF, it is customary to introduce two auxiliary techniques, namely covariance inflation and filtering. Covariance inflation compensates for the systematic underestimation of an error covariance in the EnKF. Moreover, it also makes the EnKF behave like the KF-FM. For these reasons, adopting covariance inflation in the EnKF often increases the robustness and accuracy of the filter. On the other hand, covariance filtering aims to remove spuriously large correlations between distant locations due to the effect of small ensemble size in the EnKF. Hence, adopting covariance filtering may also help an EnKF to achieve a better performance. Through some numerical experiments, we compare the performance of the stochastic EnKF with that of the ensemble transform Kalman filer (ETKF), one of the EnSRFs. We show that the ETKF consistently outperforms the stochastic EnKF. 

In Chapter~\ref{ch3:ukf} we review another type of nonlinear Kalman filter, called the scaled unscented Kalman filter (SUKF), based on the concept of the scaled unscented transform (SUT). One feature of the SUKF is that it does not require the linearization of nonlinear systems as does the extended Kalman filter. Instead, the SUKF tackles the problem of nonlinearity by producing sigma points for the purpose of approximation. This is similar to the idea of the EnKF and is convenient in implementation. We conduct an accuracy analysis for the SUKF via Taylor series expansion. In this way, we show that the SUKF can achieve better accuracy than the EnKF. For data assimilation in high dimensional systems, we propose a reduced rank version of the SUKF in order to reduce the computational cost $^{*}$. Through some numerical experiments, we examine the performance of the reduced rank SUKF and compare it with the ETKF. We show that the reduced rank SUKF outperforms the ETKF (as a representation of the EnKF) given the same amount of information. 

In Chapter~\ref{ch4:ddfs} we review another family of nonlinear Kalman filters, called the divided difference filters (DDFs), based on Stirling's Interpolation Formula. The DDFs also do not require the linearization of nonlinear systems under assimilation. Instead, like the SUKF, they generate sigma points for the purpose of approximation. For this reason, the SUKF and the DDFs are uniformly called the sigma point Kalman filters (SPKFs) or derivative-free filters in the literature. We conduct accuracy analyses on the DDFs via Taylor series expansions. For data assimilation in high dimensional systems, we also propose reduced rank versions of the DDFs in order to reduce the computational cost $^{*}$. We examine the performances of the reduced rank DDFs through some numerical experiments. A performance comparison between the reduced rank DDFs, the reduced rank SUKF and the ETKF is also presented.

In Chapter~\ref{ch5:spgsf} we consider the data assimilation problem in nonlinear/non-Gaussian systems. To this end, we introduce the Gaussian sum filter (GSF) as an approximate solution. A GSF essentially consists of a set of parallel nonlinear Kalman filters (called ``base filter'' of the GSF in this dissertation). All the aforementioned nonlinear Kalman filters, i.e., the EnKF, the reduced rank SUKF and DDFs, can be adopted as the base filters of a GSF. A potential problem of the GSF is that, in some situations, the number of Gaussian distributions in the GSF may increase very rapidly with time. To tackle this problem, we suggest conducting pdf re-approximations. We propose an auxiliary algorithm based on the concept of the unscented transform in order to implement the above strategy $^{*}$. If the GSF adopts one of the reduced rank SPKFs, such as the reduced rank SUKF or one of the reduced rank DDFs, as its base filter (the GSF implemented in this way will be called the sigma point GSF, or SPGSF for short), and if the GSF is implemented in parallel, then in principle the SPGSF can achieve almost the same computational speed as its base filter, the reduced rank SPKF. If the EnKF is chosen as the base filter, there will be extra costs in conducting pdf re-approximations. The computational speed of the EnKF-based GSF is roughly the same as those of the SPGSFs. We conduct some numerical experiments to examine the performances of the GSFs with different base filters. We show that, in general, the GSFs outperform their corresponding base filters. 

In Chapter~\ref{ch6:conclusions} we conclude the whole dissertation, and summarize the main results that we have achieved. We also discuss some outstanding problems and possible extensions of the works done in this dissertation.

\chapter{Conventional Kalman filter for linear/Gaussian systems} \label{ch1: KF linear}


\section{Overview}
The conventional Kalman filter (KF) is ``an optimal recursive data processing algorithm'' \cite{Maybeck-stochastic} for linear systems that are possibly contaminated by some Gaussian noise. Here by ``conventional'' we mean the algorithms that were originally developed in the pioneering works in the late 1950s and early 1960s by, for example, Swerling \cite{Swerling1959,Swerling1963}, Kalman \cite{Kalman-new}, Kalman and Bucy \cite{Kalman1961}, which include the scenarios where the dynamical and observation systems are either discrete or continuous \footnote{The filtering algorithm for hybrid systems, for example, a continuous dynamical system measured by a discrete observation system, can be derived in a similar way. For details see, for example, \cite[ch. 7]{Jazwinski1970} and the references therein}. The history of the conventional Kalman filter can be found in some early lecture notes \cite{Kailath1977, Sorenson1970}, and the more recent textbook \cite{Simon2006}.

In some early works, e.g. \cite{Kalman-new}, the conventional Kalman filter was derived by minimizing a quadratic cost function. This is intimately related to the least squares estimation (LSE) \cite{Kailath1977, Sorenson1970}. One advantage of adopting LSE is that it is widely studied in control and optimization theories. Thus one may apply many well-established methods in those fields to solve the data assimilation problem in various situations. However, the disadvantage would be that, 
in order to achieve the optimality in different situations, one may have to construct different proper cost functions\footnote{Here by ``proper cost function'' we mean the function that will lead to the optimal solution by minimizing it.}. However, there may lack such a systematic method that can be employed to find the proper cost functions in general situations, since the cost functions themselves would depend on the criteria of optimality in use.

Alternatively, the conventional Kalman filter can also be derived from the point of view of recursive Bayesian estimation (RBE) \cite{Ho1964}. Under the framework of RBE, the data assimilation problem is solved in terms of the posterior probability density function (pdf) of the system states conditioned on the available observations (cf. Eq.~(\ref{ch0:BRR})). The advantage of RBE is that one does not need to specify any optimality criterion when conducting recursive Bayesian estimation. Instead, it is after obtaining the posterior pdf that one makes statistical inferences, for example, estimating the mean and covariance, according to their own criteria of optimality. Thus, without involving any specific optimality criterion, the framework of RBE can be applied to various situations without any change. On the other hand, the disadvantage of RBE is that one has to compute some integrals, which are often analytically intractable. Thus some numerical methods have to be adopted to approximate the integrals, which, however, might be computationally very expensive in high dimensional systems. For this reason, the issue of how to reduce the computational cost in approximating the integrals will be a frequent topic in this dissertation, as will be seen in subsequent chapters.

The objective of this chapter is to introduce the conventional Kalman filter for linear/Gaussian systems as the starting point for studying the nonlinear Kalman filters and the Gaussian sum filters in subsequent chapters. The derivations of the conventional Kalman filter from both the points of views of LSE and RBE will be presented. In addition, two variants of the conventional Kalman filter, namely, the square root Kalman filter (SRKF) and the Kalman filter with fading memory (KF-FM), will be particularly discussed, since they will be frequently used in subsequent chapters.

\section{Problem statement and solution}
\subsection{Problem statement}
In this chapter we consider the following scenario: a linear stochastic dynamical system is driven by a Gaussian random process. The observations of this dynamical system are made by some instrument (the observation system), which is also characterized by a linear stochastic system driven by a Gaussian random process. We are interested in estimating the underlying system states at different times.

To avoid complicating our discussion, here we only study a specific class of linear systems. Later on we will consider linear systems in more general situations and give some hints for deriving the corresponding assimilation algorithms. Thus we first confine ourselves to the following class of linear systems: the dynamical system is a discrete-time first order Markov process \cite[ch. 3]{Jazwinski1970}. The dynamical and observation noise are uncorrelated, white and Gaussian with zero means. Moreover, there are no input variables existing in the dynamical system\footnote{The presence of input variables is a standard setting in many textbooks (e.g., \cite{Simon2006}), where the controllability of a dynamical system is often a concern.}. Mathematically, we can formulate the above class of linear/Gaussian systems as follows:
\begin{subequations}  \label{ch1: linear systems}
\begin{align}
\label{ch1: linear dynamical system} & \mathbf{x}_k  = \mathcal{M}_{k,k-1} \, \mathbf{x}_{k-1} + \mathbf{u}_{k}  \, ,  \\
\label{ch1: linear observation system} &  \mathbf{y}_k  = \mathcal{H}_{k} \, \mathbf{x}_{k} + \mathbf{v}_{k} \, , \\
\label{ch1: dynamical noise Gaussian} & \mathbf{u}_{k} \sim N (\mathbf{u}_{k}: \mathbf{0}, \mathbf{Q}_k ) \, ,\\
\label{ch1: observation noise Gaussian} & \mathbf{v}_{k} \sim N (\mathbf{v}_{k}: \mathbf{0}, \mathbf{R}_k ) \, ,\\
\label{ch1: dynamical noise white} & \mathbb{E} ( \mathbf{u}_{j} \mathbf{u}_{k}^T ) = \delta_{k,j}\mathbf{Q}_k \, ,\\
\label{ch1: observation noise white} & \mathbb{E} ( \mathbf{v}_{j} \mathbf{v}_{k}^T ) = \delta_{k,j}\mathbf{R}_k \, ,\\
\label{ch1: noise uncorrelated}  & \mathbb{E} ( \mathbf{u}_{i} \mathbf{v}_{j}^T ) = \mathbf{0} \quad \forall \, i, \, j \, .
\end{align}
\end{subequations}

Eqs.~(\ref{ch1: linear dynamical system}) and (\ref{ch1: linear observation system}) represent the $m$-dimensional dynamical system and the $m^{obv}$-dimensional observation system, respectively, where $\mathbf{x}_k$ denotes the $m$-dimensional system state at time $k$, and $\mathbf{y}_k$ means the $m^{obv}$-dimensional corresponding observation. On the other hand, the transition operator $\mathcal{M}_{k,k-1}$ and the observation operator $\mathcal{H}_{k}$ are $m \times m$ and $m^{obv} \times m$ matrices, respectively. They are both independent of the system states at any time.

Eqs.~(\ref{ch1: dynamical noise Gaussian})-(\ref{ch1: noise uncorrelated}) imply that the $m$-dimensional dynamical noise $\mathbf{u}_{k}$ and the $m^{obv}$-dimensional observation noise $\mathbf{v}_{k}$ are uncorrelated, white, and Gaussian with zero means. In particular, Eqs.~(\ref{ch1: dynamical noise Gaussian})-(\ref{ch1: observation noise white}) indicate that $\mathbf{u}_{k}$ and $\mathbf{v}_{k}$ follow white Gaussian processes with the covariance at time $k$ being $\mathbf{Q}_k$ and $\mathbf{R}_k$, respectively. The symbol ``$\sim$'' in Eqs.~(\ref{ch1: dynamical noise Gaussian}) and (\ref{ch1: observation noise Gaussian}) means ``following the distribution ''. The notation $N (\mathbf{x}: \mathbf{\mu}, \mathbf{\Sigma} )$ represents a Gaussian distribution with $\mathbf{x}$ being the random variable, whose mean and covariance are $\mathbf{\mu}$ and $\mathbf{\Sigma}$, respectively. Finally, $\delta_{k,j}$ denotes the Kronecker delta function, i.e.,
\begin{equation}
\delta_{k,j} = \left \{ 
\begin{array}{l}
1, \quad \text{if} \quad k=j \, , \\
0, \quad \text{if} \quad k \neq j \, .
\end{array}
\right.
\end{equation}

When there causes no confusion, we may often drop the dimension information of the matrices and vectors in subsequent derivations.

\subsection{Deriving the conventional Kalman filter from least squares estimation}\label{ch1:sec:LSE}
Here we mainly follow \cite{Bouttier-data} to derive the conventional Kalman filter from the point of view of least squares estimation. Without lost of generality, we suppose that at the $(k-1)$-th assimilation cycle one has the $m$-dimensional analysis $\mathbf{x}_{k-1}^a$ and the corresponding $m \times m$ error covariance $\mathbf{P}_{k-1}^a$. For convenience of discussion, we divide the procedures of the conventional Kalman filter into two steps: propagation (or prediction) and filtering.

\subsubsection{Propagation step}
According to Eq.~(\ref{ch1: linear dynamical system}), the expectation $\bar{\mathbf{x}}_k$ of the system state $\mathbf{x}_k$ is given by
\begin{equation}
\bar{\mathbf{x}}_k = \mathcal{M}_{k,k-1} \bar{\mathbf{x}}_{k-1} \, ,
\end{equation}
which is normally used as the estimation of the background at instant $k$. Since we do not know $\bar{\mathbf{x}}_{k-1}$, the analysis $\mathbf{x}_{k-1}^a$ will be used as an estimation instead. Thus, the $m$-dimensional background $\mathbf{x}^b_k$ is estimated as
\begin{equation}
\mathbf{x}^b_k = \mathcal{M}_{k,k-1} \mathbf{x}_{k-1}^a \, .
\end{equation}
The corresponding $m \times m$ background error covariance $\mathbf{P}^b_k$ is given by
\begin{equation} \label{ch1:sec_lse_propagation_of_analysis_cov}
\begin{split}
\mathbf{P}^b_k = & \mathbb{E} ( \mathbf{x}^b_k - \mathbf{x}_k^{tr}) ( \mathbf{x}^b_k - \mathbf{x}_k^{tr}) ^T \\
=& \mathbb{E} ( \mathcal{M}_{k,k-1} (\mathbf{x}_{k-1}^a - \mathbf{x}_{k-1}^{tr}) - \mathbf{u}_k ) ( \mathcal{M}_{k,k-1} (\mathbf{x}_{k-1}^a - \mathbf{x}_{k-1}^{tr}) -  \mathbf{u}_k ) ^T \\
=& \mathcal{M}_{k,k-1} \mathbf{P}_{k-1}^a \mathcal{M}_{k,k-1}^T + \mathbf{Q}_k \, ,
\end{split}
\end{equation}
where $\mathbf{x}_k^{tr}$ is the truth at instant $k$. Note that to derive Eq.~(\ref{ch1:sec_lse_propagation_of_analysis_cov}), we have assumed that the analysis error $\mathbf{\epsilon}_{k-1} = \mathbf{x}_{k-1}^a - \mathbf{x}_{k-1}^{tr}$ is independent of the dynamical noise $\mathbf{u}_k$.
\subsubsection{Filtering step}
After a new $m^{obv}$-dimensional observation $\mathbf{y}_k$ is available, one incorporates the new information so as to update the $m$-dimensional background $\mathbf{x}_k^b$ to the analysis $\mathbf{x}_k^a$. To this end, one needs to find an optimal $m \times m^{obv}$ weight matrix $\mathbf{K}_k$ (normally called Kalman gain), so that the analysis $\mathbf{x}_k^a$ updated according to the following rule
\begin{equation} \label{ch1:LSE_analysis_update}
\mathbf{x}_k^a = \mathbf{x}_k^b + \mathbf{K}_k \left ( \mathbf{y}_k - \mathcal{H}_k  \mathbf{x}_k^b \right )
\end{equation}
minimizes the expectation of the energy (the cost function)
\begin{equation}\label{ch1:LSE_average_energy}
J_k = \mathbb{E} \lVert \mathbf{\epsilon}^a_k \rVert^2 = \mathbb{E} \lVert \mathbf{x}_k^a - \mathbf{x}_k^{tr} \rVert^2
\end{equation}
of the analysis error $\mathbf{\epsilon}^a_k = \mathbf{x}_k^a - \mathbf{x}_k^{tr}$ \cite[p. 84]{Simon2006}. 

The reason to use Eq.~(\ref{ch1:LSE_analysis_update}) to update the background $\mathbf{x}_k^b$ is because one would normally expect the background, the analysis, and the observation to be unbiased estimations, i.e.,
\begin{equation} \label{ch1:LSE_unbiasness}
\begin{split}
&  \mathbb{E} \mathbf{\epsilon}^b_k = \mathbb{E} ( \mathbf{x}_k^b - \mathbf{x}_k^{tr} ) =0 \, , \\
& \mathbb{E} \mathbf{\epsilon}^a_k = \mathbb{E} ( \mathbf{x}_k^a - \mathbf{x}_k^{tr} ) =0 \, , \\
& \mathbb{E} \mathbf{v}_k = \mathbb{E} ( \mathbf{y}_k - \mathcal{H}_k \mathbf{x}_k^{tr} ) =0 \, , 
\end{split}
\end{equation}
where $\mathbf{\epsilon}^b_k$, $\mathbf{\epsilon}^a_k$ denote the background and analysis errors, respectively, while $\mathbf{v}_k$ is the observation noise at time $k$.

To see the rationale behind Eq.~(\ref{ch1:LSE_analysis_update}), one may first write the analysis as a linear combination of the $m$-dimensional background $\mathbf{x}_k^b$ and the $m^{obv}$-dimensional observation $\mathbf{y}_k$ such that 
\begin{equation} \label{ch1:LSE_linear_combination}
\mathbf{x}_k^a = \mathbf{C}_k \mathbf{x}_k^b + \mathbf{W}_k \mathbf{y}_k \, ,
\end{equation}
where $\mathbf{C}_k$ and $\mathbf{W}_k$ are $m \times m$ and $m \times m^{obv}$ constant matrices, respectively. Because of the unbiasedness, by Eq.~(\ref{ch1:LSE_unbiasness}) one has 
\begin{equation}
\mathbb{E} ( \mathbf{x}_k^a -  \mathbf{x}_k^{tr} ) = ( \mathbf{C}_k +\mathbf{W}_k \mathcal{H}_k -\mathbf{I}_m ) \mathbb{E} \mathbf{x}_k^{tr} = \mathbf{0} \, ,
\end{equation}
where $\mathbf{I}_m$ is the $m \times m$ identity matrix, so that $\mathbf{C}_k = \mathbf{I}_m- \mathbf{W}_k \mathcal{H}_k$ (with $\mathcal{H}_k$ being the $m^{obv} \times m$ observation operator). Substituting this identity into Eq.~(\ref{ch1:LSE_linear_combination}) and replacing $\mathbf{W}_k$ by $\mathbf{K}_k$, one obtains Eq.~(\ref{ch1:LSE_analysis_update}).

On the other hand, by definition (cf. \S~\ref{ch0_error_statistics}) the analysis error covariance 
\begin{equation}
\begin{split}
\mathbf{P}_k^a =& \mathbb{E} ( ( \mathbf{\epsilon}^a_k - \mathbb{E} \mathbf{\epsilon}^a_k ) (\mathbf{\epsilon}^a_k - \mathbb{E} \mathbf{\epsilon}^a_k )^{T}  ) \\
=& \mathbb{E} ( \mathbf{\epsilon}^a_k (\mathbf{\epsilon}^a_k )^{T}  ) \, . \\
\end{split}
\end{equation}
Thus it is clear that the cost function in Eq.~(\ref{ch1:LSE_average_energy}) is equivalent to the trace of the error covariance $\mathbf{P}_k^a$, i.e.,
\begin{equation}
J_k = \mathbb{E} \lVert \mathbf{\epsilon}^a_k \rVert^2 = \mathbb{E} ( (\mathbf{\epsilon}^a_k )^{T}  \mathbf{\epsilon}^a_k ) = \text{Tr} ( \mathbf{P}_k^a ) \, ,
\end{equation}
where the symbol $\text{Tr} ( \bullet )$ means the trace of a matrix. Consequently, the optimal state estimation problem in Eq.~(\ref{ch1: linear systems}) now becomes an optimization problem whose objective is to
\[
minimize~\text{Tr} (\mathbf{P}_k^a)~over~all~possible~weights~\mathbf{K}_k.
\]

Subtracting the truth $\mathbf{x}_k^{tr}$ from Eq. (\ref{ch1:LSE_analysis_update}), one has
\begin{equation} \label{ch1:LSE_analysis_substract}
\mathbf{x}^a_k - \mathbf{x}^{tr}_k = ( \mathbf{x}^b_k - \mathbf{x}^{tr}_k
) + \mathbf{K}_k ( (\mathbf{y}_k - \mathcal{H}_k \mathbf{x}_k^{tr}
) - ( \mathcal{H}_k \mathbf{x}^b_k - \mathcal{H}_k \mathbf{x}_k^{tr} ) ).
\end{equation}
Thus Eq.~(\ref{ch1:LSE_analysis_substract}) can be re-written as
\begin{equation} 
\mathbf{\epsilon}^a_k = \mathbf{\epsilon}^b_k + \mathbf{K}_k (
\mathbf{v}_k - \mathcal{H}_k \mathbf{\epsilon}^b_k ).
\end{equation}
Therefore one can obtain the analysis error covariance in terms of the background error covariance by noting that 
\begin{equation} \label{ch1:LSE_analysis_covariance_1}
\begin{split}
\mathbf{P}_k^a = & \mathbb{E} ( \mathbf{\epsilon}^a_k (\mathbf{\epsilon}^a_k )^{T}  ) \\
= & \mathbb{E} ( \mathbf{\epsilon}^b_k ( \mathbf{\epsilon}^b_k )^T ) - \mathbb{E} (
\mathbf{\epsilon}^b_k ( \mathbf{\epsilon}^b_k )^T )  \mathcal{H}_k^T \mathbf{K}_k^T +
\mathbf{K}_k \mathbb{E} ( \mathbf{v}_k (\mathbf{v}_k )^T
) \mathbf{K}_k^T \\
& - \mathbf{K}_k  \mathcal{H}_k \mathbb{E} ( \mathbf{\epsilon}^b_k ( \mathbf{\epsilon}^b_k )^T
) + \mathbf{K}_k  \mathcal{H}_k \mathbb{E} ( \mathbf{\epsilon}^b_k ( \mathbf{\epsilon}^b_k )^T )  \mathcal{H}_k^T \mathbf{K}_k^T \, . \\
\end{split}
\end{equation}
Note that to obtain the above result, we have assumed that the background error $\mathbf{\epsilon}^b_k$ and observation noise $\mathbf{v}_k$ are independent, so that $\mathbb{E} (
\mathbf{\epsilon}^b_k ( \mathbf{v}_k  )^T ) = \mathbb{E} (
\mathbf{v}_k ( \mathbf{\epsilon}^b_k )^T ) = \mathbf{0}$.

Also note that $\mathbf{P}_k^b = \mathbb{E} ( \mathbf{\epsilon}^b_k ( \mathbf{\epsilon}^b_k )^T )$ is the background error covariance, and $\mathbf{R}_k = \mathbb{E} (\mathbf{v}_k ( \mathbf{v}_k  )^T )$ is the covariance of the observation noise, thus one can re-write Eq.~(\ref{ch1:LSE_analysis_covariance_1}) as
\begin{equation} \label{ch1:LSE_analysis_covariance_2}
\mathbf{P}_k^a = \mathbf{P}_k^b - \mathbf{P}_k^b \mathcal{H}_k^T \mathbf{K}_k^T + \mathbf{K}_k
\mathbf{R}_k \mathbf{K}_k^T - \mathbf{K}_k \mathcal{H}_k \mathbf{P}_k^b + \mathbf{K}_k \mathcal{H}_k
\mathbf{P}_k^b \mathcal{H}_k^T \mathbf{K}_k^T \, .
\end{equation}
Therefore the trace of $\mathbf{P}_k^a$ is given by
\begin{equation} \label{DA:trace A}
\text{Tr} ( \mathbf{P}_k^a ) = \text{Tr} ( \mathbf{P}_k^b ) +  \text{Tr} ( \mathbf{K}_k \mathbf{R}_k \mathbf{K}_k^T ) - 2 \text{Tr} ( \mathbf{K}_k \mathcal{H}_k \mathbf{P}_k^b ) +  \text{Tr} ( \mathbf{K}_k \mathcal{H}_k
\mathbf{P}_k^b \mathcal{H}_k^T \mathbf{K}_k^T ) \, .
\end{equation}
Note that to derive Eq.~(\ref{DA:trace A}), we have utilized the fact that $\mathbf{P}_k^b \mathcal{H}_k^T \mathbf{K}_k^T$ is the transpose of $\mathbf{K}_k \mathcal{H}_k \mathbf{P}_k^b$, hence their traces are equivalent.  

To minimize $\text{Tr}(\mathbf{P}_k^a)$, a necessary condition for an optimal weight $\mathbf{K}_k$ is that 
$ d\text{Tr}(\mathbf{P}_k^a) / d\mathbf{K}_k =0$. For derivation, the following differential rules of matrix calculus will be useful \cite[p.~669]{Lutkepohl-introduction}: given a constant matrix $\mathbf{A}$ and a variable matrix
$\mathbf{B}$,
\begin{equation} \label{ch1:LSE_differential_rules}
\begin{split}
& \dfrac{d}{d\mathbf{B}} \text{Tr}(\mathbf{B} \mathbf{A}) =
\mathbf{A}^T \, ,\\
& \dfrac{d}{d\mathbf{B}} \text{Tr}(\mathbf{B} \mathbf{A} \mathbf{B}^T)
=  \mathbf{B} ( \mathbf{A} + \mathbf{A}^T ) \, . \\
\end{split}
\end{equation}
Applying the above rules to Eq.~(\ref{DA:trace A}) and noting that covariance matrices are symmetric, we have 
\begin{equation}
 \dfrac{d}{d\mathbf{K}_k} \text{Tr}(\mathbf{P}_k^a) = 2 ( \mathbf{K}_k \mathbf{R}_k - \mathbf{P}_k^b \mathcal{H}_k^T +  \mathbf{K}_k \mathcal{H}_k \mathbf{P}_k^b \mathcal{H}_k^T )
 = 0 \, .
\end{equation}

Therefore the optimal weight $\mathbf{K}_k^{opt}$ satisfies
\begin{equation} \label{ch1:LSE_optimal_weight}
\mathbf{K}_k^{opt} = \mathbf{P}_k^b \mathcal{H}_k^T ( \mathcal{H}_k \mathbf{P}_k^b \mathcal{H}_k^T + \mathbf{R}_k )^{-1} \, 
\end{equation}
by assuming that $\mathcal{H}_k \mathbf{P}_k^b \mathcal{H}_k^T + \mathbf{R}_k$ is invertible \footnote{Otherwise one has to adopt the generalized matrix inverse here.}.

Also note that 
\begin{equation}
\left. \dfrac{d^2}{(d \mathbf{K}_k)^2}
  \text{Tr}(\mathbf{P}_k^a) \right\vert_{\mathbf{K}_k^{opt}} = \left. 2 \dfrac{d}{\mathbf{K}_k} ( \mathbf{K}_k \mathcal{H}_k
\mathbf{P}_k^b \mathcal{H}_k^T - \mathbf{P}_k^b \mathcal{H}_k^T + \mathbf{K}_k \mathbf{R}_k  ) \right\vert_{\mathbf{K}_k^{opt}} = 2
 (\mathcal{H}_k \mathbf{P}_k^b \mathcal{H}_k^T + \mathbf{R}_k )
\end{equation} 
is positive definite, which confirms that $\text{Tr}(\mathbf{P}_k^a)$ attains its minimum at $\mathbf{K}_k^{opt}$. 

Substituting Eq.~(\ref{ch1:LSE_optimal_weight}) into Eq.~(\ref{ch1:LSE_analysis_covariance_2}) and with some algebra, it can be shown that 
\begin{equation}
\mathbf{P}_k^a \vert_{\mathbf{K}_k^{opt}} = \mathbf{P}_k^b -  \mathbf{K}_k^{opt} \mathcal{H}_k \mathbf{P}_k^b \, .
\end{equation}

\rem We note that the analysis update formula Eq.~(\ref{ch1:LSE_analysis_update}) can also be obtained by minimizing the following quadratic cost function
\begin{equation} \label{ch1:LSE_weighted_cost}
J( \mathbf{x}_k ) = \dfrac{1}{2} (\mathbf{x}_k - \mathbf{x}^b_k )^T
(\mathbf{P}_k^b )^{-1} (\mathbf{x}_k - \mathbf{x}^b_k ) + \dfrac{1}{2}
(\mathbf{y}_k - \mathcal{H}_{k}\mathbf{x}_k )^T \mathbf{R}_k^{-1}
(\mathbf{y}_k - \mathcal{H}_{k} \mathbf{x}_k ) \, .
\end{equation}
For a proof, please see Appendix~\ref{appendix:weighted LSE}. This fact will be used later in \S~\ref{ch1: KF fading} to derive the KF-FM, a variant of the conventional Kalman filter.

\subsection{Deriving the conventional Kalman filter from recursive Bayesian estimation}
Here we mainly follow \cite[\S~2.2]{Simandl2006} to derive the conventional Kalman filter from the point of view of recursive Bayesian estimation. Without loss of generality, we suppose that at instant $k-1$, one has the posterior pdf $p ( \mathbf{x}_{k-1} \vert \mathbf{Y}_{k-1} ) = N ( \mathbf{x}_{k-1}: \mathbf{x}_{k-1}^a, \mathbf{P}_{k-1}^a )$ conditioned on the observations $\mathbf{Y}_{k-1} = \left \{\mathbf{y}_{k-1}, \mathbf{y}_{k-2}, \dotsb \right \}$, where $\mathbf{x}_{k-1}^a$ and $\mathbf{P}_{k-1}^a$ are the analysis and the corresponding error covariance, respectively.

\subsubsection{Propagation step}
At the propagation step, one adopts Eq.~(\ref{BRR:prediction}) to compute the prior pdf $p ( \mathbf{x}_{k} \vert \mathbf{Y}_{k-1} )$ at instant $k$, so that
\begin{equation}\label{ch1:RBE_propagation_1}
p ( \mathbf{x}_{k} | \mathbf{Y}_{k-1} ) = \int p ( \mathbf{x}_{k} | \mathbf{x}_{k-1} )  p ( \mathbf{x}_{k-1} | \mathbf{Y}_{k-1} ) d\mathbf{x}_{k-1} \, .
\end{equation}

By Eqs.~(\ref{ch1: linear dynamical system}) and (\ref{ch1: dynamical noise Gaussian}), $p ( \mathbf{x}_{k} | \mathbf{x}_{k-1} ) = N ( \mathbf{x}_{k}: \mathcal{M}_{k,k-1} \, \mathbf{x}_{k-1}, \mathbf{Q}_{k} )$. Substituting this into Eq.~(\ref{ch1:RBE_propagation_1}), one has
\begin{equation} \label{ch1:RBE_propagation_2}
p ( \mathbf{x}_{k} | \mathbf{Y}_{k-1} ) = \int N ( \mathbf{x}_{k}: \mathcal{M}_{k,k-1} \, \mathbf{x}_{k-1}, \mathbf{Q}_{k} )   N ( \mathbf{x}_{k-1}: \mathbf{x}_{k-1}^a, \mathbf{P}_{k-1}^a ) d\mathbf{x}_{k-1} \, .
\end{equation}
With some algebra, it can be shown that, $p ( \mathbf{x}_{k} | \mathbf{Y}_{k-1} )$ follows a Gaussian distribution $N ( \mathbf{x}_{k}: \mathbf{x}_{k}^b, \mathbf{P}_{k}^b )$, where
\begin{subequations}\label{ch1:RBE_propagation_3}
\begin{align}
\label{ch1:RBE_background}& \mathbf{x}_{k}^b = \mathcal{M}_{k,k-1} \, \mathbf{x}_{k-1}^a \, , \\
\label{ch1:RBE_background_covariance}& \mathbf{P}_{k}^b = \mathcal{M}_{k,k-1} \, \mathbf{P}_{k-1}^a \, \mathcal{M}_{k,k-1}^T + \mathbf{Q}_{k}\, . 
\end{align}
\end{subequations}
The detailed deduction is provided in Appendix~\ref{appendix:KF-deduction}, also cf. \cite[\S~2.2]{Simandl2006}.

\subsubsection{Filtering step}
After a new observation $\mathbf{y}_k$ arrives, one uses Bayes' rule to update the prior pdf $p ( \mathbf{x}_{k} | \mathbf{Y}_{k-1} )$ to the posterior $p ( \mathbf{x}_{k} | \mathbf{Y}_{k} )$, so that
\begin{equation} \label{ch1:RBE_filtering_1}
\begin{split}
p ( \mathbf{x}_{k} | \mathbf{Y}_{k} ) = &  p ( \mathbf{x}_{k} | \mathbf{y}_k, \mathbf{Y}_{k-1} ) \\
=& \dfrac{ p ( \mathbf{y}_{k} | \mathbf{x}_{k} , \mathbf{Y}_{k-1}  ) p ( \mathbf{x}_{k} | \mathbf{Y}_{k-1} ) }{p ( \mathbf{y}_{k} | \mathbf{Y}_{k-1} )} \,  \\
=& \dfrac{ p ( \mathbf{y}_{k} | \mathbf{x}_{k}  ) p ( \mathbf{x}_{k} | \mathbf{Y}_{k-1} ) }{\int p ( \mathbf{y}_{k} | \mathbf{x}_{k}  ) p ( \mathbf{x}_{k} | \mathbf{Y}_{k-1} )  d\mathbf{x}_{k}} \, . \\
\end{split}
\end{equation}
The third line in Eq.~(\ref{ch1:RBE_filtering_1}) holds because $\mathbf{y}_{k}$ is considered independent of the historical observations $\mathbf{Y}_{k-1}$.

By Eqs.~(\ref{ch1: linear observation system}) and (\ref{ch1: observation noise Gaussian}), $p ( \mathbf{y}_{k} | \mathbf{x}_{k}  ) = N ( \mathbf{y}_{k}: \mathcal{H}_{k} \, \mathbf{x}_{k}, \mathbf{R}_{k} )$. Substituting it into Eq.~(\ref{ch1:RBE_filtering_1}), one has
\begin{equation} \label{ch1:RBE_filtering_2}
p ( \mathbf{x}_{k} | \mathbf{Y}_{k} ) =  \dfrac{ N ( \mathbf{y}_{k}: \mathcal{H}_{k} \, \mathbf{x}_{k}, \mathbf{R}_{k} ) N ( \mathbf{x}_{k}: \mathbf{x}_{k}^b, \mathbf{P}_{k}^b ) }{\int N ( \mathbf{y}_{k}: \mathcal{H}_{k} \, \mathbf{x}_{k}, \mathbf{R}_{k} ) N ( \mathbf{x}_{k}: \mathbf{x}_{k}^b, \mathbf{P}_{k}^b )  d\mathbf{x}_{k}} \, . 
\end{equation}
Following the same rationale in Eq.~(\ref{ch1:RBE_propagation_2}), it can be shown that 
\begin{equation} \label{ch1:RBE_filtering_3}
\int N ( \mathbf{y}_{k}: \mathcal{H}_{k} \, \mathbf{x}_{k}, \mathbf{R}_{k} ) N ( \mathbf{x}_{k}: \mathbf{x}_{k}^b, \mathbf{P}_{k}^b )  d\mathbf{x}_{k} = N ( \mathbf{y}_{k}: \mathcal{H}_{k} \, \mathbf{x}_{k}^b, \mathcal{H}_{k} \, \mathbf{P}_{k}^b \, \mathcal{H}_{k}^T + \mathbf{R}_{k} ) \, .
\end{equation}
Hence with some algebra, Eq.~(\ref{ch1:RBE_filtering_2}) is reduced to 
\begin{equation} \label{ch1:RBE_filtering_4}
\begin{split}
p ( \mathbf{x}_{k} | \mathbf{Y}_{k} ) &= \dfrac{ N ( \mathbf{y}_{k}: \mathcal{H}_{k} \, \mathbf{x}_{k}, \mathbf{R}_{k} ) N ( \mathbf{x}_{k}: \mathbf{x}_{k}^b, \mathbf{P}_{k}^b ) }{N ( \mathbf{y}_{k}: \mathcal{H}_{k} \, \mathbf{x}_{k}^b, \mathcal{H}_{k} \, \mathbf{P}_{k}^b \, \mathcal{H}_{k}^T + \mathbf{R}_{k} )} \\
&= N ( \mathbf{x}_{k}: \mathbf{x}_{k}^a, \mathbf{P}_{k}^a ) \, ,
\end{split}
\end{equation}
where 
\begin{subequations}\label{ch1:RBE_filtering_5}
\begin{align}
\label{ch1:RBE_analysis} & \mathbf{x}_k^a = \mathbf{x}_k^b + \mathbf{K}_k \left ( \mathbf{y}_k - \mathcal{H}_k  \mathbf{x}_k^b \right ) \, , \\
\label{ch1:RBE_analysis_covariance} & \mathbf{P}_k^a = \mathbf{P}_k^b -  \mathbf{K}_k \mathcal{H}_k \mathbf{P}_k^b \, ,
\end{align}
\end{subequations}
with
\begin{equation}  \label{ch1:RBE_Kalman_gain}
 \mathbf{K}_k = \mathbf{P}_k^b \mathcal{H}_k^T ( \mathcal{H}_k \mathbf{P}_k^b \mathcal{H}_k^T + \mathbf{R}_k
)^{-1} \, .
\end{equation}
The deduction of the equality between the first and second lines of Eq.~(\ref{ch1:RBE_filtering_4}) is also provided in \cite[\S~2.2]{Simandl2006}. Since $p ( \mathbf{x}_{k} | \mathbf{Y}_{k} )$ is Gaussian, the updated mean $\mathbf{x}_k^a$ and covariance $\mathbf{P}_k^a$ in Eq.~(\ref{ch1:RBE_filtering_5}) contain sufficient information for characterizing it. 


\subsection{Summary of the conventional Kalman filter algorithm}
We summarize the recursive steps of the conventional Kalman filter as follows.\\
Propagation step:
\begin{subequations}\label{ch1:KF_propagation}
\begin{align}
\label{ch1:KF_background}& \mathbf{x}_{k}^b = \mathcal{M}_{k,k-1} \, \mathbf{x}_{k-1}^a \, , \\
\label{ch1:KF_background_cov}& \mathbf{P}_{k}^b = \mathcal{M}_{k,k-1} \, \mathbf{P}_{k-1}^a \, \mathcal{M}_{k,k-1}^T + \mathbf{Q}_{k}\, . 
\end{align}
\end{subequations}
Filtering step:
\begin{subequations} \label{ch1:KF_filtering}
\begin{align}
\label{ch1:KF_gain}& \mathbf{K}_k = \mathbf{P}_k^b \mathcal{H}_k^T ( \mathcal{H}_k \mathbf{P}_k^b \mathcal{H}_k^T + \mathbf{R}_k )^{-1} \, , \\
\label{ch1:KF_analysis}& \mathbf{x}_k^a = \mathbf{x}_k^b + \mathbf{K}_k \left ( \mathbf{y}_k - \mathcal{H}_k  \mathbf{x}_k^b \right ) \, , \\
\label{ch1:KF_analysis_cov}& \mathbf{P}_k^a = \mathbf{P}_k^b -  \mathbf{K}_k \mathcal{H}_k \mathbf{P}_k^b \, . 
\end{align}
\end{subequations}

\section{Two variants of the conventional Kalman filter}

Now we introduce two variants of the conventional Kalman filter, namely the square root Kalman filter (SRKF) and the Kalman filter with fading memory (KF-FM). These two variants can benefit the performance of a filter, as will be discussed below.

\subsection{Square root Kalman filter}\label{ch1:SRKF}
The error covariance matrices, e.g., $\mathbf{P}_{k}^b$ and $\mathbf{P}_{k}^a$, should be symmetric and positive definite. However, in numerical computations, these properties may not be preserved due to the finite computational precision \cite[ch.~6]{Simon2006}. As a remedy for this problem, it is customary to use the square root Kalman filter (SRKF), which can be derived based on the conventional form.

For illustration, we first re-write the covariance matrices $\mathbf{P}_{k}^b$ and $\mathbf{P}_{k}^a$ at assimilation cycle $k$ as 
\begin{subequations}\label{ch1:sec_unnumbered_eq_sqrt_deco}
\begin{align}
& \mathbf{P}_{k}^b = \mathbf{S}_{k}^b (\mathbf{S}_{k}^b )^T \, , \\
& \mathbf{P}_{k}^a = \mathbf{S}_{k}^a (\mathbf{S}_{k}^a )^T \, ,
\end{align}
\end{subequations}
where $\mathbf{S}_{k}^b$ and $\mathbf{S}_{k}^a$ are called square root matrices of $\mathbf{P}_{k}^b$ and $\mathbf{P}_{k}^a$, respectively. We suppose that the dimensions of $\mathbf{S}_{k}^b$ and $\mathbf{S}_{k}^a$ are $m \times m_k^{sb}$ and $m \times m_k^{sa}$, respectively, where $m_k^{sb}$ and $m_k^{sa}$ may vary from cycle to cycle. In this way, in the SRKF we can propagate and update the square roots, rather than the corresponding covariance matrices, as to be shown below.

At the propagation step, in order to compute $\mathbf{S}_{k}^b$ based on $\mathbf{S}_{k-1}^a$, we substitute Eq.~(\ref{ch1:sec_unnumbered_eq_sqrt_deco}) into Eq.~(\ref{ch1:KF_background_cov}), so that
\begin{equation}\label{ch1:SRKF_propagation_cov_1}
\mathbf{S}_{k}^b (\mathbf{S}_{k}^b )^T = ( \mathcal{M}_{k,k-1} \mathbf{S}_{k-1}^a ) ( \mathcal{M}_{k,k-1} \mathbf{S}_{k-1}^a )^T + \mathbf{Q}_{k} \, . 
\end{equation}
where $\mathbf{Q}_{k}$ is the $m \times m$ covariance matrix of dynamical noise. Therefore, one can just take  $\mathbf{S}_{k}^b$ as a square root of the matrix 
\begin{equation}
\mathbf{G}_k \equiv (\mathcal{M}_{k,k-1} \mathbf{S}_{k-1}^a ) ( \mathcal{M}_{k,k-1} \mathbf{S}_{k-1}^a )^T + \mathbf{Q}_{k} \, .
\end{equation}
Note that by definition $\mathbf{P}_k^b=\mathbf{G}_k$. Here we use the notation $\mathbf{G}_k$ to represent the right hand side of Eq.~(\ref{ch1:SRKF_propagation_cov_1}), for notational convenience. If there is no dynamical noise so that $\mathbf{Q}_{k}=\mathbf{0}$, then one may simply let $\mathbf{S}_{k}^b = \mathcal{M}_{k,k-1} \mathbf{S}_{k-1}^a$. Otherwise, one may adopt the following method to compute a square root matrix of $\mathbf{G}_k$ numerically. Suppose that $\mathbf{G}_k$ is an $m \times m$ matrix, then by definition $\mathbf{G}_k$ is symmetric and positive semi-definite, so that we can perform a spectral decomposition on $\mathbf{G}_k$ \cite[\S~2.5.3]{Golub-matrix}. In doing this, $\mathbf{G}_k$ is decomposed as 
\begin{equation}
\mathbf{G}_k = \mathbf{E}_k^{G} \mathbf{D}_k^{G} (\mathbf{E}_k^{G})^T \, ,
\end{equation} 
where $\mathbf{E}_k^{G} = [\mathbf{e}_{k,1}, \dotsb, \mathbf{e}_{k,m_k^{sb}}]$ is the matrix that consists of all $m_k^{sb}$ eigenvectors $\mathbf{e}_{k,i}$ of $\mathbf{G}_k$, whose corresponding $m_k^{sb}$ eigenvalues $d_{k,i}$ ($i=1,\dotsb, m_k^{sb}$) are positive, and $\mathbf{D}_k^{G} =\text{diag}(d_{k,1},\dotsb,d_{k,m_k^{sb}})$ is the diagonal matrix whose main diagonal consists of the positive eigenvalues $d_{k,i}$ of $\mathbf{G}_k$. Therefore, the dimensions of the matrices $\mathbf{E}_k^{G}$ and $\mathbf{D}_k^{G}$ are $m \times m_k^{sb}$ and $m_k^{sb} \times m_k^{sb}$, respectively. We then define a square root $(\mathbf{D}_k^{G})^{1/2}$ of $\mathbf{D}_k^{G}$ as
\begin{equation}
(\mathbf{D}_k^{G})^{1/2} = \text{diag} (\sqrt{d_{k,1}},\dotsb,\sqrt{d_{k,m_k^{sb}}}) \, ,
\end{equation}       
i.e., $(\mathbf{D}_k^{G})^{1/2}$ is a diagonal matrix with its main diagonal consisting of the square roots of $d_{k,i}$ ($i=1,\dotsb,m_k^{sb}$), so that $(\mathbf{D}_k^{G})^{1/2}$ is also an      $m_k^{sb} \times m_k^{sb}$ matrix. Then we can let a square root matrix of $\mathbf{G}_k$ be
\begin{equation}
\mathbf{S}_{k}^b = \mathbf{E}_k^{G} (\mathbf{D}_k^{G})^{1/2} \, , 
\end{equation} 
so that $\mathbf{S}_{k}^b$ is an $m \times m_k^{sb}$ matrix. In this way, it can be verified that $\mathbf{S}_{k}^b (\mathbf{S}_{k}^b )^T = \mathbf{G}_k$, and it is guaranteed that the product $\mathbf{S}_{k}^b (\mathbf{S}_{k}^b )^T$ is positive semi-definite in numerical computations. Note that, $\mathbf{S}_{k}^b$ obtained in this way is unique with respect to the method we adopt. But $\mathbf{S}_{k}^b$ is in general not the unique square root of $\mathbf{P}_k^b$, as will be explained later.    

On the other hand, in order to compute $\mathbf{S}_{k}^a$ based on $\mathbf{S}_{k}^b$ at the filtering step, we substitute Eq.~(\ref{ch1:KF_gain}) into Eq.~(\ref{ch1:KF_analysis_cov}), so that
\begin{equation}\label{ch1:SRKF_filtering_cov_1}
\mathbf{P}_k^a = \mathbf{P}_k^b -  \mathbf{P}_k^b \mathcal{H}_k^T ( \mathcal{H}_k \mathbf{P}_k^b \mathcal{H}_k^T + \mathbf{R}_k )^{-1} \mathcal{H}_k \mathbf{P}_k^b \, .
\end{equation}
By writing the covariance matrices in terms of their square roots, we have
\begin{equation}\label{ch1:SRKF_filtering_cov_2}
\begin{split}
\mathbf{S}_{k}^a (\mathbf{S}_{k}^a )^T =& \mathbf{S}_{k}^b ( \mathbf{I}_{m_k^{sb}} -(\mathbf{S}_{k}^b )^T \mathcal{H}_k^T ( \mathcal{H}_k \mathbf{S}_{k}^b (\mathbf{S}_{k}^b )^T \mathcal{H}_k^T + \mathbf{R}_k )^{-1} \mathcal{H}_k \mathbf{S}_{k}^b ) (\mathbf{S}_{k}^b )^T \, \\
=& \mathbf{S}_{k}^b ( \mathbf{I}_{m_k^{sb}} -(\mathbf{S}_{k}^{h} )^T (\mathbf{S}_{k}^{h} (\mathbf{S}_{k}^{h} )^T + \mathbf{R}_k )^{-1} \mathbf{S}_{k}^{h} ) (\mathbf{S}_{k}^b )^T \, ,\\
\end{split}
\end{equation}
where $\mathbf{I}_{m_k^{sb}}$ is the $ m_k^{sb} \times m_k^{sb}$ identity matrix,
\begin{equation}
\mathbf{S}_{k}^{h} = \mathcal{H}_k \mathbf{S}_{k}^b
\end{equation}
represents the projection of the square root $\mathbf{S}_{k}^b$ onto the observation space. By definition $\mathbf{S}_{k}^{h}$ is an $m^{obv} \times m_k^{sb}$ matrix (with $\mathcal{H}_k$ being an $m^{obv} \times m$ matrix), while $\mathbf{R}_k$ is the   $m^{obv} \times m^{obv}$ covariance matrix of the observation noise. Thus one has the general solution of Eq.~(\ref{ch1:SRKF_filtering_cov_2}) given by \cite{Tippett-ensemble}
\begin{equation} \label{ch1_SRKF_update}
\mathbf{S}_{k}^a = \mathbf{S}_{k}^b \mathbf{Z}_{k} \mathbf{U}_k \, ,	
\end{equation}
with $\mathbf{S}_{k}^a$ being an $m \times m_k^{u}$ matrix, $\mathbf{Z}_{k}$ an $m_k^{sb} \times m_k^{sa}$ square root matrix of $\mathbf{I}_{m_k^{sb}} -(\mathbf{S}_{k}^{h} )^T (\mathbf{S}_{k}^{h} (\mathbf{S}_{k}^{h} )^T + \mathbf{R}_k )^{-1} \mathbf{S}_{k}^{h}$, and $\mathbf{U}_k$ an arbitrary $m_k^{sa} \times m_k^{u}$ matrix such that $\mathbf{U}_k\mathbf{U}_k^T = \mathbf{I}_{m_k^{sa}}$, with $m_k^{u}$ being a positive integer, and $\mathbf{I}_{m_k^{sa}}$ the $m_k^{sa} \times m_k^{sa}$ identity matrix. Note that it can be shown that $\mathbf{I}_{m_k^{sb}} -(\mathbf{S}_{k}^{h} )^T (\mathbf{S}_{k}^{h} (\mathbf{S}_{k}^{h} )^T + \mathbf{R}_k )^{-1} \mathbf{S}_{k}^{h}$ is symmetric and positive definite \cite[Thm. 6]{Livings-unbiased}, thus we can also use spectral decomposition to compute $\mathbf{Z}_{k}$, the same as the case in computing $\mathbf{S}_{k}^b$ at the propagation step. Therefore, here $m_k^{sa}$ corresponds to the number of positive eigenvalues of $\mathbf{I}_{m_k^{sb}} -(\mathbf{S}_{k}^{h} )^T (\mathbf{S}_{k}^{h} (\mathbf{S}_{k}^{h} )^T + \mathbf{R}_k )^{-1} \mathbf{S}_{k}^{h}$. Because of the positive definiteness, we have $m_k^{sa} = m_k^{sb}$. However, in general discussion, we choose to keep using the notation $m_k^{sa}$. Also note that, in principle $m_k^{u}$ can be an arbitrary number. But in certain circumstances, there might be some constraints on the choice of $m_k^{u}$, as will be seen in \S~\ref{ch2:sec_EnSRF}. In addition, the freedom in choosing the matrix $\mathbf{U}_k$ also implies that, in Eq.~(\ref{ch1_SRKF_update}) the square root $\mathbf{S}_{k}^a$ usually is not unique (although $\mathbf{Z}_{k}$ is unique with respect to the method we adopt in numerical computation). 

Analogous to the conventional Kalman filter, the steps of the square root Kalman filter can be written as: \\
Propagation step:
\begin{subequations}\label{ch1:SRKF_propagation}
\begin{align}
\label{ch1:SRKF_background}& \mathbf{x}_{k}^b = \mathcal{M}_{k,k-1} \, \mathbf{x}_{k-1}^a \, , \\
\label{ch1:SRKF_background_SR}& \mathbf{S}_{k}^b =\sqrt{(\mathcal{M}_{k,k-1} \mathbf{S}_{k-1}^a ) ( \mathcal{M}_{k,k-1} \mathbf{S}_{k-1}^a )^T + \mathbf{Q}_{k}} , \\
\label{ch1:SRKF_projection_SR} & \mathbf{S}_{k}^{h} = \mathcal{H}_k \mathbf{S}_{k}^b \, .
\end{align}
\end{subequations}
Filtering step:
\begin{subequations} \label{ch1:SRKF_filtering}
\begin{align}
\label{ch1:SRKF_gain}& \mathbf{K}_k = \mathbf{S}_k^b ( \mathbf{S}_{k}^{h} )^T ( \mathbf{S}_{k}^{h} ( \mathbf{S}_{k}^{h} )^T + \mathbf{R}_k )^{-1} \, , \\
\label{ch1:SRKF_analysis}& \mathbf{x}_k^a = \mathbf{x}_k^b + \mathbf{K}_k \left ( \mathbf{y}_k - \mathcal{H}_k  \mathbf{x}_k^b \right ) \, , \\
\label{ch1:SRKF_analysis_Z} & \mathbf{Z}_{k} = \sqrt{\mathbf{I}_{m_k^{sb}} -(\mathbf{S}_{k}^{h} )^T (\mathbf{S}_{k}^{h} (\mathbf{S}_{k}^{h} )^T + \mathbf{R}_k )^{-1} \mathbf{S}_{k}^{h}} \, , \\
\label{ch1:SRKF_analysis_SR}& \mathbf{S}_{k}^a = \mathbf{S}_{k}^b \mathbf{Z}_{k} \mathbf{U}_k \, , 
\end{align}
\end{subequations}
where in Eqs. (\ref{ch1:SRKF_background_SR}) and (\ref{ch1:SRKF_analysis_Z}) the symbol $\sqrt{\mathbf{A}}$ means a square root of the matrix $\mathbf{A}$. This can be calculated through a certain numerical scheme, for example, the spectral decomposition. For convenience, the information of the dimensions of the matrices involved in the above steps is summarized in Table~\ref{ch1:dimensions_in_SRKF}.

\begin{table*}[!t]
	\centering
	\caption{\label{ch1:dimensions_in_SRKF} Information of dimensions involved in the steps of the SRKF.}
	\begin{tabular}{p{7cm}p{7cm}}
        \hline \hline
		Number & Meaning \\
		\hline
		$m$ & Dimension of the state space \\
		\hline
		$m^{obv}$ & Dimension of the observation space \\
		\hline
		$m_k^{sb}$ & Number of positive eigenvalues of
		$( \mathcal{M}_{k,k-1} \mathbf{S}_{k-1}^a ) ( \mathcal{M}_{k,k-1} \mathbf{S}_{k-1}^a )^T + 			\mathbf{Q}_{k} $\\ 
		\hline
		$m_k^{sa}$ ($=m_k^{sb}$) & Number of positive eigenvalues of
		$\mathbf{I}_{m_k^{sb}} -(\mathbf{S}_{k}^{h} )^T (\mathbf{S}_{k}^{h} (\mathbf{S}_{k}^{h} )^T 		+ \mathbf{R}_k )^{-1} \mathbf{S}_{k}^{h}$ \\ 
		\hline \hline 
        Matrix & Dimension \\
        \hline
		$\mathcal{M}_{k,k-1}$ & $m \times m$ \\
		$\mathcal{H}_k$ & $m^{obv} \times m$ \\
		$\mathbf{Q}_k$ & $m \times m$ \\
		$\mathbf{R}_k$ & $m^{obv} \times m^{obv}$ \\
		$\mathbf{K}_k$ & $m \times m^{obv}$ \\
		$\mathbf{S}_{k}^b$ & $m \times m_{k}^{sb}$ \\
		$\mathbf{S}_{k}^h$ & $m^{obv} \times m_{k}^{sb}$ \\
		$\mathbf{Z}_{k}$ & $m_k^{sb} \times m_{k}^{sa}$ \\
		$\mathbf{U}_{k}$ & $m_k^{sa} \times m_{k}^{u}$ \\
		$\mathbf{S}_{k}^a$ & $m \times m_{k}^{u}$ \\
        \hline \hline
     \end{tabular}
\end{table*} 

\subsection{Kalman filter with fading memory}\label{ch1: KF fading}
In \S~\ref{ch1:sec:LSE} we have mentioned that, the analysis update formula Eq.~(\ref{ch1:LSE_analysis_update}) at the filtering step of the conventional Kalman filter can be obtained by minimizing the following cost function 
\begin{equation} \tag{$\ref{ch1:LSE_weighted_cost}$}
J( \mathbf{x}_k ) = \dfrac{1}{2} (\mathbf{x}_k - \mathbf{x}^b_k )^T
(\mathbf{P}_k^b )^{-1} (\mathbf{x}_k - \mathbf{x}^b_k ) + \dfrac{1}{2}
(\mathbf{y}_k - \mathcal{H}_{k}\mathbf{x}_k )^T \mathbf{R}_k^{-1}
(\mathbf{y}_k - \mathcal{H}_{k} \mathbf{x}_k ) \, .
\end{equation}
The cost function $J( \mathbf{x}_k )$ consists of two terms. The first term on the right hand side (rhs) of Eq.~(\ref{ch1:LSE_weighted_cost}) represents the information contents in the past, such as the initial condition $\mathbf{x}_0$, and the historical observations $\mathbf{Y}_{k-1} = \left \{ \mathbf{y}_{i}\right \}_{i=0}^{k-1}$ up to and including time $k-1$. The second term represents the information content of the incoming observation $\mathbf{y}_k$, which provides additional information to update the background to the analysis at time $k$. 

In practice, more often than not, the information contained in the first term on the rhs of Eq.~(\ref{ch1:LSE_weighted_cost}) may not completely reflect reality for some reasons, for example, our knowledge limit in understanding the underlying mechanism of the dynamical system, or the limit in model resolution. In contrast, the observation system might be better characterized and the observations are normally recorded with certain accuracy. In such circumstances, instead of using Eq.~(\ref{ch1:LSE_weighted_cost}) as the cost function, it may be better to put more relative weight on the second term on the rhs of Eq.~(\ref{ch1:LSE_weighted_cost}) in order to emphasize that one is more confident on the incoming observation. To this end, one can choose the following modified cost function:
\begin{equation} \label{ch1:KF_fading_cost}
J_{f}( \mathbf{x}_k ) = \dfrac{1}{2} (\mathbf{x}_k - \mathbf{x}^b_k )^T
(\mathbf{P}_k^b )^{-1} (\mathbf{x}_k - \mathbf{x}^b_k ) + \dfrac{1}{2} \, ( 1+ \delta )^2
(\mathbf{y}_k - \mathcal{H}_{k}\mathbf{x}_k )^T \mathbf{R}_k^{-1}
(\mathbf{y}_k - \mathcal{H}_{k} \mathbf{x}_k ) \, ,
\end{equation}
where $\delta \ge 0$ is a non-negative scalar constant, and is called the \sindex{covariance inflation factor} in this dissertation \footnote{The choice of this factor will be discussed in \S~\ref{ch2:sec_covariance_inflation}.}. By choosing a cost function like Eq.~(\ref{ch1:KF_fading_cost}) at each assimilation cycle, the relative weights of the historical information contents will drop faster than the situation without any covariance inflation. For this reason, the filtering algorithm derived with the cost function in Eq.~(\ref{ch1:KF_fading_cost}) (see below) is called the Kalman filter with fading memory (KF-FM) \cite[p. 208]{Simon2006}. With a fading memory, the filter can be more robust against the inaccurate information contents in the past, e.g., the errors in specifying the initial conditions, or the occasional outliers of the observations in the past.   

To derive the KF-FM, one may re-write Eq.~(\ref{ch1:KF_fading_cost}) as follows:
\begin{equation} \label{ch1:KF_fading_cost_effective}
 J_{d}( \mathbf{x}_k ) = \dfrac{1}{2} (\mathbf{x}_k - \mathbf{x}^b_k )^T ( \tilde{\mathbf{P}}_k^b )^{-1}
(\mathbf{x}_k - \mathbf{x}^b_k ) + \dfrac{1}{2} 
(\mathbf{y}_k - \mathcal{H}_{k}\mathbf{x}_k )^T \mathbf{R}_k^{-1} 
(\mathbf{y}_k - \mathcal{H}_{k} \mathbf{x}_k ) \, ,
\end{equation}
where $ J_{d}( \mathbf{x}_k ) =  ( 1+ \delta )^{-2} J_{f}( \mathbf{x}_k )$ is a discounted cost function of $J_{f}( \mathbf{x}_k )$, and $\tilde{\mathbf{P}}_k^b = ( 1+ \delta )^{2} \mathbf{P}_k^b$ is the inflated background error covariance. Thus the analysis update formula of the KF-FM is derived by minimizing $J_{d}( \mathbf{x}_k )$ in Eq.~(\ref{ch1:KF_fading_cost_effective}). In the spirit of the derivation in \S~\ref{ch1:sec:LSE}, the steps of the KF-FM are then given by:\\
Propagation step:
\begin{subequations}\label{ch1:KF_fading_propagation}
\begin{align}
\label{ch1:KF_fading_background}& \mathbf{x}_{k}^b = \mathcal{M}_{k,k-1} \, \mathbf{x}_{k-1}^a \, , \\
\label{ch1:KF_fading_background_cov}& \mathbf{P}_{k}^b = \mathcal{M}_{k,k-1} \, \mathbf{P}_{k-1}^a \, \mathcal{M}_{k,k-1}^T + \mathbf{Q}_{k}\, . 
\end{align}
\end{subequations}
Filtering step:
\begin{subequations} \label{ch1:KF_fading_filtering}
\begin{align}
\label{ch1:KF_fading_cov_inflation} & \mathbf{P}_k^b \rightarrow  ( 1+ \delta )^{2} \mathbf{P}_k^b \, , \\ 
\label{ch1:KF_fading_gain}& \mathbf{K}_k = \mathbf{P}_k^b \mathcal{H}_k^T ( \mathcal{H}_k \mathbf{P}_k^b \mathcal{H}_k^T + \mathbf{R}_k )^{-1} \, , \\
\label{ch1:KF_fading_analysis}& \mathbf{x}_k^a = \mathbf{x}_k^b + \mathbf{K}_k \left ( \mathbf{y}_k - \mathcal{H}_k  \mathbf{x}_k^b \right ) \, , \\
\label{ch1:KF_fading_analysis_cov}& \mathbf{P}_k^a = \mathbf{P}_k^b -  \mathbf{K}_k \mathcal{H}_k \mathbf{P}_k^b \, . 
\end{align}
\end{subequations}
where Eq.~(\ref{ch1:KF_fading_cov_inflation}) means that one conducts covariance inflation on $\mathbf{P}_k^b$, and uses the inflated covariance $( 1+ \delta )^{2} \mathbf{P}_k^b$ to replace $\mathbf{P}_k^b$ in subsequent computations.
\section{Summary of the chapter}\label{ch1:summary}
In this chapter we considered the data assimilation problem in a specific class of linear/Gaussian systems. The solution to the problem, which turned out to be the well-known Kalman filter, was derived from both the points of views of least squares estimation (LSE) and recursive Bayesian estimation (RBE). We also introduced two variants of the conventional Kalman filter, namely the square root Kalman filter (SRKF) and the Kalman filter with fading memory (KF-FM). The SRKF was introduced to improve the numerical precision of a filter, while the KF-FM was designed to improve its stability (or robustness). As will be shown in subsequent chapters, in practice one can implement these two variants simultaneously to improve the performance of a filter. 

Before closing this chapter, we would like to give some hints or references for deriving data assimilation algorithms for linear systems that do not fall into the category described by Eq.~(\ref{ch1: linear systems}):
\begin{itemize}
\item One can convert a higher order Markov process into a first-order one by introducing some argumented variables. This is similar to the idea of converting a higher order scalar autoregressive (AR) process into a first-order vector autoregressive (VAR) process \cite[ch. 2]{Lutkepohl-introduction}. For example, to write 
\[
\mathbf{x}_k  = \mathcal{M}_{k,k-1} \, \mathbf{x}_{k-1} + \mathcal{M}_{k,k-2} \, \mathbf{x}_{k-2} + \mathbf{u}_{k} \] 
into the form of a first-order Markov process, we define the argumented variable 
\[
\mathbf{X}_k = \begin{pmatrix} \mathbf{x}_k \\  \mathbf{x}_{k-1} \end{pmatrix} \, .
\]
Thus we have the desired form given by
\[
\mathbf{X}_k = \begin{pmatrix} \mathcal{M}_{k,k-1} & \mathcal{M}_{k,k-2} \\ \mathbf{I} & \mathbf{0} \\ \end{pmatrix} \, \mathbf{X}_{k-1} + \begin{pmatrix} \mathbf{u}_k \\  \mathbf{0} \end{pmatrix} \, ,
\]
where $\mathbf{I}$ denote the identity matrix. 

\item The conventional Kalman filter for continuous systems can be derived by letting the time steps of discrete systems tend to zero \cite[ch. 8]{Simon2006}. The filter for hybrid systems (e.g., continuous dynamical systems observed by discrete ones) can be obtained in a similar way, although one may also derive the algorithm from other points of views. For example, see \cite[ch. 7]{Jazwinski1970}.

\item In Eqs.~(\ref{ch1: linear dynamical system}) and (\ref{ch1: linear observation system}), the dynamical noise and the observation noise can also be correlated and/or Gaussian coloured. One can generalize the conventional Kalman filter to accommodate such situations. For details, see \cite[ch. 7]{Simon2006}.   
\end{itemize} 

\chapter{Ensemble Kalman filter for data assimilation} \label{ch2: EnKF}


\section{Overview}
In the previous chapter, we derived the conventional Kalman filter based on two fundamental assumptions, namely, the linearity of the dynamical and observation systems and the Gaussianity of the dynamical and observation noise. In practice, these two assumptions are often violated. Moreover, a practical aspect not mentioned previously is the computational cost, which may not be a problem for low dimensional systems, but will be an important factor in consideration when assimilating high dimensional systems like a weather forecasting model. Thus in this and the next few chapters, we will introduce some filters that are designed to tackle some, if not all, of the following problems: nonlinearity, non-Gaussianity, and high dimensionality. 

In this chapter we focus on studying the ensemble Kalman filter (EnKF) initially proposed in \cite{Evensen-sequential}. The EnKF is essentially a Monte Carlo implementation of the Kalman filter (see \cite{Butala2008} for a rigorous proof). Suppose that, at the beginning of each assimilation cycle, one has an ensemble of the background (called \em{background ensemble}), usually obtained from the previous assimilation cycle. Then, with an incoming observation, one applies the KF scheme to update each individual member of the background ensemble. To do this, the mean and error covariance of the background are approximated by the sample mean and sample covariance of the background ensemble, so that one can apply Eqs.~(\ref{ch1:KF_gain}) and (\ref{ch1:KF_analysis}) to obtain an ensemble of the analysis. The \em{analysis ensemble} is then used to estimate the mean and covariance of the underlying system states. By propagating the analysis ensemble forward through the dynamical system, one obtains a new background ensemble for the next assimilation cycle. In this way, by using only a small ensemble to evaluate the statistics (mean and covariance) at both the propagation and filtering steps, the computational cost of the filter can be reduced \cite{Evensen-sequential}.

Depending on whether to perturb the observations or not, the EnKF can be classified into two types: stochastic and deterministic \cite{Kalnay-4dvar,Tippett-ensemble}. The stochastic EnKF uses the incoming observation and the covariance matrix of the observation noise to produce an ensemble of perturbed observations, which are then used to update the background ensemble. For examples, see \cite{Burgers-analysis,Evensen-sequential,Evensen-ensemble,Evensen2006,Evensen-assimilation,Houtekamer1998}. In contrast, the deterministic EnKF, often known as the ensemble square root filter (EnSRF), does not perturb the incoming observation. Given a background ensemble, the EnSRF uses the incoming observation to update the sample mean of the background, while the analysis ensemble is taken as the sample mean plus some perturbations derived from the updated square root of the analysis error covariance. For examples, see \cite{Anderson-ensemble,Bishop-adaptive,Whitaker-ensemble}, also see the reviews in \cite{Evensen-ensemble,Evensen2006,Tippett-ensemble}. Apart from the aforementioned EnKFs, there are also some other variants in the literature. For examples, see \cite{Beezley-morphing,Sakov2007,Wang-which,Zupanski-maximum}.

In this chapter we first present the mathematical descriptions of both the stochastic and deterministic versions of the EnKF. We introduce two auxiliary techniques, namely, covariance filtering and inflation, which are useful for improving the performance and robustness of the EnKF. Finally, we use a $40$-dimensional system as the testbed to examine the effects of some parameters on the performance of the EnKF.
\section{Problem statement and a Monte Carlo approximation to the solution}\label{ch2:sec_ps}
Similar to Eq.~(\ref{ch1: linear systems}), we consider the data assimilation problem in the following scenario:
\begin{subequations}  \label{ch2:ps} %
\begin{align}
 \label{ch2:ps_dyanmical_system} & \mathbf{x}_k  = \mathcal{M}_{k,k-1} \left( \mathbf{x}_{k-1} \right) + \mathbf{u}_{k}  \, ,  \\
  \label{ch2:ps_observation_system} &  \mathbf{y}_k  = \mathcal{H}_{k} \left( \mathbf{x}_{k} \right) + \mathbf{v}_{k} \, , \\
  \label{ch2:ps_dyanmical_noise} & \mathbf{u}_{k} \sim N \left(\mathbf{u}_{k}: \mathbf{0}, \mathbf{Q}_k \right) \, ,\\
 \label{ch2:ps_observation_noise} & \mathbf{v}_{k} \sim N \left(\mathbf{v}_{k}: \mathbf{0}, \mathbf{R}_k \right) \, ,\\
  \label{ch2:ps_dyanmical_white_noise}& \mathbb{E} \left( \mathbf{u}_{j} \mathbf{u}_{k}^T \right) = \delta_{k,j}\mathbf{Q}_k \, ,\\
  \label{ch2:ps_observation_white_noise}& \mathbb{E} \left( \mathbf{v}_{j} \mathbf{v}_{k}^T \right) = \delta_{k,j}\mathbf{R}_k \, ,\\
  \label{ch2:ps_uncorrelated_noise}& \mathbb{E} \left( \mathbf{u}_{i} \mathbf{v}_{j}^T \right) = 0 \quad \forall \, i, \, j \, .
\end{align}
\end{subequations}
Note that, in general, the systems in Eq.~(\ref{ch2:ps}) are different from those in Eq.~(\ref{ch1: linear systems}), since the dynamical system Eq.~(\ref{ch2:ps_dyanmical_system}) and the observation system Eq.~(\ref{ch2:ps_observation_system}) are both possibly nonlinear, i.e., $\mathcal{M}_{k,k-1}$ and $ \mathcal{H}_{k}$ are possibly nonlinear functions. Again, we assume that the dimensions of the state space and the observation space are $m$ and $m^{obv}$, respectively. But when there causes no confusion, we may often drop the dimension information.

Eq.~(\ref{ch2:ps_dyanmical_noise}) and (\ref{ch2:ps_observation_noise}) mean that we assume both the dynamical and observation noise are Gaussian. This may not always be realistic in practice. But at the moment let us be content with this assumption. Later in Chapters~\ref{ch5:spgsf} we will address the issue of non-Gaussianity, where the main idea is to conduct pdf approximations. Also note that by Eqs.~(\ref{ch2:ps_dyanmical_white_noise}) - (\ref{ch2:ps_uncorrelated_noise}), we assume again that the dynamical and observation noise are white and uncorrelated. 

Before proceeding to introduce the details of different versions of the EnKF, we would like to give an outline of the Monte Carlo approximation to the solution of the data assimilation problem in Eq.~(\ref{ch2:ps}), from the point of view of recursive Bayesian estimation (RBE). In doing this, one may see the rationale behind the EnKF.

\subsection{Propagation step}
We suppose that at the $(k-1)$-th assimilation cycle, one has an $n$-member analysis ensemble $\mathbf{X}_{k-1}^a = \left \{  \mathbf{x}_{k-1,i}^a \right \}_{i=1}^n$, rather than the posterior pdf $p\left( \mathbf{x}_{k-1} \vert \mathbf{Y}_{k-1} \right)$. Thus to use Eq.~(\ref{BRR:prediction}) 
\begin{equation} \tagref{BRR:prediction}
p \left( \mathbf{x}_{k} | \mathbf{Y}_{k-1} \right) = \int p \left( \mathbf{x}_{k} | \mathbf{x}_{k-1} \right)  p \left( \mathbf{x}_{k-1} | \mathbf{Y}_{k-1} \right) d\mathbf{x}_{k-1}
\end{equation}
in RBE to compute the prior pdf $p\left( \mathbf{x}_{k} \vert \mathbf{Y}_{k-1} \right)$ at the next assimilation cycle, one approximates $p\left( \mathbf{x}_{k-1} \vert \mathbf{Y}_{k-1} \right)$ in terms of the analysis ensemble $ \mathbf{X}_{k-1}^a$ by \cite[ch. 7]{Simandl2006}
\begin{equation} \label{ch2:ps_analysis_pdf_MC_approx}
p\left( \mathbf{x}_{k-1} \vert \mathbf{Y}_{k-1} \right) \approx \dfrac{1}{n} \, \sum\limits_{i=1}^{n} \delta \left( \mathbf{x}_{k-1} - \mathbf{x}_{k-1,i}^a \right) ,
\end{equation}
where $\delta$ is the Dirac delta function, so that
\begin{equation} \label{ch2:ps_Dirac_delta_function}
\delta \left( \mathbf{x} \right) = \left \{ 
\begin{array}{l}
+ \infty, \quad \text{if} \quad \mathbf{x}=0 \, , \\
0, \quad \text{otherwise} \, ,
\end{array}
\right.
\end{equation} 
and
\begin{equation}
\int f \left( \mathbf{x} \right) \delta(\mathbf{x}-\mathbf{c}) d\mathbf{x} =  f \left( \mathbf{c} \right)
\end{equation}
for a function $f$ and constant $\mathbf{c}$.

Also note that by Eqs.~(\ref{ch2:ps_dyanmical_system}) and (\ref{ch2:ps_dyanmical_noise}), one has 
\begin{equation} \label{ch2:ps_transition_pdf}
p \left( \mathbf{x}_{k} | \mathbf{x}_{k-1} \right) = N \left(\mathbf{x}_{k}: \mathcal{M}_{k,k-1} \left( \mathbf{x}_{k-1} \right), \mathbf{Q}_k \right) \, .
\end{equation}
Substituting Eqs.~(\ref{ch2:ps_analysis_pdf_MC_approx}) and (\ref{ch2:ps_transition_pdf}) into Eq.~(\ref{BRR:prediction}), one has
\begin{equation} \label{ch2:ps_predictive_pdf_approx}
\begin{split}
p \left( \mathbf{x}_{k} | \mathbf{Y}_{k-1} \right) & \approx  \dfrac{1}{n} \, \sum\limits_{i=1}^{n} \int N \left(\mathbf{x}_{k}: \mathcal{M}_{k,k-1} \left( \mathbf{x}_{k-1} \right), \mathbf{Q}_k \right) \, \delta \left( \mathbf{x}_{k-1} - \mathbf{x}_{k-1,i}^a \right) d\mathbf{x}_{k-1} \\
& =  \dfrac{1}{n} \, \sum\limits_{i=1}^{n} N \left(\mathbf{x}_{k}: \mathcal{M}_{k,k-1} \left( \mathbf{x}_{k-1,i}^a \right), \mathbf{Q}_k \right) \, ,
\end{split}
\end{equation}
which is the sum of a set of Gaussian pdfs.

To evaluate the mean $\hat{\mathbf{x}}$ and covariance $\hat{\mathbf{P}}$ of a Gaussian sum pdf $p \left( \mathbf{x} \right)$ given by
\begin{equation}
p \left( \mathbf{x} \right) = \sum_{s=1}^{n} c_{s} N \left( \mathbf{x}: \hat{\mathbf{x}}_{s},  \hat{\mathbf{P}}_{s}\right) \, ,
\end{equation}
which contains a set of Gaussian pdfs $\left \{ N \left( \mathbf{x}: \hat{\mathbf{x}}_{s},  \hat{\mathbf{P}}_{s}\right) \right \}_{s=1}^n$ with the normalized weights $\left \{ c_s \right \}_{s=1}^n$ ($\sum\limits_{s=1}^n c_s =1$), the following formulae \cite[ch. 8]{Anderson-optimal} will be useful:
\begin{subequations} \label{ch2:ps_gs_statistics}
\begin{align}
\label{estimated mean} \hat{\mathbf{x}} &= \sum_{s=1}^{n} c_{s} \hat{\mathbf{x}}_{s} \, , \\
\label{estimated cov} \hat{\mathbf{P}} &= \sum_{s=1}^{n} c_{s} \left(\hat{\mathbf{P}}_{s} + \left( \hat{\mathbf{x}} -\hat{\mathbf{x}}_{s} \right) \left( \hat{\mathbf{x}} -\hat{\mathbf{x}}_{s} \right)^T \right) \, .
\end{align}
\end{subequations}
Applying Eq.~(\ref{ch2:ps_gs_statistics}) to Eq.~(\ref{ch2:ps_predictive_pdf_approx}), one has the estimated mean $\hat{\mathbf{x}}_{k}^b$ and covariance $\hat{\mathbf{P}}_{k}^b$ of the background at the $k$-th assimilation cycle given by:
\begin{subequations} \label{ch2:ps_backgroup_eva}
\begin{align}
& \mathbf{x}^b_{k,i} = \mathcal{M}_{k,k-1} \left( \mathbf{x}_{k-1,i}^a \right), \,~i=1, 2, \dotsb,n \, , \\
\label{ch2:ps_background_mean} &\hat{\mathbf{x}}_k^b = \frac{1}{n} \sum\limits_{i=1}^n \mathbf{x}^b_{k,i} \, , \\
\label{ch2:ps_background_cov} &\hat{\mathbf{P}}_k^b = \frac{1}{n} \sum\limits_{i=1}^n \left( \mathbf{x}^b_{k,i} - \hat{\mathbf{x}}_k^b \right) \left( \mathbf{x}^b_{k,i} - \hat{\mathbf{x}}_k^b \right)^T + \mathbf{Q}_k \, ,
\end{align}
\end{subequations}
where $ \mathbf{x}^b_{k,i}$ are the forecasts of the propagations of $ \mathbf{X}_{k-1}^a = \left \{  \mathbf{x}_{k-1,i}^a \right \}_{i=1}^n$.

\subsection{Filtering step}\label{ch2:sec_ps_filtering_step}
After a new observation $\mathbf{y}_k$ is available, one updates the prior pdf $p \left( \mathbf{x}_{k} | \mathbf{Y}_{k-1} \right)$ to the posterior $p \left( \mathbf{x}_{k} | \mathbf{Y}_{k} \right)$ according to Bayes' rule Eq.~(\ref{BRR:update}):
\begin{equation} \tagref{BRR:update} 
p \left( \mathbf{x}_{k} | \mathbf{Y}_{k} \right) = \dfrac{ p \left( \mathbf{y}_{k} | \mathbf{x}_{k}  \right) p \left( \mathbf{x}_{k} | \mathbf{Y}_{k-1} \right) }{\int p \left( \mathbf{y}_{k} | \mathbf{x}_{k}  \right) p \left( \mathbf{x}_{k} | \mathbf{Y}_{k-1} \right)  d\mathbf{x}_{k}} \, .
\end{equation}
By Eqs.~(\ref{ch2:ps_observation_system}) and (\ref{ch2:ps_observation_noise}), 
\begin{equation} \label{ch2:sec_unlabelled_eq_obv_noise}
p \left( \mathbf{y}_{k} | \mathbf{x}_{k}  \right) = N \left(\mathbf{y}_{k}: \mathcal{H}_{k} \left( \mathbf{x}_{k} \right), \mathbf{R}_k \right) \, .
\end{equation}
In evaluation, we approximate $p \left( \mathbf{x}_{k} | \mathbf{Y}_{k-1} \right)$ by a Gaussian pdf $N \left(\mathbf{x}_{k}: \hat{\mathbf{x}}_{k}^b, \hat{\mathbf{P}}_k^b \right)$, where $\hat{\mathbf{x}}_{k}^b$ and $\hat{\mathbf{P}}_k^b$ are the estimated mean and covariance of the background given in Eq.~(\ref{ch2:ps_backgroup_eva}). Doing this implies that we assume the underlying system state $\mathbf{x}_{k}$ follows (or can be approximated by) a Gaussian distribution. Therefore, Eq.~(\ref{BRR:update}) is reduced to 
\begin{equation} 
p \left( \mathbf{x}_{k} | \mathbf{Y}_{k} \right) = \dfrac{ N \left(\mathbf{y}_{k}: \mathcal{H}_{k} \left( \mathbf{x}_{k} \right), \mathbf{R}_k \right) N \left(\mathbf{x}_{k}: \hat{\mathbf{x}}_{k}^b, \hat{\mathbf{P}}_k^b \right)}{\int N \left(\mathbf{y}_{k}: \mathcal{H}_{k} \left( \mathbf{x}_{k} \right), \mathbf{R}_k \right) N \left(\mathbf{x}_{k}: \hat{\mathbf{x}}_{k}^b, \hat{\mathbf{P}}_k^b \right)  d\mathbf{x}_{k}} \, ,
\end{equation}

If the observation operator $\mathcal{H}_{k}$ is nonlinear, the pdf $p \left( \mathbf{x}_{k} | \mathbf{Y}_{k} \right)$ may not have a closed form as in linear/Gaussian systems (cf. Eq.~ (\ref{ch1:RBE_filtering_2}) in the previous chapter). As an approximation, one may choose to linearize $\mathcal{H}_{k}$ first, and then apply the identity in Eq.~(\ref{ch1:RBE_filtering_4}) to obtain an approximate closed form for $p \left( \mathbf{x}_{k} | \mathbf{Y}_{k} \right)$.

Concretely, one first expands $\mathcal{H}_{k}  \left( \mathbf{x}_{k} \right)$ so that
\begin{equation}
\mathcal{H}_{k}  \left( \mathbf{x}_{k} \right) = \mathcal{H}_{k}  \left( \hat{\mathbf{x}}_{k}^b \right) + \mathbf{H}_{k} \vert_{ \hat{\mathbf{x}}_{k}^b} \, \delta \mathbf{x}_{k} + \text{h.o.t} \, ,
\end{equation}
where $\mathbf{H}_{k} \vert_{ \hat{\mathbf{x}}_{k}^b}$ denotes the Jacobian matrix of $\mathcal{H}_{k}$ evaluated at $ \hat{\mathbf{x}}_{k}^b$, $\delta \mathbf{x}_{k} = \mathbf{x}_{k} - \hat{\mathbf{x}}_{k}^b$, and "$\text{h.o.t}$" represents the higher order terms of Taylor series expansion. If the perturbation $\delta \mathbf{x}_{k}$ is small, or $\mathbf{H}_{k}$ is weakly nonlinear such that the higher order derivatives of $\mathcal{H}_{k}$ evaluated at $\hat{\mathbf{x}}_{k}^b$ are small, one may discard those higher order terms and approximate $\mathcal{H}_{k}  \left( \mathbf{x}_{k} \right)$ by
\begin{equation} \label{ch2:sec_unlabelled_eq_linearization}
\begin{split}
 \mathcal{H}_{k}  \left( \mathbf{x}_{k} \right) & \approx \mathcal{H}_{k}  \left( \hat{\mathbf{x}}_{k}^b \right) + \mathbf{H}_{k} \, \delta \mathbf{x}_{k} \\
& = \mathbf{H}_{k} \mathbf{x}_k + \left(  \mathcal{H}_{k}  \left( \hat{\mathbf{x}}_{k}^b \right) - \mathbf{H}_{k} \hat{\mathbf{x}}_{k}^b \right) \, .
\end{split}
\end{equation}
For notational convenience, we dropped the localization information of the Jacobian of $\mathcal{H}_{k}$ in Eq.~(\ref{ch2:sec_unlabelled_eq_linearization}). Substituting Eq.~(\ref{ch2:sec_unlabelled_eq_linearization}) into Eq.~(\ref{ch2:sec_unlabelled_eq_obv_noise}), we have
\begin{equation}
\begin{split}
& N \left(\mathbf{y}_{k}: \mathcal{H}_{k} \left( \mathbf{x}_{k} \right), \mathbf{R}_k \right) \\
& \approx N \left(\mathbf{y}_{k}: \mathbf{H}_{k} \mathbf{x}_k + \left(  \mathcal{H}_{k}  \left( \hat{\mathbf{x}}_{k}^b \right) - \mathbf{H}_{k} \hat{\mathbf{x}}_{k}^b \right), \mathbf{R}_k \right) \\
& =  N \left(\mathbf{y}_{k}^{tr}: \mathbf{H}_{k} \mathbf{x}_k, \mathbf{R}_k \right) \, ,
\end{split}
\end{equation}
where $\mathbf{y}_{k}^{tr} = \mathbf{y}_{k} - \left(  \mathcal{H}_{k}  \left( \hat{\mathbf{x}}_{k}^b \right) - \mathbf{H}_{k} \hat{\mathbf{x}}_{k}^b \right)$ is a translation of the observation $\mathbf{y}_{k}$. Therefore we can approximate $p \left( \mathbf{x}_{k} | \mathbf{Y}_{k} \right)$ by
\begin{equation} \label{ch2:ps_analysis_pdf_approx}
\begin{split}
p \left( \mathbf{x}_{k} | \mathbf{Y}_{k} \right) & \approx \dfrac{ N \left(\mathbf{y}_{k}^{tr}: \mathbf{H}_{k} \mathbf{x}_k, \mathbf{R}_k \right) N \left(\mathbf{x}_{k}: \hat{\mathbf{x}}_{k}^b, \hat{\mathbf{P}}_k^b \right)}{\int N \left(\mathbf{y}_{k}^{tr}: \mathbf{H}_{k} \mathbf{x}_k, \mathbf{R}_k \right) N \left(\mathbf{x}_{k}: \hat{\mathbf{x}}_{k}^b, \hat{\mathbf{P}}_k^b \right) d\mathbf{x}_{k}} \\
& = N \left(\mathbf{x}_{k}: \hat{\mathbf{x}}_{k}^a, \hat{\mathbf{P}}_k^a \right) \, ,
\end{split} 
\end{equation}
where
\begin{subequations} \label{ch2:ps_analysis_eva}
\begin{align}
\label{ch2:ps_analysis_mean}  \hat{\mathbf{x}}_{k}^a & = \hat{\mathbf{x}}_{k}^b + \mathbf{K}_k \left ( \mathbf{y}_k^{tr} - \mathbf{H}_k  \hat{\mathbf{x}}_{k}^b \right )  \\
\nonumber & = \hat{\mathbf{x}}_{k}^b + \mathbf{K}_k \left ( \mathbf{y}_k - \mathcal{H}_k \left( \hat{\mathbf{x}}_{k}^b \right ) \right) \, , \\
\label{ch2:ps_analysis_cov}  \hat{\mathbf{P}}_{k}^a  & = \hat{\mathbf{P}}_{k}^b -  \mathbf{K}_k \mathbf{H}_k \hat{\mathbf{P}}_{k}^b \, , \\
\label{ch2:ps_Kalman_gain}  \mathbf{K}_k & =\hat{\mathbf{P}}_{k}^b \mathbf{H}_k^T \left( \mathbf{H}_k \hat{\mathbf{P}}_{k}^b \mathbf{H}_k^T + \mathbf{R}_k \right)^{-1} \, ,
\end{align}
\end{subequations}
are obtained in the spirit of Eq.~(\ref{ch1:RBE_filtering_4}).

To carry out ensemble forecasting at the next assimilation cycle, one also needs to generate an ensemble $ \mathbf{X}_{k}^a = \left \{  \mathbf{x}_{k,i}^a \right \}_{i=1}^n$ of the analysis as the samples of the pdf $N \left(\mathbf{x}_{k}: \hat{\mathbf{x}}_{k}^a, \hat{\mathbf{P}}_k^a \right)$. 
The approach to generating the analysis ensemble $\mathbf{X}_{k}^a$ is called \sindex{analysis scheme}. Different implementations of the EnKF may have different analysis schemes, as will be shown below. 
\section{Implementation of the ensemble Kalman filter}\label{ch2:implementation}
\subsection{Stochastic ensemble Kalman filter}
Given a background ensemble $ \mathbf{X}_{k}^b = \left \{\mathbf{x}^b_{k,i}: \mathbf{x}^b_{k,i} =  \mathcal{M}_{k,k-1} \left( \mathbf{x}_{k-1,i}^a \right) \right \}_{i=1}^n$\footnote{More precisely, $\mathbf{X}_{k}^b$ is the ensemble of the forecast of the analysis ensemble at the previous cycle. However, for brevity, we choose to call it the``background ensemble'' in this dissertation unless otherwise stated.}, in principle the computations of the sample means and covariances of the background and analysis can just follow Eqs.~(\ref{ch2:ps_backgroup_eva}) and~(\ref{ch2:ps_analysis_eva}), respectively. But in the literature there may be some different ways in computing or approximating those statistics. For example, at the propagation step, when computing the background covariance $\hat{\mathbf{P}}_k^b$, one may use the following formula
\begin{equation}
\hat{\mathbf{P}}_k^b = \frac{1}{n-1} \sum\limits_{i=1}^n \left( \mathbf{x}^b_{k,i} - \hat{\mathbf{x}}_k^b \right) \left( \mathbf{x}^b_{k,i} - \hat{\mathbf{x}}_k^b \right)^T + \mathbf{Q}_k \, ,
\end{equation}
which differs from Eq.~(\ref{ch2:ps_background_cov}) in that the factor before the summation is $1/(n-1)$, rather than $1/n$. The factor $1/n$ used in some works (e.g. \cite{Wang-which}) represents the maximum likelihood estimation of the background covariance, while the factor $1/(n-1)$ used in others (e.g. \cite{Houtekamer1998}) represents the unbiased estimation. The difference between these two estimations is not significant even for a small number $n$ (say, around $10$), thus we do not particularly favor either the criterion of maximum likelihood or unbiasedness. However, since a larger background covariance may benefit the performance of the filter (see the discussion in \S~\ref{ch2:sec_covariance_inflation}), we will normally adopt the unbiased estimator in this dissertation unless otherwise stated.

Another point worth mentioning is that, in practice, it may not be convenient to evaluate the Jacobian of a nonlinear function in a multivariate scenario. Thus in order to evaluate the Kalman gain $\mathbf{K}_k$ and the sample covariance $\hat{\mathbf{P}}_k^a$ of the analysis in Eq.~(\ref{ch2:ps_analysis_eva}), we adopt the following approximations \cite{Houtekamer-sequential}:
\begin{subequations} \label{ch2:EnKF_filtering_approx}
\begin{align}
\label{ch2:EnKF_gain_approx}& \mathbf{K}_k =\hat{\mathbf{P}}^{cr}_k \left( \hat{\mathbf{P}}^{pr}_k + \mathbf{R}_k \right)^{-1} \, , \\
\label{ch2:EnKF_analysis_cov_approx}& \hat{\mathbf{P}}_k^a = \hat{\mathbf{P}}_k^b -  \mathbf{K}_k  \left(\hat{\mathbf{P}}^{cr}_k \right)^T \, ,
\end{align}
\end{subequations}
with
\begin{subequations} \label{ch2:EnKF_cov_approx}
\begin{align}
\label{ch2:EnKF_projection_mean} & \hat{\mathbf{y}}_k =  \frac{1}{n} \sum\limits_{i=1}^n \mathcal{H}_k \left ( \mathbf{x}^b_{k,i} \right ), \\
\label{ch2:EnKF_cross_cov_approx} & \hat{\mathbf{P}}^{cr}_k =  \frac{1}{n-1} \sum\limits_{i=1}^n \left( \mathbf{x}^b_{k,i} - \hat{\mathbf{x}}_k^b \right) \left( \mathcal{H}_k \left ( \mathbf{x}^b_{k,i} \right )- \hat{\mathbf{y}}_k \right ) ^T , \\
\label{ch2:EnKF_prj_cov_approx}& \hat{\mathbf{P}}^{pr}_k = \frac{1}{n-1} \sum\limits_{i=1}^n \left( \mathcal{H}_k \left ( \mathbf{x}^b_{k,i} \right )- \hat{\mathbf{y}}_k \right) \left( \mathcal{H}_k \left ( \mathbf{x}^b_{k,i} \right)- \hat{\mathbf{y}}_k\right ) ^T \, .
\end{align}
\end{subequations}
$\hat{\mathbf{P}}^{cr}_k$ in Eq.~(\ref{ch2:EnKF_cross_cov_approx}) is the (sample) cross covariance between the background ensemble and its predicted projection onto the observation space, while $\hat{\mathbf{P}}^{pr}_k$ is the (sample) covariance of the predicted projection of the background ensemble onto the observation space. Hereafter we will call $\hat{\mathbf{P}}^{cr}_k$ and $\hat{\mathbf{P}}^{pr}_k$ cross covariance and projection covariance, respectively. Note that, Eq.~(\ref{ch2:EnKF_projection_mean}) represents an unbiased estimation of the mean of the projection of the background ensemble onto the observation space. In the literature, there may be other ways for estimation. For example, in \cite{Zupanski-maximum} Eq.~(\ref{ch2:EnKF_projection_mean}) is replaced by
\begin{equation} \label{ch2:EnKF_projection_mean_max}
\hat{\mathbf{y}}_k =  \mathcal{H}_k \left ( \hat{\mathbf{x}}^b_{k} \right ), \\
\end{equation} 
which is then used for subsequent computations in Eqs.~(\ref{ch2:EnKF_cross_cov_approx}) and (\ref{ch2:EnKF_prj_cov_approx}). In doing this, Eq.~(\ref{ch2:EnKF_projection_mean_max}) represents a maximum likelihood estimation of the mean of the projection of the background ensemble onto the observation space. If the observation operator $\mathcal{H}_k$ is linear, then Eq.~(\ref{ch2:EnKF_projection_mean}) and (\ref{ch2:EnKF_projection_mean_max}) are equivalent. Otherwise they are different in general. Thus which equation to choose may depend on the favour of the user. In this dissertation, we prefer to using Eq.~(\ref{ch2:EnKF_projection_mean}). Because in doing this, the similarity between the stochastic EnKF and the scaled unscented Kalman filter (SUKF) to be introduced later will be more clear by comparing Eq.~(\ref{ch2:EnKF_cov_approx}) with (\ref{ch3:SUKF_cov_approx_cmp}) in the next chapter.

To generate the analysis ensemble, the stochastic version of the ensemble Kalman filter (stochastic EnKF hereafter) needs to produce some surrogate observations $\mathbf{Y}_k^s = \left\{\mathbf{y}_{k,i}^s \right\}_{i=1}^n$, where $\mathbf{y}_{k,i}^s$ are the samples drawn from the Gaussian distribution with mean $\mathbf{y}_k$ and covariance $\mathbf{R}_k$. The analysis ensemble $\mathbf{X}_k^a = \left\{\mathbf{x}_{k,i}^a \right\}_{i=1}^n$ consists of the updates of the sample mean $\hat{\mathbf{x}}_k^b$ of the background according to Eq.~(\ref{ch2:ps_analysis_mean}), but with the observation therein replaced by the surrogate observations $\mathbf{Y}_k^s$. Concretely, one has
\begin{equation} 
\mathbf{x}_{k,i}^a = \hat{\mathbf{x}}_{k}^b + \mathbf{K}_k \left ( \mathbf{y}_{k,i}^{s} - \mathcal{H}_k \left(  \hat{\mathbf{x}}_{k}^b \right) \right ),~i=1,\dotsb, n \,.
\end{equation}
To summarize, the implementation of the stochastic EnKF contains the following procedures:\\
Propagation step:

\begin{subequations} \label{ch2:stochastic_EnKF_propagation}
\begin{align}
\label{ch2:stochastic_EnKF_background_propagations} & \mathbf{x}^b_{k,i} =  \mathcal{M}_{k,k-1} \left( \mathbf{x}_{k-1,i}^a \right),~i=1,\dotsb, n \, , \\
\label{ch2:stochastic_EnKF_background_mean} &\hat{\mathbf{x}}_k^b = \frac{1}{n} \sum\limits_{i=1}^n \mathbf{x}^b_{k,i} \, , \\
& \hat{\mathbf{y}}_k =  \frac{1}{n} \sum\limits_{i=1}^n \mathcal{H}_k \left ( \mathbf{x}^b_{k,i} \right ), \\
\label{ch2:stochastic_EnKF_background_cov} &\hat{\mathbf{P}}_k^b = \frac{1}{n-1} \sum\limits_{i=1}^n \left( \mathbf{x}^b_{k,i} - \hat{\mathbf{x}}_k^b \right) \left( \mathbf{x}^b_{k,i} - \hat{\mathbf{x}}_k^b \right)^T + \mathbf{Q}_k \, , \\
\label{ch2:stochastic_EnKF_cross_cov} & \hat{\mathbf{P}}^{cr}_k =  \frac{1}{n-1} \sum\limits_{i=1}^n \left( \mathbf{x}^b_{k,i} - \hat{\mathbf{x}}_k^b \right) \left( \mathcal{H}_k \left ( \mathbf{x}^b_{k,i} \right )- \hat{\mathbf{y}}_k \right ) ^T , \\
\label{ch2:stochastic_EnKF_prj_cov}& \hat{\mathbf{P}}^{pr}_k = \frac{1}{n-1} \sum\limits_{i=1}^n \left( \mathcal{H}_k \left ( \mathbf{x}^b_{k,i} \right )- \hat{\mathbf{y}}_k \right) \left( \mathcal{H}_k \left ( \mathbf{x}^b_{k,i} \right)- \hat{\mathbf{y}}_k\right ) ^T \, .
\end{align}
\end{subequations}
Filtering step:
\begin{subequations} \label{ch2:stochastic_EnKF_filtering}
\begin{align}
\label{ch2:stochastic_EnKF_gain_approx}& \mathbf{K}_k =\hat{\mathbf{P}}^{cr}_k \left( \hat{\mathbf{P}}^{pr}_k + \mathbf{R}_k \right)^{-1} \, , \\
\label{ch2:stochastic_EnKF_analysis_mean}& \hat{\mathbf{x}}_k^a = \hat{\mathbf{x}}_k^b + \mathbf{K}_k \left ( \mathbf{y}_k - \mathcal{H}_k \left( \hat{\mathbf{x}}_k^b \right) \right ) \, , \\
\label{ch2:stochastic_EnKF_analysis_cov}& \hat{\mathbf{P}}_k^a = \hat{\mathbf{P}}_k^b -  \mathbf{K}_k  \left(\hat{\mathbf{P}}^{cr}_k \right)^T \, .
\end{align}
\end{subequations}
Analysis scheme:
\begin{subequations} \label{ch2:stochastic_EnKF_analysis_scheme}
\begin{align}
\label{ch2:stochastic_EnKF_surrogate_obv} & \mathbf{y}_{k,i}^{s} \xleftarrow{\text{d.f.}} N \left( \mathbf{y}^s_k: \mathbf{y}_k, \mathbf{R}_k \right) \, , \\
\label{ch2:stochastic_EnKF_analysis_ensemble} & \mathbf{x}_{k,i}^a = \hat{\mathbf{x}}_{k}^b + \mathbf{K}_k \left ( \mathbf{y}_{k,i}^{s} - \mathcal{H}_k \left(  \hat{\mathbf{x}}_{k}^b \right) \right ),~i=1,\dotsb, n \, ,
\end{align}
\end{subequations}
where Eq.~(\ref{ch2:stochastic_EnKF_surrogate_obv}) means that the surrogate observations $\mathbf{y}_{k,i}^{s}$ are samples drawn from the Gaussian distribution $N \left( \mathbf{y}^s_k: \mathbf{y}_k, \mathbf{R}_k \right)$. Also note that in practice, it is not necessary to evaluate the covariances $\hat{\mathbf{P}}_k^b$ and $\hat{\mathbf{P}}_k^a$ in Eqs.~(\ref{ch2:stochastic_EnKF_background_cov}) and (\ref{ch2:stochastic_EnKF_analysis_cov}) for computations. However, here we still choose to list them for completeness.

\rem Sampling the Gaussian distribution $N \left( \mathbf{y}^s_k: \mathbf{y}_k, \mathbf{R}_k \right)$ normally brings some sampling errors. This causes a problem in that the sample covariance computed based on the analysis ensemble $\mathbf{X}_{k}^a = \left\{ \mathbf{x}_{k,i}^a  \right\}_{i=1}^n$ may not be the same as the targeted covariance given by Eq.~(\ref{ch2:stochastic_EnKF_analysis_cov}). Instead, it was shown in \cite{Whitaker-ensemble} that the sample covariance computed based on the analysis ensemble $\mathbf{X}_{k}^a = \left\{ \mathbf{x}_{k,i}^a  \right\}_{i=1}^n$ will underestimate the targeted covariance in Eq.~(\ref{ch2:stochastic_EnKF_analysis_cov}) due to the effect of finite ensemble size, which may cause the divergence of the EnKF. As a remedy for this problem, we will proceed to introduce a different implementation of the EnKF based on the concept of square root Kalman filter (SRKF) in \S~\ref{ch1:SRKF}. We will also introduce the idea of covariance inflation in \S~\ref{ch2:sec_covariance_inflation} in order to compensate for the covariance underestimation.

\subsection{Ensemble square root filter}\label{ch2:sec_EnSRF}
As aforementioned, generating surrogate observations will introduce sampling errors to the EnKF. To overcome this problem, a simple idea is to avoid perturbing the observations. Suppose that at each assimilation cycle, one wants to generate $n$-member ensembles for both the background and analysis. In order to generate an analysis ensemble with its sample covariance matching that in Eq.~(\ref{ch2:stochastic_EnKF_analysis_cov}), one may use the sample mean $\hat{\mathbf{x}}_k^a$ and a square root of the $m \times m$ covariance $\hat{\mathbf{P}}_k^a$ to generate the analysis ensemble $\mathbf{X}_{k}^a = \left\{ \mathbf{x}_{k,i}^a  \right\}_{i=1}^n$. This leads to the ensemble square root filter (EnSRF) in the literature \cite{Anderson-ensemble,Bishop-adaptive,Whitaker-ensemble}.  

For illustration, suppose that one has a square root $\mathbf{S}_{k}^a$ of the covariance $\hat{\mathbf{P}}_k^a$, which is updated from a square root $\mathbf{S}_{k}^b$ of the background covariance $\hat{\mathbf{P}}_k^b$ according to a certain rule. Then the analysis ensemble $\mathbf{X}_k^a$ is generated according to the following formula, 
\begin{equation} \label{ch2:EnSRF_analysis_scheme}
\mathbf{x}_{k,i}^a = \hat{\mathbf{x}}_k^a +\sqrt{n-1} \left( \mathbf{S}_k^a \right)_i, \, i=1, 2, \dotsb, n, 
\end{equation}
where $\left( \mathbf{S}_k^a \right)_i$ denotes the $i$-th column of the square root matrix $\mathbf{S}_k^a$. Note that in Eq.~(\ref{ch2:EnSRF_analysis_scheme}), the sample covariance of the analysis ensemble $\left\{ \mathbf{x}_{k,i}^a  \right\}_{i=1}^n$ is equivalent to $\hat{\mathbf{P}}_k^a$. However, its sample mean may not be equivalent to $\hat{\mathbf{x}}_k^a$ unless 
\begin{equation} \label{ch2:EnSRF_sr_constraint}
\sum\limits_{i=1}^n \left( \mathbf{S}_k^a \right)_i =0 \, .
\end{equation}   
Failing to satisfy Eq.~(\ref{ch2:EnSRF_sr_constraint}) means that, compared with the mean of the analysis evaluated by Eq.~(\ref{ch2:stochastic_EnKF_analysis_mean}), there is a bias in the sample mean of the analysis ensemble produced by Eq.~(\ref{ch2:EnSRF_analysis_scheme}), which may cause covariance underestimation, as reported in \cite{Livings-unbiased}. For this reason, the ensemble filters satisfying the constraint in Eq.~(\ref{ch2:EnSRF_sr_constraint}), called \sindex{unbiased ensemble filters} in \cite{Livings-unbiased}, is favored in this dissertation.

The propagation and filtering steps of the EnSRF may follow those of the square root Kalman filter in \S~\ref{ch1:SRKF}. Concretely, let the background ensemble $ \mathbf{X}_{k}^b = \left \{\mathbf{x}^b_{k,i} \right \}_{i=1}^n$ be the forecasts of the propagations of $ \mathbf{X}_{k-1}^a = \left \{  \mathbf{x}_{k-1,i}^a \right \}_{i=1}^n$ (cf. Eq.~ (\ref{ch2:stochastic_EnKF_background_propagations})), then the sample mean $\hat{\mathbf{x}}_{k}^b$ and covariances $\hat{\mathbf{P}}_{k}^b$ of the EnSRF are exactly the same as those in Eq.~(\ref{ch2:stochastic_EnKF_propagation}). Here we suppose that the rank of the background covariance $\hat{\mathbf{P}}_{k}^b$ is $m_k^{sb}$, which is determined by both the background ensemble $ \mathbf{X}_{k}^b$ and the covariance $\mathbf{Q}_k$ of dynamical noise. In the case that there is no dynamical noise, $m_k^{sb}=n-1$ given $n$ independent ensemble members \cite{Wang-which}. But with the existence of $\mathbf{Q}_k$, in general $m_k^{sb} \ge n-1$. Again, for numerical reasons, it is customary in the EnSRF to re-write the covariances in terms of their square roots \cite{Anderson-ensemble,Bishop-adaptive,Whitaker-ensemble}:
\begin{subequations}
\begin{align}
\label{ch2:EnSRF_background_cov} &\hat{\mathbf{P}}_k^b = \mathbf{S}_{k}^b \left( \mathbf{S}_{k}^b \right)^T= \mathbf{S}_{k}^{xb} \left( \mathbf{S}_{k}^{xb} \right)^T+\mathbf{Q}_k \, , \\
\label{ch2:EnSRF_cross_cov} & \hat{\mathbf{P}}^{cr}_k \approx  \mathbf{S}_{k}^{b}  \left( \mathbf{S}_{k}^h \right)^T \ , \\
\label{ch2:EnSRF_prj_cov}& \hat{\mathbf{P}}^{pr}_k \approx \mathbf{S}_{k}^h  \left( \mathbf{S}_{k}^h \right)^T \, ,
\end{align}
\end{subequations}    
where $\mathbf{S}_{k}^{xb}$, $\mathbf{S}_{k}^b$ and $\mathbf{S}_{k}^h$ are square root matrices satisfying
\begin{subequations}\label{ch2:EnSRF_sr}
\begin{align}
\label{ch2:EnSRF_propagation_sr}& \mathbf{S}_{k}^{xb} = \dfrac{1}{\sqrt{n-1}} \left[ \mathbf{x}^b_{k,1} - \hat{\mathbf{x}}_k^b, \dotsb, \mathbf{x}^b_{k,n} - \hat{\mathbf{x}}_k^b \right] \, , \\
\label{ch2:EnSRF_background_sr}& \mathbf{S}_{k}^{b} = \sqrt{ \mathbf{S}_{k}^{xb} \left( \mathbf{S}_{k}^{xb} \right)^T+\mathbf{Q}_k} \, , \\
\label{ch2:EnSRF_prj_sr}&  \mathbf{S}_{k}^h = \mathcal{H}_k \left ( \mathbf{S}_{k}^{b} \right ) \equiv \left[ \mathcal{H}_k ( (\mathbf{S}_{k}^{b})_1 ), \dotsb, \mathcal{H}_k ( (\mathbf{S}_{k}^{b})_{m_k^{sb}} ) \right] \, , 
\end{align}
\end{subequations}    
with $(\mathbf{S}_{k}^{b})_i$ being the $i$-th column of $\mathbf{S}_{k}^{b}$, and $\mathbf{S}_{k}^{xb}$, $\mathbf{S}_{k}^{b}$ and $\mathbf{S}_{k}^h$ are $m \times n$, $m \times m_k^{sb}$, and $m^{obv} \times m_k^{sb}$ square roots, respectively, where $m^{obv}$ represents the dimension of the observation space. Note that in Eq.~(\ref{ch2:EnSRF_background_sr}), because $\mathbf{S}_{k}^{xb} \left( \mathbf{S}_{k}^{xb} \right)^T+\mathbf{Q}_k$ is positive semi-definite, one can adopt the method discussed in \S~\ref{ch1:SRKF} to calculate $\mathbf{S}_{k}^{b}$, so that it is guaranteed that the numerical result of the product $\mathbf{S}_{k}^{b} (\mathbf{S}_{k}^{b})^T$ is positive semi-definite. In particular, if there exists no dynamical noise, then it is customary to take $\mathbf{S}_{k}^{xb}$ in Eq.~(\ref{ch2:EnSRF_propagation_sr}) as the square root of $\hat{\mathbf{P}}_{k}^b$. For examples, see \cite{Anderson-ensemble,Bishop-adaptive,Whitaker-ensemble}.  

Analogous to the situation in \S~\ref{ch1:SRKF}, the Kalman gain $\mathbf{K}_k$ and the square root $\mathbf{S}_{k}^{a}$ of the sample covariance $\hat{\mathbf{P}}^{a}_k$ are expressed in terms of $\mathbf{S}_{k}^{b}$ and $\mathbf{S}_{k}^h$, so that
\begin{subequations} \label{ch1:SRKF_filtering}
\begin{align}
\label{ch2:EnSRF_gain}& \mathbf{K}_k \approx \mathbf{S}_k^{b} \left( \mathbf{S}_{k}^{h} \right)^T \left( \mathbf{S}_{k}^{h} \left( \mathbf{S}_{k}^{h} \right)^T + \mathbf{R}_k \right)^{-1} \, , \\
\label{ch2:EnSRF_sqrt} & \mathbf{T}_{k} = \sqrt{\mathbf{I}_{m_k^{sb}} -\left(\mathbf{S}_{k}^{h} \right)^T \left(\mathbf{S}_{k}^{h} \left(\mathbf{S}_{k}^{h} \right)^T + \mathbf{R}_k \right)^{-1} \mathbf{S}_{k}^{h}} \, , \\
\label{ch2:SRKF_analysis_SR}& \mathbf{S}_{k}^a = \mathbf{S}_{k}^b \mathbf{T}_{k} \mathbf{U}_k \, , 
\end{align}
\end{subequations}
where $\mathbf{U}_k$ is a matrix satisfying $\mathbf{U}_k (\mathbf{U}_k)^T = \mathbf{I}_{m_k^{sb}}$, with $\mathbf{I}_{m_k^{sb}}$ being the $m_k^{sb}$ dimensional identity matrix (cf. \S~\ref{ch1:SRKF}). In Eq.~(\ref{ch2:EnSRF_gain}), the Kalman gain $\mathbf{K}_k$ is an $m \times m^{obv}$ matrix. The dimensions of $\mathbf{T}_{k}$, $\mathbf{S}_{k}^a$ and $\mathbf{U}_k$ will be given below. 

In the literature, there are different implementations of the EnSRF. For examples, see \cite{Anderson-ensemble,Bishop-adaptive,Whitaker-ensemble}. Essentially, these implementations  differ from one another only in the choice of the matrix $\mathbf{U}_k$ \cite{Tippett-ensemble}. For this reason, in this dissertation, we do not intend to give a detailed review of all these different versions of the EnSRF. Instead, we just choose the \sindex{ensemble transform Kalman filter} (ETKF) proposed in \cite{Bishop-adaptive} as the representative. In the ETKF \cite{Bishop-adaptive}, one chooses
\begin{equation} \label{ch2_EnSRF_transform_mtx}
\mathbf{T}_k = \mathbf{E}_k^{wpr} \left( \mathbf{D}_k^{wpr} + \mathbf{I}_{m_k^{sb}} \right)^{-1/2} \, ,
\end{equation}
where $\mathbf{T}_k$ is an $m_k^{sb} \times m_k^{sb}$ matrix. The matrices $\mathbf{E}_k^{wpr}$ and $\mathbf{D}_k^{wpr}$ are constructed in the following way. Let $\mathbf{P}^{wpr}_k$ be an $m_k^{sb} \times m_k^{sb}$ weighted projection covariance, so that
\begin{equation} \label{ch2:srp}
\mathbf{P}^{wpr}_k = \left(\mathbf{S}_k^{b} \right)^T \mathbf{H}_k^T \mathbf{R}_k^{-1} \mathbf{H}_k\mathbf{S}_k^{b} \approx \left(\mathbf{S}_k^{h} \right)^T \mathbf{R}_k^{-1} \mathbf{S}_k^{h},
\end{equation}
where $\mathbf{H}_k$ is the Jacobian of the observation operator $\mathcal{H}_k$ evaluated at $\hat{\mathbf{x}}_{k}^b$. We perform a spectral decomposition on $\mathbf{P}^{wpr}_k$ so as to obtain a set of eigenvectors $\left \{ \mathbf{e}_{k,i}\right \}_{i=1}^{m_k^{sb}}$ and the corresponding eigenvalues $\left \{ d_{k,i}\right \}_{i=1}^{m_k^{sb}}$. Then $\mathbf{E}_k^{wpr}$ is an $m_k^{sb} \times m_k^{sb}$ matrix consisting of the eigenvectors $\left \{ \mathbf{e}_{k,i}\right \}_{i=1}^{m_k^{sb}}$ so that 
\begin{equation}
\mathbf{E}_k^{wpr} = \left[ \mathbf{e}_{k,1}, \mathbf{e}_{k,2}, \dotsb, \mathbf{e}_{k,m_k^{sb}} \right] \, ,
\end{equation}   
while $\mathbf{D}_k^{wpr}$ is an $m_k^{sb} \times m_k^{sb}$ diagonal matrix consisting of the corresponding eigenvalues $\left \{ d_{k,i}\right \}_{i=1}^{m_k^{sb}}$, so that $\mathbf{D}_k^{wpr} = \text{diag}(d_{k,1}, d_{k,2}, \dotsb, d_{k,m_k^{sb}})$. It is shown in \cite{Bishop-adaptive} that, if the observation operator is linear, then 
\begin{equation} \label{ch2_EnSRF_transform_mtx_prod}
\mathbf{T}_k (\mathbf{T}_k)^T = \mathbf{I}_{m_k^{sb}} -\left(\mathbf{S}_{k}^{h} \right)^T \left(\mathbf{S}_{k}^{h} \left(\mathbf{S}_{k}^{h} \right)^T + \mathbf{R}_k \right)^{-1} \mathbf{S}_{k}^{h} \, .
\end{equation}
The details of the deduction are given in \cite{Bishop-adaptive}\footnote{As a hint, one may put   $(\mathbf{S}_k^{h})^T \mathbf{R}_k^{-1/2} = \mathbf{E}_k^{wpr} ( \mathbf{D}_k^{wpr})^{1/2}$, so that on the rhs of Eq.~(\ref{ch2_EnSRF_transform_mtx_prod}), $(\mathbf{S}_{k}^{h})^T(\mathbf{S}_{k}^{h} (\mathbf{S}_{k}^{h})^T + \mathbf{R}_k)^{-1} \mathbf{S}_{k}^{h} = \mathbf{E}_k^{wpr} \mathbf{D}_k^{wpr} ( \mathbf{D}_k^{wpr} + \mathbf{I}_{m_k^{sb}} )^{-1} (\mathbf{E}_k^{wpr})^T$. Also Note that $\mathbf{I}_{m_k^{sb}} = \mathbf{E}_k^{wpr} ( \mathbf{D}_k^{wpr} + \mathbf{I}_{m_k^{sb}} ) ( \mathbf{D}_k^{wpr} + \mathbf{I}_{m_k^{sb}} )^{-1} (\mathbf{E}_k^{wpr})^T$. Then it can be verified that $\mathbf{T}_k$ in Eq.~(\ref{ch2_EnSRF_transform_mtx}) is a solution of Eq.~(\ref{ch2_EnSRF_transform_mtx_prod}).}. Also, it can be shown that $\mathbf{T}_k$ in Eq.~(\ref{ch2_EnSRF_transform_mtx}) is a non-singular solution to Eq.~(\ref{ch2_EnSRF_transform_mtx_prod}) \cite[Thm. 6]{Livings-unbiased}, in the sense that the product $\mathbf{T}_k (\mathbf{T}_k)^T$ in Eq.~(\ref{ch2_EnSRF_transform_mtx_prod}), with $\mathbf{T}_k$ given by Eq.~(\ref{ch2_EnSRF_transform_mtx}), is positive definite.     

Originally, the ETKF proposed in \cite{Bishop-adaptive} lets $\mathbf{U}_k = \mathbf{I}_{m_k^{sb}}$ in Eq.~(\ref{ch2:SRKF_analysis_SR}), which may cause a bias in evaluation of the sample mean $\hat{\mathbf{x}}_k^a$, since there is no guarantee that the square root $\mathbf{S}_{k}^a$ obtained in this way satisfies the constraint in Eq.~(\ref{ch2:EnSRF_sr_constraint}). Here we  follow a revision of the ETKF proposed in \cite{Wang-which}, and derive the matrix $\mathbf{U}_k$ from the concept of \sindex{spherical simplex unscented transform} \cite{Julier-spherical,Julier2004}. Concretely, $\mathbf{U}_k$ is chosen as an $m_k^{sb} \times (m_k^{sb}+1)$ matrix in the following form (cf. Eq.~(C15) of \cite{Wang-which}):
\begin{equation}\label{ch2:EnSRF_spherical_orthornormal_matrix}
\begin{split}
&
\left[ \begin{array}{ccccccc}
-\dfrac{1}{\sqrt{2}} & \dfrac{1}{\sqrt{2}} & 0 & 0 & \dotsb & \dotsb & 0 \\
\vdots & \vdots & \vdots & \vdots & \vdots & \vdots & \vdots \\
-\dfrac{1}{\sqrt{i(i+1)}} & \dotsb & -\dfrac{1}{\sqrt{i(i+1)}} & \dfrac{i}{\sqrt{i(i+1)}} & 0 & \dotsb & 0 \\
\vdots & \vdots & \vdots & \vdots & \vdots & \vdots & \vdots \\
-\dfrac{1}{\sqrt{m_k^{sb}(m_k^{sb}+1)}} & \dotsb & \dotsb & \dotsb & \dotsb & -\dfrac{1}{\sqrt{m_k^{sb}(m_k^{sb}+1)}} & \dfrac{m_k^{sb}}{\sqrt{m_k^{sb}(m_k^{sb}+1)}} \\
\end{array}
\right] \, , \\ 
& \\
\end{split}
\end{equation}
which satisfies \cite{Wang-which} 
\begin{subequations}
\begin{align}
\label{ch2:EnSRF_orthornormal_matrix_constrain1} &\mathbf{U}_k \, \mathbf{1} = \mathbf{0} \, ,\\
\label{ch2:EnSRF_orthornormal_matrix_constrain2} &\mathbf{U}_k \, \mathbf{U}_k^T = \mathbf{I} \, ,
\end{align}
\end{subequations}
where $\mathbf{1}$ in Eq.~(\ref{ch2:EnSRF_orthornormal_matrix_constrain1}) denotes the $(m_k^{sb}+1) \times 1$ vector whose elements are all equivalent to $1$. In this way, the square root $\mathbf{S}_{k}^a$ in Eq.~(\ref{ch2:SRKF_analysis_SR}) is an $m \times (m_k^{sb}+1)$ matrix. In the case that there is no dynamical noise, $m_k^{sb}=n-1$. Therefore, when generating the analysis ensemble according to Eq.~(\ref{ch2:EnSRF_analysis_scheme}), one obtains $m_k^{sb} + 1=n$ members, the same as that of the background ensemble. But if there exists dynamical noise, one may have $m_k^{sb} > n-1$, so that the number $m_k^{sb} + 1$ of the analysis ensemble members is larger than $n$. In this case, in order to prevent the number of the ensemble members from growing at each cycle, one may have to reduce the number of the column vectors of $\mathbf{S}_{k}^a$. One such a possible strategy will be discussed in \S~\ref{ch3:sec_reduced_sukf} of the next chapter.

For convenience, we summarize the procedures in the EnSRF as follows:\\
Propagation step:
\begin{subequations} \label{ch2:EnSRF_propagation}
\begin{align}
& \mathbf{x}^b_{k,i} =  \mathcal{M}_{k,k-1} \left( \mathbf{x}_{k-1,i}^a \right),~i=1,\dotsb, n \, , \\
 &\hat{\mathbf{x}}_k^b = \frac{1}{n} \sum\limits_{i=1}^n \mathbf{x}^b_{k,i} \, , \\
& \mathbf{S}_{k}^{xb} = \dfrac{1}{\sqrt{n-1}} \left[ \mathbf{x}^b_{k,1} - \hat{\mathbf{x}}_k^b, \dotsb, \mathbf{x}^b_{k,n} - \hat{\mathbf{x}}_k^b \right] \, , \\
& \mathbf{S}_{k}^{b} = \sqrt{ \mathbf{S}_{k}^{xb} \left( \mathbf{S}_{k}^{xb} \right)^T+\mathbf{Q}_k} \, , \\
&  \mathbf{S}_{k}^h = \left[ \mathcal{H}_k ( (\mathbf{S}_{k}^{b})_1 ), \dotsb, \mathcal{H}_k ( (\mathbf{S}_{k}^{b})_{m_k^{sb}} ) \right] \, . 
\end{align}
\end{subequations}
Filtering step:
\begin{subequations} \label{ch2:EnSRF_filtering}
\begin{align}
&\mathbf{K}_k = \mathbf{S}_k^{b} \left( \mathbf{S}_{k}^{h} \right)^T \left( \mathbf{S}_{k}^{h} \left( \mathbf{S}_{k}^{h} \right)^T + \mathbf{R}_k \right)^{-1} \, , \\
& \hat{\mathbf{x}}_k^a = \hat{\mathbf{x}}_k^b + \mathbf{K}_k \left ( \mathbf{y}_k - \mathcal{H}_k \left( \hat{\mathbf{x}}_k^b \right) \right ) \, , \\
& \mathbf{T}_k = \mathbf{E}_k^{wpr} \left( \mathbf{D}_k^{wpr} + \mathbf{I}_{m_k^{sb}} \right)^{-1/2} \, , \\
&\mathbf{S}_{k}^a = \mathbf{S}_{k}^b \mathbf{T}_{k} \mathbf{U}_k \, .
\end{align}
\end{subequations}
Analysis scheme:
\begin{equation} \tagref{ch2:EnSRF_analysis_scheme}
\mathbf{x}_{k,i}^a = \hat{\mathbf{x}}_k^a +\sqrt{n-1} \left( \mathbf{S}_k^a \right)_i, \, i=1, 2, \dotsb, n. 
\end{equation}
For convenience, the dimension information of the matrices involved in the EnSRF is listed in Table~\ref{ch2:dimensions_in_EnSRF}. 

\begin{table*}[!t]
	\centering
	\caption{\label{ch2:dimensions_in_EnSRF} Information of dimensions involved in the ETKF.}
	\begin{tabular}{p{7cm}p{7cm}}
        \hline \hline
		Number & Meaning \\
		\hline
		$m$ & Dimension of the state space \\
		\hline
		$m^{obv}$ & Dimension of the observation space \\
		\hline
		$m_k^{sb}$ & Number of positive eigenvalues of
		$\mathbf{S}_{k}^{xb} \left( \mathbf{S}_{k}^{xb} \right)^T+\mathbf{Q}_k$\\ 
		\hline
		n  & Number of the ensemble members of both the background and the analysis \\ 
		\hline \hline 
        Matrix & Dimension \\
        \hline
		$\mathbf{Q}_k$ & $m \times m$ \\
		$\mathbf{R}_k$ & $m^{obv} \times m^{obv}$ \\
		$\mathbf{K}_k$ & $m \times m^{obv}$ \\
		$\mathbf{S}_{k}^b$ & $m \times m_{k}^{sb}$ \\
		$\mathbf{S}_{k}^h$ & $m^{obv} \times m_{k}^{sb}$ \\
		$\mathbf{T}_{k}$ & $m_k^{sb} \times m_{k}^{sb}$ \\
		$\mathbf{P}_k^{wpr}$ & $m_k^{sb} \times m_{k}^{sb}$ \\
		$\mathbf{E}_k^{wpr}$ & $m_k^{sb} \times m_{k}^{sb}$ \\
		$\mathbf{D}_k^{wpr}$ & $m_k^{sb} \times m_{k}^{sb}$ \\
		$\mathbf{U}_{k}$ & $m_k^{sb} \times (m_k^{sb}+1)$ \\
		$\mathbf{S}_{k}^a$ & $m \times (m_k^{sb}+1)$ \\
        \hline \hline
     \end{tabular}
\end{table*}

\subsection{Two auxiliary techniques to improve the performance of the ensemble Kalman filter}\label{ch2:sec_two_techniques}
Two auxiliary techniques are often adopted in many applications of the EnKF in order to improve the performance of the filter. One technique is \sindex{covariance inflation} \cite{Anderson-Monte,Ott-local,Whitaker-ensemble}, which is related to the Kalman filter with fading memory (KF-FM) in \S~\ref{ch1: KF fading}, as will be shown below. The other is \sindex{covariance filtering} (or \sindex{covariance localization}) \cite{Hamill-distance}.

\subsubsection{Covariance inflation} \label{ch2:sec_covariance_inflation}
Covariance inflation is a method that one artificially increases the error covariance of the background or the analysis at each assimilation cycle \cite{Anderson-Monte,Ott-local,Whitaker-ensemble}. The rationale behind covariance inflation may be explained from the following points of views. On one hand, when adopting the EnKF for data assimilation, the error covariance of the system states (either the background or the analysis) will be systematically underestimated due to the effect of finite ensemble size \cite{Whitaker-ensemble}. Therefore, it is natural to introduce covariance inflation for compensation. On the other hand, one may note that for data assimilation in nonlinear systems, the EnKF is only an approximate solution. Therefore, even in the ideal situation where there are no other sources of errors in the systems under assimilation, there may be still an \sindex{algorithmic error} in the EnKF, which may cause offsets in our estimations. Therefore, one may follow the argument in \S~\ref{ch1: KF fading} to artificially increase the error covariance of the background by a factor, which in effect will increase the relative weight of the incoming observation, and thus improve the robustness and accuracy of the EnKF.

In the EnKF, given a background ensemble $ \mathbf{X}_{k}^b = \left \{\mathbf{x}^b_{k,i} \right \}_{i=1}^n$ with the sample mean $\hat{\mathbf{x}}_k^b$, conducting covariance inflation is equivalent to replacing each ensemble member $\mathbf{x}^b_{k,i}$ by $\hat{\mathbf{x}}_k^b + (1+\delta) \left( \mathbf{x}^b_{k,i}- \hat{\mathbf{x}}_k^b \right)$, where $\delta$ is the \em{inflation factor}. In this way, the sample mean $\hat{\mathbf{x}}_k^b$ of the background ensemble $\mathbf{X}_{k}^b$ remains the same, but the sample error covariance is increased by a factor of $(1+\delta)^2$. Given an analysis ensemble, covariance inflation can be done in a similar way. 

How to optimally choose the value of the inflation factor $\delta$ is still an open question. In many works, e.g. \cite{Anderson-Monte,Ott-local,Whitaker-ensemble}, $\delta$ was often heuristically chosen as a constant when running a data assimilation algorithm. In a more recent work, Anderson \cite{Anderson2009} proposed a spatially and temporally adaptive method to choose $\delta$, which treats $\delta$ as the variable of a random process. Thus one can update $\delta$ at each assimilation cycle according to a data assimilation algorithm (e.g. the EnKF) by treating $\delta$ as a hidden (or unobserved) state variable of the dynamical system. However, one possible problem of this method is that normally one may not have the exact knowledge to specify the random process with respect to $\delta$. 

    
\subsubsection{Covariance filtering} \label{ch2:sec_covariance_filtering}
The error covariances of the EnKF are often evaluated based on small-size ensembles. For this reason, there may exist spuriously large correlations between distant locations in the practice \cite{Hamill-distance}. To address this problem, one may introduce a distance-dependent filter to the error covariance, so that the correlations between two distant locations are set to zero. For this purpose, the \sindex{Schur product} can be applied to an error covariance. Mathematically, the Schur product, $\mathbf{C} \equiv \mathbf{A} \circ \mathbf{B}$, of two matrices $\mathbf{A}$ and $\mathbf{B}$ with the same dimensions, is defined as the matrix with the same dimension as $\mathbf{A}$ and $\mathbf{B}$, whose components $C_{ij} = A_{ij} B_{ij}$, where $A_{ij}$, $B_{ij}$ and $C_{ij}$ are the components on the $i$-th row and the $j$-th column of the matrices $\mathbf{A}$, $\mathbf{B}$ and $\mathbf{C}$, respectively. For our problem, we suppose that $\mathbf{A}$ is an error covariance matrix, and $\mathbf{B}$ is the matrix introduced to taper $\mathbf{A}$ so as to reduce the spuriously large correlations in $\mathbf{A}$. Since $\mathbf{A}$ and the tapered matrix $\mathbf{C} \equiv \mathbf{A} \circ \mathbf{B}$ are covariance matrices, they shall both be positive semi-definite. As a result, we also require that $\mathbf{B}$ be at least positive semi-definite \cite[Lemma 3.7.1]{Bapat1997}.       

\begin{figure*}[!t]
\centering
\includegraphics[width=\textwidth]{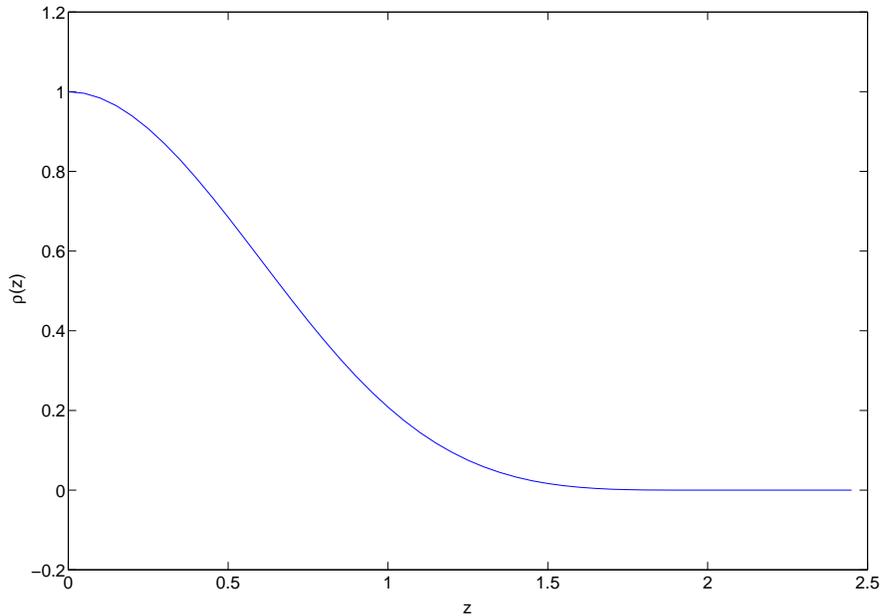}
\caption{ \label{fig:ch2_correlation_function} Shape of the function $\rho$ in Eq.~(\ref{ch2_cov_filtering_correlation_func}).}
\end{figure*}

For convenience, we call $\mathbf{B}$ the \em{taper matrix} hereafter. The construction of $\mathbf{B}$ can be done in the following way: 
\begin{equation} \label{ch2:eq_taper matrix}
B_{ij} = \rho (d_{ij}),
\end{equation}
where $d_{ij}$ is a metric measuring the difference between the locations $i$ and $j$, and $\rho$ is chosen to be \textit{a function of positive type} \cite[p. 299]{Reed1980}, so that it guarantees that the taper matrix $\mathbf{B}$ is positive definite, as a result of the Bochner's theorem \cite[p. 300]{Reed1980}. Several examples of such a function $\rho$ were discussed in \cite{Gaspari1999}. In this dissertation, we follow \cite{Houtekamer-sequential} and choose function $\rho$ in the following form:
\begin{equation} \label{ch2_cov_filtering_correlation_func}
\rho \left( z \right) = \begin{cases}
 -\dfrac{1}{4} z^5 + \dfrac{1}{2} z^3 + \dfrac{5}{8} z^3 - \dfrac{5}{3} z^2 +1 \, , & \text{if}~ 0 \le z \le 1 \, ; \\
\dfrac{1}{12} z^5 - \dfrac{1}{2} z^4 + \dfrac{5}{8} z^3 + \dfrac{5}{3} z^2 - 5 z + 4 - \dfrac{2}{3} z^{-1} \, , & \text{if}~ 1 < z \le 2 \, ;\\
 0 \, , & \text{if}~ z >2 \, . \\ 
\end{cases} 
\end{equation}  
For illustration, the shape of the function $\rho$ is plotted in Fig.~(\ref{fig:ch2_correlation_function}). As one can see there, the function $\rho$ has a ``cut-off'' effect at $z=2$, in the sense that the values of the function are set to zero for all $z>2$.    

For data assimilation in real world, $d_{ij}$ is normally a function of the distance between the available observation sites $i$ and $j$ in the three dimensional physical world. For example, see \cite{Constantinescu2007}. However, in mathematical analysis, this choice might not always be available. For example, the system states of a mathematical model may not have any physical meaning, so that we cannot observe them in the physical world. On the other hand, for a mathematical model, the (physical) distance between indices (or locations) $i$ and $j$ may also not be well defined as in the physical world. For these reasons, we follow \cite{Houtekamer-sequential} to conduct covariance filtering in the following way. 

Concretely, we suppose that $\mathbf{A}$ is the covariance of an $m$-dimensional random variable $\mathbf{x}=\left[x_1,\dotsb,x_m\right]$. For convenience, we assume that the mean $\mathbb{E}(\mathbf{x})= \mathbf{0}$, so that 
\begin{equation} \label{ch2:cov_before_tapering}
\mathbf{A} = \left [ \mathbf{r}_1^T, \dotsb, \mathbf{r}_m^T \right ]^T,
\end{equation}
where 
\begin{equation}
\mathbf{r}_i = \left[ \mathbb{E}(x_ix_1), \dotsb, \mathbb{E}(x_ix_m) \right]
\end{equation}
is the $i$-th row of $\mathbf{A}$. We define
\begin{equation} \label{ch2:distance_in_column_vectors}
d_{ij}=\lVert \mathbf{r}_{i}^T-\mathbf{r}_{j}^T \rVert_{2} / l_c \, ,
\end{equation}  
where $l_c$ is a length scale that is introduced to influence where the ``cut-off'' effect of the function $\rho$ takes place \footnote{Note that in Eq.~(\ref{ch2:distance_in_column_vectors}), by choosing the row vectors to calculate the distances $d_{ij}$, we have implicitly assumed that the number of the rows of a matrix (not necessarily square) is larger than or at least equal to the number of its columns. If this is not the case, then it is suggested to choose the column vectors to calculate the distances $d_{ij}$ instead. In this way, covariance filtering can be applied to non-square matrices like the cross covariance (when the dimension of the state space is not equal to that of the observation space, cf., for example, Eq.~(\ref{ch2:EnKF_cross_cov_approx})).}. With some algebra, it can be shown that 
\begin{equation} \label{ch2:cov_filtering_distance_in_en}
d_{ij}= \dfrac{1}{l_c} \, \sqrt{\left( \mathbb{E} ((x_i -x_j)x_1)\right)^2 + \dotsb + \left( \mathbb{E} ((x_i -x_j)x_m)\right)^2} \, .
\end{equation}
In this sense, $d_{ij}$ is a metric in measuring the statistical difference between the random variable $x_i$ and $x_j$ in the $m$-dimensional state space. In particular, if the index $i=j$ or if the length scale $l_c = \infty$ so that $d_{ij}=0$, then $\rho \left( d_{ij} \right) = 1$, which implies that in effect there is no tapering effect. But if $i \neq j$ so that $x_i$ and $x_j$ are two different random variables, $d_{ij}$ is positive and covariance filtering will take place.    

\begin{figure*}[!t] 
   \centering
   \includegraphics[width=\textwidth]{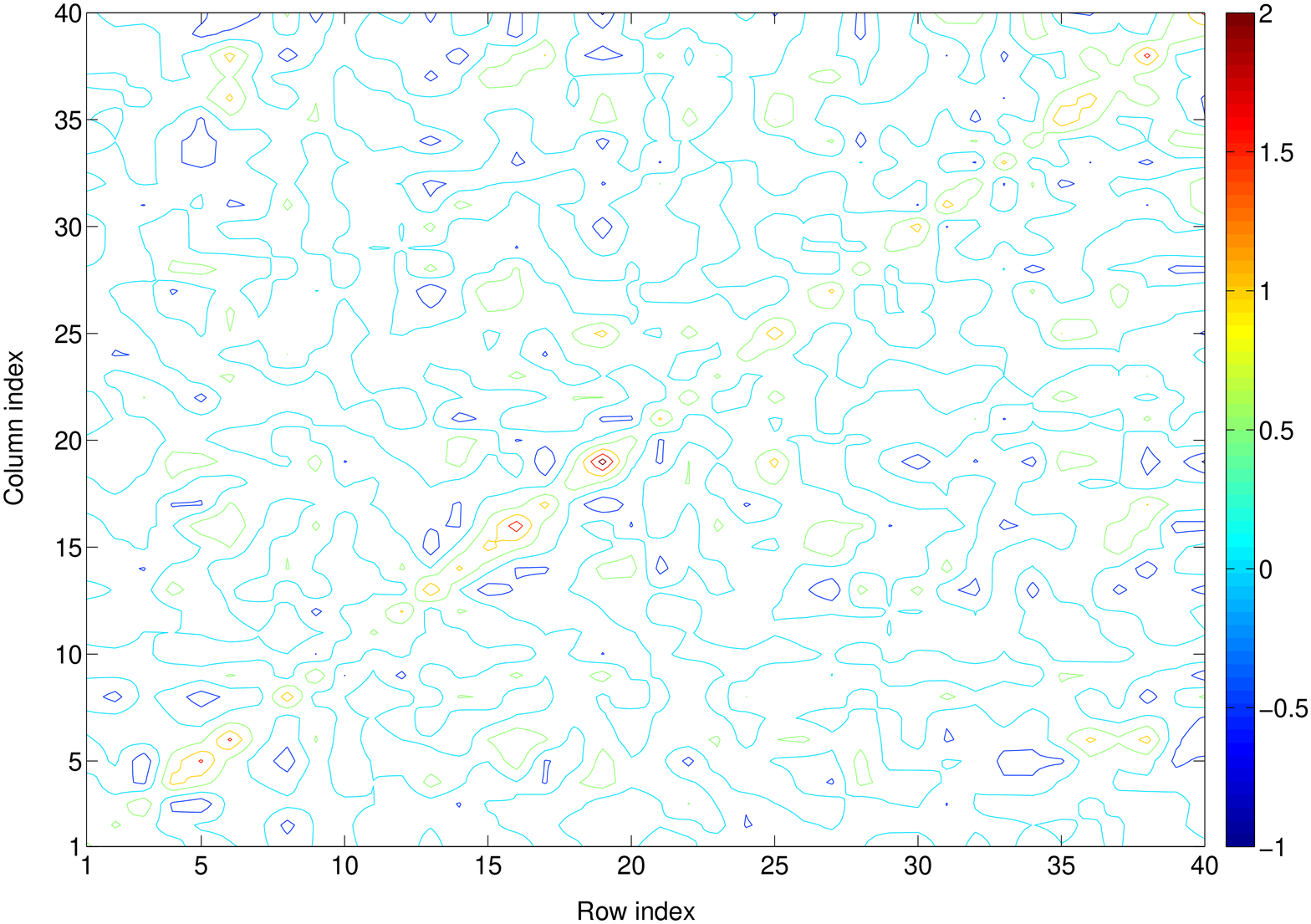} 
   \caption{ \label{fig:cov_before_schur} Contour map of the sample covariance matrix before conducting covariance filtering.}
\end{figure*}

Here we use a numerical example to illustrate the effect of covariance filtering in changing the structure of a covariance matrix. To this end, we draw $10$ samples from the $40$-dimensional normal Gaussian distribution $N(\mathbf{0},\mathbf{I}_{40})$, with $\mathbf{I}_{40}$ being the $40$-dimensional identity matrix. In consistence with the previous notations, we denote the covariance matrix calculated based on these 10 samples by $\mathbf{A}$. In Fig.~\ref{fig:cov_before_schur}, we use the interpolated contour map to represent the structure of $\mathbf{A}$, where the values of the contour levels in Fig.~\ref{fig:cov_before_schur} correspond to the values of the elements of $\mathbf{A}$. For visualisation, we use different colours to represent different values, as indicated in Fig.~\ref{fig:cov_before_schur}. As one can see, because of the effect of small samples, the sample covariance matrix $\mathbf{A}$ deviates from the $40$-dimensional identity matrix $\mathbf{I}_{40}$. In fact, by checking the eigenvalues of $\mathbf{A}$ in Fig.~\ref{fig:eigenvalues}\footnote{The eigenvalues are obtained by conducting a singular value decomposition on $\mathbf{A}$. The eigenvalues for the other matrices in Fig.~\ref{fig:eigenvalues} are obtained in the same way.}, it can be seen that $\mathbf{A}$ is singular in the sense that it only has $9$ positive eigenvalues.  
    
\begin{figure*}[!t] 
   \centering
   \includegraphics[width=\textwidth]{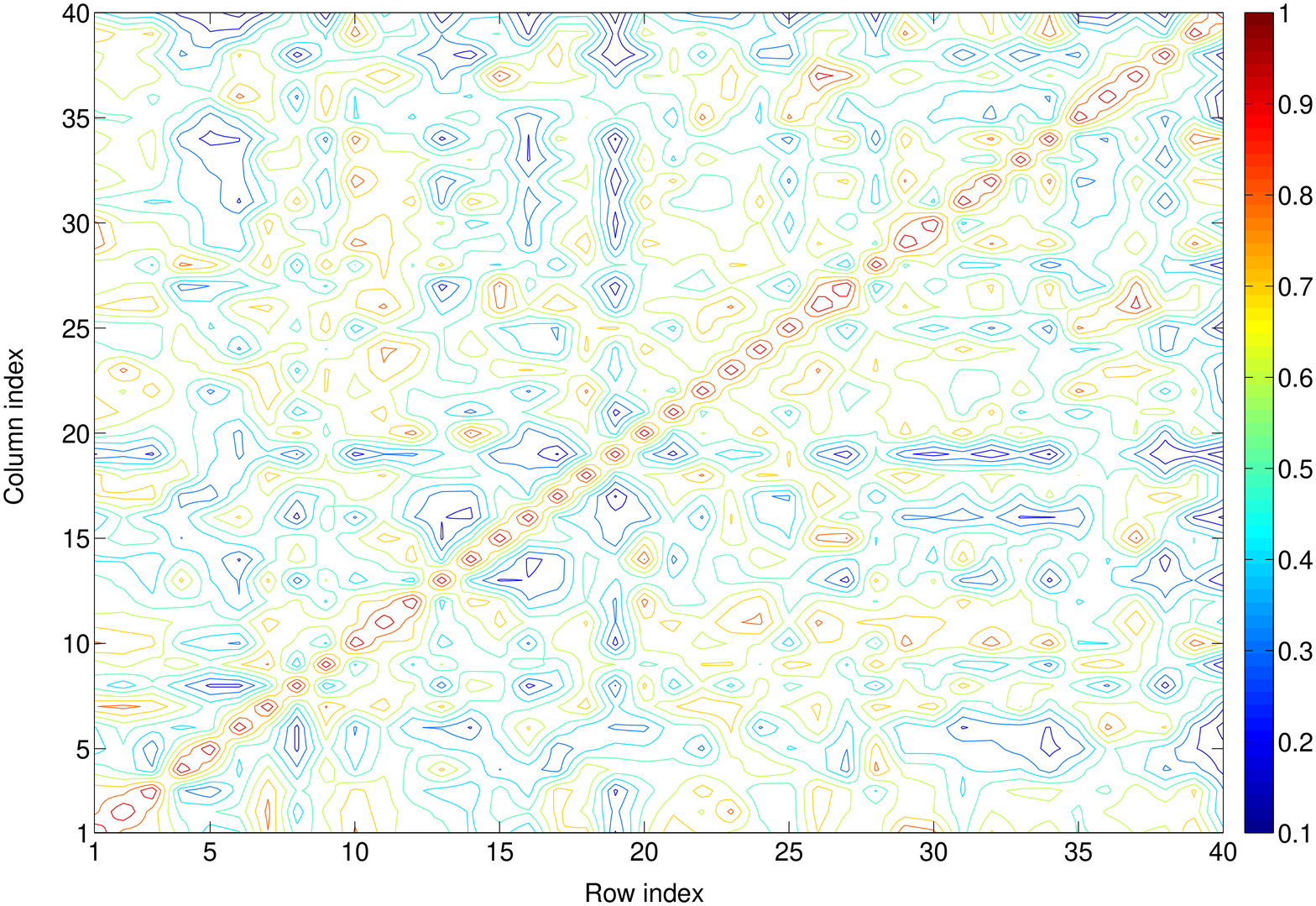} 
   \caption{ \label{fig:taper_matrix} Contour map of the taper matrix.}
\end{figure*}

The taper matrix $\mathbf{B}$ is constructed based on Eqs.~(\ref{ch2:eq_taper matrix}), (\ref{ch2_cov_filtering_correlation_func}), (\ref{ch2:cov_before_tapering}), and (\ref{ch2:distance_in_column_vectors}), where the length scale in Eq.~(\ref{ch2:distance_in_column_vectors}) is chosen as $l_c=5$. We plot the contour map of the taper matrix in Fig.~\ref{fig:taper_matrix}. As one can see, the main diagonal of $\mathbf{B}$ consists of values of $1$, where the other elements of $\mathbf{B}$ are in general less than $1$. On the other hand, we also plot the eigenvalues of $\mathbf{B}$ in Fig.~\ref{fig:eigenvalues}. Numerical experiments show that the eigenvalues of the taper matrix $\mathbf{B}$ are always positive\footnote{Some small eigenvalues of $\mathbf{B}$ are very close to zero, so it may not be distinguishable in Fig.~\ref{fig:eigenvalues}.}, which confirms that $\mathbf{B}$ is a positive definite matrix, as we expect. 

\begin{figure*}[!t] 
   \centering
   \includegraphics[width=\textwidth]{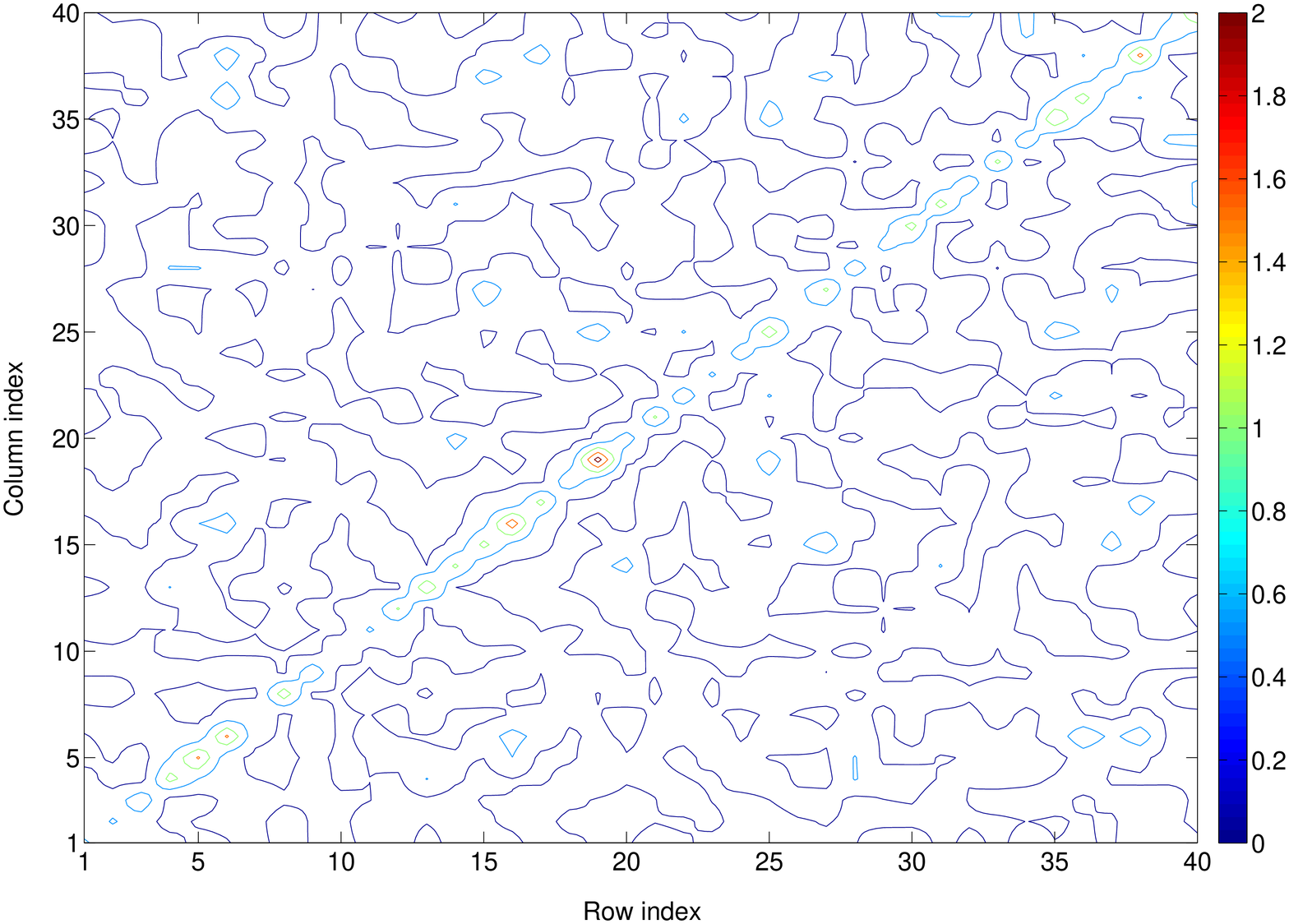} 
   \caption{ \label{fig:cov_after_schur} Contour map of the sample covariance matrix after conducting covariance filtering.}
\end{figure*}

\begin{figure*}[!t] 
   \centering
   \includegraphics[width=\textwidth]{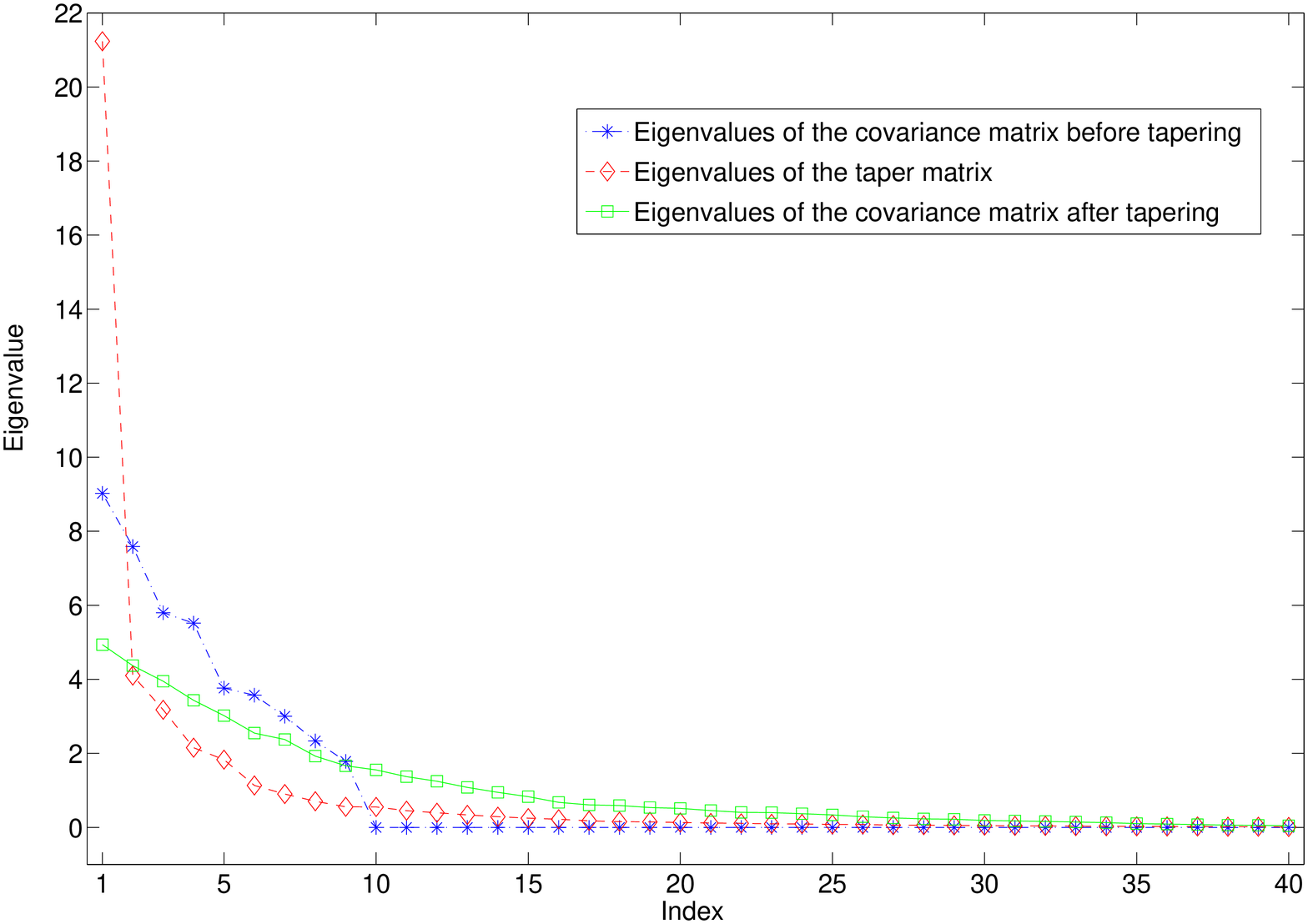} 
   \caption{ \label{fig:eigenvalues} Eigenvalues of the matrices involved in conducting covariance filtering.}
\end{figure*}

After conducting covariance filtering, we obtain the tapered sample covariance $\mathbf{C} \equiv \mathbf{A} \circ \mathbf{B}$. The contour map of $\mathbf{C}$ is plotted in Fig.~\ref{fig:cov_after_schur} \footnote{Note that in Fig.~\ref{fig:cov_after_schur}, some of the negative values are very close to $0$ (in the order of $10^{-2}$--$10^{-1}$). The negativeness of these elements is not well represented by the colour bar because of the scales of the values in plot.}. Comparing the structure of $\mathbf{A}$ in Fig.~\ref{fig:cov_before_schur} with that of $\mathbf{C}$ in Fig.~\ref{fig:cov_after_schur}, it can be seen that $\mathbf{A}$ and $\mathbf{C}$ have the same main diagonal elements. While the other elements of $\mathbf{C}$ are in general closer to $0$ than those of $\mathbf{A}$. In summary, conducting covariance filtering ``decreases the off-diagonal elements (of a covariance matrix), while keeping the (main) diagonal elements unchanged'', having the same effect as that reported in \cite{Constantinescu2007}. In this way, the spuriously large covariances between two different random variables may be reduced. In addition, as indicated in Fig.~\ref{fig:eigenvalues}, after conducting covariance filtering, the tapered sample covariance $\mathbf{C}$ becomes positive definite, since all its eigenvalues are all positive\footnote{Again, some small eigenvalues of $\mathbf{C}$ are very close to zero, so it may not be distinguishable in Fig.~\ref{fig:eigenvalues}.}.  
 
In the context of the EnKF, covariance filtering is normally conducted at the propagation step. So at the $k$-th assimilation cycle, there are three matrices, namely, the background covariance $\hat{\mathbf{P}}_k^b$, the cross covariance $\hat{\mathbf{P}}_k^{cr}$, and the projection covariance $\hat{\mathbf{P}}_k^{pr}$, that possibly need to conduct covariance filtering. The readers are referred to \cite{Constantinescu2007} for the implementation of covariance filtering in practice \footnote{Note that in the practical implementation, it often assumes that the observation system is linear, which is not a necessary assumption for our implementation}. For the reasons given previously, we do not use the observation sites in the physical world to construct the taper matrix in our analysis. Our implementation strategy is discussed below. 

First, we note that in the EnKF (including both the stochastic EnKF and the EnSRF), it is not necessary to calculate the background covariance $\hat{\mathbf{P}}_k^b$, because the subsequent procedures, such as the calculation of the Kalman gain, the updating of the background, and the generation of the analysis ensemble, do not involve $\hat{\mathbf{P}}_k^b$. Therefore it is not necessary to conduct covariance filtering on $\hat{\mathbf{P}}_k^b$. On the other hand, covariance filtering can be conducted on both the cross covariance $\hat{\mathbf{P}}_k^{cr}$ and the projection covariance $\hat{\mathbf{P}}_k^{pr}$. In the EnKF, this can be done either through Eq.~(\ref{ch2:distance_in_column_vectors}) so that the construction of the taper matrix is based on the matrix to be tapered, or through Eq.~(\ref{ch2:cov_filtering_distance_in_en}) so that the construction of the taper matrix is based on the ensemble members (but in Eq.~(\ref{ch2:cov_filtering_distance_in_en}) the expectation shall be replaced by sample mean). Which way to choose may depend on our practical consideration, i.e., which one is more efficient in a certain sense. With a moderate dimension of the dynamical system in our numerical experiments (cf. Eqs.~(\ref{ch2:ex_LE98}) and (\ref{ch2:ex_observer})), we choose to construct the taper matrix through Eq.~(\ref{ch2:distance_in_column_vectors}) since it is easier to implement the codes (in MATLAB) and runs faster (by using matrices rather than loops). 
 
Also note is that, by conducting covariance filtering, we introduce an extra parameter, the length scale $l_c$ (cf. Eq.~(\ref{ch2:distance_in_column_vectors}) or (\ref{ch2:cov_filtering_distance_in_en}) ), to the EnKF. Like the situation in choosing the optimal inflation factor, there still lacks a systematic approach to determining the optimal length scale $l_c$. Thus in practice one has to adopt some heuristic methods instead.

\section{Example: Assimilating a 40-dimensional system}
\subsection{The dynamical and observation systems} \label{ch2:sec_testbed}
For illustration, we choose the $m$-dimensional system model introduced by Lorenz and Emanuel \cite{Lorenz-predictability,Lorenz-optimal} (Lorenz-Emanuel 98, or LE 98 for short hereafter) for our numerical experiments. The LE 98 model is a simplified system for modelling atmospheric dynamics, which ``shares certain properties with many atmospheric models'' \cite{Lorenz-optimal}. Its governing equations are given by
\begin{equation} \label{ch2:ex_LE98}
\frac{dx_i}{dt} = \left( x_{i+1} - x_{i-2} \right) x_{i-1} - x_i +F, \, i=1, \dotsb, m. 
\end{equation}
The quadratic terms simulate the advection, the linear term represents the internal dissipation, while the constant $F$ acts as the external forcing \cite{Lorenz-predictability}. For consistency, the variables $x_i$'s are defined cyclically such that $x_{-1}=x_{m-1}$, $x_{0}=x_{m}$, and $x_{m+1}=x_{1}$. Note that in Eq.~(\ref{ch2:ex_LE98}), there is no dynamical noise. One may choose to add some artificial noise to the dynamical system so as to improve the performance of a data assimilation algorithm. For example, see \cite{Arulampalam2002}. In doing this, in effect one increases the background covariance, similar to the idea of covariance inflation. Since we have adopted covariance inflation in our implementation, here we choose not to introduce any artificial  noise to the dynamical system Eq.~(\ref{ch2:ex_LE98}), so that in effect we let $\mathbf{Q}_k=\mathbf{0}$. In this way, the implementation of the ETKF can be simpler, since it is more convenient to obtain the square roots of the background covariances when there is no dynamical noise (cf. the discussion in \S~\ref{ch2:sec_EnSRF}). 

We choose the observer $\mathcal{H}_k$ to be a time-invariant identity operator. Specifically, given a system state $\mathbf{x}_k=[x_{k,1},\dotsb,x_{k,m}]^T$ at the $k$-th assimilation cycle, the observations are obtained according to
\begin{equation} \label{ch2:ex_observer}
\mathbf{y}_k = \mathcal{H}_k (\mathbf{x}_k) + \mathbf{v}_k = \mathbf{x}_k + \mathbf{v}_k \, ,
\end{equation} 
where $\mathbf{v}_k$ follows the $m$-dimensional Gaussian distribution $N(\mathbf{v}_k: \mathbf{0}, \mathbf{I})$ with $\mathbf{I}$ being the identity matrix. 

Note that Eq.~(\ref{ch2:ex_LE98}) represents a set of nonlinear ordinary differential equations (ODE). Their exact solutions are intractable. Therefore it is customary to integrate the nonlinear ODEs numerically in practice. This introduces a discretization to the dynamical system, so that it becomes a first-order Markov chain, in the form of Eq.~(\ref{ch2:ps_dyanmical_system}). In doing this, the data assimilation problem falls into the scenario presented in \S~\ref{ch2:sec_ps} (by ignoring the discretization errors in the dynamical system).  

In our experiments, we set $m=40$ and $F=8$, and integrate the dynamical system Eq.~(\ref{ch2:ex_LE98}) through a fourth-order Runge-Kutta method \cite[Ch.~16]{Vetterling-numerical}. We choose the length of the integration window to be $100$ dimensionless units, and the integration time step to be $0.05$ units (corresponding to a 6-h interval in reality \cite{Lorenz-optimal}). Thus there are $2000$ assimilation cycles overall. We make the observations of the dynamical system at each assimilation cycle. 

\subsection{Two Measures of filter performance}\label{ch2:two_measures}
We adopt two statistics to measure the performance of the EnKF. One is the time-averaged relative (or normalized) rms error (\sindex{relative rmse} for short), which is defined as
\begin{equation}\label{Eq:reltiave rmse}
e_r=\frac{1}{k_{max}}\sum\limits_{k=1}^{k_{max}} \lVert\hat{\mathbf{x}}_k^{a}-{\mathbf{x}}_k^{tr}\rVert_2/ \lVert{\mathbf{x}}_k^{tr}\rVert_2,
\end{equation}
where $k_{max}$ is the maximum assimilation cycle, $\mathbf{x}_k^{tr}$ denotes the truth (the state of a control run) at the $k$-th cycle, and $\lVert \bullet \rVert_2$ means the $2$-norm. Note that $e_r$ can be interpreted as the time-averaged noise level of the trajectory $\{ \hat{\mathbf{x}}_k^{a} \}_{k=1}^{k_{max}}$ with respect to the true states. From this point of view, we can define the concept of \sindex{divergence} of the EnKF in the following sense: suppose that the relative rmse of the observations is $e_r^{obv}$, which, with the identity observation operator $\mathcal{H}_k$, is defined as 
\begin{equation}
e_r^{obv} = \frac{1}{k_{max}}\sum\limits_{k=1}^{k_{max}} \lVert\mathbf{y}_k-{\mathbf{x}}_k^{tr}\rVert_2/ \lVert{\mathbf{x}}_k^{tr}\rVert_2 = \frac{1}{k_{max}}\sum\limits_{k=1}^{k_{max}} \lVert\mathbf{v}_k \rVert_2/ \lVert{\mathbf{x}}_k^{tr}\rVert_2 \, .
\end{equation} 
If $e_r>e_r^{obv}$, then we say the EnKF is divergent because in such circumstances, the trajectory $\{ \hat{\mathbf{x}}_k^{a} \}_{k=1}^{k_{max}}$ obtained by the EnKF, on average, is more noisy than the observations, which implies that it might not make any sense to use the EnKF for data assimilation. 

Note that one may also use the time-averaged absolute rmse 
\begin{equation}
e_a=\frac{1}{k_{max}}\sum\limits_{k=1}^{k_{max}} \lVert\hat{\mathbf{x}}_k^{a}-{\mathbf{x}}_k^{tr}\rVert_2 
\end{equation}
as the measure. Since we deal with the estimation errors in a finite-dimensional space, it can be shown that the norms $\lVert \bullet \rVert_2$ and $\lVert \bullet \rVert_2 /c$ for any positive scalar constant $c$ are topologically equivalent \cite[Thm~5.36]{Hunter2001}. However, the relative rmse appears to be a more straightforward measure for indicating how good (or bad) our estimations are with respect to the true states. For this reason, we adopt the relative rmse throughout this dissertation. 

Another statistic is the time-averaged rms ratio (\sindex{rms ratio} for short), which is designed to examine the similarity between the analysis ensembles generated by the EnKF and the true states. To see this, we first introduce two types of errors with respect to an analysis ensemble $\mathbf{X}_{k}^a = \left\{ \mathbf{x}_{k,i}^a  \right\}_{i=1}^n$, in terms of
\begin{equation}
\begin{split}
& e_{k,1} = \lVert \hat{\mathbf{x}}_k^a - \mathbf{x}_{k}^{tr} \rVert_{2} \, , \\
& e_{k,2} = \dfrac{1}{n} \sum\limits_{i=1}^{n} \lVert \mathbf{x}_{k,i}^a - \mathbf{x}_{k}^{tr} \rVert_{2} \, . 
\end{split}
\end{equation}
Here, $e_{k,1}$ denotes the error of the sample mean $\hat{\mathbf{x}}_k^a$ in estimating the truth $\mathbf{x}_{k}^{tr}$, where $e_{k,2}$ means the average error of the ensemble $\mathbf{X}_{k}^a$ in estimating $\mathbf{x}_{k}^{tr}$. The time averaged rms ratio $R$ is defined as
\begin{equation} \label{R}
R = \dfrac{1}{k_{max}} \sum\limits_{k=1}^{k_{max}} e_{k,1 }/ e_{k,2} = \dfrac{1}{k_{max}} \sum\limits_{k=1}^{k_{max}} \dfrac{n\lVert \hat{\mathbf{x}}_k - \mathbf{x}_{k}^{tr} \rVert_{2}}{\sum\limits_{i=1}^{n} \lVert \mathbf{x}_{k,i} - \mathbf{x}_{k}^{tr} \rVert_{2}} \, .
\end{equation}

If the truth $\mathbf{x}_{k}^{tr}$ is statistically indistinguishable from the analysis ensemble $\mathbf{X}_{k}^a$, then it can be shown that the expectation $R_e$ of the ratio $e_{k,1}/e_{k,2}$ is given by \cite{Murphy-impact, Whitaker-ensemble}
\begin{equation}\label{R_e}
R_e =  \sqrt{\dfrac{n+1}{2n}} \, .
\end{equation}
Therefore, we say that the analysis ensemble $\mathbf{X}_{k}^a$ is statistically indistinguishable from the truth $\mathbf{x}_{k}^{tr}$ if $R$ and $R_e$ are close to one another. The relative position between $R$ and $R_e$ also qualitatively reflects the performance in estimating the error covariance, e.g., overestimation or underestimation (cf. \cite{Anderson-ensemble, Whitaker-ensemble} and the references therein). $R>R_e$ means that the covariance computed based on the ensemble $\mathbf{X}_{k}^a = \left\{ \mathbf{x}_{k,i}^a  \right\}_{i=1}^n$ underestimates the estimation error, while $R<R_e$ implies the opposite, i.e., overestimation of the estimation error \cite{Murphy-impact, Whitaker-ensemble}.

\subsection{Additional information of numerical implementations of the algorithms}
The nonlinear Kalman filters, including the EnKF in this chapter, and the reduced rank sigma point Kalman filters in chapters \ref{ch3:ukf} and \ref{ch4:ddfs}, may require the computations of the eigenvalues and eigenvectors, or the square roots, or the inverses of some matrices. Thus for clarity, here we would like to explain how we conduct these computations in our experiments. Because of the same origin of the nonlinear Kalman filters, the computation schemes discussed here are applied in the same way to the EnKF in this chapter, and the reduced rank sigma point Kalman filters in the next two chapters.

Firstly, for evaluations of the eigenvectors and eigenvalues of a symmetric matrix, we adopt the spectral decomposition in our computation by treating it as a special case of the singular value decomposition (SVD) \cite[\S~2.5.3]{Golub-matrix}.  

Next, for computations of the square roots of some covariance matrices, we note that these matrices are all positive semi-definite (cf. Eqs.~(\ref{ch2:EnSRF_propagation}), (\ref{ch3:SUKF_propagation}), (\ref{ch3:SUKF_filtering}), (\ref{ch4:DDF_propagation}), and (\ref{ch4:DDF_filtering})). Therefore, one may in general follow the scheme in \S~\ref{ch1:SRKF} and use the spectral decompositions to compute the square roots.

Finally, we note that the inverse of the matrix $\mathbf{S}_{k}^{h} \left( \mathbf{S}_{k}^{h} \right)^T + \mathbf{R}_k$ (or $\hat{\mathbf{P}}^{pr}_k + \mathbf{R}_k$) is involved when computing the Kalman gain in the nonlinear Kalman filters (cf. Eqs.~(\ref{ch2:stochastic_EnKF_filtering}),   
(\ref{ch2:EnSRF_filtering}), (\ref{ch3:SUKF_propagation}), and (\ref{ch4:DDF_propagation})). The matrix $\mathbf{S}_{k}^{h} \left( \mathbf{S}_{k}^{h} \right)^T + \mathbf{R}_k$ (or $\hat{\mathbf{P}}^{pr}_k + \mathbf{R}_k$) is normally positive definite, since this is often the case for the covariance matrix $\mathbf{R}_k$ of observation noise. For this reason, we also use the spectral decomposition to compute the inverse of $\mathbf{S}_{k}^{h} \left( \mathbf{S}_{k}^{h} \right)^T + \mathbf{R}_k$ (or $\hat{\mathbf{P}}^{pr}_k + \mathbf{R}_k$). This is based on the following fact: if $\mathbf{C}$ is a symmetric, positive definite matrix, so that by spectral decomposition we have $\mathbf{C} = \mathbf{E}^c \mathbf{D}^c (\mathbf{E}^c)^T$, where $\mathbf{E}^c$ is the matrix consisting of the eigenvectors of $\mathbf{C}$, and $\mathbf{D}^c$ is a diagonal matrix consisting of the corresponding positive eigenvalues of $\mathbf{C}$, then we have the inverse $\mathbf{C}^{-1} = \mathbf{E}^c (\mathbf{D}^c)^{-1} (\mathbf{E}^c)^T$ \cite[\S~5.5.4]{Golub-matrix}. 

\subsection{Numerical results}\label{ch2:ex_numerical_results}
Now we examine the performances of the stochastic EnKF and the ETKF through some numerical experiments.

\subsubsection{Effects of the inflation factor $\delta$ and the length scale $l_c$ on the performances of the filters}
In this experiment, we aim to examine the relative rms errors and rms ratios of the stochastic EnKF and the ETKF as functions of the covariance inflation factor $\delta$ and covariance filtering length scale $\l_c$. To this end, we let the ensemble size $n=10$, which represents the typical situation in ensemble forecasting, where the ensemble size is normally lower than the dimension of the dynamical system. Since there is no systematic method to select the optimal values of $\delta$ and $\l_c$, we choose to examine certain ranges of these two parameters. For $\delta$, we let its value start from $0$ and increase by $0.5$ each run until its value reaches $10$. For convenience, we adopt the notation $0:0.5:10$ to denote this setting. For $l_c$, we let it increase from $10$ to $400$, with a fixed increment of $20$ each run. This setting is accordingly denoted by $10:20:400$. Similar notations will be frequently adopted in subsequent chapters. Since in this and the subsequent chapters, the numerical experiments involve intense computations at different values of various intrinsic filter parameters, we choose to run the experiment once for each set of the filter parameters due to the limitation of computational resources\footnote{In practice, the typical scenario is that one has fixed observations and the freedom to choose the background ensemble. Since the ETKF is deterministic, the randomness only lies in the choice of the initial background ensemble, whose effect will be diluted as time moves on, especially with the covariance inflation technique. Similar arguments can be applied to the nonlinear Kalman filters to be introduced in Chapters~\ref{ch3:ukf} and \ref{ch4:ddfs}.}. 

\begin{figure*}[!t]
\centering
\hspace*{-0.5in} \includegraphics[width=1.15\textwidth]{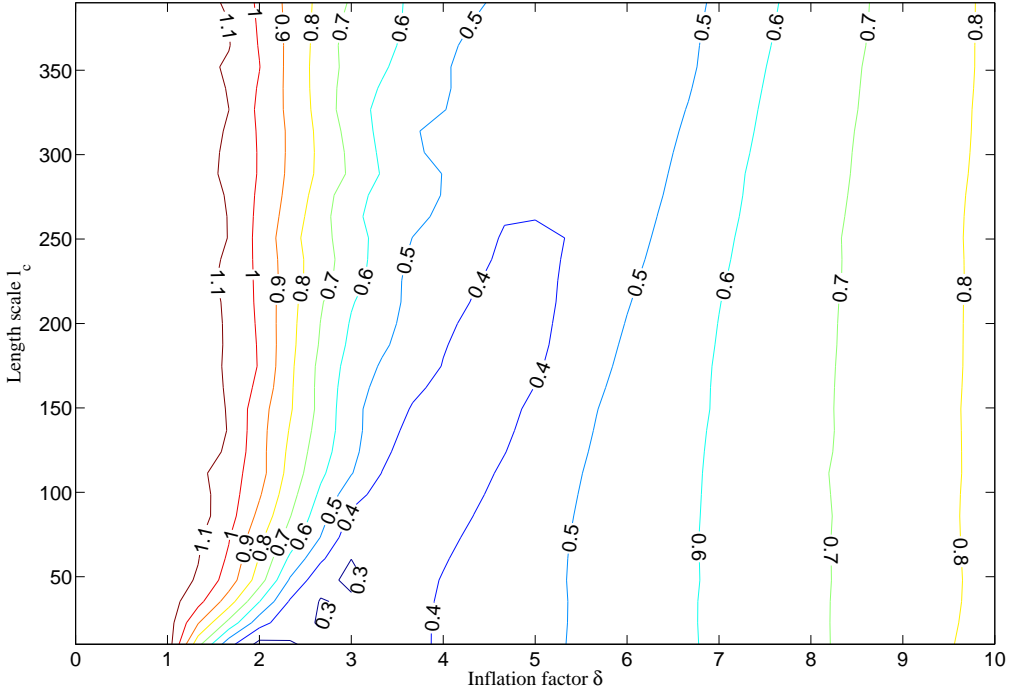} 
\caption{ \label{fig:ch2_stochastic_EnKF_delta_vs_lc_rms} The relative rmse of the stochastic EnKF as a function of the inflation factor $\delta$ and the length scale $l_c$. }
\end{figure*} 

\begin{figure*}[!t]
\centering
\hspace*{-0.5in} \includegraphics[width=1.15\textwidth]{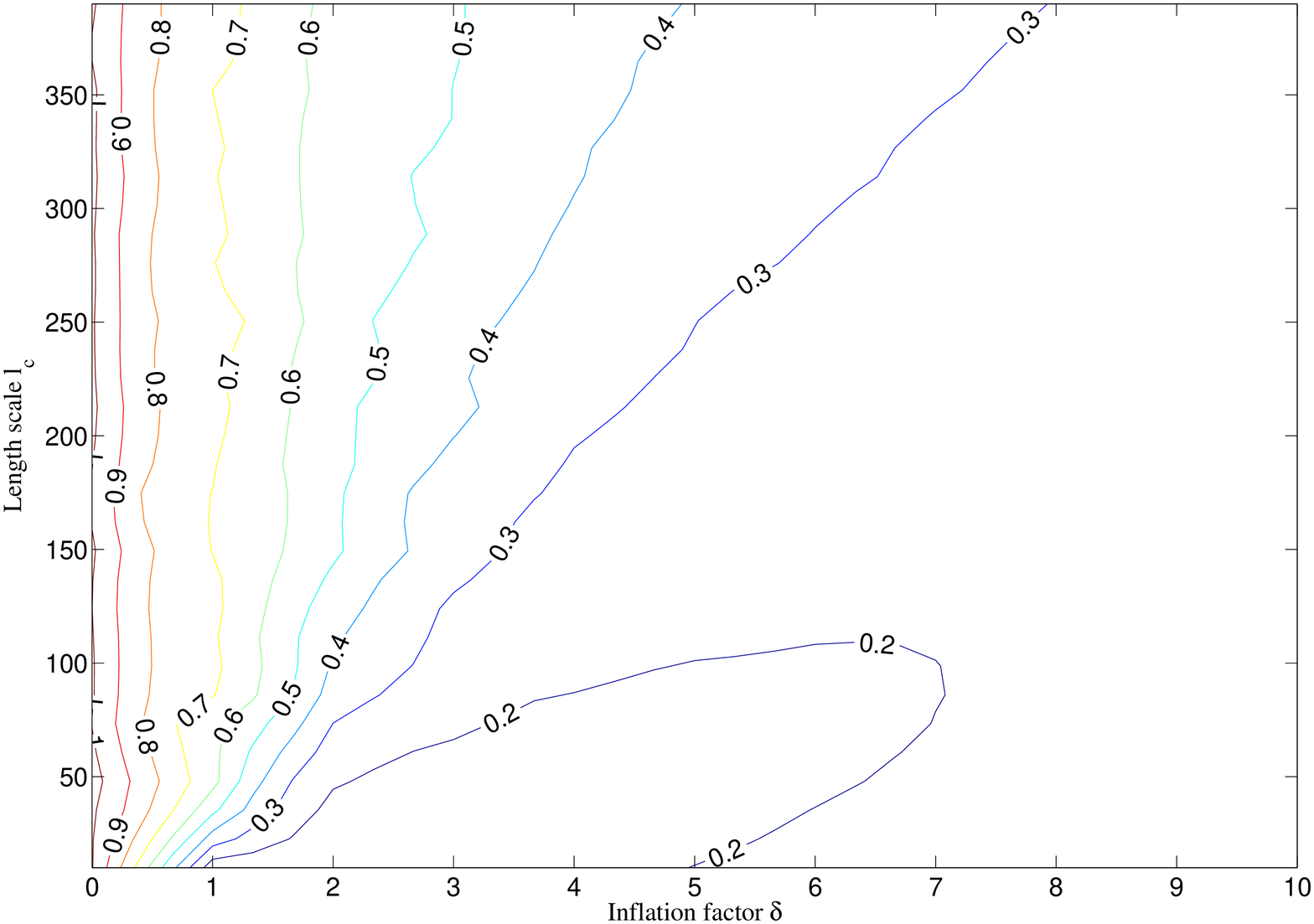} 
\caption{ \label{fig:ch2_ETKF_delta_vs_lc_rms} The relative rmse of the ETKF as a function of the inflation factor $\delta$ and the length scale $l_c$. }
\end{figure*} 

First, we plot the relative rms errors of the stochastic EnKF and the ETKF in Figs. \ref{fig:ch2_stochastic_EnKF_delta_vs_lc_rms} and \ref{fig:ch2_ETKF_delta_vs_lc_rms}, respectively. As one can see there, covariance inflation and filtering can both improve the performances of the filters given suitable values of $\delta$ and $l_c$. 

Indeed, in Figs. \ref{fig:ch2_stochastic_EnKF_delta_vs_lc_rms} and \ref{fig:ch2_ETKF_delta_vs_lc_rms}, within the ranges of the parameters tested, when fixing $\delta$, if $\delta$ is not too large (say $\delta <3$), a smaller length scale $l_c$ tends to yield lower relative rms errors for both the stochastic EnKF and the ETKF. In fact, the lowest relative rms errors of both the stochastic EnKF and the ETKF are achieved with $l_c < 110$.

On the other hand, when fixing $l_c$, the relative rmse of the stochastic EnKF exhibits a U-turn behaviour (in terms of the relative rmse) as $\delta$ increases: when $\delta$ increases from $0$, the relative rmse tends to decrease. But after $\delta$ becomes larger than a certain value, further increasing $\delta$ will instead cause a larger relative rmse. Following \cite{Luo-sgkf}, we explain the U-turn phenomenon as follows: when there is no covariance inflation, i.e., $\delta=0$, it can be shown that the error covariance of the EnKF is systematically underestimated \cite{Whitaker-ensemble}. This implies that we are over-confident about the background. Consequently, the analysis to be updated will rely too much on the background, which may cause a relatively large relative rmse, since the information content from the incoming observation will possibly be underrepresented. On the other hand, increasing $\delta$ will make the error covariance of the background become larger. This implies that we are more uncertain about the background. Thus if $\delta$ gets too large, the analysis to be updated will rely too much on the incoming observation, which may also cause a relatively large relative rmse, since the information content from our prior knowledge (the background) will possibly be underrepresented. In contrast, a moderate inflation factor $\delta$, as a trade-off between being too large and too small, will instead get a lower relative rmse. For the ETKF, one can also find the U-turn behaviour in Fig.~\ref{fig:ch2_ETKF_delta_vs_lc_rms}, which can be explained in a similar way.

A comparison between Figs.~\ref{fig:ch2_stochastic_EnKF_delta_vs_lc_rms} and \ref{fig:ch2_ETKF_delta_vs_lc_rms} reveals that, given the same $\delta$ and $l_c$, the relative rmse of the ETKF is always lower than that of the stochastic EnKF. This is consistent with the result reported in \cite{Whitaker-ensemble}. In fact, the relative rmse (i.e., noise level) of the observations in our experiment is around $0.22$. Thus from Fig.~\ref{fig:ch2_stochastic_EnKF_delta_vs_lc_rms}, one can see that the stochastic EnKF is always divergent within the ranges of the parameters tested, since its relative rmse is larger than $0.22$ everywhere. In contrast, from Fig.~\ref{fig:ch2_ETKF_delta_vs_lc_rms}, one can see that there exist some areas, for example, the one surrounded by the the horizontal axis and the contour level curve marked by the value of $0.2$, where the ETKF is not divergent in the sense that its relative rmse is no larger than $0.2$.    

\begin{figure*}[!t]
\centering
\hspace*{-0.5in} \includegraphics[width=1.15\textwidth]{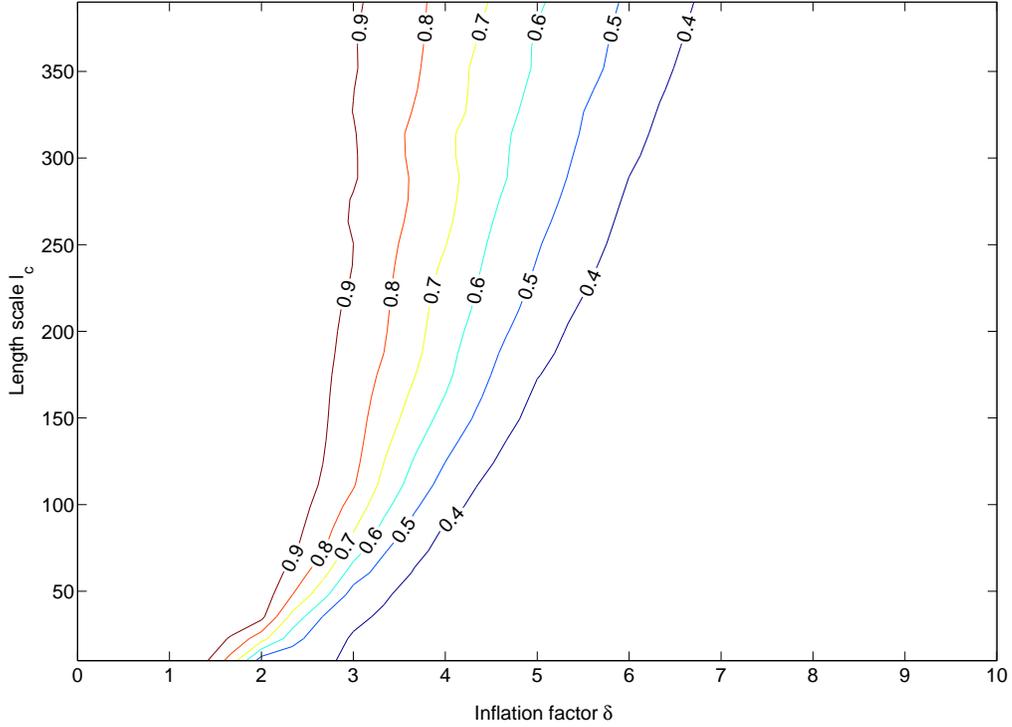} 
\caption{ \label{fig:ch2_stochastic_EnKF_delta_vs_lc_ratio} The rms ratio of the stochastic EnKF as a function of the inflation factor $\delta$ and the length scale $l_c$. }
\end{figure*} 

\begin{figure*}[!t]
\centering
\hspace*{-0.5in} \includegraphics[width=1.15\textwidth]{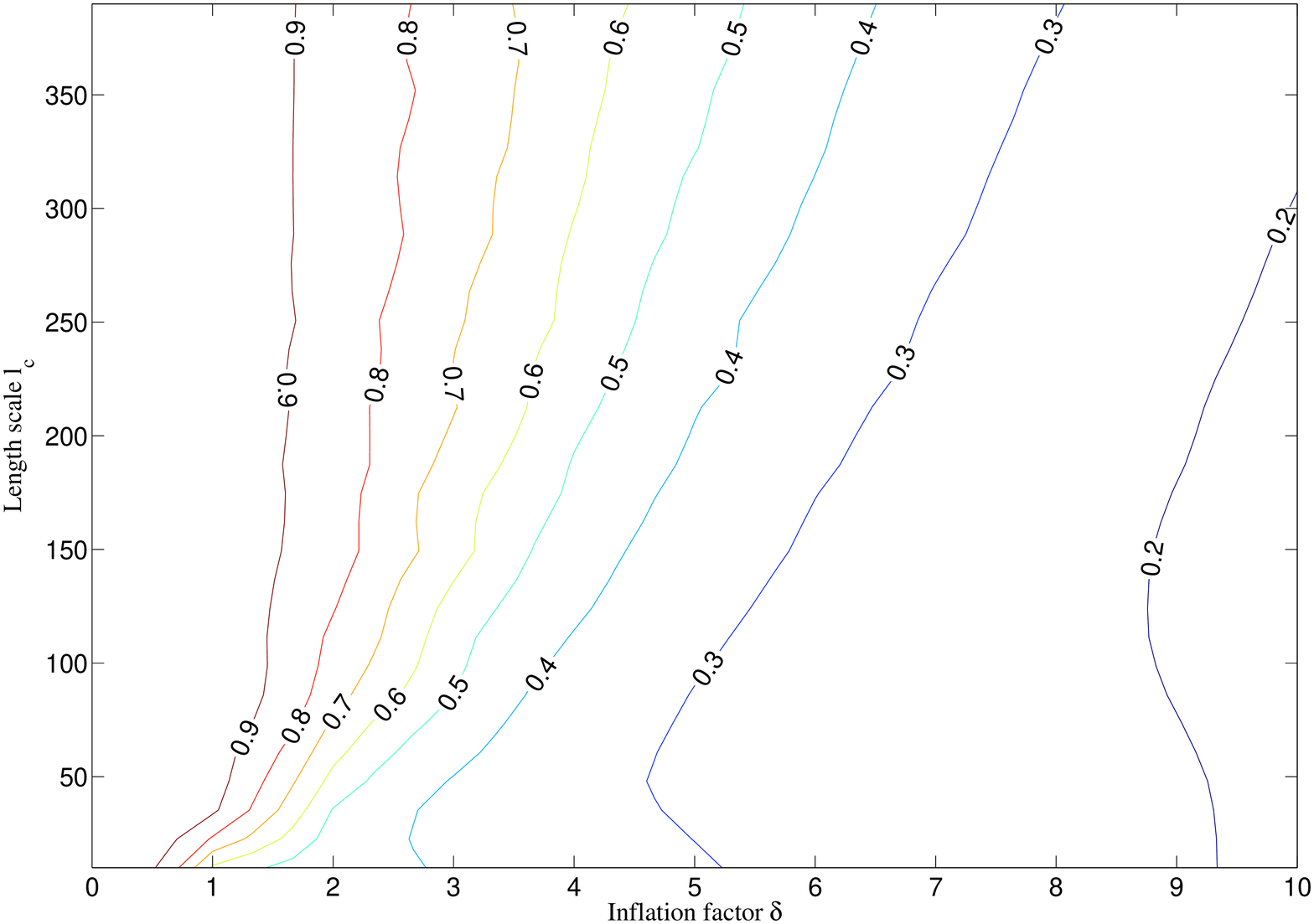} 
\caption{ \label{fig:ch2_ETKF_delta_vs_lc_ratio} The rms ratio of the ETKF as a function of the inflation factor $\delta$ and the length scale $l_c$. }
\end{figure*} 
 
Next, we plot the rms ratios of the stochastic EnKF and the ETKF in Figs.~\ref{fig:ch2_stochastic_EnKF_delta_vs_lc_ratio} and \ref{fig:ch2_ETKF_delta_vs_lc_ratio}, respectively. From them, one can see that, when fixing $l_c$, the rms ratio is a monotonically decreasing function of $\delta$, while when fixing $\delta$, the rmse ratio is roughly a monotonically increasing function of $l_c$, with some violations in the ETKF. For example, if we fix $\delta=5$ in Fig.~\ref{fig:ch2_ETKF_delta_vs_lc_ratio}, the rms ratio will decrease as $l_c$ increases from $10$ to $50$, and then start to increase after $l_c$ is over $50$. Since the ensemble size $n=10$, we have the expectation of the rms ratio $R_e \approx 0.74$ according to Eq.~(\ref{R_e}). Therefore, if the analysis ensemble is statistically indistinguishable from the truth, the rms ratio $R$ should be close to $R_e \approx 0.74$. In this sense, both the stochastic EnKF and the ETKF can generate analysis ensembles that are indistinguishable from the corresponding truths, provided that $\delta$ and $l_c$ are taken within the strips between the ratio values of $0.7$ and $0.8$ such that $R \approx 0.74$. However, in order to obtain better performances in terms of the relative rms errors, both the stochastic EnKF and the ETKF should take $\delta$ and $l_c$ outside of the aforementioned strips, so that the corresponding rms ratios are lower than the expectation $0.74$. According to the discussion in \S~\ref{ch2:two_measures} (also cf. \cite{Anderson-ensemble, Whitaker-ensemble}), this implies that the error covariances of the analysis ensembles are over-estimated. Thus these experiment results confirm the benefit of conducting covariance inflation. 

\subsubsection{Effect of the ensemble size on the performances of the filters}

Now we examine the relative rms errors and rms ratios of the stochastic EnKF and the ETKF as functions of the ensemble size $n$. To this end, we let $\delta=3$
and $l_c=50$ for both the stochastic EnKF and the ETKF. The ensemble size $n$ increases from $5$ to $40$ with a fixed increment of $1$ each run. This setting is denoted by $5:1:40$.

\begin{figure*}[!t]
\centering
\hspace*{-0.5in} \includegraphics[width=1.15\textwidth]{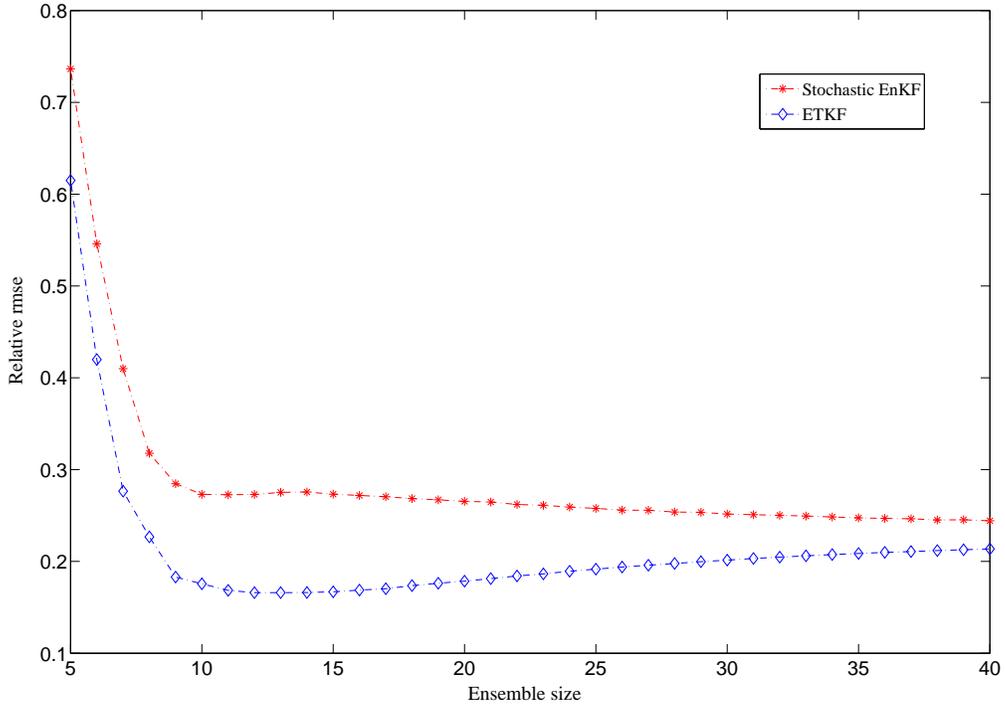} 
\caption{ \label{fig:ch2_ETKF_sEnKF_nEn_rms} The relative rms errors of the stochastic EnKF and the ETKF as functions of the ensemble size $n$. }
\end{figure*} 
 
We plot the relative rms errors of the stochastic EnKF and the ETKF in Fig.~\ref{fig:ch2_ETKF_sEnKF_nEn_rms}. For the stochastic EnKF, the relative rmse drops rapidly when $n$ increases from $5$ to $10$. After that, further increasing $n$ does not reduce the relative rmse significantly. 
For the ETKF, the situation is slightly different. The relative rmse also drops rapidly when $n$ starts from $n=5$. However, as $n$ increases, the relative rmse exhibits the U-turn behaviour, with the lowest relative rmse attained at $n=13$. For $n>13$, the ETKF with a smaller ensemble size (say $n=20$) performs better than the ETKF with a larger one (say $n=40$). A possible explanation of this phenomenon is postponed to \S~\ref{ch3:sec_experiments_threshold_bounds} in the next chapter, since we feel this phenomenon is better explained there.

Comparing the curves in Fig.~\ref{fig:ch2_ETKF_sEnKF_nEn_rms}, one can see that, with the same ensemble size $n$, the ETKF always outperforms the stochastic EnKF. The stochastic EnKF is divergent for all the ensemble sizes tested, in the sense that the corresponding rms errors are always larger than $0.22$ (the relative rmse in the observations). In contrast, the ETKF is not-divergent for $n \ge 9$.  

\begin{figure*}[!t]
\centering
\hspace*{-0.5in} \includegraphics[width=1.15\textwidth]{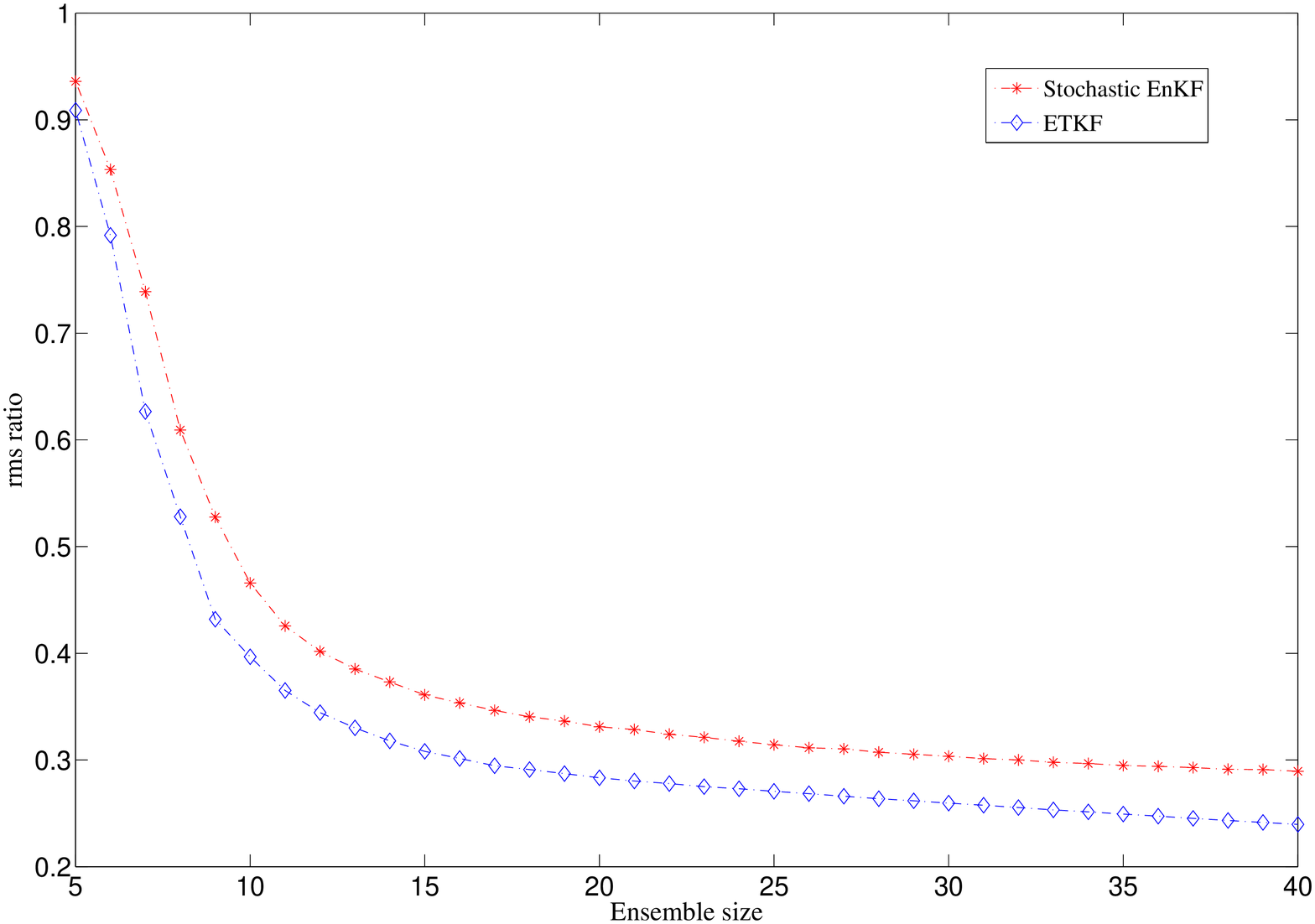} 
\caption{ \label{fig:ch2_ETKF_sEnKF_nEn_ratio} The rms ratios of the stochastic EnKF and the ETKF as functions of the ensemble size $n$. }
\end{figure*} 

We also plot the time-averaged rms ratios of the stochastic EnKF and the ETKF in Fig.~\ref{fig:ch2_ETKF_sEnKF_nEn_ratio}. For both the filters, their rms ratios decrease monotonically as the ensemble sizes $n$ increase. To make the rms ratios close to $0.74$, the ensemble sizes of both the filters should be less than $8$. Increasing $n$ will lead to smaller rms ratios and thus cause over-estimations of the error covariances. This, however, will benefit the performances of the filters in the sense that the corresponding relative rms errors are lower.   

\section{Summary of the chapter}

In this chapter we considered the data assimilation problem in nonlinear (discrete) systems, where both the dynamical and observation noise are assumed to follow some Gaussian distributions. We used recursive Bayesian estimation (RBE) to derive an approximate Monte Carlo solution to the data assimilation problem, which turned out to be consistent with the ensemble Kalman filter (EnKF) in the literature. We noted that for the validity of the deduction, apart from the Gaussianity assumptions for both the dynamical and observation noise, it is also necessary to assume that the system states also follow some Gaussian distributions.  

Depending on whether to perturb the observations or not, the EnKF can be classified as two types: the stochastic EnKF and the ensemble square root filter (EnSRF). Our numerical results showed that the ensemble transform Kalman filter (ETKF), as a representative of the EnSRFs, consistently outperformed the stochastic EnKF.

In order to improve the performance of the EnKF, we introduced two auxiliary techniques, namely covariance inflation and filtering. We also discussed the connection between the covariance inflation technique and the Kalman filter with fading memory. Through some numerical experiments, we illustrated the benefits of adopting these two auxiliary techniques in the EnKF.

\chapter{Unscented and scaled unscented Kalman filters for data assimilation} \label{ch3:ukf}

\section{Overview} \label{ch3:overview}
The conventional Kalman filter (KF) is simple but general for linear/Gaussian systems. However, when it comes to nonlinear or non-Gaussian systems, its optimality is often lost. Since the appearance of the KF, lots of works were dedicated to extending the KF to nonlinear and/or non-Gaussian systems. In fact, apart from the ensemble Kalman filter (EnKF) introduced in the previous chapter, there are some other types of extensions, for example, the extended Kalman filter (EKF) and its variants, the iterated EKF and higher order EKF \cite[ch. 13]{Simon2006}, the \sindex{unscented Kalman filter} (UKF) \cite{Julier2000} and its generalization, the \sindex{scaled unscented Kalman filter} (SUKF) \cite{Julier2004}, and the \sindex{divided difference filters} (DDFs) \cite{Ito-gaussian,Noergaard-Advances,Norgaard--new}. All these filters are intimately related to the conventional KF. For this reason, we call them nonlinear Kalman filters in this dissertation. Note that these nonlinear Kalman filters are not designed for the data assimilation problem in nonlinear/non-Gaussian systems. Instead, they yield approximate solutions for nonlinear/Gaussian systems. Here by ``nonlinear/Gaussian'', we mean that not only are the dynamical and observation noise Gaussian, but also the underlying system states, as we have pointed out in \S~\ref{ch2:sec_ps}. 

Similar to the derivation of the EnKF in \S~\ref{ch2:sec_ps}, one may also derive other nonlinear KFs from the point of view of recursive Bayesian estimation (RBE). As a result, one can also split the procedures of a nonlinear KF into the propagation (or prediction) step and the filtering step. For all of the nonlinear KFs, the main operations at the filtering step are the same, which update the mean and covariance of the background to the corresponding statistics of the analysis, in the same way as the conventional Kalman filter does. Thus in general, it is the approach to dealing with the nonlinearity at the propagation step that distinguishes different types of nonlinear KFs. 

\begin{figure*}[!t]
\centering
\begin{tikzpicture}[auto,node distance=1.8cm,>=latex']
        \tikzstyle{every node}=
        [%
          fill=green!50!black!20,%
          draw=green!50!black,%
          minimum size=8mm,%
          thick%
        ]

         \path (1,8) node[draw, shape=rectangle, scale=1.5] (a) { \text{\centering $\mathcal{F}$}};
         \node (b) [draw, shape=circle, scale=1, left of=a, node distance=2.5cm, text width = 1cm, text centered] {
                    \text{$\mathbf{x}$}};
         \node (c) [draw, shape=circle, scale=1, right of=a, node distance=2.5cm, text width = 1cm, text centered]
                    {\text{$\mathbf{\eta}$}};

         \path [thick,shorten >=1pt,-stealth']
                        (b) edge (a)
                        (a) edge (c);	  
\end{tikzpicture}
\caption{\label{ch3:fig_problem_recast} The recast problem at the propagation step of a nonlinear KF.}
\end{figure*}
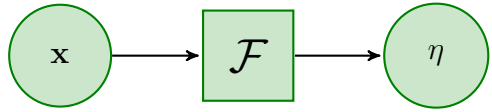

Note that under the assumption of Gaussianity, in order to estimate the pdf of a Gaussian distribution, it is sufficient to estimate its mean and covariance. Thus for a nonlinear Kalman filter, the pdf approximation problem at the propagation step Eq.~(\ref{BRR:prediction}) of RBE is equivalent to the problem in estimating the mean and covariance of the background, which itself can be recast as the estimation problem in the following scenario: as shown in Fig.~\ref{ch3:fig_problem_recast}, we suppose that there is a Gaussian random variable $\mathbf{x}$ with mean $\bar{\mathbf{x}}$ and covariance $\mathbf{P}_x$, which is transformed by a nonlinear function $\mathcal{F}$ into another random variable $\mathbf{\eta}$, so that $\mathbf{\eta} = \mathcal{F} \left( \mathbf{x} \right)$\footnote{The more general scenario, where $\mathbf{\eta} = \mathcal{F} \left( \mathbf{x}, \mathbf{u} \right)$, with $\mathbf{u}$ being Gaussian noise, is discussed in Appendix~\ref{appendix:accuracy analysis}.}. Our objective is to estimate the mean $\bar{\mathbf{\eta}}$ and covariance $\mathbf{P}_{\eta}$ of the transformed random variable $\mathbf{\eta}$.  

To solve the recast problem, the idea of the EnKF, as introduced in the previous chapter, is to generate some samples of system states and propagate them forward. Then the mean $\bar{\mathbf{\eta}}$ and covariance $\mathbf{P}_{\eta}$ of the transformed random variable $\mathbf{\eta}$ are estimated as the sample mean and sample covariance of the propagations.   

Apart from the EnKF, there are a few other methods to tackle the recast problem. One such method is the (first order) extended Kalman filter (EKF) \footnote{For convenience, hereafter whenever we say extended Kalman filter, we mean the first order approximation by default.}. The idea of the EKF is to expand the nonlinear function $\mathcal{F}$ around the mean $\bar{\mathbf{x}}$ up to first order in a Taylor series expansion. For example, let $\mathbf{x} = \bar{\mathbf{x}} + \delta \mathbf{x}$, where $\delta \mathbf{x}$ represents a small perturbation, then   
\begin{equation} \label{ch3:overview_Taylor_expansion}
\mathcal{F} \left( \bar{\mathbf{x}} + \delta \mathbf{x} \right) = \mathcal{F} \left( \bar{\mathbf{x}}\right) + \mathbf{F}|_{\bar{\mathbf{x}}} \, \delta \mathbf{x} + o \left( \delta \mathbf{x} \right) \, ,
\end{equation}
where $\mathbf{F}|_{\bar{\mathbf{x}}}$ denotes the Jacobian matrix of $\mathcal{F}$ evaluated at $\bar{\mathbf{x}}$, and $o \left( \delta \mathbf{x} \right)$ represents the higher order terms in the expansion. Thus if $\delta \mathbf{x}$ is sufficiently small, or if the system under assimilation is weakly nonlinear, so that higher order derivatives of the function $\mathcal{F}$ are relatively small compared with the Jacobian $\mathbf{F}|_{\bar{\mathbf{x}}}$, then $o \left( \delta \mathbf{x} \right)$ can be neglected in computation. To this end, let $\delta \mathbf{\eta} = \mathcal{F} \left( \bar{\mathbf{x}} + \delta \mathbf{x} \right) - \mathcal{F} \left( \bar{\mathbf{x}}\right)$, then the nonlinear system in Eq.~(\ref{ch3:overview_Taylor_expansion}) is approximated by a linear one 
\begin{equation} \label{ch3:overview_Taylor_approx_1st_order}
\delta \mathbf{\eta} \approx \mathbf{F}|_{\bar{\mathbf{x}}} \, \delta \mathbf{x} 
\end{equation}
by neglecting $o \left( \delta \mathbf{x} \right)$. Thus the conventional Kalman filter introduced in Chapter~\ref{ch1: KF linear} can be used to assimilate the approximate system Eq.~(\ref{ch3:overview_Taylor_approx_1st_order}). 

In order to implement the EKF (or its higher order variants, see \cite[Ch.~13]{Simon2006}), one has to evaluate the derivative(s), e.g., Jacobian or even Hessian, of the nonlinear function $\mathcal{F}$, which is often inconvenient in implementation. For this reason, we will not investigate the performance of the EKF in this dissertation \footnote{Another reason is that, it can been shown analytically that the unscented Kalman filter to be introduced later systematically outperforms the EKF \cite{Julier2000}. The higher order variants of the EKF may perform better than the EKF itself. However, the complication in evaluating higher order derivatives prevents their spread in practice.}. Instead, we will introduce two other types of nonlinear KFs developed in recent years, namely the unscented Kalman filter (UKF) \cite{Julier2000} and its generalization, the scaled unscented Kalman filter (SUKF) \cite{Julier2004}, and the divided difference filters (DDFs) \cite{Ito-gaussian,Noergaard-Advances,Norgaard--new}, for the data assimilation problem in nonlinear/Gaussian systems. One advantage of these filters is that they avoid the necessity of evaluating the derivatives of a nonlinear function. Instead, they all produce some specially chosen system states, called \sindex{sigma points}, for the purpose of approximation. For this reason, they are uniformly called \sindex{sigma point Kalman filters} (SPKFs) \cite{Merwe2004} or derivative-free filters \cite{Simandl2006}.

More details of the UKF, the SUKF and the DDFs will be presented in this and the next chapters. As a summary, we provide in Table~\ref{ch3:table_classification_nKFs} brief descriptions of how some of the nonlinear KFs handle the nonlinearity at the propagation step.   
\begin{table*}[!t]
	\centering
	\caption{\label{ch3:table_classification_nKFs} Different ways of the nonlinear Kalman filters in dealing with 	 		the nonlinearity at the propagation step.}
	\begin{tabular}{p{7cm}p{7cm}}
        \hline \hline
        Filter & Idea\\
        \hline
		Extended Kalman filter  & Linearizing the nonlinear function $\mathcal{F}$  \\
        \hline
        Ensemble Kalman filter  & Taking average over the propagations of ensemble members (see 	 		Ch.~\ref{ch2: EnKF}) \\
		\hline
        Unscented Kalman filter; Scaled unscented Kalman filter & Taking weighted average over the
		propagations of sigma points (similar to the ensemble Kalman filter) \\
		\hline
        Divided difference filters  & Interpolating the nonlinear function $\mathcal{F}$ by Stirling's 
		interpolation formula (similar to the extended Kalman filter)\\
        \hline \hline
     \end{tabular}
\end{table*} 

The remainder of this chapter is organized as follows: although the problem in study is the same as that in \S~\ref{ch2:sec_ps}, we choose to re-state it in \S~\ref{ch3:sec_ps} for completeness. In \S~\ref{ch3:sec_ut}, we introduce the unscented transform (UT) as the approximate solution to the recast problem in Fig.~\ref{ch3:fig_problem_recast}. We then proceed to introduce the scaled unscented transform (SUT) in \S~\ref{ch3:sec_sut} as the generalization of the UT. Applying the SUT to the propagation step of RBE leads to the scaled unscented Kalman filter (SUKF), as will be seen in \S~\ref{ch3:sec_UKF}. To apply the SUKF to assimilate high dimensional systems, we introduce the reduced rank SUKF in \S~\ref{ch3:sec_reduced_sukf}. In \S~\ref{ch3:sec_ex} we use the $40$-dimensional Lorenz-Emanuel system as the testbed to illustrate the details in implementing the reduced rank SUKF, and to investigate the effects of the intrinsic filter parameters on the performance of the SUKF. We draw our conclusion for this chapter in \S~\ref{ch3:sec_summary}.

\section{Problem statement} \label{ch3:sec_ps}
We are interested in the data assimilation problem in the same family of nonlinear/Gaussian systems as described in Eq.~(\ref{ch2:ps}), i.e.,
\begin{subequations}  
\begin{align}
 \tagref{ch2:ps_dyanmical_system} & \mathbf{x}_k  = \mathcal{M}_{k,k-1} \left( \mathbf{x}_{k-1} \right) + \mathbf{u}_{k}  \, ,  \\
  \tagref{ch2:ps_observation_system} &  \mathbf{y}_k  = \mathcal{H}_{k} \left( \mathbf{x}_{k} \right) + \mathbf{v}_{k} \, , \\
  \tagref{ch2:ps_dyanmical_noise} & \mathbf{u}_{k} \sim N \left(\mathbf{u}_{k}: \mathbf{0}, \mathbf{Q}_k \right) \, ,\\
 \tagref{ch2:ps_observation_noise} & \mathbf{v}_{k} \sim N \left(\mathbf{v}_{k}: \mathbf{0}, \mathbf{R}_k \right) \, ,\\
  \tagref{ch2:ps_dyanmical_white_noise}& \mathbb{E} \left( \mathbf{u}_{j} \mathbf{u}_{k}^T \right) = \delta_{k,j}\mathbf{Q}_k \, ,\\
  \tagref{ch2:ps_observation_white_noise}& \mathbb{E} \left( \mathbf{v}_{j} \mathbf{v}_{k}^T \right) = \delta_{k,j}\mathbf{R}_k \, ,\\
  \tagref{ch2:ps_uncorrelated_noise}& \mathbb{E} \left( \mathbf{u}_{i} \mathbf{v}_{j}^T \right) = 0 \quad \forall \, i, \, j \, .
\end{align}
\end{subequations}

The approximate solutions to the above problem, in terms of the sigma point Kalman filters (SPKFs), are given in this chapter (for the UKF and the SUKF), and the next chapter (for the DDFs), respectively. As we have pointed out previously, the nonlinear KFs differ from each other mainly at the propagation step, where the problem can be recast as the estimation problem in Fig.~\ref{ch3:fig_problem_recast}. Therefore, in this chapter, we first discuss how to solve the recast problem through the unscented transform (UT) and its generalization, the scaled unscented transform (SUT). Incorporating the UT and the SUT into the propagation step of RBE leads to the unscented Kalman filter (UKF) and the scaled unscented Kalman filter (SUKF), respectively. 

\section{Unscented transform} \label{ch3:sec_ut}
\subsection{Basic idea } \label{ch3:sec_idea_of_ut}
The idea of the unscented transform (UT) is based on the intuition that ``it is easier to approximate a probability distribution than it is to approximate an arbitrary nonlinear function or transformation'' \cite{Julier2000}. To see this, we use a continuous nonlinear transform $\mathbf{\eta} = \mathcal{F} \left( \mathbf{x} \right)$ for illustration. Suppose that $\mathbf{x}$ is an $m$-dimensional random variable (not necessarily Gaussian), with mean $\bar{\mathbf{x}}$ and covariance $\mathbf{P}_x$, and that $\mathbf{S}_x$ is an $m \times L$ square root of $\mathbf{P}_x$ such that $\mathbf{P}_x = \mathbf{S}_x \left( \mathbf{S}_x \right)^T$. We generate a set of $2L+1$ specially chosen states $\{ \mathcal{X}_i \}_{i=0}^{2L}$, called \sindex{sigma points}, with respect to the triplet $\left(\lambda, \bar{\mathbf{x}}, \mathbf{S}_x \right)$, according to the following formula:
\begin{equation} \label{ch3:ut_sigma_points}
\begin{split}
& \mathcal{X}_0 = \bar{\mathbf{x}},\\
& \mathcal{X}_i = \bar{\mathbf{x}} + \sqrt{L+\lambda} \left(  \mathbf{S}_x \right)_i, \, i=1, 2, \dotsb, L ,\\
& \mathcal{X}_i = \bar{\mathbf{x}} - \sqrt{L+\lambda} \left(  \mathbf{S}_x \right)_{i-L}, \, i=L+1, L+2, \dotsb, 2L ,\\
\end{split}
\end{equation}
where $\left(\mathbf{S}_x \right)_i$ denotes the $i$-th column of the square root matrix $\mathbf{S}_x$, and $\lambda$ is an adjustable parameter satisfying some constraints (see \S~\ref{ch3:sec_ex_implementation_issue}). The reason to introduce $\lambda$ is that it can influence higher order moments (e.g., kurtosis) of the set $\{ \mathcal{X}_i \}_{i=0}^{2L}$. Thus one may use this flexibility to reduce the approximation error of sigma points in higher order moment matching \cite{Julier2000}, as will be shown later. 

We also allocate a set of weights $\{ W_i \}_{i=0}^{2L}$
\begin{equation} \label{ch3:ut_weights}
\begin{split}
&W_0 = \frac{\lambda}{L+\lambda},\\
&W_i = \frac{1}{2\left(L+\lambda  \right)}, \, i=1,2,\dotsb, 2L, \\
\end{split}
\end{equation}
to sigma points $\{ \mathcal{X}_i \}_{i=0}^{2L}$. In this way, it can be verified that the weighted sample mean and sample covariance of the set $\{ \mathcal{X}_i \}_{i=0}^{2L}$, given by  
\begin{equation}  \label{ch3:ut_weighted_stat}
\begin{split}
&\hat{\mathcal{X}} = \sum\limits_{i=0}^{2L} W_i \mathcal{X}_i \, , \\
& \hat{\mathbf{P}}_{\mathcal{X}} = \sum\limits_{i=0}^{2L} W_i \left(\mathcal{X}_i-\hat{\mathcal{X}} \right) \left(\mathcal{X}_i-\hat{\mathcal{X}} \right)^T \, , \\
\end{split}
\end{equation}
match the mean $\bar{\mathbf{x}}$ and covariance $\mathbf{P}_{x}$ of the random variable $\mathbf{x}$, respectively. If $\mathbf{x}$ follows a Gaussian distribution, then it is suggested that $\lambda$ be chosen as $\lambda =3 - L$, so that the kurtoses of $\{ \mathcal{X}_i \}_{i=0}^{2L}$ can match as many as possible of those of the random variable $\mathbf{x}$ \cite{Julier2000,Julier2004}.  

Because of the symmetry in sigma points, the rank of the matrix $\hat{\mathbf{P}}_{\mathcal{X}}$ is $L$. To avoid rank deficiency in $\hat{\mathbf{P}}_{\mathcal{X}}$, it is suggested that the number of sigma points be larger than twice the dimension of the vector $\mathbf{x}$, or equivalently, $L \ge m$ \cite{Julier2000}. However, for high dimensional systems, this restriction should be relaxed in order to reduce the computational cost, as will be discussed later.

The weights $\{ W_i \}_{i=0}^{2L}$ satisfy the normalization condition $\sum\limits_{i=0}^{2L} W_i =1$. However, they might be inconsistent with the conventional interpretation of the weights of samples of a distribution. This is because $\lambda$ can be negative, so that $W_0$ is also negative. The negativeness of $W_0$ also causes another problem, in that the sample covariance $\hat{\mathbf{P}}_{\eta}$ (cf. Eq.~(\ref{ch3:ut_transformed_cov})) of the transformed sigma points may not always be positive semi-definite. 

One remedy to the above problem is to simply let $\lambda$ always be non-negative. In particular, by letting $\lambda = 0$ such that $W_0=0$, one in effect propagates (or transforms) sigma points $\{ \mathcal{X}_i \}_{i=0}^{2L}$ forward except for the center $\mathcal{X}_0$. This scheme, i.e., excluding the center $\mathcal{X}_0$ of sigma points, is known as \sindex{positive-negative pairs} (PNP) in the literature of data assimilation (cf. \cite{Wang-which} and the reference therein). In this sense, the PNP scheme can be deemed as a special case of the UT. In both the UT and PNP schemes, sigma points, with or without the center, will have the same sample mean $\hat{\mathcal{X}}$ and sample covariance $\hat{\mathbf{P}}_{\mathcal{X}}$, regardless of the choice of $\lambda$. The advantage of adopting the UT, however, lies in the fact that, in the UT there is an additional parameter $\lambda$, which provides an extra freedom to influence higher order moments (e.g. kurtosis) of sigma points \cite[Appendix \Rmnum{2}]{Julier2000}. Moreover, from Eq.~(\ref{ch3:ut_sigma_points}) one can also see that $\lambda$ also affects the distances of the other sigma points to the center $\mathcal{X}_0$ (but without affecting their sample mean and sample covariance). This may be desirable in some situations, where one wants to explore the dynamics of a nonlinear transform in different scales through an ensemble of system states, but does not wish to change the sample mean and covariance of the ensemble. We will come back to the issue of positive semi-definiteness later and present another remedy following the work \cite{Julier2004}. 

To estimate the mean and covariance of the transformed random variable $\mathbf{\eta}$, we first conduct the nonlinear transform on the set of sigma points $\{ \mathcal{X}_i \}_{i=0}^{2L}$ so as to obtain a set of transformed sigma points $\{ \mathcal{Y}_i:  \mathcal{Y}_i = \mathcal{F} \left( \mathcal{X}_i \right) \}_{i=0}^{2L}$. As the estimations of $\bar{\mathbf{\eta}}$ and $\mathbf{P}_{\eta}$, the sample mean $\hat{\mathbf{\eta}}$ and sample covariance $\hat{\mathbf{P}}_{\eta}$ are given by  
\begin{subequations} \label{ch3:ut_transformed_stat}
\begin{align}
\label{ch3:ut_transformed_mean} &\hat{\mathbf{\eta}} = \sum\limits_{i=0}^{2L} W_i \, \mathcal{Y}_i \, ,  \\
 \label{ch3:ut_transformed_cov} &\hat{\mathbf{P}}_{\eta} =  \sum\limits_{i=0}^{2L} W_i \, \left( \mathcal{Y}_i - \hat{\mathbf{\eta}} \right) \left(  \mathcal{Y}_i - \hat{\mathbf{\eta}} \right ) ^T 
 + \beta \left( \mathcal{Y}_0 - \hat{\mathbf{\eta}} \right) \left(  \mathcal{Y}_0 - \hat{\mathbf{\eta}} \right ) ^T \, ,
\end{align}
\end{subequations}
where the scalar $\beta$ is also an adjustable parameter. In the original work \cite{Julier2000}, the term $\beta \left( \mathcal{Y}_0 - \hat{\mathbf{\eta}} \right) \left(  \mathcal{Y}_0 - \hat{\mathbf{\eta}} \right ) ^T$ does not appear on the right hand side (rhs) of Eq.~(\ref{ch3:ut_transformed_cov}). However, introducing this additional term has the following benefits: firstly, it can reduce the approximation error. For example, it was shown in \cite{Julier2004} that, if $\mathbf{x}$ follows a Gaussian distribution, $\beta=2$ yields a better covariance estimation than $\beta=0$. Secondly, since the weight $W_0$ can be negative, it is not guaranteed that the first term $\sum\limits_{i=0}^{2L} W_i \, \left( \mathcal{Y}_i - \hat{\mathbf{\eta}} \right) \left(  \mathcal{Y}_i - \hat{\mathbf{\eta}} \right ) ^T $ on the rhs of Eq.~(\ref{ch3:ut_transformed_cov}) is positive semi-definite. However, by adding the second term, the effective weight of the transformed sigma point $\mathcal{Y}_0$ in Eq.~(\ref{ch3:ut_transformed_cov}) becomes $W_0 + \beta$. Thus by choosing an appropriate value for $\beta$, we can provide some compensation so that the sample covariance $\hat{\mathbf{P}}_{\eta}$ is guaranteed to be positive semi-definite. Finally, a positive value of $\beta$ increases the error covariance $\hat{\mathbf{P}}_{\eta}$. This is similar to the covariance inflation technique introduced in \S~\ref{ch2:sec_covariance_inflation}. Thus it may improve the performance of a filter given a suitable value of $\beta$. 

For convenience, we summarize the main procedures in the UT as follows:\\
Generation of sigma points:
\begin{equation} \tagref{ch3:ut_sigma_points}
\begin{split}
& \mathcal{X}_0 = \bar{\mathbf{x}},\\
& \mathcal{X}_i = \bar{\mathbf{x}} + \sqrt{L+\lambda} \left(  \mathbf{S}_x \right)_i, \, i=1, 2, \dotsb, L ,\\
& \mathcal{X}_i = \bar{\mathbf{x}} - \sqrt{L+\lambda} \left(  \mathbf{S}_x \right)_{i-L}, \, i=L+1, L+2, \dotsb, 2L .\\
\end{split}
\end{equation}
Allocation of associated weights:
\begin{equation} \tagref{ch3:ut_weights}
\begin{split}
&W_0 = \frac{\lambda}{L+\lambda},\\
&W_i = \frac{1}{2\left(L+\lambda  \right)}, \, i=1,2,\dotsb, 2L. \\
\end{split}
\end{equation}
Estimations of the mean and covariance of the transformed random variable $\mathbf{\eta}$:
\begin{subequations} 
\begin{align}
\nonumber &  \mathcal{Y}_i = \mathcal{F} \left( \mathcal{X}_i \right),  \, i=0,1,\dotsb, 2L, \\
\tagref{ch3:ut_transformed_mean} &\hat{\mathbf{\eta}} = \sum\limits_{i=0}^{2L} W_i \, \mathcal{Y}_i \, ,  \\
\tagref{ch3:ut_transformed_cov} &\hat{\mathbf{P}}_{\eta} =  \sum\limits_{i=0}^{2L} W_i \, \left( \mathcal{Y}_i - \hat{\mathbf{\eta}} \right) \left(  \mathcal{Y}_i - \hat{\mathbf{\eta}} \right ) ^T 
 + \beta \left( \mathcal{Y}_0 - \hat{\mathbf{\eta}} \right) \left(  \mathcal{Y}_0 - \hat{\mathbf{\eta}} \right ) ^T \, .
\end{align}
\end{subequations}
\subsection{Accuracy analysis} \label{ch3:sec_ut_accuracy_analysis}
We now conduct an accuracy analysis of the UT, which mainly follows the ideas in \cite{Julier2000,Merwe2004}.

For our purpose, we first consider the Taylor series expansion of a nonlinear function $\mathcal{F}$ around $\bar{\mathbf{x}}$, the mean of the $m$-dimensional Gaussian random variable $\mathbf{x} \sim N \left( \mathbf{x}: \bar{\mathbf{x}}, \mathbf{P}_x \right)$. Let $\mathbf{x} = \bar{\mathbf{x}} + \delta \mathbf{x}$ such that $\delta \mathbf{x} \sim N \left( \delta \mathbf{x}: \mathbf{0}, \mathbf{P}_x \right)$, then
       
\begin{equation} \label{ch3:sut_Taylor_expansion}
\mathbf{\eta} = \mathcal{F} (\bar{\mathbf{x}}+\delta \mathbf{x}) = \mathcal{F} (\bar{\mathbf{x}}) + \mathbf{D}_{\delta \mathbf{x}} \mathcal{F} + \frac{\mathbf{D}_{\delta \mathbf{x}}^2 \mathcal{F}}{2!} + \dotsb .
\end{equation}
Let $\nabla \equiv \left( \dfrac{\partial}{\partial x_1}, \dotsb, \dfrac{\partial}{\partial x_m} \right)^T$ be the gradient operator, then the operator
\begin{equation} \label{ch3:sut_operator_D}
\mathbf{D}_{\delta \mathbf{x}} \equiv \delta \mathbf{x}^T \nabla = \sum\limits_{i=1}^{m} \delta x_i \frac{\partial}{\partial x_i}
\end{equation}
acts on $\mathcal{F}$ on a component-by-component basis \cite{Julier-general}. For example,
\begin{equation}
\begin{split}
&\mathbf{D}_{\delta \mathbf{x}} \mathcal{F} = \sum\limits_{i=1}^{m}  \delta x_i \, \frac{\partial \mathcal{F}}{\partial x_i} \Bigl \vert_{ \bar{\mathbf{x}}} \bigr. \, ,\\
\end{split}
\end{equation}
where 
\begin{equation}  
\frac{\partial \mathcal{F}}{\partial x_i} = \left( \frac{\partial F_1}{\partial x_i}, \frac{\partial F_2}{\partial x_i}, \dotsb, \frac{\partial F_k}{\partial x_i} \right)^T \, , \\
\end{equation}
given that $\mathcal{F} = \left( F_1, F_2, \dotsb, F_k \right)^T$ is a $k$-dimensional vector function. Since all of the derivatives of $\mathcal{F}$ in the expansion are evaluated at $\bar{\mathbf{x}}$, for ease of notation, hereafter we may often drop the localization information. For example, we use $\partial \mathcal{F}/ \partial x_i$ to represent $\partial \mathcal{F}/ \partial x_i \vert_{ \bar{\mathbf{x}}}$ when it causes no confusion.  

Note that, the operator $\nabla$ only acts on the nonlinear function $\mathcal{F}$ or itself, but not on the perturbation $\delta \mathbf{x}$. This point will be useful in our deduction.
For example, to compute $\bar{\mathbf{\eta}} 
= \mathbb{E} \left( \mathbf{\eta}\right)$ according to Eq.~(\ref{ch3:sut_Taylor_expansion}), one has to evaluate the expectation of the second order term $\mathbb{E} \left( \mathbf{D}_{\delta \mathbf{x}}^2 \mathcal{F} \right)$, which, by definition, can be re-written as
\begin{equation}
\begin{split}
\mathbb{E} \left( \mathbf{D}_{\delta \mathbf{x}}^2 \mathcal{F} \right) &= \mathbb{E} \left[ \left( \delta \mathbf{x}^T \nabla\right) \left( \delta \mathbf{x}^T \nabla\right) \mathcal{F}  \right] \\
& = \mathbb{E} \left[ \nabla^T \left( \delta \mathbf{x} \delta \mathbf{x}^T \right) \nabla \mathcal{F}  \right] \\
& = \nabla^T \mathbb{E} \left( \delta \mathbf{x} \delta \mathbf{x}^T \right) \nabla \mathcal{F}  \\
& = \nabla^T \mathbf{P}_x \nabla \mathcal{F} \, .  \\
\end{split}
\end{equation}   
To interpret the result in the above equation, one may treat $\nabla$ as a constant vector, which acts on $\nabla$ itself and $\mathcal{F}$ only, but not on the covariance $\mathbf{P}_x$. For illustration, we consider a two dimensional case, where $\delta \mathbf{x} = \left( \delta x_1, \delta x_2 \right)^T$, $\nabla = \left( \dfrac{\partial}{\partial x_1}, \dfrac{\partial}{\partial x_2} \right)^T$, and 
\begin{equation}
\mathbf{P}_x = 
\begin{pmatrix}
P_x^{11} & P_x^{12} \\
P_x^{21} & P_x^{22} \\
\end{pmatrix} \, ,
\end{equation}   
with $P_x^{11} = \mathbb{E} \left( \delta x_1^2\right)$, $P_x^{22} = \mathbb{E} \left( \delta x_2^2\right)$, and 
$P_x^{12} = \mathbb{E} \left( \delta x_1 \delta x_2 \right) = P_x^{21} = \mathbb{E} \left( \delta x_2 \delta x_1 \right)$. In this case, we have
\begin{equation}
\begin{split}
\nabla^T \mathbf{P}_x \nabla \mathcal{F} & = \left( \dfrac{\partial}{\partial x_1},~\dfrac{\partial}{\partial x_2} \right) 
\begin{pmatrix}
P_x^{11} & P_x^{12} \\
P_x^{21} & P_x^{22} \\
\end{pmatrix}
\begin{pmatrix}
 \dfrac{\partial}{\partial x_1}\\
 \dfrac{\partial}{\partial x_2}\\
\end{pmatrix}
\mathcal{F} \\
& = \left( P_x^{11} \dfrac{\partial}{\partial x_1} + P_x^{21} \dfrac{\partial}{\partial x_2},~P_x^{12} \dfrac{\partial}{\partial x_1} + P_x^{22} \dfrac{\partial}{\partial x_2}  \right) \begin{pmatrix}
 \dfrac{\partial}{\partial x_1}\\
 \dfrac{\partial}{\partial x_2}\\
\end{pmatrix}
\mathcal{F} \\
& = \left( P_x^{11} \dfrac{\partial^2 \mathcal{F}}{\partial x_1^2} + 2 P_x^{12} \dfrac{\partial^2 \mathcal{F}}{\partial x_1\partial x_2} + P_x^{22} \dfrac{\partial^2 \mathcal{F}}{\partial x_2^2}  \right)\Bigl \vert_{ \bar{\mathbf{x}}} \bigr. \, ,
\end{split}
\end{equation}
which is consistent with the result in \cite{Julier2004}.

Applying the above principle and noting that $\mathbb{E} \left( \delta \mathbf{x} \right) = \mathbf{0}$, the mean and covariance of the random variable $\mathbf{\eta}$ in Eq.~(\ref{ch3:sut_Taylor_expansion}) are then given by 
\begin{subequations} \label{ch3:sut_y_stat_in_expansion}
\begin{align}
 \label{ch3:sut_y_mean_in_expansion} \bar{\mathbf{\eta}} & = \mathbb{E} \left( \mathbf{\eta}\right) \\
\nonumber &= \mathcal{F} (\bar{\mathbf{x}}) + \frac{1}{2} \left( \nabla^T \mathbf{P}_x \nabla \right) \mathcal{F}  + \frac{1}{6} \, \mathbb{E} \left( \mathbf{D}_{\delta \mathbf{x}}^3 \mathcal{F} \right) + \dotsb \, ,\\
 \label{ch3:sut_y_cov_in_expansion} \mathbf{P}_{\eta} &= \mathbb{E} \left[ \left( \mathbf{\eta} - \bar{\mathbf{\eta}}\right) \left( \mathbf{\eta} - \bar{\mathbf{\eta}}\right)^T \right] \\
\nonumber & =   \left( \nabla \mathcal{F} \right)^T \mathbf{P}_x \left( \nabla \mathcal{F} \right) + \mathbb{E} \left[ \frac{ \mathbf{D}_{\delta \mathbf{x}} \mathcal{F} \left( \mathbf{D}_{\delta \mathbf{x}}^3 \mathcal{F} \right)^T}{6}   \right. \\
\nonumber &\quad  +  \frac{ \mathbf{D}_{\delta \mathbf{x}}^2 \mathcal{F} \left( \mathbf{D}_{\delta \mathbf{x}}^2 \mathcal{F} \right)^T}{4} \left.  +  \frac{ \mathbf{D}_{\delta \mathbf{x}}^3 \mathcal{F} \left( \mathbf{D}_{\delta \mathbf{x}} \mathcal{F} \right)^T}{6} \right]    \\
\nonumber &\quad - \left[  \left( \frac{ \nabla^T \mathbf{P}_{x}  \nabla }{2}
\right) \mathcal{F} \right] \left[  \left( \frac{ \nabla^T
\mathbf{P}_{x}  \nabla }{2}  \right) \mathcal{F} \right]^T + \dotsb \, .
\end{align}
\end{subequations} 

In particular, if both the pdf $p \left( \delta \mathbf{x} \right)$ of $\delta \mathbf{x}$ and its support are symmetric about the origin \footnote{ $p \left( \delta \mathbf{x} \right)$ is not necessarily Gaussian. For example, it can be a uniform distribution on the interval $[-1, 1 ]$ in one dimensional case.}, the odd order (central) moments of the $m$-dimensional random variable $\delta \mathbf{x} = \left( \delta x_1, \dotsb, \delta x_m \right)^T$ are all zero, i.e.,
\begin{equation}
\mathbb{E} \left[ \prod\limits_{i=1}^{m} \left( \delta x_i \right)^{p_i} \right] =0 
\end{equation}
if the summation $p=\sum\limits_{i=1}^{m} p_i$ of the non-negative integers $p_i \ge 0$ is an odd integer. In this case, it can be shown that
\begin{equation}
\begin{split}
\mathbb{E} \left( \mathbf{D}_{\delta \mathbf{x}}^3 \mathcal{F} \right) & = \mathbb{E} \left( \delta \mathbf{x}^T \nabla \delta \mathbf{x}^T \nabla \delta \mathbf{x}^T \nabla \mathcal{F} \right) \\
& = \nabla^T \mathbb{E} \left( \delta \mathbf{x} \delta \mathbf{x}^T \nabla \delta \mathbf{x}^T \right) \nabla \mathcal{F} \\
& = \mathbf{0} \, .
\end{split}
\end{equation}
Thus Eq.~(\ref{ch3:sut_y_mean_in_expansion}) can be further reduced to
\begin{equation} \label{ch3:sut_symmetric_y_mean_in_expansion}
\bar{\mathbf{\eta}} = \mathcal{F} (\bar{\mathbf{x}}) + \frac{1}{2} \left( \nabla^T \mathbf{P}_x \nabla \right) \mathcal{F}  + \frac{1}{4!} \, \mathbb{E} \left( \mathbf{D}_{\delta \mathbf{x}}^4 \mathcal{F} \right) + \dotsb \, .
\end{equation} 

To analyze the accuracy of the UT, we also need to expand the transformed sigma points $\{ \mathcal{Y}_i:  \mathcal{Y}_i = \mathcal{F} \left( \mathcal{X}_i \right) \}_{i=0}^{2L}$ around $\mathcal{X}_0 = \bar{\mathbf{x}}$. Let $\delta \mathcal{X}_i = \mathcal{X}_i - \bar{\mathbf{x}}$, then according to Eq.~(\ref{ch3:ut_sigma_points}),
\begin{equation}
\delta \mathcal{X}_i =
\begin{cases}
\mathbf{0} \, , & \text{for~} i=0 \, , \\
\sqrt{L+\lambda} \left( \mathbf{S}_x\right)_i \, , & \text{for~} i=1,\dotsb,L \, , \\
-\sqrt{L+\lambda} \left( \mathbf{S}_x\right)_{i-L} \, , & \text{for~} i=L+1,\dotsb,2L \, . \\ 
\end{cases}  
\end{equation}

Substituting $\delta \mathcal{X}_i$ into Eq.~(\ref{ch3:sut_Taylor_expansion}), the sample mean $\hat{\mathbf{\eta}}$ of the transformed sigma points are then given by
\begin{equation} \label{ch3:sut_y_mean_in_expansion_de1}
\begin{split}
\hat{\mathbf{\eta}} & = \sum\limits_{i=0}^{2L} W_i \, \mathcal{Y}_i   \\
& = \sum\limits_{i=0}^{2L} W_i \left[ \mathcal{F} (\bar{\mathbf{x}}) + \delta \mathcal{X}_i^T \nabla \mathcal{F} + \dfrac{1}{2} \nabla^T \left( \delta \mathcal{X}_i \delta \mathcal{X}_i^T \right) \nabla \mathcal{F} \right. \\ 
& \quad \left. + \dfrac{1}{6}  \nabla^T \left( \delta \mathcal{X}_i \delta \mathcal{X}_i^T \nabla \delta \mathcal{X}_i^T \right) \nabla \mathcal{F} + \dfrac{1}{4!} \mathbf{D}_{\delta \mathcal{X}_i}^4 \mathcal{F} + \dotsb \right] \\
& = \left( \sum\limits_{i=0}^{2L} W_i \right) \mathcal{F} (\bar{\mathbf{x}}) + \left( \sum\limits_{i=0}^{2L} W_i \delta \mathcal{X}_i \right)^T \nabla \mathcal{F} + \dfrac{1}{2} \nabla^T \left( \sum\limits_{i=0}^{2L} W_i \delta \mathcal{X}_i \delta \mathcal{X}_i^T \right) \nabla \mathcal{F}  \\
& \quad + \dfrac{1}{6}  \nabla^T \left( \sum\limits_{i=0}^{2L} W_i \delta \mathcal{X}_i \delta \mathcal{X}_i^T \nabla \delta \mathcal{X}_i^T \right) \nabla \mathcal{F} + \dfrac{1}{4!} \sum\limits_{i=0}^{2L} W_i \mathbf{D}_{\delta \mathcal{X}_i}^4 \mathcal{F} + \dotsb \, .
\end{split}
\end{equation}
From Eqs. (\ref{ch3:ut_sigma_points}) - (\ref{ch3:ut_weighted_stat}), it is evident that 
\begin{equation}
\begin{split}
& \sum\limits_{i=0}^{2L} W_i = 1 \, ; \\
& \sum\limits_{i=0}^{2L} W_i \delta \mathcal{X}_i = \mathbf{0} \, ; \\
&  \sum\limits_{i=0}^{2L} W_i \delta \mathcal{X}_i \delta \mathcal{X}_i^T = \mathbf{P}_x \, .
\end{split}
\end{equation}
Moreover, because of the symmetry in sigma points, we have 
\begin{equation}
\delta \mathcal{X}_i \delta \mathcal{X}_i^T \nabla \delta \mathcal{X}_i^T + \delta \mathcal{X}_{L+i} \delta \mathcal{X}_{L+i}^T \nabla \delta \mathcal{X}_{L+i}^T =0,~\text{for}~ i=1,\dotsb,L \, .
\end{equation}
Thus it is clear that 
\begin{equation}
\sum\limits_{i=0}^{2L} W_i \delta \mathcal{X}_i \delta \mathcal{X}_i^T \nabla \delta \mathcal{X}_i^T =0 \, .
\end{equation}
Substituting the above identities into Eq.~(\ref{ch3:sut_y_mean_in_expansion_de1}), we have
\begin{equation} \label{ch3:sut_y_mean_in_expansion_de2}
\begin{split}
\hat{\mathbf{\eta}} & = \sum\limits_{i=0}^{2L} W_i \, \mathcal{Y}_i   \\
& = \mathcal{F} (\bar{\mathbf{x}}) + \dfrac{1}{2} \nabla^T \mathbf{P}_x \nabla \mathcal{F} + \dfrac{1}{4!} \sum\limits_{i=0}^{2L} W_i \mathbf{D}_{\delta \mathcal{X}_i}^4 \mathcal{F} + \dotsb \, .
\end{split}
\end{equation}

Thus by comparing the UT estimation $\hat{\mathbf{\eta}}$ in Eq.~(\ref{ch3:sut_y_mean_in_expansion_de2}) with the expectation $\bar{\mathbf{\eta}}$ in Eq.~(\ref{ch3:sut_symmetric_y_mean_in_expansion}), it can be seen that the expansion of $\hat{\mathbf{\eta}}$ matches that of $\bar{\mathbf{\eta}}$ up to third order term (i.e., the term $\mathbb{E} \left( \mathbf{D}_{\delta \mathbf{x}}^3 \mathcal{F} \right)$ in Eq.~(\ref{ch3:sut_y_mean_in_expansion})), and in general differs from $\bar{\mathbf{\eta}}$ at fourth order. However, in some special situations, for example, if $\mathcal{F}$ is a nonlinear function with fourth and higher order derivatives being zero, then $\hat{\mathbf{\eta}}$ is equal to $\bar{\mathbf{\eta}}$.    

In contrast, in the extended Kalman filter (EKF), an unbiased estimation $\hat{\mathbf{\eta}}$ based on the linearization scheme in Eq.~(\ref{ch3:overview_Taylor_expansion}) is given by $\hat{\mathbf{\eta}} = \mathcal{F} (\bar{\mathbf{x}})$, which matches the expansion of $\bar{\mathbf{\eta}}$ in Eq.~(\ref{ch3:sut_symmetric_y_mean_in_expansion}) only up to first order ($\mathbb{E} \left( \mathbf{D}_{\delta \mathbf{x}} \mathcal{F} \right)$), and is therefore less accurate than the UT in this sense.          

The same arguments can be applied to study the accuracy of the covariance estimation $\hat{\mathbf{P}}_{\eta}$. To this end, we first consider the estimation without the compensation term $\beta \left( \mathcal{Y}_0 - \hat{\mathbf{\eta}} \right) \left(  \mathcal{Y}_0 - \hat{\mathbf{\eta}} \right ) ^T$ in Eq.~(\ref{ch3:ut_transformed_cov}):
\begin{equation} \label{ch3:ut_y_ut_cov_in_expansion}
\begin{split}
\hat{\mathbf{P}}_{\eta} &=  \sum\limits_{i=0}^{2L} W_i \, \left( \mathcal{Y}_i - \hat{\mathbf{\eta}} \right) \left(  \mathcal{Y}_i - \hat{\mathbf{\eta}} \right ) ^T \\
&=   \left( \nabla \mathcal{F} \right)^T \mathbf{P}_{x}  \left( \nabla \mathcal{F} \right) + \sum_{i=0}^{2L} W_i \left[ \dfrac{\mathbf{D}_{\delta \mathcal{X}_i} \mathcal{F}  \left( \mathbf{D}_{\delta \mathcal{X}_i}^3 \mathcal{F} \right) ^T}{6}  \right. \\
&\quad \left. + \dfrac{\mathbf{D}_{\delta \mathcal{X}_i}^2 \mathcal{F}  \left( \mathbf{D}_{\delta \mathcal{X}_i}^2 \mathcal{F} \right) ^T}{4} + \dfrac{\mathbf{D}_{\delta \mathcal{X}_i}^3 \mathcal{F}  \left( \mathbf{D}_{\delta \mathcal{X}_i} \mathcal{F} \right) ^T}{6}  \right] \\
&\quad - \left[  \left( \frac{ \nabla^T \mathbf{P}_{x}  \nabla }{2}  \right) \mathcal{F} \right] \left[  \left( \frac{ \nabla^T \mathbf{P}_{x}  \nabla }{2}  \right) \mathcal{F} \right]^T  + \dotsb \, .
\end{split} 
\end{equation}
Comparing $\hat{\mathbf{P}}_{\eta}$ with its expectation $\mathbf{P}_{\eta}$ in Eq.~(\ref{ch3:sut_y_cov_in_expansion}), one can see that the expansion of $\hat{\mathbf{P}}_{\eta}$ matches the two terms in the expansion of $\mathbf{P}_{\eta}$ that contain $\left( \nabla \mathcal{F} \right)^T \mathbf{P}_{x}  \left( \nabla \mathcal{F} \right)$ and $\left[  \left( \nabla^T \mathbf{P}_{x}  \nabla  \right) \mathcal{F} \right] \left[  \left( \nabla^T \mathbf{P}_{x}  \nabla \right) \mathcal{F} \right]^T$. But the other terms in the expansion of $\hat{\mathbf{P}}_{\eta}$ might differ from those in the expansion of $\mathbf{P}_{\eta}$.       

Similarly, in the EKF, the estimation $\hat{\mathbf{P}}_{\eta}$ based on the linearization scheme in Eq.~(\ref{ch3:overview_Taylor_expansion}) is given by $\hat{\mathbf{P}}_{\eta} = \left( \nabla \mathcal{F} \right)^T \mathbf{P}_{x}  \left( \nabla \mathcal{F} \right)$, which matches only the first term in the expansion of $\mathbf{P}_{\eta}$ in Eq.~(\ref{ch3:sut_y_cov_in_expansion}), and thus is also less accurate than the estimation in Eq.~(\ref{ch3:ut_y_ut_cov_in_expansion}) obtained through the UT. 

It is also possible to apply similar arguments in this chapter to analyze the accuracies of the mean and covariance estimations of the ensemble Kalman filter (EnKF). This is done in Appendix~\ref{appendix:accuracy analysis}. The analytical results there indicate that, under the assumption of Gaussianity, estimations based on the UT can avoid some sample errors and bias that appear in the EnKF due to the effect of finite ensemble size. 

The benefits of introducing the additional term  $\beta \left( \mathcal{Y}_0 - \hat{\mathbf{\eta}} \right) \left(  \mathcal{Y}_0 - \hat{\mathbf{\eta}} \right ) ^T$ will be discussed again in the next section.

\section{Scaled unscented transform}\label{ch3:sec_sut}

From Eq.~(\ref{ch3:ut_sigma_points}) we see that, $\lambda$ is a parameter that affects the distances of sigma points to their center. With nonlinearity in the transform function $\mathcal{F}$, $\lambda$ is also a parameter that affects the higher order moments of the transformed sigma points. The scaled unscented transform (SUT) extends this idea by further introducing a scale parameter $\alpha$ to the UT \cite{Julier2004}. 

To see this, we first construct an auxiliary random variable 
\begin{equation} \label{ch3:sut_new_rv_z}
\mathbf{z} = \mathcal{F} (\bar{\mathbf{x}}) + \dfrac{\mathcal{F} (\bar{\mathbf{x}} + \alpha \, \delta \mathbf{x}) - \mathcal{F} (\bar{\mathbf{x}})}{\mu} \, ,
\end{equation}
where $\alpha$ and $\mu$ are two free parameters ($\mu \neq 0$). Eq.~(\ref{ch3:sut_new_rv_z}) is similar to the Taylor series expansion in Eq.~(\ref{ch3:sut_Taylor_expansion}) up to first order, in that the second term on the rhs of Eq.~(\ref{ch3:sut_new_rv_z}) can be considered as a divided difference approximation to the term $\mathbf{D}_{\delta \mathbf{x}} \mathcal{F}$ in Eq.~(\ref{ch3:sut_Taylor_expansion}). Compared with the idea of linearization in Eq.~(\ref{ch3:overview_Taylor_expansion}) to construct the extended Kalman filter (EKF), the advantage of taking the form in Eq.~(\ref{ch3:sut_new_rv_z}) is that there is no need to evaluate the derivatives of $\mathcal{F}$. What we need to do next is to evaluate the mean and covariance of the transformed random variable $\mathbf{\eta}=\mathcal{F} (\bar{\mathbf{x}} + \delta \mathbf{x})$ based on the auxiliary variable $\mathbf{z}$.   

In analysis, we also expand $\mathcal{F} (\bar{\mathbf{x}} + \alpha \, \delta \mathbf{x})$ around $\bar{\mathbf{x}}$ such that \cite{Julier2004}
\begin{equation}
\mathbf{z} = \mathcal{F} (\bar{\mathbf{x}}) + \dfrac{\alpha}{\mu} \, \mathbf{D}_{\delta \mathbf{x}} \mathcal{F} + \dfrac{\alpha^2}{\mu} \, \frac{\mathbf{D}_{\delta \mathbf{x}}^2 \mathcal{F}}{2!} + \dfrac{\alpha^3}{\mu} \, \frac{\mathbf{D}_{\delta \mathbf{x}}^3 \mathcal{F}}{3!} + \dotsb .
\end{equation}      
Then the mean $\bar{\mathbf{z}}$ and covariance $\mathbf{P}_z^{*} = \mu \, \mathbf{P}_z$ are given by
\begin{subequations} \label{ch3:sut_z_stat_in_expansion}
\begin{align}
 \label{ch3:sut_z_mean_in_expansion} \bar{\mathbf{z}} & = \mathbb{E} \left( \mathbf{z} \right) \\
\nonumber & = \mathcal{F} (\bar{\mathbf{x}}) + \frac{1}{2} \, \dfrac{\alpha^2}{\mu} \, \left( \nabla^T \mathbf{P}_x \nabla \mathcal{F} \right) + \frac{1}{6} \, \dfrac{\alpha^3}{\mu} \mathbb{E} \left(\mathbf{D}_{\delta \mathbf{x}}^3 \mathcal{F} \right) + \dotsb \, ,\\
 \label{ch3:sut_z_cov_in_expansion} \mathbf{P}_z^{*} &= \mu \, \mathbb{E} \left[ \left( \mathbf{z} - \bar{\mathbf{z}}\right) \left( \mathbf{z} - \bar{\mathbf{z}}\right)^T \right] \\
\nonumber & = \dfrac{\alpha^2}{\mu} \left( \nabla \mathcal{F} \right)^T \mathbf{P}_{x}  \left( \nabla \mathcal{F} \right) + \dfrac{\alpha^3}{2\mu} \left\{ \mathbb{E} \left[ \mathbf{D}_{\delta \mathbf{x}} \mathcal{F} \left( \mathbf{D}_{\delta \mathbf{x}}^2 \mathcal{F} \right)^T \right] + \mathbb{E} \left[ \mathbf{D}_{\delta \mathbf{x}}^2 \mathcal{F} \left( \mathbf{D}_{\delta \mathbf{x}} \mathcal{F} \right)^T \right] \right\} \\
\nonumber & \quad + \dfrac{\alpha^4}{6\mu} \left\{ \mathbb{E} \left[ \mathbf{D}_{\delta \mathbf{x}} \mathcal{F} \left( \mathbf{D}_{\delta \mathbf{x}}^3 \mathcal{F} \right)^T \right] + \mathbb{E} \left[ \mathbf{D}_{\delta \mathbf{x}}^3 \mathcal{F} \left( \mathbf{D}_{\delta \mathbf{x}} \mathcal{F} \right)^T \right] \right\} \\
\nonumber & \quad + \dfrac{\alpha^4}{4\mu} \left\{ \mathbb{E} \left[ \mathbf{D}_{\delta \mathbf{x}}^2 \mathcal{F} \left( \mathbf{D}_{\delta \mathbf{x}}^2 \mathcal{F} \right)^T \right] -  \left( \nabla^T \mathbf{P}_x \nabla \mathcal{F} \right) \left( \nabla^T \mathbf{P}_x \nabla \mathcal{F} \right)^T \right\} + \dotsb \, .
\end{align}
\end{subequations}

Comparing Eq.~(\ref{ch3:sut_z_stat_in_expansion}) with Eq.~(\ref{ch3:sut_y_stat_in_expansion}), it is evident that when $\alpha=\mu=1$, we have $\bar{\mathbf{\eta}} = \bar{\mathbf{z}}$ and $\mathbf{P}_{\eta} = \mathbf{P}_z^{*}$ \footnote{$\delta \mathbf{x}$ is assumed to follow the Gaussian distribution $N (\delta \mathbf{x}: \mathbf{0}, \mathbf{P}_x)$, hence in the second line of Eq.~(\ref{ch3:sut_z_cov_in_expansion}), we have $\mathbb{E} \left[ \mathbf{D}_{\delta \mathbf{x}} \mathcal{F} \left( \mathbf{D}_{\delta \mathbf{x}}^2 \mathcal{F} \right)^T \right] = \mathbb{E} \left[ \mathbf{D}_{\delta \mathbf{x}}^2 \mathcal{F} \left( \mathbf{D}_{\delta \mathbf{x}} \mathcal{F}  \right)^T \right]=0$.}. If $\alpha \neq 1$, one may use $\bar{\mathbf{z}}$ and $\mathbf{P}_z^{*}$ as the approximations to $\bar{\mathbf{\eta}}$ and $\mathbf{P}_{\eta}$, respectively. To this end, a natural choice is to let $\mu = \alpha^2$. Thus under the assumption that $\delta \mathbf{x}$ is Gaussian such that its odd moments are all zero, Eq.~(\ref{ch3:sut_z_stat_in_expansion}) is reduced to 
\begin{subequations} \label{ch3:sut_reduced_z_stat_in_expansion}
\begin{align}
 \label{ch3:sut_reduced_z_mean_in_expansion} \bar{\mathbf{z}} & = \mathbb{E} \left( \mathbf{z} \right) \\
\nonumber & = \mathcal{F} (\bar{\mathbf{x}}) + \frac{1}{2} \left( \nabla^T \mathbf{P}_x \nabla \mathcal{F} \right) + \frac{\alpha^2}{4!} \mathbb{E} \left(\mathbf{D}_{\delta \mathbf{x}}^4 \mathcal{F} \right) + \dotsb \, ,\\
 \label{ch3:sut_reduced_z_cov_in_expansion} \mathbf{P}_z^{*} &= \alpha^2 \, \mathbb{E} \left[ \left( \mathbf{z} - \bar{\mathbf{z}}\right) \left( \mathbf{z} - \bar{\mathbf{z}}\right)^T \right] \\
\nonumber & = \left( \nabla \mathcal{F} \right)^T \mathbf{P}_{x}  \left( \nabla \mathcal{F} \right) + \dfrac{\alpha^2}{6} \left\{ \mathbb{E} \left[ \mathbf{D}_{\delta \mathbf{x}} \mathcal{F} \left( \mathbf{D}_{\delta \mathbf{x}}^3 \mathcal{F} \right)^T \right] + \mathbb{E} \left[ \mathbf{D}_{\delta \mathbf{x}}^3 \mathcal{F} \left( \mathbf{D}_{\delta \mathbf{x}} \mathcal{F} \right)^T \right] \right\} \\
\nonumber & \quad + \dfrac{\alpha^2}{4} \left\{ \mathbb{E} \left[ \mathbf{D}_{\delta \mathbf{x}}^2 \mathcal{F} \left( \mathbf{D}_{\delta \mathbf{x}}^2 \mathcal{F} \right)^T \right] -  \left( \nabla^T \mathbf{P}_x \nabla \mathcal{F} \right) \left( \nabla^T \mathbf{P}_x \nabla \mathcal{F} \right)^T \right\} + \dotsb \, .
\end{align}
\end{subequations} 
With this choice, $\bar{\mathbf{z}}$ and $\mathbf{P}_z^{*}$ agree with $\bar{\mathbf{\eta}}$ and $\mathbf{P}_{\eta}$ up to second order (moment) term (i.e. the term that contains only one $\mathbf{P}_x$). Other higher order terms scale with the parameter $\alpha$.

When $\alpha^2 <1$, $\mathbf{P}_z^{*}$ underestimates $\mathbf{P}_{\eta}$, with $\alpha^2 \rightarrow 0$ being the extreme situation (equivalent to the covariance estimation in the EKF). In contrast, $\alpha^2 >1$ means that $\mathbf{P}_z^{*}$ overestimates $\mathbf{P}_{\eta}$. This is similar to the covariance inflation technique in \S~\ref{ch2:sec_covariance_inflation}, and thus is desirable provided that $\alpha^2$ is not too large. The difference $\Delta \mathbf{P} = \mathbf{P}_{\eta} - \mathbf{P}_z^{*}$ between $\mathbf{P}_{\eta}$ in Eq.~(\ref{ch3:sut_y_cov_in_expansion}) and $\mathbf{P}_z^{*}$ in the ``worst'' case $\alpha^2 \rightarrow 0$, is therefore given by 
\begin{equation} \label{ch3:sec_unnumbered_eq_1}
\begin{split}
\Delta \mathbf{P} & = \mathbb{E} \left[ \frac{ \mathbf{D}_{\delta \mathbf{x}} \mathcal{F} \left( \mathbf{D}_{\delta \mathbf{x}}^3 \mathcal{F} \right)^T}{6} +  \frac{ \mathbf{D}_{\delta \mathbf{x}}^2 \mathcal{F} \left( \mathbf{D}_{\delta \mathbf{x}}^2 \mathcal{F} \right)^T}{4} +  \frac{ \mathbf{D}_{\delta \mathbf{x}}^3 \mathcal{F} \left( \mathbf{D}_{\delta \mathbf{x}} \mathcal{F} \right)^T}{6} \right]    \\
&\quad - \left[  \left( \frac{ \nabla^T \mathbf{P}_{x}  \nabla }{2}
\right) \mathcal{F} \right] \left[  \left( \frac{ \nabla^T
\mathbf{P}_{x}  \nabla }{2}  \right) \mathcal{F} \right]^T + \dotsb \, .
\end{split}
\end{equation}   
To further reduce the approximation error $\Delta \mathbf{P}$, a simple idea is to introduce an extra term, $\beta \left( \bar{\mathbf{z}} - \mathcal{F} (\bar{\mathbf{x}})\right) \left( \mathbb{E} \left( \mathbf{z} \right) - \mathcal{F} (\bar{\mathbf{x}})\right)^T$ in the expression of $\mathbf{P}_z^{*}$. From Eq.~(\ref{ch3:sut_reduced_z_mean_in_expansion}), 
\begin{equation}
\beta \left( \bar{\mathbf{z}} - \mathcal{F} (\bar{\mathbf{x}})\right) \left( \bar{\mathbf{z}} - \mathcal{F} (\bar{\mathbf{x}})\right)^T = \dfrac{\beta}{4} \left( \nabla^T \mathbf{P}_x \nabla \mathcal{F} \right) \left( \nabla^T \mathbf{P}_x \nabla \mathcal{F} \right)^T + \dotsb \, .
\end{equation} 
Thus by choosing a proper value of $\beta$ (which itself depends on the distribution of $\mathbf{x}$), one may reduce the difference $\Delta \mathbf{P}$ in Eq.~(\ref{ch3:sec_unnumbered_eq_1}). For example, if $\delta \mathbf{x}$ follows a univariate normal distribution $N(\delta \mathbf{x}: 0, 1)$, then it can be shown that \cite{Julier2004}
\begin{equation}
\mathbb{E} \left( \mathbf{D}_{\delta \mathbf{x}}^2 \mathcal{F} \left( \mathbf{D}_{\delta \mathbf{x}}^2 \mathcal{F} \right)^T \right) = 3 \left( \nabla^T \mathbf{P}_{x}  \nabla \mathcal{F} \right) \left( \nabla^T
\mathbf{P}_{x}  \nabla \mathcal{F} \right)^T \, . 
\end{equation}
Thus choosing $\beta=2$ will reduce the approximation error in the fourth order (moment) terms, so that
\begin{equation}
\begin{split}
\Delta \mathbf{P} & = \mathbf{P}_{\eta} - \mathbf{P}_z^{*} - \beta \left( \bar{\mathbf{z}} - \mathcal{F} (\bar{\mathbf{x}})\right) \left( \bar{\mathbf{z}} - \mathcal{F} (\bar{\mathbf{x}})\right)^T \\
& = \mathbb{E} \left[ \frac{ \mathbf{D}_{\delta \mathbf{x}} \mathcal{F} \left( \mathbf{D}_{\delta \mathbf{x}}^3 \mathcal{F} \right)^T}{6} + \frac{ \mathbf{D}_{\delta \mathbf{x}}^3 \mathcal{F} \left( \mathbf{D}_{\delta \mathbf{x}} \mathcal{F} \right)^T}{6} \right] + \dotsb \, .
\end{split}
\end{equation}
Other schemes to reduce $\Delta \mathbf{P}$ should also be possible. However, they might be more complicated in implementations.   

In practice, to estimate $\bar{\mathbf{z}}$ and $\mathbf{P}_z^{*}$, we apply the UT introduced in the previous section. To this end, we first generate a set of sigma points $\{ \mathcal{X}_i \}_{i=0}^{2L}$ with respect to the quartet $\left( \alpha, \lambda, \bar{\mathbf{x}}, \mathbf{S}_x \right)$ such that
\begin{equation} \label{ch3:sut_sigma_points}
\begin{split}
& \mathcal{X}_0 = \bar{\mathbf{x}},\\
& \mathcal{X}_i = \bar{\mathbf{x}} + \alpha \sqrt{L+\lambda} \left(  \mathbf{S}_x \right)_i, \, i=1, 2, \dotsb, L ,\\
& \mathcal{X}_i = \bar{\mathbf{x}} - \alpha \sqrt{L+\lambda} \left(  \mathbf{S}_x \right)_{i-L}, \, i=L+1, L+2, \dotsb, 2L \, .
\end{split}
\end{equation}
According to Eq.~(\ref{ch3:sut_new_rv_z}), the transformed sigma points $\{\mathcal{Z}_i \}_{i=0}^{2L}$ are given by
\begin{equation}
\mathcal{Z}_i =
\begin{cases}
\mathcal{F} \left( \bar{\mathbf{x}} \right), & i=0 \, ; \\
\mathcal{Z}_0 + \dfrac{\mathcal{F} \left( \mathcal{X}_i \right) - \mathcal{Z}_0 }{\alpha^2}, & i=1, \dotsb, 2L \, . \\
\end{cases}
\end{equation}
The estimated mean $\hat{\mathbf{z}}$ and covariance $\hat{\mathbf{P}}_z^{*}$ (with the compensation term), are given by 
\begin{subequations} \label{ch3:sut_z_transformed_stat}
\begin{align}
\label{ch3:sut_z_transformed_mean} &\hat{\mathbf{z}} = \sum\limits_{i=0}^{2L} W_i \, \mathcal{Z}_i \, ,  \\
\label{ch3:ut_z_transformed_cov} &\hat{\mathbf{P}}_{z}^{*} =  \sum\limits_{i=0}^{2L} W_i \, \left( \mathcal{Z}_i - \hat{\mathbf{z}} \right) \left(  \mathcal{Z}_i - \hat{\mathbf{z}} \right ) ^T + \beta \left( \hat{\mathbf{z}} - \mathcal{F} (\bar{\mathbf{x}})\right) \left( \hat{\mathbf{z}} - \mathcal{F} (\bar{\mathbf{x}})\right)^T  \, , 
\end{align}
\end{subequations}
where $W_i$ are the weights determined by Eq.~(\ref{ch3:ut_weights}).

The sample mean $\hat{\mathbf{z}}$ and sample covariance $\hat{\mathbf{P}}_{z}^{*}$ obtained in Eq.~(\ref{ch3:sut_z_transformed_stat}) will then be used as the approximations to $\bar{\mathbf{\eta}}$ and $\mathbf{P}_{\eta}$, respectively. However, one possible problem is that $\hat{\mathbf{z}}$ and $\hat{\mathbf{P}}_{z}^{*}$ are expressed in terms of the set of sigma points $\{\mathcal{Z}_i \}_{i=0}^{2L}$, whose physical meanings might be hard to interpret. Thus it is preferable to express $\hat{\mathbf{z}}$ and $\hat{\mathbf{P}}_{z}^{*}$ in terms of the transformed sigma points $\{ \mathcal{Y}_i: \mathcal{Y}_i = \mathcal{F} \left( \mathcal{X}_i \right) \}_{i=0}^{2L}$ of the original nonlinear system. To this end, we note that $\mathcal{Z}_i = \mathcal{F} \left( \bar{\mathbf{x}} \right) + \dfrac{1}{\alpha^2} \left( \mathcal{Y}_i - \mathcal{F} \left( \bar{\mathbf{x}} \right) \right) $ is just a linear transformation of $\mathcal{Y}_i$. Thus we have \cite{Julier-scaled,Julier2004}
\begin{subequations} \label{ch3:sut_y_transformed_stat}
\begin{align}
\label{ch3:sut_y_transformed_mean} & \hat{\mathbf{\eta}} = \hat{\mathbf{z}} = \sum\limits_{i=0}^{2L} W_i^{s} \, \mathcal{Y}_i \, ,  \\
\label{ch3:ut_y_transformed_cov} & \hat{\mathbf{P}}_{\eta} = \hat{\mathbf{P}}_{z}^{*} =  \sum\limits_{i=0}^{2L} W_i^s \, \left( \mathcal{Y}_i - \hat{\mathbf{\eta}} \right) \left(  \mathcal{Y}_i - \hat{\mathbf{\eta}} \right ) ^T + \left( 1+ \beta - \alpha^2 \right) \left( \mathcal{Y}_0 - \hat{\mathbf{\eta}} \right) \left( \mathcal{Y}_0 - \hat{\mathbf{\eta}} \right)^T  \, , 
\end{align}
\end{subequations}       
where the weights, $W_i^{s}$, are given by
\begin{equation} \label{ch3:sut_weights}
\begin{split}
\begin{split}
&W_0^s = \frac{\lambda}{\alpha^2(L+\lambda)} + 1-\dfrac{1}{\alpha^2} \, ,\\
&W_i^s = \frac{1}{2 \alpha^2 \left(L+\lambda  \right)} \,,~i=1,2,\dotsb, 2L \, . \\
\end{split}
\end{split}
\end{equation}
It can be verified that the sigma points in Eq.~(\ref{ch3:sut_sigma_points}), associated with the weights $W_i^{s}$, also capture the mean $\bar{\mathbf{x}}$ and covariance $\mathbf{P}_x$ of the random variable $\mathbf{x}$, so that
\begin{equation}  \label{ch3:sut_weighted_stat}
\begin{split}
& \sum\limits_{i=0}^{2L} W_i^s = 1 \, ; \\
& \hat{\mathcal{X}} = \sum\limits_{i=0}^{2L} W_i^s \mathcal{X}_i = \bar{\mathbf{x}} \, ; \\
& \hat{\mathbf{P}}_{\mathcal{X}} = \sum\limits_{i=0}^{2L} W_i^s \left(\mathcal{X}_i-\hat{\mathcal{X}} \right) \left(\mathcal{X}_i-\hat{\mathcal{X}} \right)^T = \mathbf{P}_x \, . \\
\end{split}
\end{equation}
Note that in Eq.~(\ref{ch3:ut_y_transformed_cov}) there exists an extra term $\left( 1 - \alpha^2 \right) \left( \mathcal{Y}_0 - \hat{\mathbf{\eta}} \right) \left( \mathcal{Y}_0 - \hat{\mathbf{\eta}} \right)^T$, which is due to the introduction of the scale parameter $\alpha$. This does not appear in the unscented transform (UT), where $\alpha=1$.  

We summarize the main procedures of the SUT as follows:\\
Generation of sigma points:
\begin{equation} \tagref{ch3:sut_sigma_points}
\begin{split}
& \mathcal{X}_0 = \bar{\mathbf{x}},\\
& \mathcal{X}_i = \bar{\mathbf{x}} + \alpha \sqrt{L+\lambda} \left(  \mathbf{S}_x \right)_i, \, i=1, 2, \dotsb, L ,\\
& \mathcal{X}_i = \bar{\mathbf{x}} - \alpha \sqrt{L+\lambda} \left(  \mathbf{S}_x \right)_{i-L}, \, i=L+1, L+2, \dotsb, 2L .\\
\end{split}
\end{equation}
Allocation of associated weights:
\begin{equation} \tagref{ch3:sut_weights}
\begin{split}
\begin{split}
&W_0^s = \frac{\lambda}{\alpha^2(L+\lambda)} + 1-\dfrac{1}{\alpha^2} \, ,\\
&W_i^s = \frac{1}{2 \alpha^2 \left(L+\lambda  \right)} \,,~i=1,2,\dotsb, 2L \, . \\
\end{split}
\end{split}
\end{equation}
Estimations of the mean and covariance of the transformed random variable $\mathbf{\eta}$:
\begin{subequations} 
\begin{align}
\nonumber &  \mathcal{Y}_i = \mathcal{F} \left( \mathcal{X}_i \right),  \, i=0,1,\dotsb, 2L, \\
\tagref{ch3:sut_y_transformed_mean} & \hat{\mathbf{\eta}} = \sum\limits_{i=0}^{2L} W_i^{s} \, \mathcal{Y}_i \, ,  \\
\tagref{ch3:ut_y_transformed_cov} & \hat{\mathbf{P}}_{\eta} =  \sum\limits_{i=0}^{2L} W_i^s \, \left( \mathcal{Y}_i - \hat{\mathbf{\eta}} \right) \left(  \mathcal{Y}_i - \hat{\mathbf{\eta}} \right ) ^T + \left( 1+ \beta - \alpha^2 \right) \left( \mathcal{Y}_0 - \hat{\mathbf{\eta}} \right) \left( \mathcal{Y}_0 - \hat{\mathbf{\eta}} \right)^T  \, . 
\end{align}
\end{subequations}  

The accuracy analysis of the SUT can be conducted in a similar way to that in \S~\ref{ch3:sec_ut_accuracy_analysis}, but with the weights $W_i$ of the UT therein replaced by the weights $W_i^s$ of the SUT. Thus here we do not repeat it.

Since the UT can be considered as a special case of the SUT, hereafter we will use the SUT in general discussions and drop the superscripts in the weights of the SUT.

\section{Scaled Unscented Kalman filter as the approximate solution} \label{ch3:sec_UKF}
Applying the SUT to the propagation step of RBE leads to the \sindex{scaled unscented Kalman filter} (SUKF). Without loss of generality, we assume that at instant $k-1$, a set of sigma points $\left \{ \mathcal{X}_{k-1,i}^a \right \}_{i=0}^{2L_{k-1}}$ with respect to the quartet $\left( \alpha, \lambda, \hat{\mathbf{x}}_{k-1}^a,  {\mathbf{S}}_{k-1}^{xa} \right)$ is available, where $\alpha$ and $\lambda$ are the same parameters as in Eq.~(\ref{ch3:sut_sigma_points}), $\hat{\mathbf{x}}_{k-1}^a$ is the analysis mean, and ${\mathbf{S}}_{k-1}^{xa}$ is a square root matrix (with $L_{k-1}$ column vectors) of the analysis error covariance $\hat{\mathbf{P}}_{k-1}^a$. We also assume that the associated weights are $\left \{ W_{k-1,i}\right \}_{i=0}^{2L_{k-1}}$.     

Following \cite{Julier2000,Julier2004,Merwe2004,VanMerwe2001}, the main procedures of the SUKF are also split into the propagation and filtering steps.

\subsection{Propagation step} \label{ch3:sec_SUKF_progagation_step}
The ensemble mean $\hat{\mathbf{x}}_{k}^b$ and covariance $ \hat{\mathbf{P}}_k^b$ are evaluated according to the following formulae:
\begin{subequations} 
\begin{align}
& \mathbf{x}_{k,i}^b =  \mathcal{M}_{k,k+1} \left( \mathcal{X}_{k-1,i}^a \right), i=0,\dotsb,2L_{k-1} \, , \\
\label{sut mean} & \hat{\mathbf{x}}_{k}^b = \sum\limits_{i=0}^{2L_{k-1}} W_{k-1,i} \mathbf{x}_{k,i}^b \, , \\
 \label{sut cov} & \hat{\mathbf{P}}_k^b=\sum\limits_{i=0}^{2L_{k-1}} W_{k-1,i} \left( \mathbf{x}_{k,i}^b - \hat{\mathbf{x}}_{k}^b \right) \left( \mathbf{x}_{k,i}^b - \hat{\mathbf{x}}_{k}^b \right)^T \\
\nonumber &\qquad + \left(1+\beta-\alpha^2 \right) \left( \mathbf{x}_{k,0}^b - \hat{\mathbf{x}}_{k}^b \right) \left( \mathbf{x}_{k,0}^b -  \hat{\mathbf{x}}_{k}^b \right)^T + \mathbf{Q}_k. 
\end{align}
\end{subequations}

To compute the Kalman gain $\mathbf{K}_k$, it is customary to first compute the cross covariance $\hat{\mathbf{P}}^{cr}_k$ and the projection covariance $\hat{\mathbf{P}}^{pr}_k$ \cite{Julier2000,Julier2004}, given by
\begin{subequations} \label{ch3:SUKF_cov_approx_cmp}
\begin{align}
\label{sut obv mean} \hat{\mathbf{y}}_{k} = & \sum\limits_{i=0}^{2L_{k-1}} W_{k-1,i} \mathcal{H}_k \left( \mathbf{x}_{k,i}^b \right), \\
\label{sut cross covariance} \hat{\mathbf{P}}^{cr}_k = & \sum\limits_{i=0}^{2L_{k-1}} W_{k-1,i} \left( \mathbf{x}_{k,i}^b - \hat{\mathbf{x}}_{k}^b \right) \left( \mathcal{H}_k \left( \mathbf{x}_{k,i}^b\right) - \hat{\mathbf{y}}_{k} \right)^T \\
\nonumber &+ \left(1+\beta-\alpha^2 \right) \left( \mathbf{x}_{k,0}^b - \hat{\mathbf{x}}_{k}^b \right) \left( \mathcal{H}_k \left( \mathbf{x}_{k,0}^b\right) - \hat{\mathbf{y}}_{k} \right)^T, \\
 \label{sut projection covariance} \hat{\mathbf{P}}^{pr}_k =  & \sum\limits_{i=0}^{2L_{k-1}} W_{k-1,i} \left( \mathcal{H}_k \left( \mathbf{x}_{k,i}^b\right) - \hat{\mathbf{y}}_{k} \right) \left( \mathcal{H}_k \left( \mathbf{x}_{k,i}^b\right) - \hat{\mathbf{y}}_{k} \right)^T \\
\nonumber &+ \left(1+\beta-\alpha^2 \right) \left( \mathcal{H}_k \left( \mathbf{x}_{k,0}^b\right) - \hat{\mathbf{y}}_{k} \right) \left( \mathcal{H}_k \left( \mathbf{x}_{k,0}^b\right) - \hat{\mathbf{y}}_{k} \right)^T.
\end{align}
\end{subequations}

As in the ensemble square root filter (EnSRF), we re-write the above covariances in terms of some square root matrices. To this end, we introduce two square roots, $\mathbf{S}^{x}_k$ and $\mathbf{S}^{h}_k$, which are defined as 
\begin{subequations}
\begin{align}
\label{sut SRX} \mathbf{S}^{x}_k = & \left[ \sqrt{W_{k-1,0}^{\alpha \beta}} \left( \mathbf{x}_{k,0}^b - \hat{\mathbf{x}}_{k}^b \right), \sqrt{W_{k-1,1}} \left( \mathbf{x}_{k,1}^b - \hat{\mathbf{x}}_{k}^b \right), \dotsb, \sqrt{W_{k-1,2L_{k-1}}} \left( \mathbf{x}_{k,2L_{k-1}}^b - \hat{\mathbf{x}}_{k}^b \right) \right], \\
\nonumber  \label{sut SRH} \mathbf{S}^{h}_k = & \left[ \sqrt{W_{k-1,0}^{\alpha \beta}} \left( \mathcal{H}_k \left ( \mathbf{x}_{k,0}^b \right )- \hat{\mathbf{y}}_{k} \right), \sqrt{W_{k-1,1}} \left( \mathcal{H}_k \left ( \mathbf{x}_{k,1}^b \right )- \hat{\mathbf{y}}_{k} \right), \right.      \\
& \left. \dotsb, \sqrt{W_{k-1,2L_{k-1}}} \left( \mathcal{H}_k \left ( \mathbf{x}_{k,2L_{k-1}}^b \right )- \hat{\mathbf{y}}_{k} \right) \right],
\end{align}
\end{subequations}
where $W_{k-1,0}^{\alpha \beta} = W_{k-1,0}+1+\beta-\alpha^2$. Then the covariances can be re-written as
\begin{subequations}
\begin{align}
\label{sut SQ cov} & \hat{\mathbf{P}}_k^b = \mathbf{S}^{xb}_k \left( \mathbf{S}^{xb}_k \right)^T= \mathbf{S}^{x}_k \left( \mathbf{S}^{x}_k \right)^T + \mathbf{Q}_k,\\
\label{sut SQ cross cov} & \hat{\mathbf{P}}^{cr}_k =  \mathbf{S}^{x}_k \left( \mathbf{S}^{h}_k \right)^T, \\
\label{sut SQ projection cov} & \hat{\mathbf{P}}^{pr}_k =   \mathbf{S}^{h}_k \left( \mathbf{S}^{h}_k \right)^T, 
\end{align}
\end{subequations}
where $\mathbf{S}^{xb}_k$ is a square root of $\hat{\mathbf{P}}_k^b$, which can be obtained by letting $ \mathbf{S}^{xb}_k = \sqrt{ \mathbf{S}^{x}_k \left( \mathbf{S}^{x}_k \right)^T + \mathbf{Q}_k }$, following the numerical scheme in \S~\ref{ch1:SRKF} \footnote{In the context of the SUKF, there is actually no need to compute $\mathbf{S}^{xb}_k$ \cite{Julier2000,Julier2004}. This, however, is not true for the divided difference filters, as will be seen in the next chapter.}. In particular, if there is no dynamical noise, then it is customary to let $\mathbf{S}^{xb}_k = \mathbf{S}^{x}_k$. 

Finally, the Kalman gain  $\mathbf{K}_k$ can be calculated in terms of the square roots such that
\begin{equation} \label{Kalman gain}
\begin{split}
\mathbf{K}_k & = \hat{\mathbf{P}}^{cr}_k  \left( \hat{\mathbf{P}}^{pr}_k + \mathbf{R}_k\right)^{-1} \\
&= \mathbf{S}^{x}_k \left( \mathbf{S}^{h}_k \right)^T  \left( \mathbf{S}^{h}_k \left( \mathbf{S}^{h}_k \right)^T + \mathbf{R}_k \right)^{-1}.
\end{split}
\end{equation}

\subsection{Filtering step}  \label{ch3:sec_SUKF_filtering_step}

Once a new observation is available, one updates the sample mean and sample covariance of the background, so that
\begin{subequations}
\begin{align}
\label{analysis mean} & \hat{\mathbf{x}}_k^{a} =  \hat{\mathbf{x}}_k^{b} + \mathbf{K}_k  \left( \mathbf{y}_k - \mathcal{H}_k \left( \hat{\mathbf{x}}_k^{b} \right) \right),\\
\label{analysis cov} & \hat{\mathbf{P}}_k^a = \hat{\mathbf{P}}_k^b -  \mathbf{K}_k \left( \hat{\mathbf{P}}^{cr}_k \right)^T .
\end{align}
\end{subequations}
To obtain a square root $\mathbf{S}^{xa}_k$ of $\hat{\mathbf{P}}_k^a$, one may adopt the numerical scheme in \S~\ref{ch1:SRKF} to factorize the updated covariance as $\hat{\mathbf{P}}_k^a=\mathbf{S}^{xa}_k \left( \mathbf{S}^{xa}_k \right)^T$. 

Having the updated sample mean $\hat{\mathbf{x}}_k^{a}$ and a square root $\mathbf{S}^{xa}_k$, one can generate a new set of sigma points $\{ \mathcal{X}_{k,i}^a \}_{i=0}^{2L_{k}}$ at instant $k$ with respect to the quartet $\left(\alpha,\lambda, \hat{\mathbf{x}}_{k}^a, \mathbf{S}_{k}^{xa}\right)$, and compute the associated weights $\{ W_{k,i}\}_{i=0}^{2L_{k}}$. Then by propagating $\{ \mathcal{X}_{k,i}^a \}_{i=0}^{2L_{k}}$ forward, one can start a new assimilation cycle at instant $k+1$. 

\subsection{Summary of the scaled unscented Kalman filter}
We summarize the main procedures of the SUKF as follows:\\ \\
Propagation step:
\begin{subequations} \label{ch3:SUKF_propagation}
\begin{align}
& \mathbf{x}^b_{k,i} =  \mathcal{M}_{k,k-1} \left( \mathcal{X}_{k-1,i}^a \right),~i=0,\dotsb, 2L_{k-1} \, , \\
 &\hat{\mathbf{x}}_k^b = \sum\limits_{i=0}^{2L_{k-1}} W_{k-1,i} \mathbf{x}^b_{k,i} \, , \\
& \hat{\mathbf{y}}_{k} = \sum\limits_{i=0}^{2L_{k-1}} W_{k-1,i} \mathcal{H}_k \left( \mathbf{x}_{k,i}^b \right), \\
& \mathbf{S}^{x}_k = \left[ \sqrt{W_{k-1,0}^{\alpha \beta}} \left( \mathbf{x}_{k,0}^b - \hat{\mathbf{x}}_{k}^b \right), \sqrt{W_{k-1,1}} \left( \mathbf{x}_{k,1}^b - \hat{\mathbf{x}}_{k}^b \right), \right. \\
\nonumber & \qquad \left. \dotsb, \sqrt{W_{k-1,2L_{k-1}}} \left( \mathbf{x}_{k,2L_{k-1}}^b - \hat{\mathbf{x}}_{k}^b \right) \right], \\
& \mathbf{S}^{xb}_k = \sqrt{ \mathbf{S}^{x}_k \left( \mathbf{S}^{x}_k \right)^T + \mathbf{Q}_k } \, , \\
& \mathbf{S}^{h}_k = \left[ \sqrt{W_{k-1,0}^{\alpha \beta}} \left( \mathcal{H}_k \left ( \mathbf{x}_{k,0}^b \right )- \hat{\mathbf{y}}_{k} \right), \sqrt{W_{k-1,1}} \left( \mathcal{H}_k \left ( \mathbf{x}_{k,1}^b \right )- \hat{\mathbf{y}}_{k} \right), \right.      \\
\nonumber & \qquad \left. \dotsb, \sqrt{W_{k-1,2L_{k-1}}} \left( \mathcal{H}_k \left ( \mathbf{x}_{k,2L_{k-1}}^b \right )- \hat{\mathbf{y}}_{k} \right) \right] \, , \\
& \hat{\mathbf{P}}_k^b = \mathbf{S}^{xb}_k \left( \mathbf{S}^{xb}_k \right)^T \, ,\\ 
& \hat{\mathbf{P}}^{cr}_k =  \mathbf{S}^{x}_k \left( \mathbf{S}^{h}_k \right)^T \, , \\
&\mathbf{K}_k = \mathbf{S}_k^x \left( \mathbf{S}_{k}^{h} \right)^T \left( \mathbf{S}_{k}^{h} \left( \mathbf{S}_{k}^{h} \right)^T + \mathbf{R}_k \right)^{-1} \, . 
\end{align}
\end{subequations} \\ \\
Filtering step:
\begin{subequations} \label{ch3:SUKF_filtering}
\begin{align}
& \hat{\mathbf{x}}_k^a = \hat{\mathbf{x}}_k^b + \mathbf{K}_k \left ( \mathbf{y}_k - \mathcal{H}_k \left( \hat{\mathbf{x}}_k^b \right) \right ) \, , \\
& \hat{\mathbf{P}}_k^a = \hat{\mathbf{P}}_k^b -  \mathbf{K}_k \left( \hat{\mathbf{P}}^{cr}_k \right)^T \, , \\
&\mathbf{S}_{k}^{xa} = \sqrt{\hat{\mathbf{P}}_k^a} \, .
\end{align}
\end{subequations} \\ 
Analysis scheme:\\
\indent Sigma points:
\begin{equation} 
\begin{split}
& \mathcal{X}_{k,0}^a = \hat{\mathbf{x}}_k^a \, ,\\
& \mathcal{X}_{k,i}^a = \hat{\mathbf{x}}_k^a + \alpha \sqrt{L_{k}+\lambda} \left(  \mathbf{S}_k^{xa} \right)_i, \, i=1, 2, \dotsb, L_{k} \, ,\\
& \mathcal{X}_{k,i}^a = \hat{\mathbf{x}}_k^a - \alpha \sqrt{L_{k}+\lambda} \left(  \mathbf{S}_k^{xa} \right)_{i-L_{k}}, \, i=L_{k}+1, L_{k}+2, \dotsb, 2L_{k} \, .\\
\end{split}
\end{equation} \\
\indent Associated weights:
\begin{equation} 
\begin{split}
\begin{split}
&W_{k,0} = \frac{\lambda}{\alpha^2(L_{k}+\lambda)} + 1-\dfrac{1}{\alpha^2} \, ,\\
&W_{k,i} = \frac{1}{2 \alpha^2 \left(L_{k}+\lambda  \right)} \,,~i=1,2,\dotsb, 2L_{k} \, . \\
\end{split}
\end{split}
\end{equation}

\section{Reduced rank scaled unscented Kalman filter for high dimensional systems} \label{ch3:sec_reduced_sukf}
In practice, one may not wish to apply the SUKF directly to high dimensional systems. To see this, recall that in the generation of sigma points, one requirement is that the number of sigma points be larger than twice the dimension of the system under assimilation in order to avoid rank deficiency. This may be infeasible, and sometimes actually unnecessary \footnote{Because some of the system states may be correlated, so that the covariance of the system states itself may not be a full rank matrix.}, for data assimilation in high dimensional systems. Therefore, some modifications have to be introduced to the SUKF in high dimensional systems. To do this, we follow the idea in \cite{Luo-ensemble}. We perform a truncated singular value decomposition (SVD) on a covariance matrix, as described below, to generate sigma points with controlled numbers. The SUKF producing sigma points in this way will be called the reduced rank SUKF, which will be used in subsequent numerical experiments. For convenience, we may sometimes use ``SUKF'' to mean the ``reduced rank SUKF'' when it causes no confusion.  

Without loss of generality, we assume that at time instant $k-1$, we have a set of $(2l_{k-1}+1)$ sigma points, in terms of $\mathcal{X}_{k-1}^a = \left \{\mathcal{X}_{k-1,i}^a \right\}_{i=0}^{2l_{k-1}}$ with the corresponding weights $\mathcal{X}_{k-1}^a = \left \{W_{k-1,i} \right\}_{i=0}^{2l_{k-1}}$, where the choice of $l_{k-1}$ will be discussed later.

\subsection{Propagation and filtering steps}
The procedures at the propagation and filtering steps of the reduced rank SUKF are the same as those of the SUKF. We first define a set of forecasts of the propagated sigma points
 \begin{equation}\label{forecasts DA}
\mathbf{X}_k^b = \left \{ \mathbf{x}_{k,i}^b: \mathbf{x}_{k,i}^b =  \mathcal{M}_{k,k-1} \left( \mathcal{X}_{k-1,i}^a\right), i=0,\dotsb,2l_{k-1} \right \},
\end{equation}
based upon which the background sample mean $\hat{\mathbf{x}}_k^b$, sample covariance $\hat{\mathbf{P}}_k^b$  and the Kalman gain $\mathbf{K}_k$ can be computed according to the formulae in \S~\ref{ch3:sec_SUKF_progagation_step}, but with $L_{k-1}$ therein replaced by $l_{k-1}$. Then the analysis mean $\hat{\mathbf{x}}_k^a$ and covariance $\hat{\mathbf{P}}_k^a$ are updated using the formulae in \S~\ref{ch3:sec_SUKF_filtering_step}. 

\subsection{Analysis scheme}
To generate a set of sigma points $\mathcal{X}_{k}^a = \left \{\mathcal{X}_{k,0}^a, \dotsb,\mathcal{X}_{k,2l_{k}}^a \right \}$ with controlled number, the truncated singular value decomposition (SVD) is conducted on $\hat{\mathbf{P}}_k^a$.  Let $\hat{\mathbf{P}}_{k}^{a}$ be an $m \times m$ matrix, then it can be decomposed as
\begin{equation}
\hat{\mathbf{P}}_{k}^{a} = \mathbf{E}_k^a \mathbf{D}_k^a \left(\mathbf{E}_k^a \right)^T,
\end{equation}
where $\mathbf{D}_K^a = \text{diag} (\sigma_{k,1}^2, \dotsb, \sigma_{k,m}^2)$ is a diagonal matrix consisting of the eigenvalues $\sigma_{k,i}^2$ of $\hat{\mathbf{P}}_k^a$, which are arranged in descending order, i.e., $\sigma_{k,i}^2 \ge \sigma_{k,j}^2 \ge 0$ for $i>j$, and $\mathbf{E}_k^a = \left[\mathbf{e}_{k,1}, \dotsb,  \mathbf{e}_{k,m} \right] $  is the matrix consisting of the corresponding eigenvectors $\mathbf{e}_{k,i}$. A new set of sigma points $\mathcal{X}_{k}^a = \left \{\mathcal{X}_{k,0}^a, \dotsb,\mathcal{X}_{k,2l_{k}}^a \right \}$ is then generated as follows:
\begin{equation} \label{ch3:reduced_sut_sigma_point}
\begin{split}
& \mathcal{X}_{k,0}^a = \hat{\mathbf{x}}_k^{a},\\
& \mathcal{X}_{k,i}^a =  \hat{\mathbf{x}}^{a}_k + \alpha \left( l_k + \lambda \right)^{1/2} \sigma_{k,i} \mathbf{e}_{k,i} \, , ~ i=1, \dotsb, l_k ,\\
& \mathcal{X}_{k,i}^a =  \hat{\mathbf{x}}^{a}_k - \alpha \left( l_k + \lambda \right)^{1/2} \sigma_{k,i-l_k} \mathbf{e}_{k,i-l_k} \, , ~ i=l_k+1, \dotsb, 2l_k ,\\
\end{split}
\end{equation}
where $l_k$ is an integer to be specified. Note that using the new sigma points as the analysis ensemble, the sample mean of sigma points is equal to $\hat{\mathbf{x}}_k^{a}$. Thus the SUKF is an unbiased ensemble filter according to the definition in \cite{Livings-unbiased} (also see the discussion in \S~\ref{ch2:sec_EnSRF}). 

It is worth noting that, Eq.~(\ref{ch3:reduced_sut_sigma_point}) only requires the first $l_k$ pairs of  eigenvalues and eigenvectors, rather than the full spectrum. Therefore, to reduce the computational cost in high dimensional problems, some fast SVD algorithms, e.g.,  the Lanczos or block Lanczos algorithm (cf. \cite{Cullum1985} and \cite[ch 9]{Golub-matrix}), can be adopted to compute the first $l_k$ pairs of eigenvalues and eigenvectors only (for example, see \cite{Treebushny-construction}). This may reduce the computational cost of the SUKF in high dimensional systems \footnote{To see this, we consider a simple scenario, where the $m$-dimensional dynamical system is given by $\mathbf{x}_{k+1} = \mathbf{A} \, \mathbf{x}_{k}$. Here $\mathbf{A}$ is supposed to be a full rank matrix (otherwise the model size can be reduced). Then the computational complexity of propagating one state point forward is in the order of $m^2$, denoted by $\mathcal{O}(m^2)$. Therefore for the full rank SUKF, the computational complexity of propagating all sigma points forward is $\mathcal{O}(m^3)$. In contrast, for the reduced rank SUKF, by using the Lanczos algorithm (or its variants) to compute the eigenvalues and eigenvectors, the computational complexity of one iteration is at most $\mathcal{O}(m^2)$, or even less if $\mathbf{A}$ is a sparse matrix \cite[p.~35]{Cullum1985}. Thus to evaluate the first $l_k$ pairs of eigenvalues and eigenvectors, the computational complexity is $l_k \times \bar{n}^{it} \times \mathcal{O}(m^2)$, where $\bar{n}^{it}$ is the average number of iterations in executing the Lanczos algorithm. The computational complexity of evolving $2l_k+1$ sigma points forward is $(2l_k+1) \times \mathcal{O}(m^2)$. Therefore, for the reduced rank SUKF, the overall computational complexity of generating sigma points and propagating them forward is $ [ l_k \times (\bar{n}^{it}+2) +1 ]\times \mathcal{O}(m^2)$, which can be (much) less than $\mathcal{O}(m^3)$ in some high dimensional systems, e.g., a weather forecasting model with millions of state variables (or even more), while the sizes of $l_k$ and $\bar{n}^{it}$ may be in the orders of $10^2$ and $10^3$, respectively, or even less (as an example, see \cite{Zhang1998} for the convergence of a Lanczos algorithm).}.

For convenience, we call $l_k$ the \sindex{truncation number} (at time $k$), which influences the performance of the reduced rank SUKF. To see this, we let 
\begin{equation}
\tilde{\mathbf{P}}_k^a = \sum\limits_{i=1}^{l_k} \sigma_{k,i}^2 \mathbf{e}_{k,i} \left( \mathbf{e}_{k,i} \right)^T \, ,
\end{equation}
which can be considered as an approximation to the matrix 
\begin{equation}
\hat{\mathbf{P}}_k^a = \sum\limits_{i=1}^{m} \sigma_{k,i}^2 \mathbf{e}_{k,i} \left( \mathbf{e}_{k,i} \right)^T \, . 
\end{equation}
If $l_k$ is too small, some important information of  $\hat{\mathbf{P}}_k^a$, in terms of $\sigma_{k,i}^2 \mathbf{e}_{k,i} \left( \mathbf{e}_{k,i} \right)^T$ for $i > l_k$, will be lost. However, as the computational cost is also a concern, it is not desirable for $l_k$ to get too large. Moreover, in many situations, if $l_k$ is large enough, $\sigma_{k,l_k}^2$ may be already very small compared with the leading eigenvalues. Thus the improvement obtained by further increasing $l_k$ becomes negligible. In this sense, one may choose a moderate value for $l_k$ to achieve a tradeoff between accuracy and efficiency. In our implementation, to prevent $l_k$ getting too large or too small, we also pre-specify some upper and lower bounds, denoted by $l_l$ and $l_u$ respectively, to guarantee that $l_k$ falls within an acceptable range $l_l \le l_k \le l_u$ \cite{Luo-ensemble} \footnote{The choice of $l_l$ and $l_u$ itself may depend on our experience and needs in running the system under assimilation, and hence is case-dependent in general.}. 

Another point to note is that, the approximate matrix $\tilde{\mathbf{P}}_k^a$ based on the set of sigma points $\mathcal{X}_{k}^a = \left \{\mathcal{X}_{k,0}^a, \dotsb,\mathcal{X}_{k,2l_{k}}^a \right \}$ is of rank $l_k$ because of the symmetry in sigma points. But at the next assimilation cycle, the background covariance $\hat{\mathbf{P}}_{k+1}^b$ evaluated based on the propagated sigma points $\mathbf{X}_{k+1}^b = \left \{\mathcal{M}_{k+1,k}(\mathcal{X}_{k,0}^a), \dotsb,\mathcal{M}_{k+1,k}(\mathcal{X}_{k,2l_{k}}^a) \right \}$ may have a rank higher than $l_k$ for a nonlinear transition operator $\mathcal{M}_{k+1,k}$. This is because, in the set $\mathcal{X}_{k}^a = \left \{\mathcal{X}_{k,0}^a, \dotsb,\mathcal{X}_{k,2l_{k}}^a \right \}$ there exists redundant information due to the symmetry in sigma points. But after propagation, the symmetry will normally be broken thanks to the nonlinearity of $\mathcal{M}_{k+1,k}$. The set $\mathbf{X}_{k+1}^b = \left \{\mathcal{M}_{k+1,k}(\mathcal{X}_{k,0}^a), \dotsb,\mathcal{M}_{k+1,k}(\mathcal{X}_{k,2l_{k}}^a) \right \}$ may thus explore more information of $\mathcal{M}_{k+1,k}$ than any one of its (strict) subsets does. Therefore the rank of $\hat{\mathbf{P}}_{k+1}^b$ can be higher than that of $\tilde{\mathbf{P}}_k^a$. On the other hand, as illustrated in \S~\ref{ch2:sec_covariance_filtering} (cf. Fig.~\ref{fig:eigenvalues}), by conducting covariance filtering one can in effect increase the rank of a sample covariance. For these two reasons, one may conduct SVDs on the analysis covariances without worrying about the deficiency of their ranks.

In principle, the choice of the truncation number $l_k$ may be determined by the geometry of a dynamical system in phase space. Take a dynamical system with a chaotic attractor as an example, the attractor dimension (e.g., the Hausdorff dimension) may be substantially lower than the (topological) dimension of the dynamical system. Suppose that at the $k$-th assimilation cycle, the local dimension of the trajectory around the analysis $\hat{\mathbf{x}}_k^a$ is $d_k$, then in principle one can choose $l_k$ to be around $\min(d_k, l_u)$, where $\min(a,b)$ means the minimum between $a$ and $b$, and $l_u$ is an acceptable upper bound of $l_k$ for practical computation. Therefore, if $d_k$ is not too large ($d_k < l_u$), one can let $l_k $ be close to $d_k$ so that the number of sigma points is about $2d_k+1$. In this case, the number of sigma points is not too large, but the approximation matrix $\tilde{\mathbf{P}}_k^a$ captures the structure of  $\hat{\mathbf{P}}_k^a$ well such that $\tilde{\mathbf{P}}_k^a \approx \hat{\mathbf{P}}_k^a$. However, if the local attractor dimension still appears too large ($d_k > l_u$) for the purpose of computation, the upper bound $l_u$ will work to prevent the number of sigma points ($2l_u+1$) getting too large, but at the cost of deteriorating the quality of covariance approximation.

In practice, it is infeasible to compute the local attractor dimension $d_k$ at each assimilation cycle. One may instead use some ad hoc criterion to choose the value of $l_k$. In our implementation, we let $l_k$ be an integer such that
\begin{equation} \label{enukf:threshold}
\begin{split}
& \sigma_{k,i}^2 > \text{trace} \left( \hat{\mathbf{P}}_k^a \right) / \Gamma_k \, ,~ i=1, \dotsb, l_k \, ,\\
& \sigma_{k,i}^2 \le \text{trace} \left( \hat{\mathbf{P}}_k^a \right) / \Gamma_k \, ,~  i>l_k +1 \, ,
\end{split}
\end{equation}
where $\Gamma_k$ is the threshold at the $k$-th cycle (we will discuss how to choose $\Gamma_k$ later). This is equivalent to saying that we generate sigma points based on the eigenvectors whose corresponding eigenvalues are larger than a specified tolerance. 

Under the assumption of Gaussianity, it can be verified that the perturbations of sigma points with respect to their center $\mathcal{X}_{k,0}^a$, in terms of $\alpha (l_k+\lambda)^{1/2} \sigma_{k,i} \mathbf{e}_{k,i}$ for $i=1,\dotsb, l_k$, are equally likely in the sense that their probabilities, in terms of  
\begin{equation}
p(\delta \mathbf{x}) = \left( 2 \pi \right)^{m/2} (\det \hat{\mathbf{P}}_k^a)^{-1/2} \text{exp} \left\{ -\dfrac{1}{2} \left( \delta \mathbf{x}\right)^T \left( \hat{\mathbf{P}}_k^a \right)^{-1} \left( \delta \mathbf{x}\right) \right\} \, , 
\end{equation}
are the same (also see the discussions in \cite{Wang-which}), where $\det \bullet$ means the determinant of a matrix. Therefore it is natural to assign an identical weight to all the perturbations. Consequently, in the spirit of Eq.~(\ref{ch3:sut_weights}), the weights associated with sigma points are allocated as follows:
\begin{equation} \label{ch3:reduced_sut_weights}
\begin{split}
&W_{k,0} = \frac{\lambda}{\alpha^2(l_k+\lambda)} + 1-\dfrac{1}{\alpha^2},\\
&W_{k,i} = \frac{1}{2 \alpha^2 \left(l_k+\lambda  \right)}, \,  ~ i=1,2,\dotsb, 2l_k. \\
\end{split}
\end{equation}

In summary, the analysis scheme of the reduced rank SUKF is given as follows:\\
Generation of sigma points:
\begin{equation} \tagref{ch3:reduced_sut_sigma_point}
\begin{split}
& \mathcal{X}_{k,0}^a = \hat{\mathbf{x}}_k^{a},\\
& \mathcal{X}_{k,i}^a =  \hat{\mathbf{x}}^{a}_k + \alpha \left( l_k + \lambda \right)^{1/2} \sigma_{k,i} \mathbf{e}_{k,i}, \,~ i=1, \dotsb, l_k ,\\
& \mathcal{X}_{k,i}^a =  \hat{\mathbf{x}}^{a}_k - \alpha \left( l_k + \lambda \right)^{1/2} \sigma_{k,i-l_k} \mathbf{e}_{k,i-l_k}, \, ~ i=l_k+1, \dotsb, 2l_k ,\\
\end{split}
\end{equation}
Allocation of associated weights:
\begin{equation} \label{ch3:reduced_sut_weights}
\begin{split}
&W_{k,0} = \frac{\lambda}{\alpha^2(l_k+\lambda)} + 1-\dfrac{1}{\alpha^2},\\
&W_{k,i} = \frac{1}{2 \alpha^2 \left(l_k+\lambda  \right)}, \,~ i=1,2,\dotsb, 2l_k. \\
\end{split}
\end{equation}

\section{Example: Assimilating the 40-dimensional Lorenz-Emanuel 98 system} \label{ch3:sec_ex}

\subsection{The testbed and the measures of filter performance} \label{ch3:sec_testbed}
The dynamical and observation systems are the same as those in \S~\ref{ch2:sec_testbed}. That is, the dynamical system (LE 98) is governed by
\begin{equation} \tagref{ch2:ex_LE98}
\frac{dx_i}{dt} = \left( x_{i+1} - x_{i-2} \right) x_{i-1} - x_i + 8, \,~\text{for}~ i=1, \dotsb, 40 \, , 
\end{equation}
while the observation system is
\begin{equation} \tagref{ch2:ex_observer}
\mathbf{y}_k = \mathbf{x}_k + \mathbf{v}_k \, ,
\end{equation} 
where $\mathbf{v}_k$ follows the Gaussian distribution $N \left( \mathbf{v}_k: \mathbf{0}, \mathbf{I} \right)$. 

We integrate the dynamical system Eq.~(\ref{ch2:ex_LE98}) through a fourth-order Runge-Kutta method \cite[Ch.~16]{Vetterling-numerical}. We choose the length of the integration window to be $100$ dimensionless units, and the integration time step to be $0.05$. For notational convenience, we denote this setting by $0:0.05:100$. Similar notations will also be used later. We make the observations of the dynamical system at every integration step.

We also adopt the relative rmse $e_r$ and rms ratio $R$ defined in \S~\ref{ch2:two_measures} as the measures of filter performance. In the context of the reduced rank SUKF, they are given by
\begin{equation}\tagref{Eq:reltiave rmse}
e_r=\frac{1}{k_{max}}\sum\limits_{k=1}^{k_{max}} \lVert\hat{\mathbf{x}}_k^{a}-{\mathbf{x}}_k^{tr}\rVert_2/ \lVert{\mathbf{x}}_k^{tr}\rVert_2 
\end{equation}
and
\begin{equation} \label{ch3:R}
R = \dfrac{1}{k_{max}} \sum\limits_{k=1}^{k_{max}} \dfrac{ (2l_k +1) \, \lVert \hat{\mathbf{x}}_k^a - \mathbf{x}_{k}^{tr} \rVert_{2}}{\sum\limits_{i=1}^{2l_k +1} \lVert \mathcal{X}_{k,i}^a - \mathbf{x}_{k}^{tr} \rVert_{2}} \, ,
\end{equation}
respectively.

According to Eq.~(\ref{R_e}), the expectation $R_e$ of the rms ratio is 
\[
R_e = \sqrt{(l_{eff}+1)/(2l_{eff}+1)} 
\]
by letting $n=2l_{eff}+1$ in Eq.~(\ref{R_e}), where $l_{eff}$ is the ``effective'' truncation number over the whole assimilation window. Hence, if the true system states are statistically indistinguishable from the corresponding sigma points, the values of $R$ and $R_e$ will be close to one another. Note that $R_e \approx 0.71$ for a sufficiently large $l_{eff}$. For simplicity, we let $l_{eff}$ equal the average of the truncation number $\bar{l}$, i.e., $l_{eff} = \bar{l} = \sum_{i=1}^{k_{max}} l_k / k_{max}$. Again, $R>R_e$ means that a sample covariance of sigma points underestimates the error in state estimation, while $R<R_e$ overestimates the error in state estimation \cite{Murphy-impact, Whitaker-ensemble}.

\subsection{Some issues in implementation}\label{ch3:sec_ex_implementation_issue}

\subsubsection{Positive semi-definiteness of the covariance matrices}
One issue in implementing the SUKF is to guarantee the positive semi-definiteness of the covariance matrices. To this end, first of all we require $\l_k + \lambda >0$ so that in Eq.~(\ref{ch3:reduced_sut_sigma_point}) the square root of $\l_k + \lambda$ is real. Also note, when computing the covariances (cf. Eqs. ~(\ref{sut cov}), (\ref{sut cross covariance}) and (\ref{sut projection covariance})), the effective weight of $ \mathbf{x}_{k+1,0}^b$ is $W_{k,0}+1+\beta-\alpha^2$ ($\beta \ge 0$). So we also require that $W_{k,0}+1+\beta-\alpha^2 \ge 0$, which, together with Eq.~(\ref{ch3:reduced_sut_weights}), is equivalent to saying
\begin{equation}
 \frac{\lambda}{\alpha^2(l_k+\lambda)} + 1-\dfrac{1}{\alpha^2} + 1+\beta-\alpha^2 \ge 0 \, .
\end{equation}
$l_k$ may take different values at different assimilation cycles. However, since $l_k$ is bounded such that $0<l_l \le l_k \le l_u$, with some algebra, one can obtain the sufficient conditions 
\begin{equation} 
\begin{split}
& \lambda \ge - l_l + \dfrac{l_l}{\left( 2+\beta\right)^2} \; ,\\
& \alpha \ge \sqrt{ 2+\beta -\sqrt{ \left( 2+\beta \right)^2 - \dfrac{l_l}{l_l+\lambda} } } \, ,\\
&  \alpha \le \sqrt{ 2+\beta + \sqrt{ \left( 2+\beta \right)^2 - \dfrac{l_l}{l_l+\lambda} }} \, ,
\end{split}
\end{equation}
which guarantee the positive semi-definiteness of the covariances.

\subsubsection{The choice of the threshold $\Gamma_k$}
The choice of the threshold $\Gamma_k$ follows the work \cite{Luo-ensemble}. We begin by specifying a threshold $\Gamma_1$ at the first assimilation cycle. If $\Gamma_1$ is a proper value such that the corresponding truncation number $l_1$ satisfies $l_l \le l_1 \le l_u$, then we keep $\Gamma_1$ and at the next cycle we start with $\Gamma_2=\Gamma_1$. If $\Gamma_1$ is too small, so that $l_1<l_l$, then we increase it by replacing $\Gamma_1$ by $1.1 \Gamma_1 + 200$. We continue the replacement until $l_1$ falls into the specified range, or the number of replacement operations reaches $30$ (in which case we simply put $l_1=l_l$, regardless of the value of $\Gamma_1$). Similarly, if $\Gamma_1$ is too large, so that $l_1>l_u$, then we decrease it by replacing $\Gamma_1$ with $\Gamma_1/1.1 - 200$. We continue the replacement until $l_1$ falls in the specified range, or the number of the operations reaches $30$ (in which case we simply put $l_1=l_u$). After the adjustment, at the next cycle we start with $\Gamma_2=\Gamma_1$ and adjust it (if necessary) to let $l_2$ fall into the specified range, and so on.

\subsection{Numerical experiments and results} \label{ch3:sec_numerical_results}

\subsubsection{Effects of the inflation factor $\delta$ and the length scale $l_c$ on the performance of the reduced rank SUKF} \label{ch3:sec_ex_delta_vs_lc}

\begin{figure*}[!t]
\centering
\hspace*{-0.5in} \includegraphics[width=1.15\textwidth]{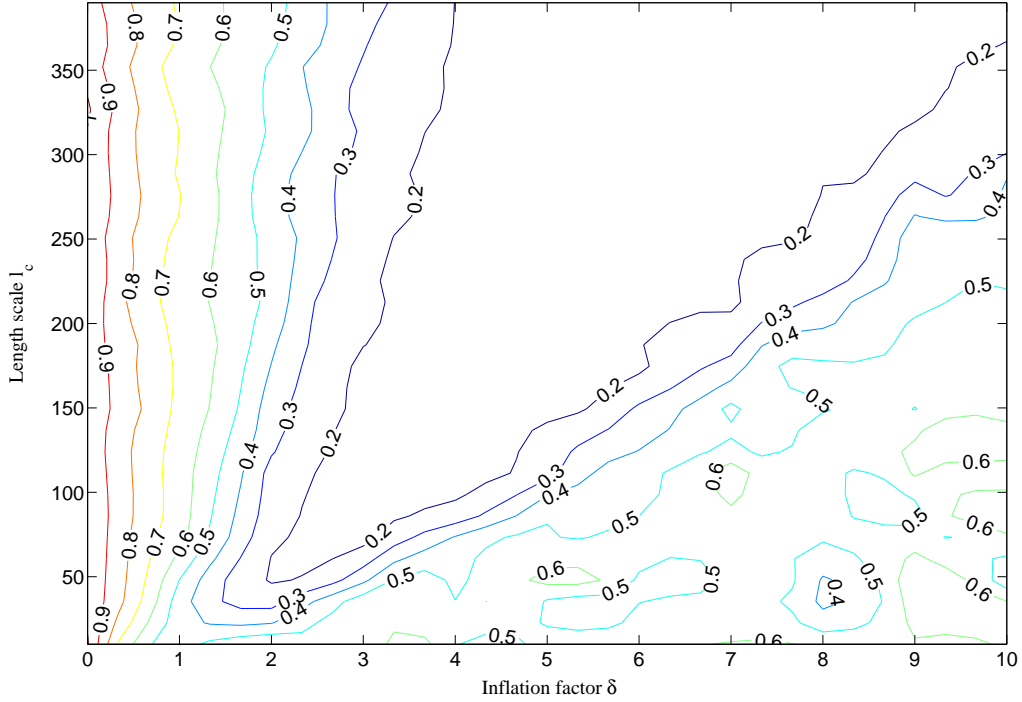} 
\caption{ \label{fig:ch3_SUKF_delta_vs_lc_rms} The relative rmse of the SUKF as a function of the inflation factor $\delta$ and the length scale $l_c$. }
\end{figure*} 

To improve the performance of the reduced rank SUKF, we also adopt the covariance inflation and filtering techniques introduced in \S~\ref{ch2:sec_two_techniques}. Here we first examine the effects of the inflation factor $\delta$ and the length scale $l_c$ on the performance of the SUKF. The parameters in the experiments are set as follows: the inflation factor $\delta$ increases from $0$ to $10$, with a fixed increment of $0.5$ each time. We denote this setting by $0:0.5:10$. The length scale $\l_c$ increases from $10$ to $400$, with a fixed increment of $20$ each time. This setting is thus denoted by $10:20:400$. Other (fixed) parameter values are $\alpha=1$, $\beta=2$, $\lambda=-2$, lower bound $l_l = 3$, upper bound $l_u=6$, and the threshold at the first assimilation cycle $\Gamma_1=1000$. 

To begin the assimilation, we randomly choose an initial condition to start a control run, and so obtain the true trajectory within the specified assimilation window. We then add some Gaussian noise drawn from the distribution $N \left(\mathbf{v}_k: \mathbf{0}, \mathbf{I} \right)$ to the true trajectory to generate the observations. The noise level (relative rmse) of the observations $e_r^{obv} \approx 0.22$. To start the SUKF, we also generate $6$ randomly perturbed initial conditions \footnote{Given the truth $\mathbf{x}_{1}^{th}$ at the first assimilation cycle, these $6$ different initial conditions are just the samples drawn from the distribution $N \left(\mathbf{v}_k: \mathbf{x}_{1}^{th}, \mathbf{I} \right)$.} as the background ensemble at the first assimilation cycle. This represents a typical scenario in data assimilation, where the ensemble size of the background is often (much) smaller than the dimension of the dynamical system. Note that, at the first cycle, there are no sigma points propagated from the previous cycle. Thus at the first assimilation cycle, we use the ensemble transform Kalman filter (ETKF) introduced in the previous chapter to update the sample mean and covariance of the background to the corresponding statistics of the analysis, and then generate sigma points accordingly. After propagating sigma points forward, the SUKF can start running recursively from the second assimilation cycle. 

First we examine the performance of the SUKF in terms of the relative rmse. We plot the relative rmse of the SUKF as a function of the inflation factor $\delta$ and the length scale $l_c$ in Fig.~\ref{fig:ch3_SUKF_delta_vs_lc_rms}. As one can see, when fixing $l_c$, the relative rmse of the SUKF also exhibits the U-turn behaviour as $\delta$ increases. This phenomenon was already explained in \S~\ref{ch2:ex_numerical_results}. On the other hand, when fixing $\delta$, and provided that $\delta$ is not large (say, $\delta <2$), the relative rmse is roughly insensitive to the change of $l_c$. For $2<\delta<4$, the relative rmse exhibits the U-turn behaviour as $l_c$ increases. For $\delta>4$, the relative rmse of the SUKF tends to decrease overall as $l_c$ increases.

\begin{figure*}[!t]
\centering
\hspace*{-0.5in} \includegraphics[width=1.15\textwidth]{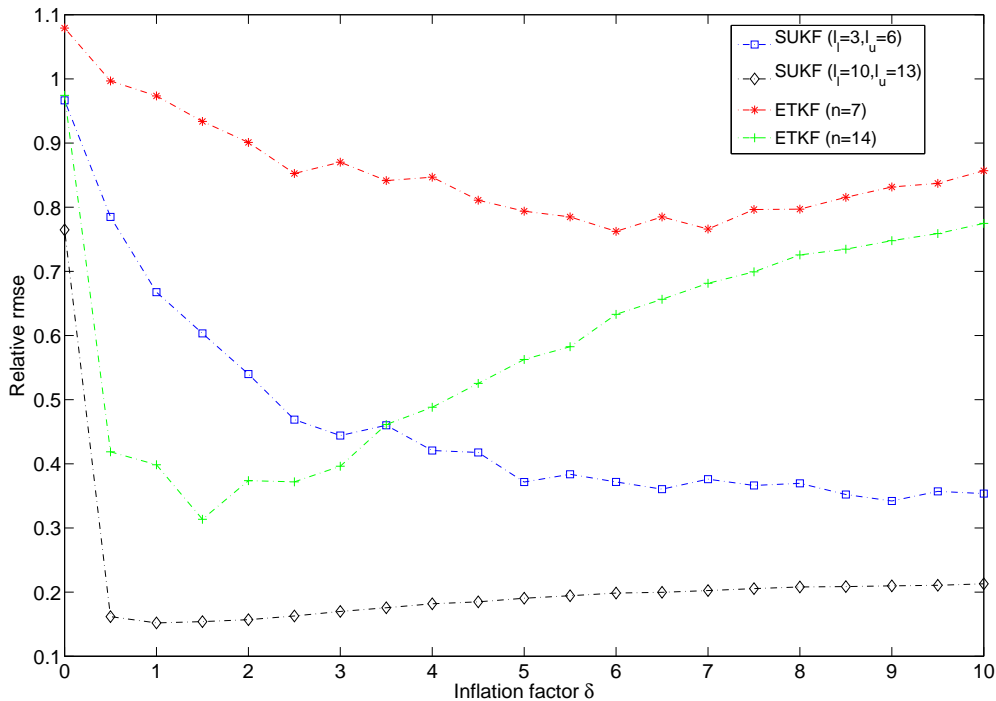} 
\caption{ \label{fig:EnUKF_rmse_vs_delta_no_inflation} The relative rms errors of the SUKF and the ETKF as functions of the covariance inflation factor $\delta$, but without any covariance filtering ($l_c = \infty$). Here the experiment setting of the SUKF is almost the same as that specified at the beginning of \S~\ref{ch3:sec_ex_delta_vs_lc}, except for that in one experiment (corresponding to the dash-dotted line in blue marked by squares), the initial ensemble size of the background is $n=6$, with the lower bound $l_l=3$ and the upper bound $l_u=6$, while in another experiment (corresponding to the dash-dotted line in black marked by diamonds), the initial ensemble size of the background is $n=10$, with $l_l=10$ and $l_u=13$. The experiment setting of the ETKF is also almost the same as that in~\S~\ref{ch2:ex_numerical_results}, but with the initial ensemble size of the background $n=7$ in one experiment (corresponding to the dash-dotted line in red marked by asterisks), and $n=14$ in another experiment (corresponding to the dash-dotted line in green marked by plus signs).}
\end{figure*} 

Comparing Fig.~\ref{fig:ch2_ETKF_delta_vs_lc_rms} with Fig.~\ref{fig:ch3_SUKF_delta_vs_lc_rms}, the SUKF does not consistently outperform the ETKF. This might be due to the following reasons. 
\begin{enumerate}[(a)]
\item The states of a nonlinear system do not strictly follow a Gaussian distribution, which violates the Gaussianity assumption in nonlinear Kalman filters (see \S~\ref{ch2:sec_ps_filtering_step}); 
\item The covariance filtering technique works better for the ETKF than for the SUKF (see the discussion below);
\item The amount of information in use (also see the discussion below). 
\end{enumerate}

In fact, a closer examination on Fig.~\ref{fig:ch2_ETKF_delta_vs_lc_rms} indicates that, the SUKF appears to be ``less dependent'' on the covariance filtering technique than the ETKF, in the sense that, to achieve a lower relative rmse (e.g. $e_r<0.2$), the length scale $l_c$ of the SUKF tends to be larger than that of the ETKF \footnote{From the discussion in \S~\ref{ch2:sec_covariance_filtering}, one can see that, given a covariance matrix $\mathbf{P}$, the components of the correlation matrix $\mathbf{\Phi}$ will be closer to $1$ for a larger length scale $l_c$, which means that the Schur product $\mathbf{P} \circ \mathbf{\Phi}$ is closer to the original matrix $\mathbf{P}$. In the extreme situation such that $l_c = + \infty$, one has $\mathbf{P} \circ \mathbf{\Phi} = \mathbf{P}$, which means that introducing covariance filtering does not change the covariance matrix $\mathbf{P}$ at all.}. The SUKF also appears to have a broader region than the ETKF where the filter does not diverge (i.e. $e_r < e_r^{obv} \approx 0.22$). 

A possible explanation of the above difference between the ETKF and the SUKF may be given based on the accuracy analysis in Appendix~\ref{appendix:accuracy analysis}, where we show that in the EnKF (including the ETKF), because of the effect of finite ensemble size, some spurious modes and bias exist in the estimation of a sample covariance. Thus the covariance filtering technique works well to reduce the effect by choosing relatively small length scales. In contrast, because of the symmetry in sigma points, those spurious modes and bias in the EnKF do not appear in the SUKF. Thus there is no need to change a sample covariance of the SUKF as much as that of the ETKF. Hence larger length scales will work better for the SUKF. 

In Fig.~\ref{fig:EnUKF_rmse_vs_delta_no_inflation}, we examine the situation where there is no covariance filtering conducted on both the SUKF and the ETKF. Note that in the SUKF, given $2l_k+1$ sigma points, only the first $l_k+1$ sigma points contain useful information because of the symmetry in sigma points (the information from the last $l_k$ sigma points are redundant). However, by propagating all sigma points forward through a nonlinear function, $2l_k+1$ propagated sigma points may explore more information about the nonlinear function compared with the choice of propagating $l_k+1$ sigma points forward only. Indeed, from Fig.~\ref{fig:EnUKF_rmse_vs_delta_no_inflation} one can see that, if the upper bound $l_u$ of the SUKF is equal to the ensemble size $n$ of the ETKF minus one, so that either $l_u+1=n=7$ or $l_u+1=n=14$ in Fig.~\ref{fig:EnUKF_rmse_vs_delta_no_inflation}, the SUKF always outperforms the ETKF. On the other hand, if the ensemble size $n$ in the ETKF is about equal to twice the upper bound $l_u$ plus one, for example, $n=14$ and $l_u=6$ in Fig.~\ref{fig:EnUKF_rmse_vs_delta_no_inflation}, the performance of the SUKF with $l_u=6$ is still comparable to that of the ETKF with $n=14$. Therefore in some situations, if it is inconvenient or expensive to produce background ensembles, the SUKF may be adopted to improve the performance of data assimilation.

\begin{figure*}[!t]
\centering
\hspace*{-0.5in} \includegraphics[width=1.15\textwidth]{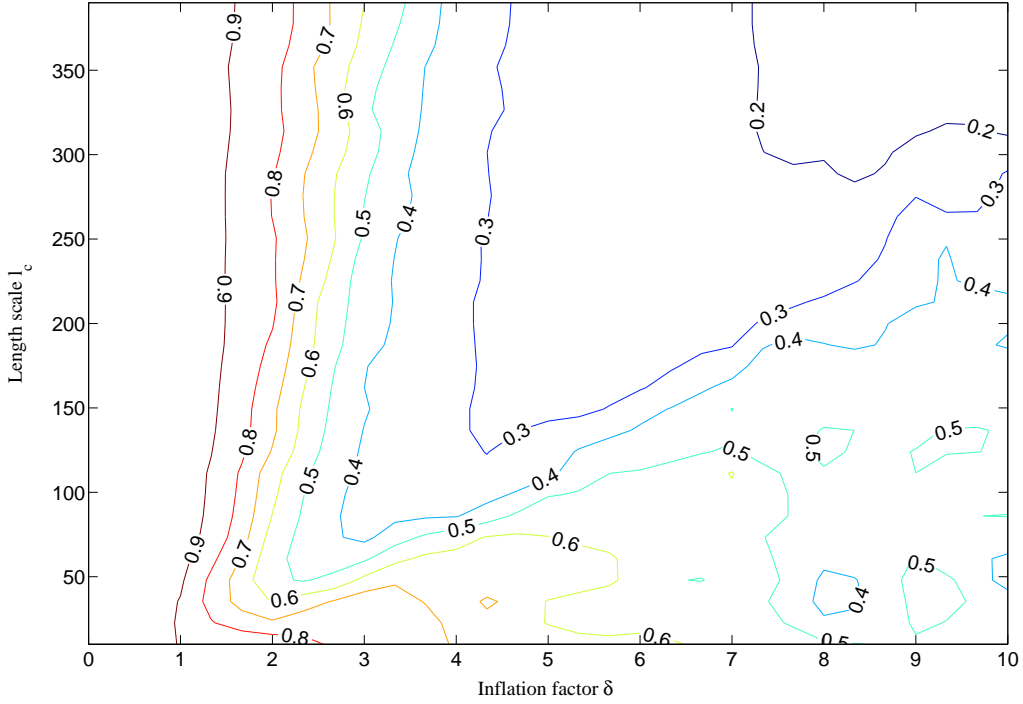} 
\caption{ \label{fig:ch3_SUKF_delta_vs_lc_ratio} The rms ratio of the SUKF as a function of the inflation factor $\delta$ and the length scale $l_c$. }
\end{figure*} 

Next we examine the rms ratio of the SUKF. We plot the rms ratio of the SUKF as a function of $\delta$ and $l_c$ in Fig.~\ref{fig:ch3_SUKF_delta_vs_lc_ratio}. As one can see there, when fixing $l_c$, if $l_c$ is not too large (say $l_c < 30$), the rms ratio $R$ tends to decrease as $\delta$ increases. If $l_c$ is relatively large (say $l_c > 30$), the rms ratio $R$ exhibits the $U$-turn behaviour as $\delta$ increases. On the other hand, when fixing $\delta$, if $\delta$ is not too large (say $\delta <1$), the rms ratio appears insensitive to the change of $l_c$. But as $\delta$ increases above $2$, the rms ratio $R$ also exhibits the $U$-turn behaviour. To make the analysis ensemble (sigma points) indistinguishable from the truth (i.e. $R \approx 0.71$), one should take the parameter values of $\delta$ and $l_c$ within the strip between the contour levels of $0.7$ and $0.8$. However, overestimation of the analysis covariance (i.e. $R<0.71$) can in fact improve the performance of the SUKF in the sense that it can achieve lower relative rms errors, the same  as the phenomenon observed in the ETKF (cf. Fig.~\ref{fig:ch2_ETKF_delta_vs_lc_ratio}).         

\subsubsection{Effect of the scale factor $\alpha$ on the performance of the reduced rank SUKF}

\begin{figure*}[!t]
\centering
\hspace*{-0.5in} \includegraphics[width=1.15\textwidth]{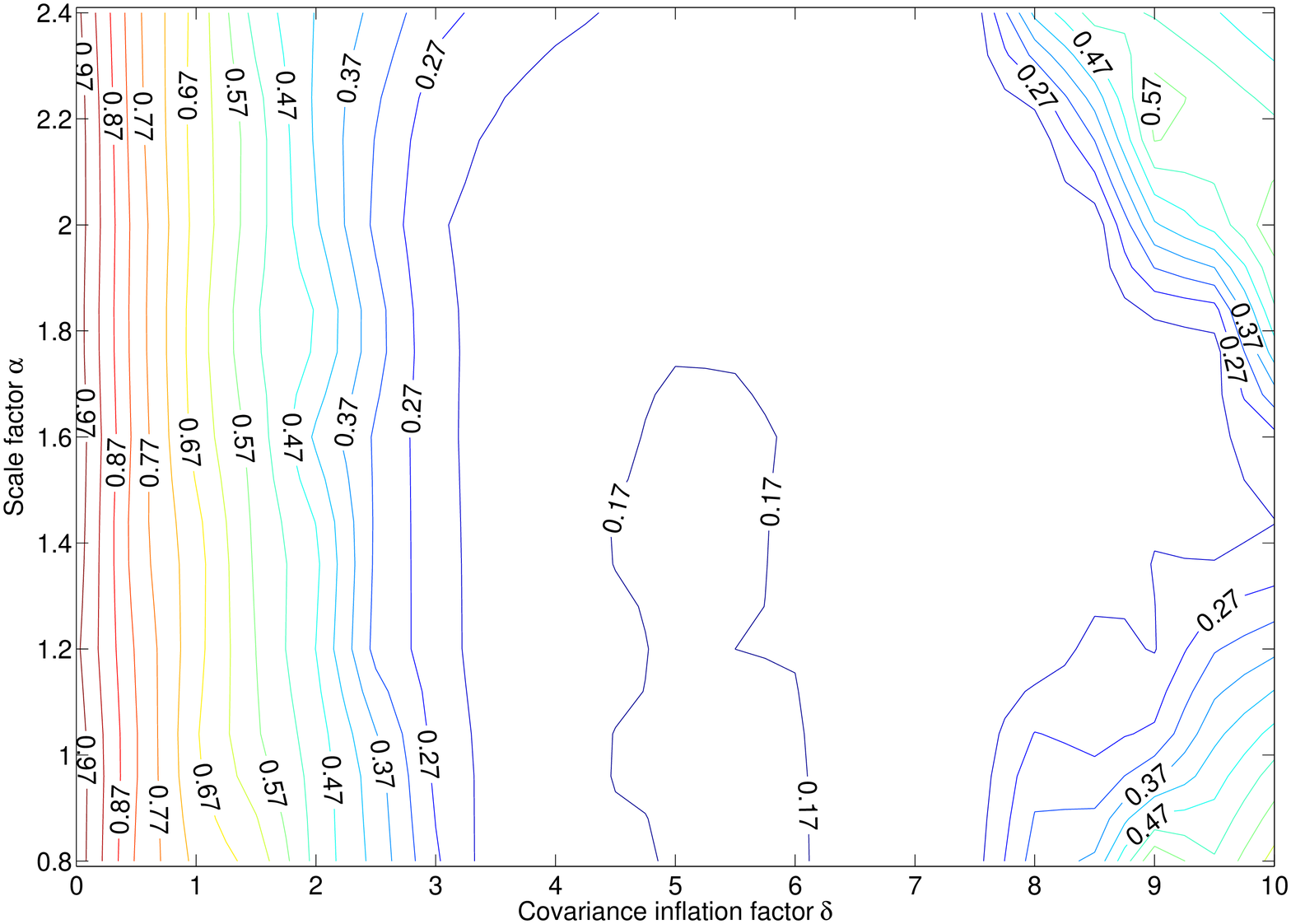} 
\caption{ \label{fig:sukf_alpha_vs_delta_rms} The relative rmse of the SUKF as a function of the scale factor $\alpha$ and the inflation factor $\delta$. }
\end{figure*} 

Now we examine the effect of the scale factor $\alpha$ on the performance of the SUKF. For a more thorough examination, we also include the covariance inflation factor $\delta$ as another variable parameter, although in the previous experiments we have already studied its effect. The scale factor $\alpha$ and the inflation factor $\delta$ take values from the sets $0.8:0.2:2.4$ and $0:0.5:10$, respectively. The values of the other parameters are: $\beta=2$, $\lambda=-2$, length scale $l_c=240$, lower bound $l_l = 3$, upper bound $l_u=6$, and the threshold at the first assimilation cycle is $\Gamma_1=1000$.  

We first plot the relative rmse of the SUKF as a function of $\alpha$ and $\delta$ in Fig.~\ref{fig:sukf_alpha_vs_delta_rms}. When fixing $\delta$, and if $\delta$ is not too large (say, $\delta < 3$), the relative rmse is insensitive to the change of $\alpha$ \footnote{This phenomenon has also been found in other experiments, see, for examples, Figs.~\ref{fig:ch2_stochastic_EnKF_delta_vs_lc_rms}, \ref{fig:ch2_ETKF_delta_vs_lc_rms} and~\ref{fig:ch3_SUKF_delta_vs_lc_rms}, where the common feature is that, when the covariance inflation factor $\delta$ is small, the relative rmse of the filter (either the EnKF or the SUKF) is roughly insensitive to the change of the other parameter in test. One possible explanation to this phenomenon is that, when $\delta$ is small, the error covariance of the background is underestimated, so that the background will dominate the computations of the sample mean and covariance of the analysis, while the influence of the incoming observation is not significant. Therefore, the relative rmse does not change too much for relatively small inflation factors.}. If $\delta$ is large (say, $\delta > 8$), then the relative rmse exhibits the U-turn behaviour as $\alpha$ increases, which can be explained from the following point of view. Comparing the true error covariance, given by Eq.~(\ref{ch3:sut_y_cov_in_expansion}), with the estimated error covariance of the SUT, given by Eq.~(\ref{ch3:sut_reduced_z_cov_in_expansion}), we see that $\alpha$ plays a role similar to that of the covariance inflation factor $\delta$. If $\alpha <1$, the error covariance of the SUT is underestimated. If $\alpha > 1$, the error covariance of the SUT is overestimated, which can therefore improve the performance of the SUKF, provided that $\alpha$ is not too large.          

\begin{figure*}[!t]
\centering
\hspace*{-0.5in} \includegraphics[width=1.15\textwidth]{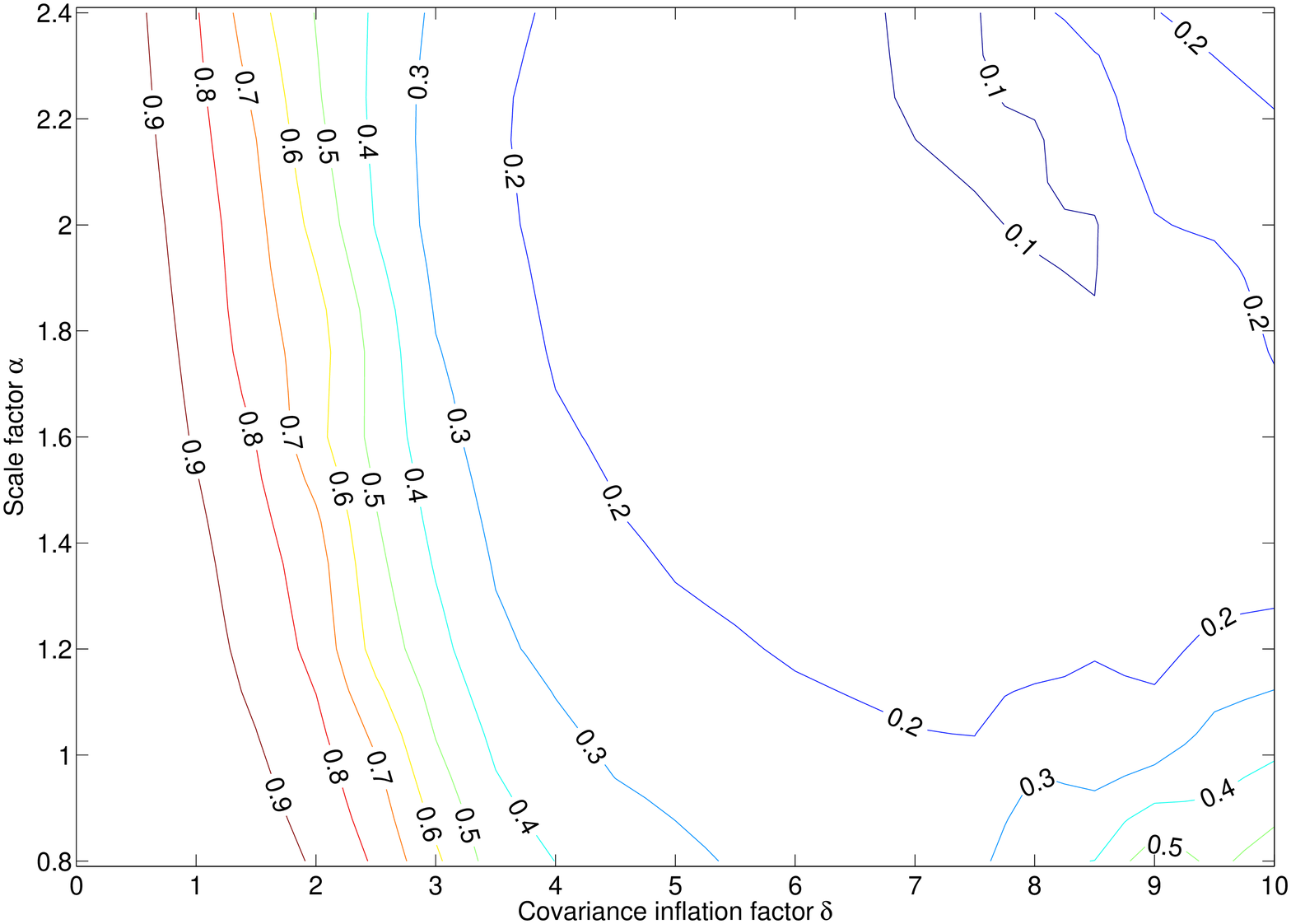} 
\caption{ \label{fig:sukf_alpha_vs_delta_ratio} The rms ratio of the SUKF as a function of the scale factor $\alpha$ and the inflation factor $\delta$. }
\end{figure*} 

Next we examine the rms ratio of the SUKF, which we plot as a function of $\delta$ and $\alpha$ in Fig.~\ref{fig:sukf_alpha_vs_delta_ratio}. When fixing $\delta$, and if $\delta$ is not too large (say $\delta <2$), the rms ratio will decrease as $\alpha$ increases. If $\delta$ is larger (say $\delta >4$), the rms ratio $R$ also exhibits a $U$-turn behaviour as $\alpha$ increases. To make sigma points indistinguishable from the truth (i.e., $R \approx 0.71$), one should take the parameter values of $\delta$ and $\alpha$ within the strip between the contour levels of $0.7$ and $0.8$. However, overestimation of the analysis covariance (i.e., $R<0.71$) can also improve the performance of the SUKF, just as in the previous experiments.        

\subsubsection{Effects of the threshold $\Gamma_1$ and the bounds $l_l$, $l_u$ on the performance of the reduced rank SUKF} \label{ch3:sec_experiments_threshold_bounds}

Here the experiments are designed to examine the effects of the threshold $\Gamma_1$ and the bounds $l_l$, $l_u$ on the performance of the SUKF. Since the rms ratio is only a qualitative measure of filter performance (e.g., underestimation or overestimation of the error covariance), we will henceforth only use the relative rmse to examine the performance of the SUKF.

\begin{figure*}[!t]
\centering
\hspace*{-0.5in} \includegraphics[width=1.15\textwidth]{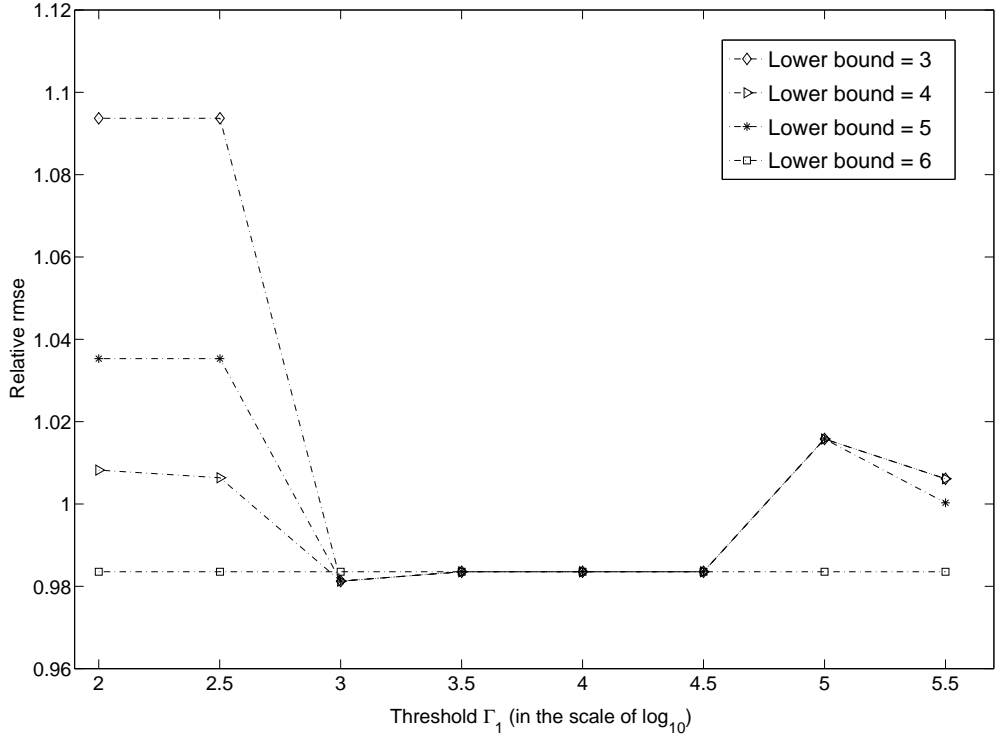} 
\caption{ \label{fig:ukf_RRmse_vs_LOGthreshold} The relative rmse of the SUKF as a function of the threshold $\Gamma_1$ (in the scale of $\log_{10}$) with different lower bounds $l_l$.}
\end{figure*} 

In the first experiment, we let the covariance inflation factor $\delta=0$, the length scale $l_c=240$, the initial ensemble size $n=4$, and take $\alpha=1$, $\beta=2$ and $\lambda=-2$. We fix the upper bound $l_u=6$, but vary the lower bound such that $l_l$ takes values from the set $3:1:6$. We also vary the threshold $\Gamma_1$ such that the logarithmic function $\log_{10}\Gamma_1$ takes values from the set $2:0.5:5.5$ \footnote{This range represents the moderate values of $\Gamma_1$ in our choice so as to make the truncation numbers $l_k$ neither too large nor too small.}. 

We show the numerical results in Fig.~\ref{fig:ukf_RRmse_vs_LOGthreshold} . Intuitively, the larger the threshold $\Gamma_1$ and the bound $l_l$, the larger the truncation number $l_k$ tends to be, which, however, does not guarantee a better performance in terms of the relative rmse. Indeed, in Fig. \ref{fig:ukf_RRmse_vs_LOGthreshold}, the optimal threshold $\log_{10}\Gamma_1=3$ is the same for lower bounds of $l_l = 3, 4,5$\footnote{$l_l=l_u=6$ means $l_k=6$ at every cycle, so the threshold $\Gamma_1$ does not affect the value of $l_k$ in this case.}, while thresholds larger than this value will result in larger relative rms errors. For the lower bound $l_l=6$, its relative rms errors are smaller than, or at least approximately equal to those of the bounds $l_l = 3, 4,5$ in most cases. However, for $\log_{10}\Gamma_1=3$, the relative rmse for $l_l=6$ is higher than the other cases. To explain this phenomenon, we conjecture that, too small a truncation number $l_k$ is not likely to achieve a performance as good as a modest value because it means poor quality of covariance approximation. In contrast, too large a truncation number $l_k$ also does not necessarily achieve a better performance than a modest value. This is because, if a covariance of the system states is not a full rank matrix, too large a truncation number may introduce some spurious structures from the null space of SVD into sigma points, which are then treated as equally likely as the other sigma points, and propagated forward to the next cycle. The effect of the spurious structures may be accumulated and eventually deteriorate the overall performance. 

\begin{figure*}[!t]
\centering
\hspace*{-0.5in} \includegraphics[width=1.15\textwidth]{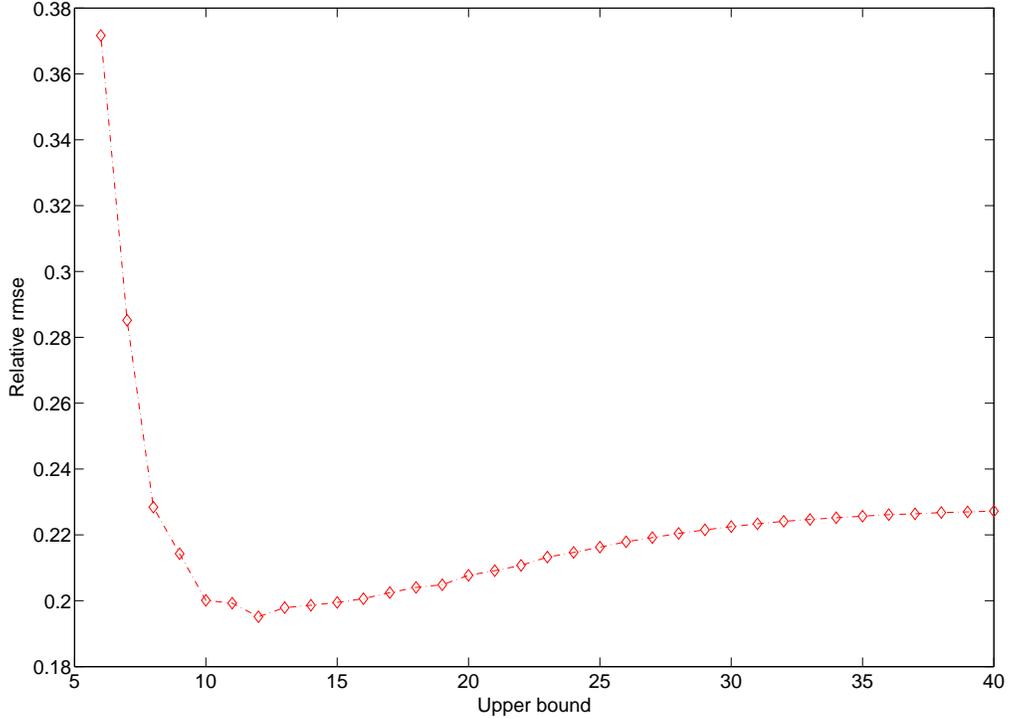} 
\caption{ \label{fig:EnUKF_rms_upper_bound} The relative rmse of the SUKF as a function of the upper bound $l_u$.}
\end{figure*} 

In the second experiment, we let the covariance inflation factor $\delta=6$, 
the length scale $l_c=\infty$ (no covariance filtering), the initial ensemble size $n=10$, the initial threshold $\Gamma_1=1000$, and we take $\alpha=1$, $\beta=2$ and $\lambda=-2$. We fix the lower bound $l_l=3$, but take the values of the upper bound $l_u$ from the set $6:1:40$. 

Fig.~\ref{fig:EnUKF_rms_upper_bound} show the relative rmse as a function of the upper bound $l_u$. As one can see, the relative rmse also exhibits the U-turn behaviour as $l_u$ increases, as for the ETKF in Fig.~\ref{fig:ch2_ETKF_sEnKF_nEn_rms}. A possible explanation of this phenomenon may be the same as the argument in the first experiment, that is, some of sigma points are actually obtained from the null space of SVD, which cannot be evaluated and propagated as equally as the other sigma points, otherwise spurious structures will be introduced so as to deteriorate the performance of the SUKF. Similar arguments can be applied to explain the U-turn behaviour of the ETKF in Fig.~\ref{fig:ch2_ETKF_sEnKF_nEn_rms}, since the square root of an error covariance, although not necessarily obtained through a SVD, is also involved in the ETKF. 

\subsubsection{Effects of the parameters $\lambda$ and $\beta$ on the performance of the reduced rank SUKF}\label{ch3:sec_experiments_lambda_beta}

Finally we examine the effects of the parameters $\lambda$ and $\beta$. In the experiments, we let $\delta=0$, $l_c=240$, $\alpha=1$, $l_l=3$, $l_u=6$, $\Gamma_1=1000$, and we take the initial ensemble size $n=4$. We consider four different scenarios with $\beta=0,2,4,6$ respectively\footnote{To guarantee the positive semi-definiteness of sample covariances, the values of $\lambda$ will depend on the choice of $\beta$. Thus it is inconvenient to plot a contour plot with the relative rmse as a function of $\beta$ and $\lambda$. For this reason, we only single out four values of $\beta$ for study.}, and compute $20$ values of $\lambda$ in each case. To guarantee the positive semi-definiteness of the sample covariances, we start with $\lambda=-\beta l_l / (1+\beta)$, and increase $\lambda$ by $\Delta \lambda =1$ each time. In particular, when $\beta=0$ and $\lambda=0$, the effective weight $W_{k,0} + \beta$ of the ensemble mean $\hat{\mathbf{x}}_{k}^a$ equals zero for any $k$. Therefore, in this case, the SUKF can be considered as the EnKF equipped with the analysis scheme of positive-negative pairs (PNP) (cf \cite{Wang-which} and the references therein).

\begin{figure*}[!t] 
     \centering
	 \subfigure[Relative rmse vs $\lambda$ with $\beta=0$]{
        \label{fig:ukf_rmse_vs_lambda_beta0}
        \includegraphics[width=0.45\textwidth]{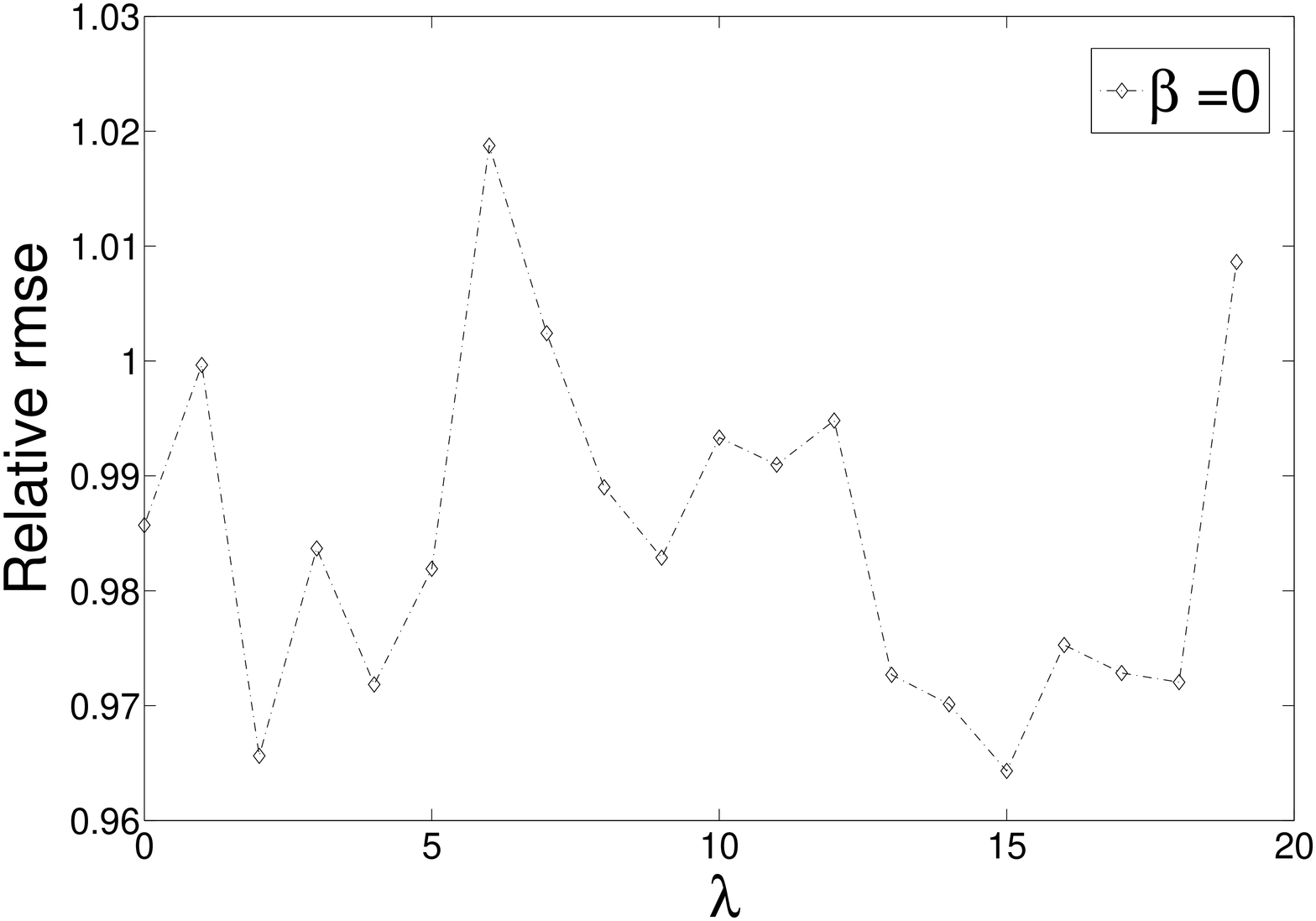}} 
	 \hspace{.3in}
     \subfigure[Relative rmse vs $\lambda$ with $\beta=2$]{
          \label{fig:ukf_rmse_vs_lambda_beta2}
          \includegraphics[width=0.45\textwidth]{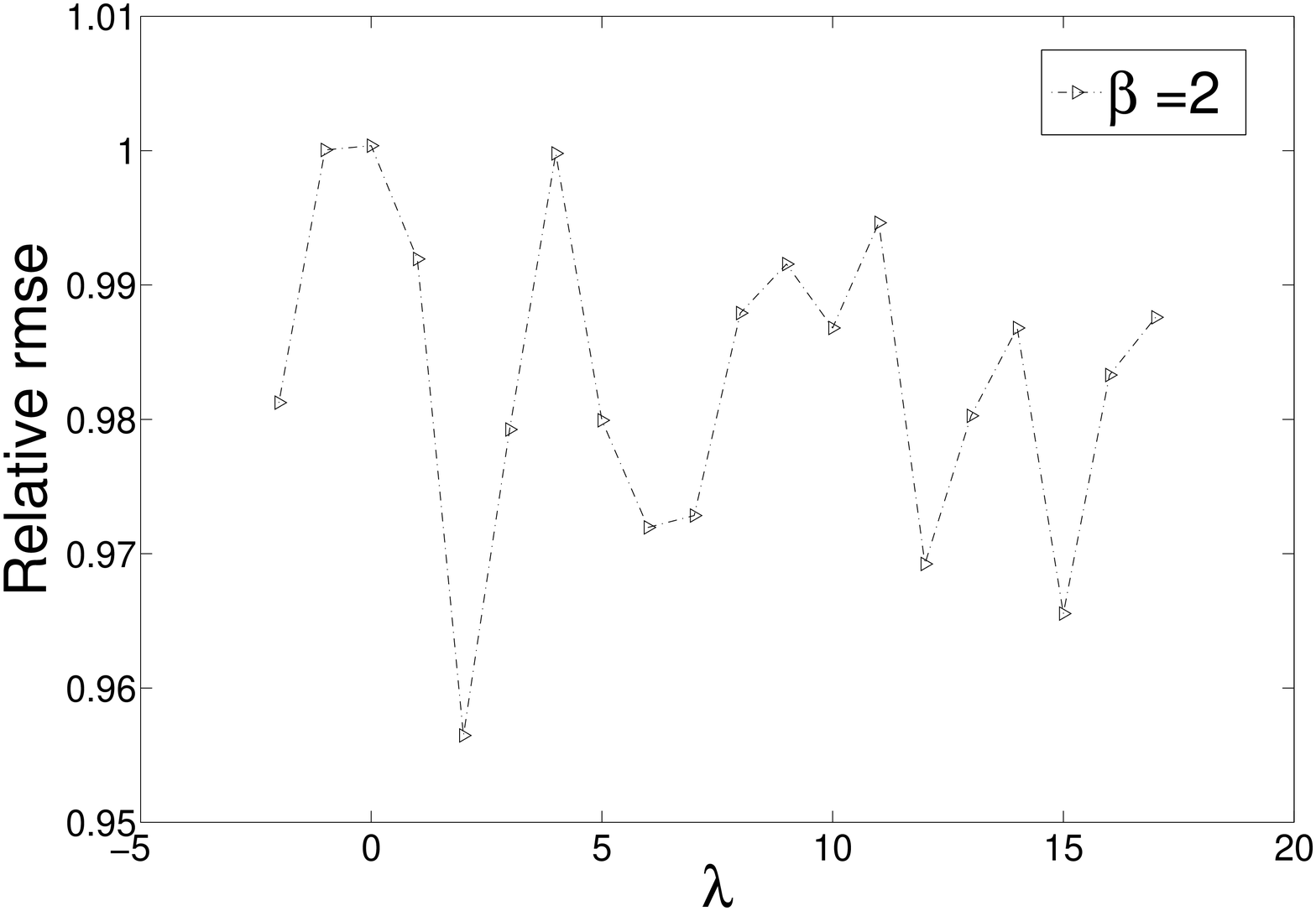}}
    \hspace{.3in}
	\subfigure[Relative rmse vs $\lambda$ with $\beta=4$]{
        \label{fig:ukf_rmse_vs_lambda_beta4}
        \includegraphics[width=0.45\textwidth]{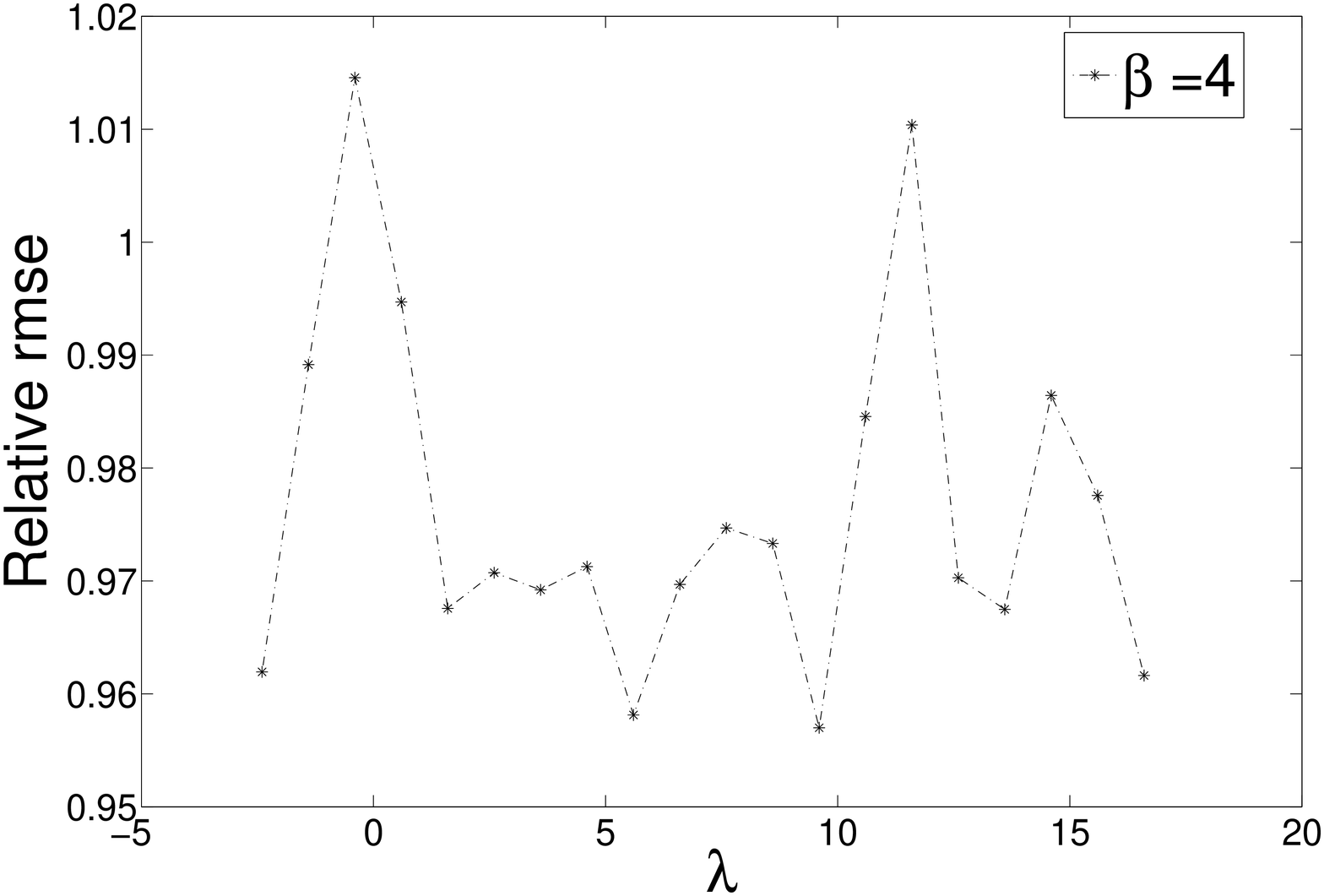}}
	 \hspace{.3in}
     \subfigure[Relative rmse vs $\lambda$ with $\beta=6$]{
          \label{fig:ukf_rmse_vs_lambda_beta6}
          \includegraphics[width=0.45\textwidth]{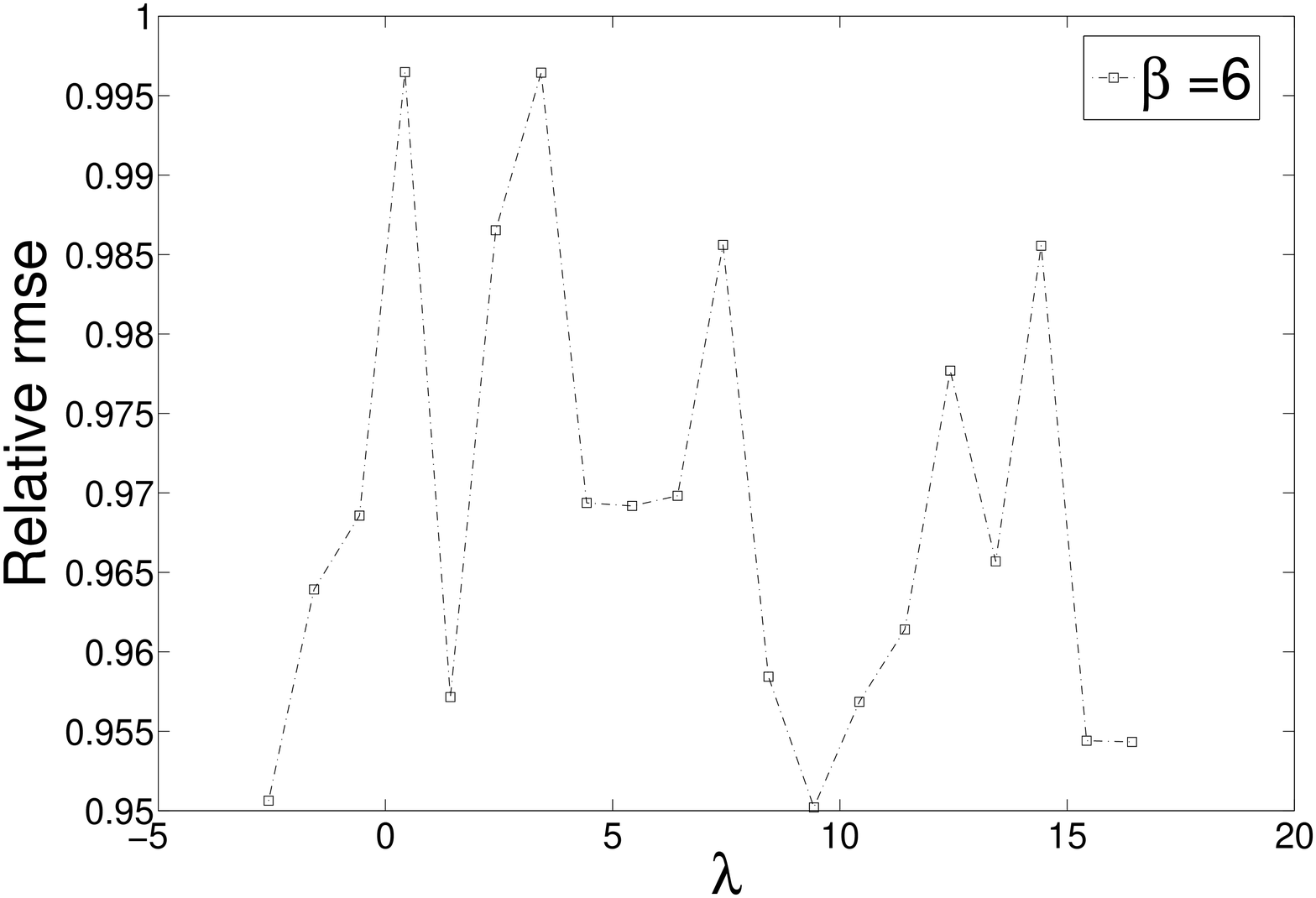}}
     \hspace{.3in}
     \caption{ \label{fig:ukf beta} Effects of the parameters $\beta$ and $\lambda$ on the performance of the SUKF.}
\end{figure*}

We plot the numerical results in Fig.~\ref{fig:ukf beta}. As one can see, when $\beta$ increases from $0$ to $6$, the minimum relative rmse for a given value of $\beta$ decreases. This may be interpreted as follows: as pointed out in \S~\ref{ch3:sec_idea_of_ut}, a positive value of $\beta$ will increase the error covariance, which is similar to the covariance inflation technique introduced in \S~\ref{ch2:sec_covariance_inflation}, and so a larger value of $\beta$ tends to result in a smaller relative rmse, provided that $\beta$ is not too large (otherwise the U-turn behaviour may appear). 


However, for each fixed $\beta$, there is no clear trend indicating the optimal value of $\lambda$. A larger value of $\lambda$ does not imply a smaller relative rmse, or vice verse. As an explanation of this phenomenon, we note that, with the other parameters being fixed, $\lambda$ determines the relative weights between the sample mean and the other sigma points (cf. Eq.~(\ref{ch3:reduced_sut_weights})). If the underlying system is linear, then in principle we can compute the optimal relative weights between the sample mean and the other sigma points (under the assumption of Gaussianity), and so determine the optimal value of $\lambda$. Nevertheless, the existence of nonlinearity may make the problem intractable. For nonlinear systems, the optimal relative weights (hence $\lambda$) may vary from cycle to cycle. However, to search for the optimal parameter $\lambda$ at each assimilation cycle will be computationally expensive. Thus in our experiments, we chose to fix $\lambda$ within the same assimilation window, so that the same value of $\lambda$ is used at each assimilation cycle. In doing this, the fixed $\lambda$ cannot capture the variation of its optimal values at different assimilation cycles, therefore it may be difficult to find a clear trend of its optimal value in Fig.~(\ref{fig:ukf beta}).   

\section{Summary of the chapter}\label{ch3:sec_summary}

In this chapter we introduced the basic idea of the unscented transform (UT) and its extension, the scaled unscented transform (SUT). We conducted an accuracy analysis for the UT via Taylor series expansions. We also showed in Appendix~\ref{appendix:accuracy analysis} that, under the assumption of Gaussianity, the UT can achieve better accuracy than the EnKF (including the ETKF).

Incorporating the UT or the SUT into the propagation step of recursive Bayesian estimation (RBE) will lead to the unscented Kalman filter (UKF) or the scaled unscented Kalman filter (SUKF), respectively. In practice, however, one may not wish to apply the UKF or the SUKF directly to high dimensional systems, since the computational cost in those circumstances will become very expensive. To this end, we introduced the reduced rank SUKF to reduce the computational cost.

For illustration, we took the $40$-dimensional LE 98 system as the testbed to demonstrate the details in implementing the reduced rank SUKF. We also investigated the effects of the intrinsic parameters (e.g., $\alpha$, $\beta$, $\lambda$ etc) on the performance of the filter. Currently, there are no theoretical grounds that can be used to determine the optimal values of these filter parameters in general situations. The experiments conducted in this chapter may provide some insights into how these parameters affect the performance of the reduced rank SUKF.   

\chapter{Divided difference filters for data assimilation} \label{ch4:ddfs}

\section{Overview} \label{ch4:sec_overview}

The divided difference filters (DDFs) are similar to the extended Kalman filter (EKF). At the propagation step, the DDFs also involve a local expansion of a nonlinear function, not via a Taylor series expansion as in the EKF, but through Stirling's interpolation formula. The advantage of adopting this formula is that the computation does not involve the derivatives of a nonlinear function. Instead, one uses divided differences for approximation and thus can avoid the difficulty in the EKF. By adopting Stirling's interpolation formula, one also needs to generate sigma points as in the scaled unscented Kalman filter (SUKF). Thus although the DDFs and the SUKF are derived from different points of view, they are similar to one another in many aspects, as will be shown later. 

This chapter is organized as follows. In \S~\ref{ch4:sec_ps} we state the problem of interest. Then we proceed to introduce Stirling's interpolation formula in \S~\ref{ch4:sec_stirling} as the approximate solution to the recast problem in Fig.~\ref{ch3:fig_problem_recast}. Incorporating this formula into the propagation step of recursive Bayesian estimation (RBE) leads to the DDFs, as will be introduced in \S~\ref{ch4:sec_ddfs}. To reduce the computational cost, we introduce the reduced rank DDFs in \S~\ref{ch4:sec_reduced_rank_ddfs}. In \S~\ref{ch4:sec_experiments} we use the $40$-dimensional Lorenz-Emanuel 98 model as the testbed to illustrate the details in implementing the reduced rank DDFs, and investigate the effects of filter parameters on the performance of the DDFs. Finally, we draw our conclusions for this chapter in \S~\ref{ch4:sec_summary}.  

\section{Problem statement} \label{ch4:sec_ps}

Consider the data assimilation problem in the systems described by Eq.~(\ref{ch2:ps}):
\begin{subequations}  
\begin{align}
 \tagref{ch2:ps_dyanmical_system} & \mathbf{x}_k  = \mathcal{M}_{k,k-1} \left( \mathbf{x}_{k-1} \right) + \mathbf{u}_{k}  \, ,  \\
  \tagref{ch2:ps_observation_system} &  \mathbf{y}_k  = \mathcal{H}_{k} \left( \mathbf{x}_{k} \right) + \mathbf{v}_{k} \, , \\
  \tagref{ch2:ps_dyanmical_noise} & \mathbf{u}_{k} \sim N \left(\mathbf{u}_{k}: \mathbf{0}, \mathbf{Q}_k \right) \, ,\\
 \tagref{ch2:ps_observation_noise} & \mathbf{v}_{k} \sim N \left(\mathbf{v}_{k}: \mathbf{0}, \mathbf{R}_k \right) \, ,\\
  \tagref{ch2:ps_dyanmical_white_noise}& \mathbb{E} \left( \mathbf{u}_{j} \mathbf{u}_{k}^T \right) = \delta_{k,j}\mathbf{Q}_k \, ,\\
  \tagref{ch2:ps_observation_white_noise}& \mathbb{E} \left( \mathbf{v}_{j} \mathbf{v}_{k}^T \right) = \delta_{k,j}\mathbf{R}_k \, ,\\
  \tagref{ch2:ps_uncorrelated_noise}& \mathbb{E} \left( \mathbf{u}_{i} \mathbf{v}_{j}^T \right) = 0 \quad \forall \, i, \, j \, .
\end{align}
\end{subequations}

We first discuss how to solve the recast problem in Fig.~\ref{ch3:fig_problem_recast} through Stirling's interpolation formula. Then we apply this formula to the propagation step of RBE to derive the DDFs. 

\section{Stirling's Interpolation Formula} \label{ch4:sec_stirling}

\subsection{Basic idea}
The ideas and analyses presented here and in \S~\ref{ch4:sec_ddfs} mainly follow the works \cite{Ito-gaussian,Noergaard-Advances,Norgaard--new}. 

We first re-state the estimation problem in Fig.~\ref{ch3:fig_problem_recast}. Let $\mathbf{x}$ be an $m$-dimensional Gaussian random variable such that $\mathbf{x} \sim N \left( \mathbf{x}: \bar{\mathbf{x}}, \mathbf{P}_x \right)$. We transform $\mathbf{x}$ by a nonlinear function $\mathcal{F}$ to give a transformed random variable $\mathbf{\eta} = \mathcal{F} \left( \mathbf{x} \right)$. Our objective is to estimate the mean $\bar{\mathbf{\eta}}$ and covariance $\mathbf{P}_{\eta}$ of $\mathbf{\eta}$.  

In Chapter~\ref{ch3:ukf} we have mentioned the extended Kalman filter (EKF), derived via a Taylor series expansion of $\mathcal{F}$. Alternatively, one can choose to expand $\mathcal{F}$ through Stirling's interpolation polynomials \cite{Noergaard-Advances,Norgaard--new}. For example, a second-order approximation can be conducted based on the formula \cite[Eq. (11-13)]{Noergaard-Advances}
\begin{equation} \label{ch3:SI-2}
\mathbf{\eta} = \mathcal{F} \left( \bar{\mathbf{x}} +  \delta \mathbf{x}\right) \approx \mathcal{F} \left( \bar{\mathbf{x}}\right) + \mathcal{D}_{ \delta \mathbf{e}}  \mathcal{F} \left( \bar{\mathbf{x}}\right) + \dfrac{1}{2} \mathcal{D}_{ \delta \mathbf{e}}^2  \mathcal{F} \left( \bar{\mathbf{x}}\right),
\end{equation}
where $\mathcal{D}_{ \delta \mathbf{e}}$ and $\mathcal{D}_{ \delta \mathbf{e}}^2$ are the divided difference operators defined through the following operations \cite{Noergaard-Advances}: 
\begin{subequations} \label{ch3:operator_def_1}
\begin{align}
 \mathcal{D}_{ \delta \mathbf{e}} \mathcal{F} \left( \bar{\mathbf{x}}\right) = & \dfrac{1}{h} \left( \sum_{i=1}^{L} \delta e_i \mathcal{P}_i (h/2) \mathcal{N}_i (h/2) \right) \mathcal{F} \left( \bar{\mathbf{x}}\right), \\
  \mathcal{D}^2_{ \delta \mathbf{e}} \mathcal{F} \left( \bar{\mathbf{x}}\right) =& \dfrac{1}{h^2} \left(\sum_{i=1}^{L} \left(\delta e_i \right)^2 \left[ \mathcal{N}_i (h/2) \right]^2  \right. \\
\nonumber & \left. + \sum_{i=1}^{L} \sum_{j=1,j \neq i}^{L} \delta e_i \delta e_j \left[ \mathcal{P}_i (h/2) \mathcal{N}_i (h/2) \right] \left[ \mathcal{P}_j (h/2) \mathcal{N}_j (h/2) \right] \right) \mathcal{F} \left( \bar{\mathbf{x}}\right).
\end{align}
\end{subequations}
Here $\mathcal{P}_i (h/2)$ and $\mathcal{N}_i (h/2)$ are operators satisfying
\begin{equation} \label{ch3:operator_def_2}
\begin{split}
 \mathcal{P}_i (h/2) \mathcal{F} \left( \bar{\mathbf{x}}\right) = & \dfrac{1}{2} \left( \mathcal{F} \left( \bar{\mathbf{x}} + \dfrac{h}{2} \left(\mathbf{S}_x \right)_i \right) + \mathcal{F} \left( \bar{\mathbf{x}} - \dfrac{h}{2} \left(\mathbf{S}_x \right)_i \right) \right) , \\
 \mathcal{N}_i (h/2) \mathcal{F} \left( \bar{\mathbf{x}}\right) = & \mathcal{F} \left( \bar{\mathbf{x}} + \dfrac{h}{2} \left(\mathbf{S}_x \right)_i \right)  -   \mathcal{F} \left( \bar{\mathbf{x}} - \dfrac{h}{2} \left(\mathbf{S}_x \right)_i \right) ,
\end{split}
\end{equation}
with the parameter $h$ being the interval length of interpolation, and $\left( \mathbf{S}_x \right)_i$ the $i$-th column of the $m \times L$ square root matrix $\mathbf{S}_x$ of $\mathbf{P}_x$. And $\delta e_i$ denotes the $i$-th element of the Gaussian random variable $\delta \mathbf{e} \sim N \left( \delta \mathbf{e}: \mathbf{0}, \mathbf{I} \right)$, where $\mathbf{I}$ is the $L \times L$ identity matrix. Therefore,
\begin{equation}
 \mathbb{E} \left( \delta e_i \right)=0, ~\mathbb{E} \left( \delta e_i^2 \right)=1~\text{for}~ i =1, \dotsb, L \, .
\end{equation} 
Moreover, $\delta \mathbf{x} = \mathbf{S}_x \delta \mathbf{e}$ follows the Gaussian distribution $N \left( \delta \mathbf{x}: \mathbf{0}, \mathbf{P}_x \right)$.  

With some algebra, it can be shown that
\begin{equation}
\begin{split}
& \left[ \mathcal{P}_i (h/2) \right]^2 \mathcal{F} \left( \bar{\mathbf{x}}\right) = \dfrac{1}{2} \left[ \mathcal{P}_i (h) + 1 \right] \mathcal{F} \left( \bar{\mathbf{x}}\right) \, ; \\
& \left[ \mathcal{N}_i (h/2) \right]^2 \mathcal{F} \left( \bar{\mathbf{x}}\right) = 2 \left[ \mathcal{P}_i (h) - 1 \right] \mathcal{F} \left( \bar{\mathbf{x}}\right) \, ; \\
& \left[ \mathcal{P}_i (h/2) \mathcal{N}_i (h/2) \right] \mathcal{F} \left( \bar{\mathbf{x}}\right) = \left[ \mathcal{N}_i (h/2) \mathcal{P}_i (h/2) \right] \mathcal{F} \left( \bar{\mathbf{x}}\right) = \dfrac{1}{2} \mathcal{N}_i (h) \, \mathcal{F} \left( \bar{\mathbf{x}}\right) \, . \\
\end{split}
\end{equation}
Thus Eq.~(\ref{ch3:operator_def_1}) becomes
\begin{subequations} 
\begin{align}
 \mathcal{D}_{ \delta \mathbf{e}} \mathcal{F} \left( \bar{\mathbf{x}}\right) = & \dfrac{1}{2h} \left( \delta \mathbf{e} \right)^T \mathcal{N} (h) \mathcal{F} \left( \bar{\mathbf{x}}\right), \\
 \label{ch3:operator_2nd_order} \mathcal{D}^2_{ \delta \mathbf{e}} \mathcal{F} \left( \bar{\mathbf{x}}\right) =& \dfrac{1}{4h^2} \left(8 \, \sum_{i=1}^{L} \left(\delta e_i \right)^2 \left[ \mathcal{P}_i (h) -1 \right]  \right. \\
\nonumber & \left. + \sum_{i=1}^{L} \sum_{j=1,j \neq i}^{L} \delta e_i \delta e_j \mathcal{N}_i (h) \mathcal{N}_j (h) \right) \mathcal{F} \left( \bar{\mathbf{x}}\right) \, ,
\end{align}
\end{subequations}
where $\mathcal{N} (h) \equiv \left[ \mathcal{N}_1 (h), \dotsb, \mathcal{N}_L (h)\right]^T $. Note that in Eq.~(\ref{ch3:operator_2nd_order}), evaluating the terms $\sum_{i=1}^{L} \sum_{j=1,j \neq i}^{L} \delta e_i \delta e_j \mathcal{N}_i (h) \mathcal{N}_j (h) \mathcal{F} \left( \bar{\mathbf{x}}\right)$ requires one to generate $L(L-1)$ additional system states (apart from the sigma points to be introduced later), which would be prohibitive if the system dimension $m$ is large and we require $L \ge m$ to avoid rank deficiency in the sample covariance. Thus to reduce the computational cost, we discard these terms following \cite{Noergaard-Advances}. Therefore, Eq.~(\ref{ch3:operator_2nd_order}) is reduced to 
\begin{equation}
\mathcal{D}^2_{ \delta \mathbf{e}} \mathcal{F} \left( \bar{\mathbf{x}}\right) \approx \dfrac{2}{h^2}  \sum_{i=1}^{L} \left(\delta e_i \right)^2 \left[ \mathcal{P}_i (h) -1 \right] \mathcal{F} \left( \bar{\mathbf{x}}\right) \, .
\end{equation}  

As for the scaled unscented transform (SUT), we also need to generate a set of special system states $\left\{ \mathcal{X}_i \right\}_{i=0}^{2L}$ ($L \ge m $):
\begin{equation} \label{ch4:ddf_sigma_points}
\begin{split}
& \mathcal{X}_0 = \bar{\mathbf{x}},\\
& \mathcal{X}_i = \bar{\mathbf{x}} + h \left(  \mathbf{S}_x \right)_i, \, i=1, 2, \dotsb, L ,\\
& \mathcal{X}_i = \bar{\mathbf{x}} - h \left(  \mathbf{S}_x \right)_{i-L}, \, i=L+1, L+2, \dotsb, 2L ,\\
\end{split}
\end{equation} 
which are also called sigma points. But note that here sigma points are generated for the purpose of function interpolation, while in the SUT, sigma points are particularly chosen to capture certain moments of the distribution of $\mathbf{x}$. 
 
Let the transformed sigma points be $ \left\{ \mathcal{Y}_i : \mathcal{Y}_i = \mathcal{F} \left( \mathcal{X}_i\right) \right\}_{i=0}^{2L}$. In what follows we introduce three different approximation schemes.

\subsubsection{First order divided difference approximation}

In the first order divided difference (DD1) approximation scheme, the nonlinear function $\mathcal{F}$ is approximated by \cite{Noergaard-Advances,Norgaard--new}:
\begin{equation} 
\begin{split}
\mathbf{\eta} &= \mathcal{F} \left( \bar{\mathbf{x}} +  \delta \mathbf{x} \right) \\
&\approx \mathcal{F} \left( \bar{\mathbf{x}}\right) + \mathcal{D}_{ \delta \mathbf{e}}  \mathcal{F} \left( \bar{\mathbf{x}}\right) \\
&= \mathcal{F} \left( \bar{\mathbf{x}}\right) + \dfrac{1}{2h} \left( \delta \mathbf{e} \right)^T \mathcal{N} (h) \mathcal{F} \left( \bar{\mathbf{x}}\right) \, . 
\end{split}
\end{equation}

Therefore the estimated mean $\hat{\mathbf{\eta}}$ is given by
\begin{equation}
\hat{\mathbf{\eta}} = \mathbb{E} \left[ \mathcal{F} \left( \bar{\mathbf{x}}\right) + \mathcal{D}_{ \delta \mathbf{e}}  \mathcal{F} \left( \bar{\mathbf{x}}\right) \right] = \mathcal{F} \left( \bar{\mathbf{x}} \right) = \mathcal{Y}_0 \, .\\
\end{equation}

Similarly, the estimated covariance 
\begin{equation}
\begin{split}
\hat{\mathbf{P}}_{\eta} &= \mathbb{E} \left( \mathbf{\eta} - \hat{\mathbf{\eta}} \right)\left( \mathbf{\eta} - \hat{\mathbf{\eta}} \right)^T \\
& = \dfrac{1}{4h^2} \mathbb{E} \left[ \left( \delta \mathbf{e} \right)^T \mathcal{N} (h) \mathcal{F} \left( \bar{\mathbf{x}}\right) \right]  \left[ \left( \delta \mathbf{e} \right)^T \mathcal{N} (h) \mathcal{F} \left( \bar{\mathbf{x}}\right) \right]^T \\
& = \dfrac{1}{4h^2} \left[ \mathcal{N} (h) \mathcal{F} \left( \bar{\mathbf{x}}\right) \right]^T \mathbb{E} \left[ \left( \delta \mathbf{e} \right) \, \left( \delta \mathbf{e} \right)^T  \right]  \left[  \mathcal{N} (h) \mathcal{F} \left( \bar{\mathbf{x}}\right) \right] \\
& = \dfrac{1}{4h^2} \left[ \mathcal{N} (h) \mathcal{F} \left( \bar{\mathbf{x}}\right) \right]^T \left[  \mathcal{N} (h) \mathcal{F} \left( \bar{\mathbf{x}}\right) \right] \\
& = \dfrac{1}{4h^2} \sum\limits_{i=1}^{L}  \left[ \mathcal{N}_i (h) \mathcal{F} \left( \bar{\mathbf{x}}\right) \right] \left[ \mathcal{N}_i (h) \mathcal{F} \left( \bar{\mathbf{x}}\right) \right]^T \\
& = \dfrac{1}{4h^2} \sum\limits_{i=1}^{L}  \left[ \mathcal{F} \left( \bar{\mathbf{x}} + h \left( \mathbf{S}_x \right)_i \right) - \mathcal{F} \left( \bar{\mathbf{x}} - h \left( \mathbf{S}_x \right)_i \right) \right] \left[ \mathcal{F} \left( \bar{\mathbf{x}} + h \left( \mathbf{S}_x \right)_i \right) - \mathcal{F} \left( \bar{\mathbf{x}} - h \left( \mathbf{S}_x \right)_i \right) \right]^T \\
& =   \dfrac{1}{4h^2} \sum\limits_{i=1}^{L} \left( \mathcal{Y}_i -\mathcal{Y}_{L+i} \right) \left(  \mathcal{Y}_i -\mathcal{Y}_{L+i} \right)^T. 
\end{split}
\end{equation}

For convenience, it is customary in practice to also compute the cross covariance (for evaluation of the Kalman gain in the DDFs), which is given by
\begin{equation} \label{ch3:stirling_interpolation_1st_order_cross_cov}
\begin{split}
\hat{\mathbf{P}}_{x\eta} & = \mathbb{E} \left[ \delta \mathbf{x} \left( \mathbf{\eta} - \hat{\mathbf{\eta}} \right)   ^T \right]\\
& \approx \dfrac{1}{2h} \, \mathbb{E} \left[ \mathbf{S}_x \delta \mathbf{e} \left( \delta \mathbf{e} \right)^T \mathcal{N} (h) \mathcal{F} \left( \bar{\mathbf{x}}\right) \right] \\
& = \dfrac{1}{2h} \, \mathbf{S}_x \mathcal{N} (h) \mathcal{F} \left( \bar{\mathbf{x}}\right) \\
& = \dfrac{1}{2h} \, \sum\limits_{i=1}^{L} \left( \mathbf{S}_x \right)_i \left( \mathcal{Y}_i -\mathcal{Y}_{L+i} \right)^T \, .
\end{split}
\end{equation} 

To summarize, in the DD1 approximation scheme, the solution to the recast problem is given by
\begin{subequations}  \label{ch4:ddf_1st_order_estimations}
\begin{align}
 \label{ch4:ddf_1st_order_mean} &\hat{\mathbf{\eta}} = \mathcal{F} \left( \bar{\mathbf{x}} \right)= \mathcal{Y}_0 \, ;\\
 \label{ch4:ddf_1st_order_cov} & \hat{\mathbf{P}}_{\eta}=   \dfrac{1}{4h^2} \sum\limits_{i=1}^{L} \left( \mathcal{Y}_i -\mathcal{Y}_{L+i} \right) \left(  \mathcal{Y}_i -\mathcal{Y}_{L+i} \right)^T \, ; \\ 
\label{ch4:ddf_1st_order_cross_cov} &\hat{\mathbf{P}}_{x\eta} = \dfrac{1}{2h} \, \sum\limits_{i=1}^{L} \left( \mathbf{S}_x \right)_i \left( \mathcal{Y}_i -\mathcal{Y}_{L+i} \right)^T \, .
\end{align}
\end{subequations}

\subsubsection{Second order divided difference approximation}
In the second order divided difference (DD2) approximation scheme, the nonlinear function $\mathcal{F}$ is approximated by \cite{Noergaard-Advances,Norgaard--new}:
\begin{equation} 
\begin{split}
\mathbf{\eta} &= \mathcal{F} \left( \bar{\mathbf{x}} +  \delta \mathbf{x} \right) \\
&\approx \mathcal{F} \left( \bar{\mathbf{x}}\right) + \dfrac{1}{2h} \left( \delta \mathbf{e} \right)^T \mathcal{N} (h) \mathcal{F} \left( \bar{\mathbf{x}}\right) + \dfrac{1}{h^2} \sum_{i=1}^{L} \left(\delta e_i \right)^2 \left[ \mathcal{P}_i (h) -1 \right] \mathcal{F} \left( \bar{\mathbf{x}} \right) \, .
\end{split}
\end{equation}

Thus we have the estimated mean 
\begin{equation}
\begin{split}
\hat{\mathbf{\eta}} & = \mathcal{F} \left( \bar{\mathbf{x}}\right) + \dfrac{1}{h^2} \sum_{i=1}^{L} \mathbb{E} \left[ \left(\delta e_i \right)^2 \right] \left[ \mathcal{P}_i (h) -1 \right] \mathcal{F} \left( \bar{\mathbf{x}} \right)\\
& = \dfrac{h^2-L}{h^2} \, \mathcal{F} \left( \bar{\mathbf{x}}\right) +  \dfrac{1}{h^2} \sum_{i=1}^{L} \mathcal{P}_i (h) \mathcal{F} \left( \bar{\mathbf{x}} \right) \\
& = \dfrac{h^2-L}{h^2} \, \mathcal{Y}_0 +  \dfrac{1}{2 h^2} \sum_{i=1}^{L} \left( \mathcal{Y}_i + \mathcal{Y}_{L+i} \right) \, .\\
\end{split}
\end{equation}

To estimate the covariance, we have 
\begin{equation}
\hat{\mathbf{P}}_{\eta} = \mathbb{E} \left( \mathbf{\eta} -\hat{\mathbf{\eta}}\right) \left( \mathbf{\eta} -\hat{\mathbf{\eta}}\right)^T \, .
\end{equation}
To facilitate the evaluations, one may note that 
\begin{equation}
\begin{split}
\hat{\mathbf{P}}_{\eta} &= \mathbb{E} \left( \mathbf{\eta} -\hat{\mathbf{\eta}}\right) \left( \mathbf{\eta} -\hat{\mathbf{\eta}}\right)^T \\
&= \mathbb{E} \left( \mathbf{\eta} - \mathcal{F} \left( \bar{\mathbf{x}}\right)\right) \left( \mathbf{\eta} -\mathcal{F} \left( \bar{\mathbf{x}}\right) \right)^T - \left( \hat{\mathbf{\eta}} - \mathcal{F} \left( \bar{\mathbf{x}}\right) \right) \left( \hat{\mathbf{\eta}} - \mathcal{F} \left( \bar{\mathbf{x}}\right) \right)^T \\
& = \dfrac{1}{4h^2} \left( \mathcal{N} (h) \mathcal{F} \left( \bar{\mathbf{x}}\right) \right)^T \left( \mathcal{N} (h) \mathcal{F} \left( \bar{\mathbf{x}}\right) \right) \\
& \quad + \dfrac{1}{h^4} \sum\limits_{i=1}^{L} \left( \mathbb{E} (\delta e_i)^4 \right) \left[ \left(\mathcal{P}_i (h) -1 \right) \mathcal{F} \left( \bar{\mathbf{x}} \right) \right] \left[ \left(\mathcal{P}_i (h) -1 \right) \mathcal{F} \left( \bar{\mathbf{x}} \right) \right]^T \\
& \quad + \dfrac{1}{h^4} \sum\limits_{i=1}^{L} \sum\limits_{j=1, j \neq i}^{L} \left( \mathbb{E} (\delta e_i)^2 (\delta e_j)^2 \right) \left[ \left(\mathcal{P}_i (h) -1 \right) \mathcal{F} \left( \bar{\mathbf{x}} \right) \right] \left[ \left(\mathcal{P}_j (h) -1 \right) \mathcal{F} \left( \bar{\mathbf{x}} \right) \right]^T \\ 
& \quad -  \dfrac{1}{h^4} \sum\limits_{i=1}^{L} \left( \mathbb{E} (\delta e_i)^2 \right)^2 \left[ \left(\mathcal{P}_i (h) -1 \right) \mathcal{F} \left( \bar{\mathbf{x}} \right) \right] \left[ \left(\mathcal{P}_i (h) -1 \right) \mathcal{F} \left( \bar{\mathbf{x}} \right) \right]^T \\
& \quad - \dfrac{1}{h^4} \sum\limits_{i=1}^{L} \sum\limits_{j=1, j \neq i}^{L} \left( \mathbb{E} (\delta e_i)^2\right) \left( \mathbb{E} (\delta e_j)^2\right) \left[ \left(\mathcal{P}_i (h) -1 \right) \mathcal{F} \left( \bar{\mathbf{x}} \right) \right] \left[ \left(\mathcal{P}_j (h) -1 \right) \mathcal{F} \left( \bar{\mathbf{x}} \right) \right]^T \, .
\end{split}
\end{equation}
Note that, to derive the above equation, we utilize the fact that the third order moments $\mathbb{E} (\delta e_i \delta e_j \delta e_k ) \equiv 0$ for arbitrary admissible indices $i$, $j$, and $k$, since we assume $\delta \mathbf{e} \sim N \left( \delta \mathbf{e}: \mathbf{0}, \mathbf{I} \right)$. Moreover, we also have
\begin{equation}
\mathbb{E} (\delta e_i)^2 (\delta e_j)^2 = \left( \mathbb{E} (\delta e_i)^2 \right) \left( \mathbb{E}(\delta e_j)^2 \right) = 1, \text{for}~ i \neq j \, .
\end{equation}
Note that in \cite{Noergaard-Advances, Norgaard--new}, the author chose to parameterize the term $\mathbb{E} (\delta e_i)^4$ and set $\mathbb{E} (\delta e_i)^4 = h^2$. It is in this respect that the DD2 approximation differs from the central (divided) difference (CD) approximation, as will be seen later. 

Following the choice in \cite{Noergaard-Advances, Norgaard--new}, we have
\begin{equation}
\begin{split}
\hat{\mathbf{P}}_{\eta} &= \dfrac{1}{4h^2} \left( \mathcal{N} (h) \mathcal{F} \left( \bar{\mathbf{x}}\right) \right)^T \left( \mathcal{N} (h) \mathcal{F} \left( \bar{\mathbf{x}}\right) \right) \\
& \quad + \dfrac{h^2 -1}{h^4} \sum\limits_{i=1}^{L} \left[ \left(\mathcal{P}_i (h) -1 \right) \mathcal{F} \left( \bar{\mathbf{x}} \right) \right] \left[ \left(\mathcal{P}_i (h) -1 \right) \mathcal{F} \left( \bar{\mathbf{x}} \right) \right]^T \\
& = \dfrac{1}{4h^2} \sum\limits_{i=1}^{L} \left( \mathcal{Y}_i -\mathcal{Y}_{L+i} \right) \left(  \mathcal{Y}_i -\mathcal{Y}_{L+i} \right)^T \\
& \quad+ \dfrac{h^2-1}{4h^4} \sum\limits_{i=1}^{L} \left( \mathcal{Y}_i + \mathcal{Y}_{L+i} -  2 \mathcal{Y}_0 \right) \left( \mathcal{Y}_i + \mathcal{Y}_{L+i} -  2 \mathcal{Y}_0 \right)^T  \, .
\end{split}
\end{equation}
Note that to guarantee the positive semi-definiteness of $\hat{\mathbf{P}}_{\eta}$, a sufficient condition is that $h \ge 1$.

Similarly, we have the estimated cross covariance
\begin{equation}
\begin{split}
\hat{\mathbf{P}}_{x\eta} & = \mathbb{E} \left[ \delta \mathbf{x} \left( \mathbf{\eta} - \hat{\mathbf{\eta}} \right)   ^T \right]\\
& \approx \dfrac{1}{2h} \, \mathbb{E} \left[ \mathbf{S}_x \delta \mathbf{e} \left( \delta \mathbf{e} \right)^T \mathcal{N} (h) \mathcal{F} \left( \bar{\mathbf{x}}\right) \right] \\
& = \dfrac{1}{2h} \, \mathbf{S}_x \mathcal{N} (h) \mathcal{F} \left( \bar{\mathbf{x}}\right) \\
& = \dfrac{1}{2h} \, \sum\limits_{i=1}^{L} \left( \mathbf{S}_x \right)_i \left( \mathcal{Y}_i -\mathcal{Y}_{L+i} \right)^T \, ,
\end{split}
\end{equation} 
which is the same as that of the first order approximation (cf. Eq.~(\ref{ch3:stirling_interpolation_1st_order_cross_cov})).

To summarize, in the DD2 approximation scheme, the solution to the recast problem is given by
\begin{subequations}  \label{ch4:ddf_2st_order_estimations}
\begin{align}
\label{ch4:ddf_2st_order_mean} &  \hat{\mathbf{\eta}} =  \dfrac{h^2-L}{h^2} \mathcal{Y}_0 + \dfrac{1}{2h^2} \sum\limits_{i=1}^{2L}\mathcal{Y}_i \, ;\\
\label{ch4:ddf_2st_order_cov} &  \hat{\mathbf{P}}_{\eta} =  \dfrac{1}{4h^2} \sum\limits_{i=1}^{L} \left( \mathcal{Y}_i -\mathcal{Y}_{L+i} \right) \left(  \mathcal{Y}_i -\mathcal{Y}_{L+i} \right)^T   \\
\nonumber & \quad \qquad  + \dfrac{h^2-1}{4h^4} \sum\limits_{i=1}^{L} \left( \mathcal{Y}_i + \mathcal{Y}_{L+i} -  2 \mathcal{Y}_0 \right) \left( \mathcal{Y}_i + \mathcal{Y}_{L+i} -  2 \mathcal{Y}_0 \right)^T \, ; \\
\label{ch4:ddf_2st_order_cross_cov} & \hat{\mathbf{P}}_{x\eta}  = \dfrac{1}{2h} \, \sum\limits_{i=1}^{L} \left( \mathbf{S}_x \right)_i \left( \mathcal{Y}_i -\mathcal{Y}_{L+i} \right)^T \, .
\end{align}
\end{subequations}

\subsubsection{Central (divided) difference approximation}
The central (divided) difference (CD) approximation scheme \cite{Ito-gaussian} is almost the same as the DD2 approximation scheme, except that it does not parameterize the fourth-order moment $\mathbb{E} (\delta e_i)^4$. Instead, it takes $\mathbb{E} (\delta e_i)^4 = 3$, as is the case for the Gaussian distribution   
$N \left( \delta \mathbf{e}: \mathbf{0}, \mathbf{I} \right)$. Thus we do not repeat the derivation. Instead, we summarize the main results as follows:
\begin{subequations}   \label{ch4:ddf_cdf_estimations}
\begin{align}
\label{ch4:ddf_cdf_mean} & \hat{\mathbf{\eta}} = \dfrac{h^2-L}{h^2} \mathcal{Y}_0 + \dfrac{1}{2h^2} \sum\limits_{i=1}^{2L}\mathcal{Y}_i \, ;\\
\label{ch4:ddf_cdf_cov} & \hat{\mathbf{P}}_{\eta}= \dfrac{1}{4h^2} \sum\limits_{i=1}^{L} \left( \mathcal{Y}_i -\mathcal{Y}_{L+i} \right) \left(  \mathcal{Y}_i -\mathcal{Y}_{L+i} \right)^T  \\
\nonumber & \quad \qquad + \dfrac{1}{2h^4} \sum\limits_{i=1}^{L} \left( \mathcal{Y}_i + \mathcal{Y}_{L+i} -  2 \mathcal{Y}_0 \right) \left( \mathcal{Y}_i + \mathcal{Y}_{L+i} -  2 \mathcal{Y}_0 \right)^T \, ; \\
\label{ch4:ddf_cdf_cross_cov} & \hat{\mathbf{P}}_{x\eta}  = \dfrac{1}{2h} \, \sum\limits_{i=1}^{L} \left( \mathbf{S}_x \right)_i \left( \mathcal{Y}_i -\mathcal{Y}_{L+i} \right)^T \, .
\end{align}
\end{subequations}

\subsection{Accuracy analysis} \label{sec:SIF_accuracy_analysis}
Following Chapter~\ref{ch3:ukf}, we conduct accuracy analyses for the divided difference approximation schemes. To this end, we first define perturbations $\left \{ \delta \mathcal{X}_i \right \}_{i=0}^{2L}$ of sigma points around $\bar{\mathbf{x}}$ according to Eq.~(\ref{ch4:ddf_sigma_points}):
\begin{equation}
\delta \mathcal{X}_i = 
\begin{cases}
0, & i=0 \, ;\\
h \left( \mathbf{S}_x \right)_i, & i =1, \dotsb, L \, ; \\
- h \left( \mathbf{S}_x \right)_{i-L}, & i =L+1, \dotsb, 2L \, . \\
\end{cases}
\end{equation}
Expanding $\mathcal{F}$ around $\bar{\mathbf{x}}$ gives (cf. Eq.~(\ref{ch3:sut_y_mean_in_expansion_de1})) 
\begin{equation}
\begin{split}
\mathcal{Y}_i &= \mathcal{F} \left(\bar{\mathbf{x}} + \delta \mathcal{X}_i \right) \\
& =  \mathcal{F} (\bar{\mathbf{x}}) + \mathbf{D}_{\delta \mathcal{X}_i} \mathcal{F} + \frac{\mathbf{D}_{\delta \mathcal{X}_i}^2 \mathcal{F}}{2!} +  \frac{\mathbf{D}_{\delta \mathcal{X}_i}^3 \mathcal{F}}{3!} + \dotsb \\
& = \mathcal{F} (\bar{\mathbf{x}}) + \delta \mathcal{X}_i^T \nabla \mathcal{F} + \dfrac{1}{2} \nabla^T \left( \delta \mathcal{X}_i \delta \mathcal{X}_i^T \right) \nabla \mathcal{F} \\
& \quad + \dfrac{1}{6}  \nabla^T \left( \delta \mathcal{X}_i \delta \mathcal{X}_i^T \nabla \delta \mathcal{X}_i^T \right) \nabla \mathcal{F} + \dotsb \, .
\end{split} 
\end{equation}

On the other hand, under the assumption that $\mathbf{x} \sim N \left( \mathbf{x}: \bar{\mathbf{x}}, \mathbf{P}_x \right)$, the mean and covariance of the transformed random variable $ \mathbf{\eta} = \mathcal{F} (\mathbf{x})$ are given by 
\begin{subequations} 
\begin{align}
 \tagref{ch3:sut_y_mean_in_expansion} \bar{\mathbf{\eta}} & = \mathbb{E} \left( \mathbf{\eta}\right) \\
\nonumber &= \mathcal{F} (\bar{\mathbf{x}}) + \frac{1}{2} \left( \nabla^T \mathbf{P}_x \nabla \right) \mathcal{F}  + \frac{1}{4!} \, \mathbb{E} \left( \mathbf{D}_{\delta \mathbf{x}}^4 \mathcal{F} \right) + \dotsb \, ,\\
 \tagref{ch3:sut_y_cov_in_expansion} \mathbf{P}_{\eta} &= \mathbb{E} \left[ \left( \mathbf{\eta} - \bar{\mathbf{\eta}}\right) \left( \mathbf{\eta} - \bar{\mathbf{\eta}}\right)^T \right] \\
\nonumber & =   \left( \nabla \mathcal{F} \right)^T \mathbf{P}_x \left( \nabla \mathcal{F} \right) + \mathbb{E} \left[ \frac{ \mathbf{D}_{\delta \mathbf{x}} \mathcal{F} \left( \mathbf{D}_{\delta \mathbf{x}}^3 \mathcal{F} \right)^T}{6}   \right. \\
\nonumber &\quad  +  \frac{ \mathbf{D}_{\delta \mathbf{x}}^2 \mathcal{F} \left( \mathbf{D}_{\delta \mathbf{x}}^2 \mathcal{F} \right)^T}{4} \left.  +  \frac{ \mathbf{D}_{\delta \mathbf{x}}^3 \mathcal{F} \left( \mathbf{D}_{\delta \mathbf{x}} \mathcal{F} \right)^T}{6} \right]    \\
\nonumber &\quad - \left[  \left( \frac{ \nabla^T \mathbf{P}_{x}  \nabla }{2}
\right) \mathcal{F} \right] \left[  \left( \frac{ \nabla^T
\mathbf{P}_{x}  \nabla }{2}  \right) \mathcal{F} \right]^T + \dotsb \, .
\end{align}
\end{subequations}  

\subsubsection{Accuracy of first order approximation}

Compared with Eq.~(\ref{ch3:sut_y_mean_in_expansion}), it is clear that the first order estimation of the mean in Eq.~(\ref{ch4:ddf_1st_order_mean}) is carried out only up to first order in the Taylor series expansion, which is zero in both equations under the assumption of Gaussianity.

On the other hand, note that
\begin{equation}
\mathcal{Y}_i - \mathcal{Y}_{i+L} = 2 \mathbf{D}_{\delta \mathcal{X}_i} \mathcal{F} +  \frac{1}{3} \mathbf{D}_{\delta \mathcal{X}_i}^3 \mathcal{F} + \dotsb \, ,~\text{for}~ i = 1, \dotsb, L \, ,
\end{equation}
where the even-order derivative terms vanish because of the symmetry in $\left \{ \delta \mathcal{X}_i \right \}_{i=0}^{2L}$. Thus we have
\begin{equation} \label{ch4:accuracy_dd1_cov}
\begin{split}
\hat{\mathbf{P}}_{\eta} & =   \dfrac{1}{4h^2} \sum\limits_{i=1}^{L} \left( \mathcal{Y}_i -\mathcal{Y}_{L+i} \right) \left(  \mathcal{Y}_i -\mathcal{Y}_{L+i} \right)^T \\
& = \dfrac{1}{h^2} \sum\limits_{i=1}^{L} \left\{ \mathbf{D}_{\delta \mathcal{X}_i} \mathcal{F} \left( \mathbf{D}_{\delta \mathcal{X}_i} \mathcal{F} \right)^T + \dfrac{1}{6} \mathbf{D}_{\delta \mathcal{X}_i} \mathcal{F} \left( \mathbf{D}_{\delta \mathcal{X}_i}^3 \mathcal{F}\right)^T + \dfrac{1}{6} \mathbf{D}_{\delta \mathcal{X}_i}^3 \mathcal{F} \left( \mathbf{D}_{\delta \mathcal{X}_i} \mathcal{F}\right)^T \right\} + \dotsb \\
& = \dfrac{1}{2h^2} \sum\limits_{i=0}^{2L} \left\{ \mathbf{D}_{\delta \mathcal{X}_i} \mathcal{F} \left( \mathbf{D}_{\delta \mathcal{X}_i} \mathcal{F} \right)^T + \dfrac{1}{6} \mathbf{D}_{\delta \mathcal{X}_i} \mathcal{F} \left( \mathbf{D}_{\delta \mathcal{X}_i}^3 \mathcal{F}\right)^T + \dfrac{1}{6} \mathbf{D}_{\delta \mathcal{X}_i}^3 \mathcal{F} \left( \mathbf{D}_{\delta \mathcal{X}_i} \mathcal{F}\right)^T \right\} + \dotsb \\
& = \left( \nabla \mathcal{F} \right)^T \mathbf{P}_x \left( \nabla \mathcal{F} \right) + \dfrac{1}{2h^2} \sum\limits_{i=0}^{2L} \left\{ \dfrac{1}{6} \mathbf{D}_{\delta \mathcal{X}_i} \mathcal{F} \left( \mathbf{D}_{\delta \mathcal{X}_i}^3 \mathcal{F}\right)^T + \dfrac{1}{6} \mathbf{D}_{\delta \mathcal{X}_i}^3 \mathcal{F} \left( \mathbf{D}_{\delta \mathcal{X}_i} \mathcal{F}\right)^T \right\} + \dotsb \\
\end{split}
\end{equation}  
Note that in the final line of the above equation, the terms
\[
\dfrac{1}{2h^2} \sum\limits_{i=0}^{2L} \left\{ \dfrac{1}{6} \mathbf{D}_{\delta \mathcal{X}_i} \mathcal{F} \left( \mathbf{D}_{\delta \mathcal{X}_i}^3 \mathcal{F}\right)^T + \dfrac{1}{6} \mathbf{D}_{\delta \mathcal{X}_i}^3 \mathcal{F} \left( \mathbf{D}_{\delta \mathcal{X}_i} \mathcal{F}\right)^T \right\} 
\]
can be considered as the estimation of the terms 
\[
\mathbb{E} \left[ \frac{ \mathbf{D}_{\delta \mathbf{x}} \mathcal{F} \left( \mathbf{D}_{\delta \mathbf{x}}^3 \mathcal{F} \right)^T}{6} +  \frac{ \mathbf{D}_{\delta \mathbf{x}}^3 \mathcal{F} \left( \mathbf{D}_{\delta \mathbf{x}} \mathcal{F} \right)^T}{6} \right]
\]
in Eq.~(\ref{ch3:sut_y_cov_in_expansion}). However, an estimation of the part 
\[
\mathbb{E} \left[ \frac{ \mathbf{D}_{\delta \mathbf{x}}^2 \mathcal{F} \left( \mathbf{D}_{\delta \mathbf{x}}^2 \mathcal{F} \right)^T}{4} \right] - \left[  \left( \frac{ \nabla^T \mathbf{P}_{x}  \nabla }{2}
\right) \mathcal{F} \right] \left[  \left( \frac{ \nabla^T
\mathbf{P}_{x}  \nabla }{2}  \right) \mathcal{F} \right]^T 
\] 
in Eq.~(\ref{ch3:sut_y_cov_in_expansion}) is missing. 

\subsubsection{Accuracy of second order approximations}
In order to analyze the accuracy of the mean estimation of the second order approximations, one may note the equivalence between the mean estimation of the unscented transform (UT) and those in Eqs.~(\ref{ch4:ddf_2st_order_mean}) and (\ref{ch4:ddf_cdf_mean}) \cite{Noergaard-Advances,Norgaard--new}. To see this, let $h=\sqrt{L+\lambda}$, with $\lambda$ being the free parameter of the UT (cf. Eq.~(\ref{ch3:ut_sigma_points})), and treat the set $\left\{\dfrac{h^2-L}{h^2}, \dfrac{1}{2h^2}, \dotsb, \dfrac{1}{2h^2} \right\}$ as the weights of the propagated sigma points $\left\{ \mathcal{Y}_0, \mathcal{Y}_1, \dotsb, \mathcal{Y}_{2L} \right \}$. Then it can be shown that Eqs.~(\ref{ch4:ddf_2st_order_mean}) and (\ref{ch4:ddf_cdf_mean}) are equivalent to Eq.~(\ref{ch3:ut_transformed_mean}). Thus the accuracy analysis of the mean estimations of the DD2 and CD approximations just follows that of the UT in Eq.~(\ref{ch3:sut_y_mean_in_expansion_de2}).

To analyze the accuracy of covariance estimations of the DD2 and CD approximations, we temporally parameterize the fourth order moment $\mathbb{E} \left( \delta e_i \right)^4 = \sigma_4$, and let the covariance estimation be
\begin{equation} \label{ch4:accuracy_dd2_cov_1}
\begin{split}
  \hat{\mathbf{P}}_{\eta} & =  \dfrac{1}{4h^2} \sum\limits_{i=1}^{L} \left( \mathcal{Y}_i -\mathcal{Y}_{L+i} \right) \left(  \mathcal{Y}_i -\mathcal{Y}_{L+i} \right)^T   \\
 & \quad  + \dfrac{\sigma_4-1}{4h^4} \sum\limits_{i=1}^{L} \left( \mathcal{Y}_i + \mathcal{Y}_{L+i} -  2 \mathcal{Y}_0 \right) \left( \mathcal{Y}_i + \mathcal{Y}_{L+i} -  2 \mathcal{Y}_0 \right)^T \, . \\
\end{split}
\end{equation}
By Taylor series expansion, one has 
\begin{equation}
\mathcal{Y}_i + \mathcal{Y}_{L+i} -  2 \mathcal{Y}_0 = \mathbf{D}_{\delta \mathcal{X}_i}^2 \mathcal{F} +  \frac{2}{4!} \mathbf{D}_{\delta \mathcal{X}_i}^4 \mathcal{F} + \dotsb \, ,~\text{for}~ i = 1, \dotsb, L \, .
\end{equation}
Moreover, note that $\delta \mathcal{X}_0 = \mathbf{0}$, therefore $\mathbf{D}_{\delta \mathcal{X}_0}^j \mathcal{F} = 0$ for all $j \ge 1 $. Because of the symmetry in $\delta \mathcal{X}_i$, one has
\begin{equation} \label{ch4:sec_unnumbered_eq_1}
\begin{split}
\sum\limits_{i=1}^{L} \mathbf{D}_{\delta \mathcal{X}_i}^2 \mathcal{F} \left( \mathbf{D}_{\delta \mathcal{X}_i}^2 \mathcal{F} \right)^T & = \dfrac{1}{2} \sum\limits_{i=0}^{2L} \mathbf{D}_{\delta \mathcal{X}_i}^2 \mathcal{F} \left( \mathbf{D}_{\delta \mathcal{X}_i}^2 \mathcal{F} \right)^T \\
 & = h^4 \left( \nabla^T \mathbf{P}_{x}  \nabla
\mathcal{F}\right) \left( \nabla^T \mathbf{P}_{x}  \nabla
\mathcal{F}\right)^T + \mathbf{\Delta}\, ,
\end{split}
\end{equation}
where
\begin{equation}
\mathbf{\Delta} = \sum\limits_{i=1}^{L} \mathbf{D}_{\delta \mathcal{X}_i}^2 \mathcal{F} \left( \mathbf{D}_{\delta \mathcal{X}_i}^2 \mathcal{F} \right)^T - h^4 \left( \nabla^T \mathbf{P}_{x}  \nabla
\mathcal{F}\right) \left( \nabla^T \mathbf{P}_{x}  \nabla
\mathcal{F}\right)^T \, .
\end{equation}
Since the set of sigma points $\left \{ \mathcal{X}_i \right \}_{i=0}^{2L}$ in general cannot capture all of the fourth order moments of the random variable $\mathbf{x}$, the term $\mathbf{\Delta}$ may not vanish.  

Substituting Eq.~(\ref{ch4:sec_unnumbered_eq_1}) and Eq.~(\ref{ch4:accuracy_dd1_cov}) into Eq.~(\ref{ch4:accuracy_dd2_cov_1}), we have
\begin{equation} \label{ch4:accuracy_dd2_cov_2}
\begin{split}
  \hat{\mathbf{P}}_{\eta} & =  \dfrac{1}{4h^2} \sum\limits_{i=1}^{L} \left( \mathcal{Y}_i -\mathcal{Y}_{L+i} \right) \left(  \mathcal{Y}_i -\mathcal{Y}_{L+i} \right)^T   \\
 & \quad  + \dfrac{\sigma_4-1}{4h^4} \sum\limits_{i=1}^{L} \left( \mathcal{Y}_i + \mathcal{Y}_{L+i} -  2 \mathcal{Y}_0 \right) \left( \mathcal{Y}_i + \mathcal{Y}_{L+i} -  2 \mathcal{Y}_0 \right)^T  \\
& = \left( \nabla \mathcal{F} \right)^T \mathbf{P}_x \left( \nabla \mathcal{F} \right) + \dfrac{1}{2h^2} \sum\limits_{i=0}^{2L} \left\{ \dfrac{1}{6} \mathbf{D}_{\delta \mathcal{X}_i} \mathcal{F} \left( \mathbf{D}_{\delta \mathcal{X}_i}^3 \mathcal{F}\right)^T + \dfrac{1}{6} \mathbf{D}_{\delta \mathcal{X}_i}^3 \mathcal{F} \left( \mathbf{D}_{\delta \mathcal{X}_i} \mathcal{F}\right)^T \right\}  \\
& \quad + \dfrac{\sigma_4 - 1}{4h^4}\sum\limits_{i=1}^{L} \mathbf{D}_{\delta \mathcal{X}_i}^2 \mathcal{F} \left( \mathbf{D}_{\delta \mathcal{X}_i}^2 \mathcal{F} \right)^T + \dotsb \\
& = \left( \nabla \mathcal{F} \right)^T \mathbf{P}_x \left( \nabla \mathcal{F} \right) + \dfrac{1}{2h^2} \sum\limits_{i=0}^{2L} \left\{ \dfrac{\mathbf{D}_{\delta \mathcal{X}_i} \mathcal{F} \left( \mathbf{D}_{\delta \mathcal{X}_i}^3 \mathcal{F}\right)^T}{6}  \right. \\
& \left. \quad + \dfrac{\sigma_4 }{h^2} \dfrac{\mathbf{D}_{\delta \mathcal{X}_i}^2 \mathcal{F} \left( \mathbf{D}_{\delta \mathcal{X}_i}^2 \mathcal{F} \right)^T}{4}  + \dfrac{\mathbf{D}_{\delta \mathcal{X}_i}^3 \mathcal{F} \left( \mathbf{D}_{\delta \mathcal{X}_i} \mathcal{F}\right)^T}{6}  \right\}  \\
& \quad  - \left[  \left( \frac{ \nabla^T \mathbf{P}_{x}  \nabla }{2} \right) \mathcal{F} \right] \left[  \left( \frac{ \nabla^T
\mathbf{P}_{x}  \nabla }{2}  \right) \mathcal{F} \right]^T + \dotsb \, , 
\end{split}
\end{equation} 
where 
\[
\dfrac{1}{2h^2} \sum\limits_{i=0}^{2L} \left\{ \dfrac{\mathbf{D}_{\delta \mathcal{X}_i} \mathcal{F} \left( \mathbf{D}_{\delta \mathcal{X}_i}^3 \mathcal{F}\right)^T}{6} + \dfrac{\sigma_4 }{h^2} \dfrac{\mathbf{D}_{\delta \mathcal{X}_i}^2 \mathcal{F} \left( \mathbf{D}_{\delta \mathcal{X}_i}^2 \mathcal{F} \right)^T}{4}  + \dfrac{\mathbf{D}_{\delta \mathcal{X}_i}^3 \mathcal{F} \left( \mathbf{D}_{\delta \mathcal{X}_i} \mathcal{F}\right)^T}{6} \right \} 
\]
can be considered as the estimation of 
\[
\mathbb{E} \left[ \frac{ \mathbf{D}_{\delta \mathbf{x}} \mathcal{F} \left( \mathbf{D}_{\delta \mathbf{x}}^3 \mathcal{F} \right)^T}{6}   +  \frac{ \mathbf{D}_{\delta \mathbf{x}}^2 \mathcal{F} \left( \mathbf{D}_{\delta \mathbf{x}}^2 \mathcal{F} \right)^T}{4} +  \frac{ \mathbf{D}_{\delta \mathbf{x}}^3 \mathcal{F} \left( \mathbf{D}_{\delta \mathbf{x}} \mathcal{F} \right)^T}{6} \right]    
\]
in Eq.~(\ref{ch3:sut_y_cov_in_expansion}).

If one takes $\sigma_4 = h^2$, $h^2=L+\lambda$, and $W_i = 1/2(L+\lambda) = 1/2h^2$ for $i=1,\dotsb,2L$, then it can be shown that the covariance estimation of the DD2 approximation matches that of the UT in Eq.~(\ref{ch3:ut_y_ut_cov_in_expansion}) (with $\beta =0$) for the terms \em{presented} on the rhs of Eq.~(\ref{ch4:accuracy_dd2_cov_2}) \footnote{But the omitted terms on the rhs will not be completely the same. In fact, it can be verified that the covariance estimation Eq.~(\ref{ch4:ddf_2st_order_cov}) of the DD2 approximation in general is not equal to the covariance estimation Eq.~(\ref{ch3:ut_transformed_cov}) of the UT, even when $\beta=0$.}. On the other hand, if one lets $\sigma_4 = 3$, then, in general, there would be a deviation of the CD approximation from the UT estimation in the term $\mathbf{D}_{\delta \mathcal{X}_i}^2 \mathcal{F} \left( \mathbf{D}_{\delta \mathcal{X}_i}^2 \mathcal{F} \right)^T$.    

\section{Divided difference filters as the approximate solutions}\label{ch4:sec_ddfs}
Incorporating the divided difference approximations into the propagation step of RBE leads to the corresponding divided difference filters (DDFs). Without loss of generality, we assume that at time instant $k-1$, one has obtained the analysis sample mean $\hat{\mathbf{x}}^a_{k-1}$ and a square root $\mathbf{S}^{xa}_{k-1}$ of the error covariance $\hat{\mathbf{P}}^a_{k-1}$. Based on these, a set of $2L_{k-1}+1$ ($L_{k-1} \ge m$) sigma points with respect to the triplet $\left(h, \hat{\mathbf{x}}^a_{k-1}, \mathbf{S}^{xa}_{k-1}\right)$ can be generated in the spirit of Eq.~(\ref{ch4:ddf_sigma_points}), so that
\begin{equation} \label{spkf sigma points}
\begin{split}
& \mathcal{X}_{k-1,0}^a = \hat{\mathbf{x}}^a_{k-1},\\
& \mathcal{X}_{k-1,i}^a = \hat{\mathbf{x}}^a_{k-1} + h \left( \mathbf{S}^{xa}_{k-1} \right)_i, \, i=1, 2, \dotsb, L_{k-1} ,\\
& \mathcal{X}_{k-1,i}^a =  \hat{\mathbf{x}}^a_{k-1} - h \left( \mathbf{S}^{xa}_{k-1} \right)_{i-L_{k-1}}, \, i=L_{k-1}+1, L_{k-1}+2, \dotsb, 2L_{k-1}. \\
\end{split}
\end{equation}

After generating sigma points at $k-1$, one propagates them forward through the system model. Let the ensemble of forecasts of the propagations be 
\begin{equation}\label{forecasts of propagations}
\mathbf{X}_k^b = \left \{ \mathbf{x}_{k,i}^b: \mathbf{x}_{k,i}^b =  \mathcal{M}_{k,k+1} \left( \mathcal{X}_{k-1,i}^a \right), i=0,\dotsb,2L_{k-1} \right \} \, .
\end{equation}
Then the ensemble mean $\hat{\mathbf{x}}_{k}^b$ and covariance $ \hat{\mathbf{P}}_k^b$ of the background can be estimated in a way consistent with the chosen approximation method, as will be shown below. For convenience, we also split the procedures of the DDFs into the propagation (or prediction) step and the filtering step.

\clearpage
\bigskip
\begin{sidewaystable*}[!h]
\centering
\caption{\label{SQRT in spkfs} Square roots at the propagation steps of sigma point Kalman filters.}
\begin{tabular}{p{1.5cm}ll}
\hline \hline
 & \multicolumn{1}{c}{Square Roots}   &  \multicolumn{1}{c}{Remarks}  \\
\hline 
\multirow{3}{*}{SUKF} &  $ \mathbf{S}^{x}_k = \left[ \sqrt{W_{k,0}^{\alpha \beta}} \left( \mathbf{x}_{k,0}^b - \hat{\mathbf{x}}_{k}^b \right), \sqrt{W_{k,1}} \left( \mathbf{x}_{k,1}^b - \hat{\mathbf{x}}_{k}^b \right), \dotsb,  \sqrt{W_{k,2L_{k-1}}} \left( \mathbf{x}_{k,2L_{k-1}}^b - \hat{\mathbf{x}}_{k}^b \right) \right]$   & \multirow{3}{*}{$W_{k,0}^{\alpha \beta} = W_{k,0}+1+\beta-\alpha^2$}  \\[+0.15cm]
          &  $ \mathbf{S}^{h}_k = \left[ \sqrt{W_{k,0}^{\alpha \beta}} \left( \mathcal{H}_k \left ( \mathbf{x}_{k,0}^b \right )-  \hat{\mathbf{y}}_{k} \right), \sqrt{W_{k,1}} \left( \mathcal{H}_k \left ( \mathbf{x}_{k,1}^b \right )- \hat{\mathbf{y}}_{k} \right), \right. $ &     \\
		& $ \qquad \quad \left. \dotsb, \sqrt{W_{k,2L_{k-1}}} \left( \mathcal{H}_k \left ( \mathbf{x}_{k,2L_{k-1}}^b \right )-  \hat{\mathbf{y}}_{k} \right) \right]$ &  \\ [+0.15cm]
\hline  
 \multirow{2}{*}{DD1} & $ \mathbf{S}^{x}_k = \dfrac{1}{2h} \left[  \mathbf{x}_{k,1}^b - \mathbf{x}_{k,L_{k-1}+1}^b, \dotsb ,  
 \mathbf{x}_{k,L_{k-1}}^b - \mathbf{x}_{k,2L_{k-1}}^b \right]$ &  \multirow{2}{*}{$-$} \\[+0.15cm]
& $ \mathbf{S}^{h}_k = \mathbf{S}^{h1}_k  = \dfrac{1}{2h} \left[\mathbf{y}_{k,1}^b - \mathbf{y}_{k,L_{k}+1}^b, \dotsb,  \mathbf{y}_{k,L_{k}}^b - \mathbf{y}_{k,2L_{k}}^b  \right]$ &  \\[+0.15cm]
\hline 
\multirow{6}{*}{DD2} & $ \mathbf{S}^{x}_k  =\left[  \mathbf{S}^{x1}_k,  \mathbf{S}^{x2}_k \right]$ & \\[+0.15cm]  
& $\qquad  \mathbf{S}^{x1}_k = \dfrac{1}{2h} \left[  \mathbf{x}_{k,1}^b - \mathbf{x}_{k,L_{k-1}+1}^b, \dotsb,  \mathbf{x}_{k,L_{k-1}}^b - \mathbf{x}_{k,2L_{k-1}}^b \right]$  & \multirow{6}{*}{$-$} \\[+0.15cm]
&  $\qquad \mathbf{S}^{x2}_k = \dfrac{\sqrt{h^2-1}}{2h^2} \left[  \mathbf{x}_{k,1}^b + \mathbf{x}_{k,L_{k-1}+1}^b - 2\mathbf{x}_{k,0}^b, \dotsb,  \mathbf{x}_{k,L_{k-1}}^b + \mathbf{x}_{k,2L_{k-1}}^b - 2\mathbf{x}_{k,0}^b \right]$ & \\[+0.15cm]
&  $ \mathbf{S}^{h}_k  =\left[  \mathbf{S}^{h1}_k,  \mathbf{S}^{h2}_k \right]$ & \\[+0.15cm]   
& $ \qquad \mathbf{S}^{h1}_k = \dfrac{1}{2h} \left[\mathbf{y}_{k,1}^b - \mathbf{y}_{k,L_{k}+1}^b, \dotsb,  \mathbf{y}_{k,L_{k}}^b - \mathbf{y}_{k,2L_{k}}^b  \right]$ & \\[+0.15cm]
&  $ \qquad  \mathbf{S}^{h2}_k = \dfrac{\sqrt{h^2-1}}{2h^2} \left[\mathbf{y}_{k,1}^b + \mathbf{y}_{k,L_{k}+1}^b - 2 \mathbf{y}_{k,0}^b, \dotsb, \mathbf{y}_{k,L_{k}+1}^b + \mathbf{y}_{k,2L_{k}}^b - 2 \mathbf{y}_{k,0}^b \right]$ & \\[+0.15cm]
\hline 
\multirow{6}{*}{CDF} & $ \mathbf{S}^{x}_k  =\left[  \mathbf{S}^{x1}_k,  \mathbf{S}^{x2}_k \right]$ & \\[+0.15cm]  
& $\qquad \mathbf{S}^{x1}_k = \dfrac{1}{2h} \left[  \mathbf{x}_{k,1}^b - \mathbf{x}_{k,L_{k-1}+1}^b, \dotsb,  \mathbf{x}_{k,L_{k-1}}^b - \mathbf{x}_{k,2L_{k-1}}^b \right]$  & \multirow{6}{*}{$-$} \\[+0.15cm]
&  $\qquad \mathbf{S}^{x2}_k = \dfrac{\sqrt{2}}{2h^2} \left[  \mathbf{x}_{k,1}^b + \mathbf{x}_{k,L_{k-1}+1}^b - 2\mathbf{x}_{k,0}^b, \dotsb,  \mathbf{x}_{k,L_{k-1}}^b + \mathbf{x}_{k,2L_{k-1}}^b - 2\mathbf{x}_{k,0}^b \right]$ & \\[+0.15cm]
&  $ \mathbf{S}^{h}_k  =\left[  \mathbf{S}^{h1}_k,  \mathbf{S}^{h2}_k \right]$ & \\[+0.15cm]   
& $ \qquad \mathbf{S}^{h1}_k = \dfrac{1}{2h} \left[\mathbf{y}_{k,1}^b - \mathbf{y}_{k,L_{k}+1}^b, \dotsb,  \mathbf{y}_{k,L_{k}}^b - \mathbf{y}_{k,2L_{k}}^b \right]$ & \\[+0.15cm]
&  $ \qquad  \mathbf{S}^{h2}_k = \dfrac{\sqrt{2}}{2h^2} \left[\mathbf{y}_{k,1}^b + \mathbf{y}_{k,L_{k}+1}^b - 2 \mathbf{y}_{k,0}^b, \dotsb, \mathbf{y}_{k,L_{k}+1}^b + \mathbf{y}_{k,2L_{k}}^b - 2 \mathbf{y}_{k,0}^b \right]$ & \\[+0.15cm]
\hline \hline
\end{tabular}
\end{sidewaystable*}

\clearpage

\subsection{Propagation step} \label{ch4:sec_ddf_progagation_step}
In the DDFs, the ensemble mean $\hat{\mathbf{x}}_{k}^b$ and covariance $ \hat{\mathbf{P}}_k^b$ are evaluated according to the chosen approximation method. Specifically,

\begin{itemize}
\item[-] For the DD1 filter 
\begin{subequations} 
\begin{align}
\label{dd1 mean} \hat{\mathbf{x}}_{k}^b = & \mathbf{x}_{k,0}^b, \\
 \label{dd1 cov} \hat{\mathbf{P}}_k^b= &  \dfrac{1}{4h^2} \sum\limits_{i=1}^{L_{k-1}} \left( \mathbf{x}_{k,i}^b - \mathbf{x}_{k,L_{k-1}+i}^b \right) \left(  \mathbf{x}_{k,i}^b - \mathbf{x}_{k,L_{k-1}+i}^b \right)^T + \mathbf{Q}_k.
\end{align}
\end{subequations}
\item[-] For the DD2 filter 
\begin{subequations}  
\begin{align}
 \label{dd2 mean} \hat{\mathbf{x}}_{k}^b =& \dfrac{h^2-L_{k-1}}{h^2}  \mathbf{x}_{k,0}^b + \dfrac{1}{2h^2} \sum\limits_{i=1}^{2L_{k-1}}\mathbf{x}_{k,i}^b,\\
 \label{dd2 cov} \hat{\mathbf{P}}_k^b = & \dfrac{1}{4h^2} \sum\limits_{i=1}^{L_{k-1}} \left( \mathbf{x}_{k,i}^b - \mathbf{x}_{k,L_{k-1}+i}^b \right) \left( \mathbf{x}_{k,i}^b - \mathbf{x}_{k,L_{k-1}+i}^b \right)^T  \\
\nonumber & + \dfrac{h^2-1}{4h^4} \sum\limits_{i=1}^{L_{k-1}} \left( \mathbf{x}_{k,i}^b + \mathbf{x}_{k,L_{k-1}+i}^b -  2 \mathbf{x}_{k,0}^b \right) \left(  \mathbf{x}_{k,i}^b + \mathbf{x}_{k,L_{k-1}+i}^b -  2 \mathbf{x}_{k,0}^b \right)^T + \mathbf{Q}_k .
\end{align}
\end{subequations}
\item[-] For the central difference filter (CDF) 
\begin{subequations}  
\begin{align}
 \label{cdf mean} \hat{\mathbf{x}}_{k}^b =& \dfrac{h^2-L_{k-1}}{h^2}  \mathbf{x}_{k,0}^b + \dfrac{1}{2h^2} \sum\limits_{i=1}^{2L_{k-1}}\mathbf{x}_{k,i}^b,\\
 \label{cdf cov} \hat{\mathbf{P}}_k^b = & \dfrac{1}{4h^2} \sum\limits_{i=1}^{L_{k-1}} \left( \mathbf{x}_{k,i}^b - \mathbf{x}_{k,L_{k-1}+i}^b \right) \left( \mathbf{x}_{k,i}^b - \mathbf{x}_{k,L_{k-1}+i}^b \right)^T  \\
\nonumber & + \dfrac{1}{2h^4} \sum\limits_{i=1}^{L_{k-1}} \left( \mathbf{x}_{k,i}^b + \mathbf{x}_{k,L_{k-1}+i}^b -  2 \mathbf{x}_{k,0}^b \right)  \left(  \mathbf{x}_{k,i}^b + \mathbf{x}_{k,L_{k-1}+i}^b -  2 \mathbf{x}_{k,0}^b \right)^T + \mathbf{Q}_k .
\end{align}
\end{subequations}
\end{itemize}

For each DDF one can also re-write its background covariance $\hat{\mathbf{P}}_k^b$ in terms of some square roots, so that
\begin{equation} \label{ch4:ddf_cov_in_sqrt}
\hat{\mathbf{P}}_k^b= \mathbf{S}^{xb}_k \left( \mathbf{S}^{xb}_k \right)^T = \mathbf{S}^{x}_k \left( \mathbf{S}^{x}_k \right)^T + \mathbf{Q}_k, 
\end{equation}
where the square root $\mathbf{S}^{x}_k$ of each DDF is listed in Table~\ref{SQRT in spkfs} for convenience (also with the square roots of the SUKF listed there for comparison). To compute the square root $\mathbf{S}^{xb}_k$, one can let $ \mathbf{S}^{xb}_k = \sqrt{ \mathbf{S}^{x}_k \left( \mathbf{S}^{x}_k \right)^T + \mathbf{Q}_k }$ , following the numerical scheme in \S~\ref{ch1:SRKF}. 

If the observation operator $\mathcal{H}_{k}$ is nonlinear, then a divided difference approximation has to be conducted on $\mathcal{H}_{k}$ once again. This is because the background ensemble $\mathbf{X}_k^b$ in Eq.~(\ref{forecasts of propagations}) is in general no longer symmetric about the sample mean $\hat{\mathbf{x}}_{k}^b$ as are sigma points $\left\{ \mathcal{X}_{k-1,i}^a \right\}_{i=1}$ in the previous assimilation cycle. Thus one has to regenerate sigma points with respect to the triplet $\left(h, \hat{\mathbf{x}}_{k}^b, \mathbf{S}^{xb}_k \right)$ in order to conduct a divided difference approximation to estimate the cross and projection covariances of the background ensemble (cf. Eq.~(\ref{ddf covariance}) below) \footnote{In contrast, the SUKF does not regenerate sigma points at the propagation step because it is based on statistical approximation. Each member of the background ensemble $\mathbf{X}_k^b$ is already assigned a weight for the purpose of approximation when sigma points are generated in the previous cycle. Thus it is not necessary to require that the ensemble members of $\mathbf{X}_k^b$ be symmetric about $\hat{\mathbf{x}}_{k}^b$.}.

Let $L_{k}$ be the number of the column vectors of $\mathbf{S}^{xb}_k$. Then one generates another set of sigma points, $\mathcal{X}_k^b=\left\{ \mathcal{X}_{k,0}^b, \dotsb, \mathcal{X}_{k,2L_{k}}^b \right\}$, with respect to $\left(h, \hat{\mathbf{x}}_{k}^b, \mathbf{S}^{xb}_k \right)$ as follows:
\begin{equation}
\begin{split}
& \mathcal{X}_{k,0}^b = \hat{\mathbf{x}}^b_{k},\\
& \mathcal{X}_{k,i}^b =  \hat{\mathbf{x}}^b_{k} + h \left( \mathbf{S}^{xb}_k \right)_i, \, i=1, \dotsb, L_{k} ,\\
& \mathcal{X}_{k,i}^b =   \hat{\mathbf{x}}^b_{k} - h \left( \mathbf{S}^{xb}_k \right)_{i-L_{k}}, \, i=L_{k}+1, \dotsb, 2L_{k}. \\
\end{split}
\end{equation}

Similarly, we can define a set of forecasts of the projections of the above sigma points 
\begin{equation}\label{forecasts of projections}
\mathbf{Y}_k^b = \left \{ \mathbf{y}_{k,i}^b: \mathbf{y}_{k,i}^b =  \mathcal{H}_{k} \left( \mathcal{X}_{k,i}^b \right), i=0,\dotsb,2L_{k} \right \}.
\end{equation}
Then the cross and projection covariances of all the DDFs, in terms of square roots, can be computed as follows:
\begin{subequations} \label{ddf covariance}
\begin{align}
\label{ddf cross covariance} \hat{\mathbf{P}}^{cr}_k = & \mathbf{S}^{xb}_k \left( \mathbf{S}^{h1}_k \right)^T, \\
\label{ddf projection covariance} \hat{\mathbf{P}}^{pr}_k =  & \mathbf{S}^{h}_k \left( \mathbf{S}^{h}_k \right)^T, 
\end{align}
\end{subequations}
where 
\begin{equation}
\mathbf{S}^{h1}_k = \dfrac{1}{2h} \left[\mathbf{y}_{k,1}^b - \mathbf{y}_{k,L_{k}+1}^b, \dotsb,  \mathbf{y}_{k,L_{k}}^b - \mathbf{y}_{k,2L_{k}}^b  \right]
\end{equation}
is the same for all the DDFs, but $\mathbf{S}^{h}_k$ has different forms, which are again summarized in Table~\ref{SQRT in spkfs}. 

Finally, the Kalman gain $\mathbf{K}_k$ is given by
\begin{equation} \label{Kalman gain}
\begin{split}
\mathbf{K}_k &= \hat{\mathbf{P}}^{cr}_k  \left( \hat{\mathbf{P}}^{pr}_k + \mathbf{R}_k \right)^{-1} \\
& = \mathbf{S}^{xb}_k \left( \mathbf{S}^{h1}_k \right)^T  \left( \mathbf{S}^{h}_k \left( \mathbf{S}^{h}_k \right)^T + \mathbf{R}_k \right)^{-1}.
\end{split}
\end{equation}

\subsection{Filtering step}  \label{ch4:sec_ddf_filtering_step}

At the filtering step, the procedures of the DDFs are the same as those of the SUKF. One first computes the updated sample mean and covariance through the following formulae
\begin{subequations}
\begin{align}
\tag{$\ref{analysis mean}$} & \hat{\mathbf{x}}_k^{a} =  \hat{\mathbf{x}}_k^{b} + \mathbf{K}_k  \left( \mathbf{y}_k - \mathcal{H}_k \left( \hat{\mathbf{x}}_k^{b} \right) \right),\\
\tag{$\ref{analysis cov}$} & \hat{\mathbf{P}}_k^a = \hat{\mathbf{P}}_k^b -  \mathbf{K}_k \left( \hat{\mathbf{P}}^{cr}_k \right)^T.
\end{align}
\end{subequations}
By adopting a certain algorithm to compute a square root $\mathbf{S}^{xa}_k$ of $\hat{\mathbf{P}}_k^a$, one generates a new set of sigma points, now denoted by $\mathcal{X}_k^a=\left\{ \mathcal{X}_{k,0}^a, \mathcal{X}_{k,1}^a, \dotsb  \right\}$, in the spirit of Eq.~(\ref{spkf sigma points}). Propagating these sigma points forward, one starts a new assimilation cycle at instant $k+1$. 

\subsection{Summary of the divided difference filters}

We summarize the main procedures of the DDFs as follows:\\ \\
Propagation step:
\begin{subequations} \label{ch4:DDF_propagation}
\begin{align}
& \mathbf{X}^b_{k} = \left\{ \mathbf{x}^b_{k,i}: \mathbf{x}^b_{k,i} =  \mathcal{M}_{k,k-1} \left( \mathcal{X}_{k-1,i}^a \right) \right \}_{i=0}^{2L_{k-1}} \, , \\
 \label{ch4:DDF_propagation_generic_mean} & \left[ \hat{\mathbf{x}}_k^b , \mathbf{S}^{x}_k, \hat{\mathbf{P}}_k^b \right ]= \text{ddf}\left( h,\mathbf{X}^b_{k}, \mathbf{Q}_k \right) \, , \\
& \mathbf{S}^{xb}_k = \sqrt{ \mathbf{S}^{x}_k \left( \mathbf{S}^{x}_k \right)^T + \mathbf{Q}_k }\, , \\
\label{ch4:DDF_propagation_sigma_points} & \mathcal{X}^b_{k} = \left\{ \mathcal{X}^b_{k,i}: \mathcal{X}^b_{k,i} =  \sigma \left( h, \hat{\mathbf{x}}_k^b, \mathbf{S}^{xb}_k \right) \right \}_{i=0}^{2L_{k}} \, , \\
&\mathbf{Y}_k^b = \left \{ \mathbf{y}_{k,i}^b: \mathbf{y}_{k,i}^b =  \mathcal{H}_{k} \left( \mathcal{X}_{k,i}^b \right)\right \}_{i=0}^{2L_{k}} \, , \\
\label{ch4:DDF_propagation_generic_sqrt} & \left[ \mathbf{S}^{h1}_k, \mathbf{S}^{h}_k \right ]= \text{ddf}\left( h, \mathbf{Y}_k^b, \mathbf{R}_k \right) \, , \\
& \hat{\mathbf{P}}^{cr}_k = \mathbf{S}^{xb}_k \left( \mathbf{S}^{h1}_k \right)^T \, , \\
& \hat{\mathbf{P}}^{pr}_k =   \mathbf{S}^{h}_k \left( \mathbf{S}^{h}_k \right)^T \, , \\
&\mathbf{K}_k = \mathbf{S}_k^{xb} \left( \mathbf{S}_{k}^{h1} \right)^T \left( \mathbf{S}_{k}^{h} \left( \mathbf{S}_{k}^{h} \right)^T + \mathbf{R}_k \right)^{-1} \, , 
\end{align}
\end{subequations}
where Eqs.~(\ref{ch4:DDF_propagation_generic_mean}) and (\ref{ch4:DDF_propagation_generic_sqrt}) mean that $\hat{\mathbf{x}}_k^b$, $\mathbf{S}^{x}_k$, $\hat{\mathbf{P}}_k^b$, $\mathbf{S}^{h1}_k$ and $\mathbf{S}^{h}_k$ are computed according to the divided difference approximation scheme in use, while Eq.~(\ref{ch4:DDF_propagation_sigma_points}) means that the sigma points are generated with respect to the triplet $\left( h, \hat{\mathbf{x}}_k^b, \mathbf{S}^{xb}_k \right)$.  \\ \\
Filtering step:
\begin{subequations} \label{ch4:DDF_filtering}
\begin{align}
& \hat{\mathbf{x}}_k^a = \hat{\mathbf{x}}_k^b + \mathbf{K}_k \left ( \mathbf{y}_k - \mathcal{H}_k \left( \hat{\mathbf{x}}_k^b \right) \right ) \, , \\
& \hat{\mathbf{P}}_k^a = \hat{\mathbf{P}}_k^b -  \mathbf{K}_k \left( \hat{\mathbf{P}}^{cr}_k \right)^T \, , \\
&\mathbf{S}_{k}^{xa} = \sqrt{\hat{\mathbf{P}}_k^a} \, .
\end{align}
\end{subequations} \\ \\
Analysis scheme:\\
Sigma points:
\begin{equation} 
\begin{split}
\mathcal{X}^a_{k} = \left\{ \mathcal{X}^a_{k,i}: \mathcal{X}^a_{k,i} =  \sigma \left( h, \hat{\mathbf{x}}_k^a, \mathbf{S}^{xa}_k \right), i =1, 2, \dotsb \right \} \, . \\
\end{split}
\end{equation}

\section{Reduced rank divided difference filters for high dimensional systems} \label{ch4:sec_reduced_rank_ddfs}
For the DDFs, the modification scheme is similar to that of the SUKF. The difference for the DDFs lies in that, one has to generate sigma points twice, rather than only once as in the SUKF. Moreover, as to be shown below, in a DDF, a truncated singular value decomposition (SVD) is conducted at the propagation step for the generation of sigma points, rather than at the filtering step as in the SUKF. For convenience of discussion, we also assume that at time instant $k-1$, one has obtained a set of sigma points $\mathcal{X}_{k-1}^a = \left \{\mathcal{X}_{k-1,0}^a, \dotsb,\mathcal{X}_{k-1,2l_{k-1}}^a \right \}$.
\subsection{Propagation step}
We define a set of forecasts of the propagated sigma points by
 \begin{equation}\label{forecasts DA}
\mathbf{X}_k^b = \left \{ \mathbf{x}_{k,i}^b: \mathbf{x}_{k,i}^b =  \mathcal{M}_{k,k+1} \left( \mathcal{X}_{k-1,i}^a\right), i=0,\dotsb,2l_{k-1} \right \} \, ,
\end{equation}
based upon which the ensemble mean $\hat{\mathbf{x}}_{k}^{b}$ and covariance $\hat{\mathbf{P}}_{k}^{b}$ can be computed accordingly, depending on which approximation scheme is chosen (cf. \S~\ref{ch4:sec_ddf_progagation_step}). 

Let $\hat{\mathbf{P}}_{k}^{b}$ be decomposed as
\begin{equation}
\hat{\mathbf{P}}_{k}^{b} = \mathbf{E}_k^b \mathbf{D}_k^b \left(\mathbf{E}_k^b \right)^T,
\end{equation}
where $\mathbf{D}_k^b = \text{diag} (\sigma_{k,1}^2, \dotsb, \sigma_{k,m}^2)$ is a diagonal matrix of eigenvalues $\sigma_{k,i}^2$ of $\hat{\mathbf{P}}_k^b$, and $\mathbf{E}_K^b = \left[\mathbf{e}_{k,1}, \dotsb,  \mathbf{e}_{k,m} \right] $  is the matrix consisting of the corresponding eigenvectors $\mathbf{e}_{k,i}$. Then, a new set of $2l_k+1$ sigma points, denoted by  $\mathcal{X}_{k}^b = \left \{\mathcal{X}_{k,0}^b, \dotsb,\mathcal{X}_{k,2l_{k}}^b \right \}$, can be generated as follows:
\begin{equation} \label{spkf DA: sigma point generation}
\begin{split}
& \mathcal{X}_{k,0}^b = \hat{\mathbf{x}}_k^{b},\\
& \mathcal{X}_{k,i}^b =  \hat{\mathbf{x}}^{b}_k + h \sigma_{k,i} \mathbf{e}_{k,i}, \,~ i=1, \dotsb, l_k ,\\
& \mathcal{X}_{k,i}^b =  \hat{\mathbf{x}}^{b}_k - h \sigma_{k,i-l_k} \mathbf{e}_{k,i-l_k}, \,~ i=l_k+1, \dotsb, 2l_k ,\\
\end{split}
\end{equation}
where $l_k$ is an integer satisfying 
\begin{equation} \label{enukf:threshold}
\begin{split}
& \sigma_{k,i}^2 > \text{trace} \left( \hat{\mathbf{P}}_k^b \right) / \Gamma_k , \, ~ i=1, \dotsb, l_k \, ,\\
& \sigma_{k,i}^2 \le \text{trace} \left( \hat{\mathbf{P}}_k^b \right) / \Gamma_k , \, ~  i>l_k +1 \, ,
\end{split}
\end{equation}
with $\Gamma_k$ being a pre-specified threshold (the values of $\Gamma_k$ are chosen in the same way as in \S~\ref{ch3:sec_ex_implementation_issue}). Moreover, we also specify upper and lower bounds, $l_u$ and $l_l$ respectively, to prevent $l_k$ from getting too large or too small.

By projecting the sigma points in Eq.~(\ref{spkf DA: sigma point generation}) onto the observation space, we have the forecasts of the projections
\begin{equation}\label{DA: forecasts of propagations}
\mathbf{Y}_k^b = \left \{ \mathbf{y}_{k,i}^b: \mathbf{y}_{k,i}^b =  \mathcal{H}_{k} \left( \mathcal{X}_{k,i}^b \right), i=0,\dotsb,2l_k \right \}.
\end{equation}

Based on sigma points and their projection forecasts, we can obtain some approximate square roots. Specifically, we take $\tilde{\mathbf{S}}_k^{xb} = \left[\sigma_{k,1} \mathbf{e}_{k,1}, \dotsb, \sigma_{k,l_k} \mathbf{e}_{k,l_k} \right]$ as an approximate square root of $\hat{\mathbf{P}}_k^b$. The approximate square roots to $\mathbf{S}_k^{h1}$ and $\mathbf{S}_k^{h}$, denoted by $\tilde{\mathbf{S}}_k^{h1}$ and $\tilde{\mathbf{S}}_k^{h}$ respectively, are computed according to the formulae in Table~\ref{SQRT in spkfs}, but with $L_{k-1}$ and $L_{k}$ therein replaced by $l_{k-1}$ and $l_k$, respectively. The corresponding approximate cross and projection covariances are given by
\begin{subequations} 
\begin{align}
\label{DA: cross covariance} \tilde{\mathbf{P}}^{cr}_k = & \tilde{\mathbf{S}}_k^{xb} \left( \tilde{\mathbf{S}}_k^{h1} \right)^T, \\
\label{DA: projection covariance} \tilde{\mathbf{P}}^{pr}_k =  & \tilde{\mathbf{S}}_k^{h} \left( \tilde{\mathbf{S}}_k^{h} \right)^T.
\end{align}
\end{subequations}
Consequently, the Kalman gain is approximated by 
\begin{equation} \label{Kalman gain}
\begin{split}
\tilde{\mathbf{K}}_k &= \tilde{\mathbf{P}}^{cr}_k \left(\tilde{\mathbf{P}}^{pr}_k + \mathbf{R}_k \right)^{-1} \\
&=\tilde{\mathbf{S}}_k^{xb}  \left( \tilde{\mathbf{S}}_k^{h1} \right)^T  \left( \tilde{\mathbf{S}}_k^{h} \left( \tilde{\mathbf{S}}_k^{h} \right)^T + \mathbf{R}_k \right)^{-1}.
\end{split}
\end{equation}

\subsection{Filtering step}
When a new observation $\mathbf{y}_k$ is available, the ensemble mean is updated as follows:
\begin{equation}
\label{DA: analysis mean}  \hat{\mathbf{x}}_k^{a} =  \hat{\mathbf{x}}_k^{b} + \tilde{\mathbf{K}}_k  \left( \mathbf{y}_k - \mathcal{H}_k \left( \hat{\mathbf{x}}_k^{b} \right) \right).
\end{equation}

To obtain a square root of the updated covariance $\hat{\mathbf{P}}_{k}^{a}$ , in principle, one may compute the covariance first, and then perform a matrix factorization through a certain numerical algorithm. To reduce the computational cost, however, we follow the idea in the ensemble square root filter (EnSRF) \cite{Anderson-ensemble,Bishop-adaptive,Tippett-ensemble,Whitaker-ensemble} to update $\tilde{\mathbf{S}}_k^{xb}$ to $\tilde{\mathbf{S}}_k^{xa}$ directly. For example, using the ensemble transform Kalman filter (ETKF) \cite{Bishop-adaptive}, one may update the square root $\tilde{\mathbf{S}}_k^{xb}$ via
\begin{equation} \label{DA:SR update}
\tilde{\mathbf{S}}_k^{xa} = \tilde{\mathbf{S}}_k^{xb} \mathbf{T}_k \mathbf{U}_k,
\end{equation}
where $\mathbf{U}_k$ is the centering matrix in Eq.~(\ref{ch2:EnSRF_spherical_orthornormal_matrix}), and $\mathbf{T}_k$ is the transformation matrix given by (cf. \S~\ref{ch2:sec_EnSRF}):
\begin{equation}
\mathbf{T}_k = \mathbf{E}_k^{wpr} \left( \mathbf{D}_k^{wpr} + \mathbf{I} \right)^{-1/2},
\end{equation}
with $\mathbf{I}$ being the identity matrix. $ \mathbf{E}_k^{wpr}$ and $ \mathbf{D}_k^{wpr}$ are the eigenvector matrix and the corresponding diagonal matrix of eigenvalues, respectively, of the weighted projection matrix $\tilde{\mathbf{P}}^{wpr}_k$, which is defined by
\begin{equation} \label{srp}
\tilde{\mathbf{P}}^{wpr}_k = \left( \tilde{\mathbf{S}}_k^{h1} \right)^T \mathbf{R}_k^{-1} \tilde{\mathbf{S}}_k^{h1} =\mathbf{E}_k^{wpr} \mathbf{D}_k^{wpr} \left(\mathbf{E}_k^{wpr}\right)^T.
\end{equation}
Note that, by using the square root $\tilde{\mathbf{S}}_k^{h1}$,  Eq.~(\ref{srp}) is equivalent to the original form in \cite{Bishop-adaptive} if the observation operator $\mathcal{H}_k$ is linear, but it avoids evaluating the Jacobian matrix of $\mathcal{H}_k$ when $\mathcal{H}_k$ is nonlinear. 

After obtaining $\hat{\mathbf{x}}_k^{a}$ and $\tilde{\mathbf{S}}_k^{xa}$, one produces $2l_k +1$ sigma points with respect to $\left(h, \hat{\mathbf{x}}_k^{a}, \tilde{\mathbf{S}}_k^{xa} \right)$ and then propagates them forward to start a new assimilation cycle.

\section{Example: Assimilating the 40-dimensional Lorenz-Emanuel 98 system} \label{ch4:sec_experiments}

\subsection{The testbed and the measures of filter performance} \label{ch4:sec_testbed}
The testbed and the measures of filter performance are the same as those in \S~\ref{ch3:sec_testbed}. The dynamical system (LE 98 model) is governed by
\begin{equation} \tagref{ch2:ex_LE98}
\frac{dx_i}{dt} = \left( x_{i+1} - x_{i-2} \right) x_{i-1} - x_i + 8, \,~ i=1, \dotsb, 40 \, , 
\end{equation}
while the observation system is
\begin{equation} \tagref{ch2:ex_observer}
\mathbf{y}_k = \mathbf{x}_k + \mathbf{v}_k \, ,
\end{equation} 
where $\mathbf{v}_k$ follows the Gaussian distribution $N \left( \mathbf{v}_k: \mathbf{0}, \mathbf{I} \right)$. 

We integrate the dynamical system Eq.~(\ref{ch2:ex_LE98}) by a fourth-order Runge-Kutta method \cite[Ch.~16]{Vetterling-numerical}, and choose the length of the integration window to be $0:0.05:100$. We make the observations at every integration step.

The measures of filter performance are the time averaged relative rmse and rms ratio, given by
\begin{equation}\tagref{Eq:reltiave rmse}
e_r=\frac{1}{k_{max}}\sum\limits_{k=1}^{k_{max}} \lVert\hat{\mathbf{x}}_k^{a}-{\mathbf{x}}_k^{tr}\rVert_2/ \lVert{\mathbf{x}}_k^{tr}\rVert_2 
\end{equation}
and
\begin{equation} \tagref{ch3:R}
R = \dfrac{1}{k_{max}} \sum\limits_{k=1}^{k_{max}} \dfrac{ (2l_k +1) \, \lVert \hat{\mathbf{x}}_k^a - \mathbf{x}_{k}^{tr} \rVert_{2}}{\sum\limits_{i=1}^{2l_k +1} \lVert \mathcal{X}_{k,i}^a - \mathbf{x}_{k}^{tr} \rVert_{2}} \, ,
\end{equation}
respectively. Again, according to Eq.~(\ref{R_e}), the expectation of the rms ratio 
\[
R_e = \sqrt{(l_{eff}+1)/(2l_{eff}+1)} \, ,
\]
with $l_{eff}$ being the ``effective'' truncation number over the whole assimilation window. $R_e$ is about $0.71$ for a large $l_{eff}$.

\subsection{Numerical results} \label{ch4:sec_ex_result}

\subsubsection{Effects of the inflation factor $\delta$ and the length scale $l_c$ on the performances of the reduced rank DDFs} \label{ch4:sec_ex_result_delta_vs_lc}

We also adopt the covariance inflation and filtering techniques in our experiments to improve the performances of the reduced rank DDFs. To examine the effects of the inflation factor $\delta$ and the length scale $l_c$ on the performances of the filters, we let $\delta$ and $\l_c$ take values from the sets $0:0.5:10$ and $10:20:400$, respectively. The values of the other parameters are: interval length $h=3$, lower bound $l_l = 3$, upper bound $l_u=6$, and the threshold at the first assimilation cycle $\Gamma_1=1000$. 

We choose an initial condition at random to start a control run, and thus obtain a trajectory of the true states within the specified assimilation window. We then add some Gaussian noise drawn from the distribution $N \left(\mathbf{v}_k: \mathbf{0}, \mathbf{I} \right)$ to the true trajectory to generate the observations. The noise level (relative rmse) $e_r^{obv} \approx 0.22$. We also follow the procedure in \S~\ref{ch3:sec_ex_delta_vs_lc} to initialize the reduced rank DDFs by generating $6$ randomly perturbed initial conditions as the background ensemble at the first assimilation cycle. Then we use the ensemble transform Kalman filter (ETKF) to update the background to the analysis, and so generate sigma points. After propagating the sigma points forward, the DDFs can start running recursively from the second cycle. 

\begin{figure*}[!t]
\centering
\hspace*{-0.5in} \includegraphics[width=1.15\textwidth]{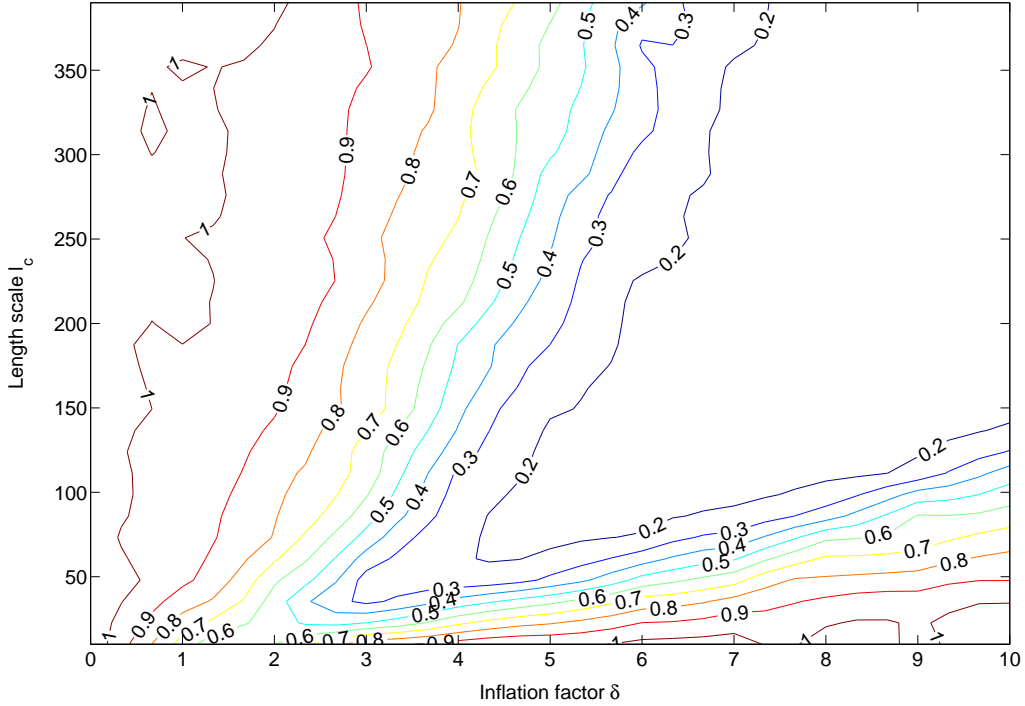} 
\caption{ \label{fig:ch4_DD1_delta_vs_lc_rms} The relative rmse of the DD1 filter as a function of the inflation factor $\delta$ and the length scale $l_c$. }
\end{figure*} 

\begin{figure*}[!t]
\centering
\hspace*{-0.5in} \includegraphics[width=1.15\textwidth]{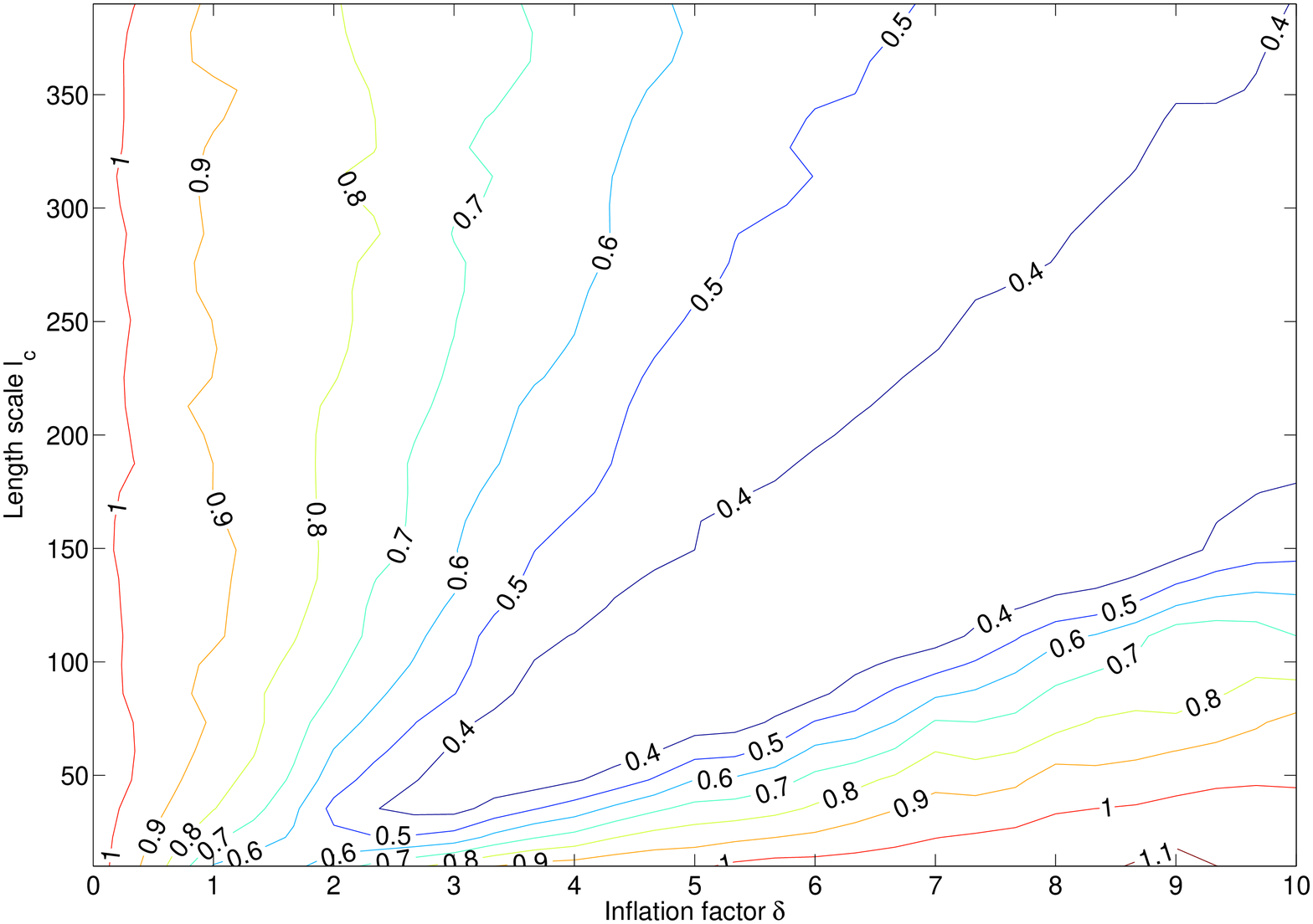} 
\caption{ \label{fig:ch4_DD2_delta_vs_lc_rms} The relative rmse of the DD2 filter as a function of the inflation factor $\delta$ and the length scale $l_c$. }
\end{figure*} 

\begin{figure*}[!t]
\centering
\hspace*{-0.5in} \includegraphics[width=1.15\textwidth]{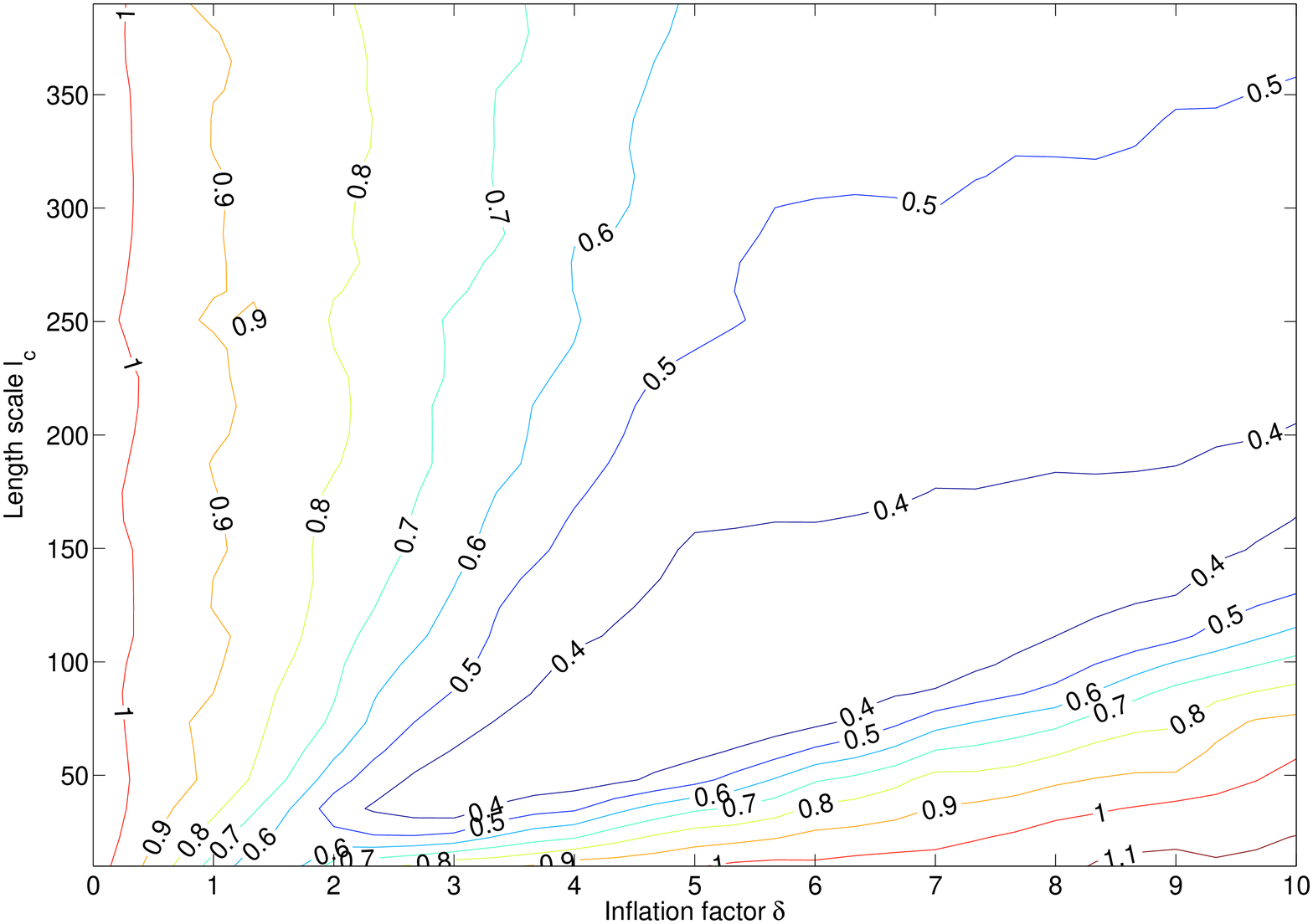} 
\caption{ \label{fig:ch4_CDF_delta_vs_lc_rms} The relative rmse of the CDF as a function of the inflation factor $\delta$ and the length scale $l_c$. }
\end{figure*} 

In our experiments, we first examine the performances of the DDFs in terms of the relative rms errors. To this end, we plot the relative rms errors of the DD1, DD2 filters and the CDF as functions of $\delta$ and $l_c$ in Figs.~\ref{fig:ch4_DD1_delta_vs_lc_rms}, \ref{fig:ch4_DD2_delta_vs_lc_rms} and \ref{fig:ch4_CDF_delta_vs_lc_rms}, respectively. In all these figures, when fixing $\delta$, if $\delta$ is relatively small (e.g., $\delta=2$ for the DD1 filter), the relative rms errors are roughly monotonically increasing functions of $l_c$. If $\delta$ is relatively large (e.g., $\delta=5$ for the DD1 filter), then the relative rms errors of all three DDFs exhibit the U-turn behaviour as $l_c$ increases. On the other hand, when fixing $l_c$, if $l_c$ is relatively large (e.g., $l_c = 250$ for the DD1 filter), then the relative rms errors are roughly monotonically decreasing functions of $\delta$. If $l_c$ is relatively small (e.g., $l_c = 100$ for the DD1 filter), however, then the relative rms errors of all three DDFs also exhibit the U-turn behaviour as $\delta$ increases.      

A comparison between the DD2 filter and the CDF shows that these two filters have similar performances, with the DD2 filter being slightly better than the CDF for some parameter values (e.g., $\delta=8$ and $l_c=200$). A comparison between the DD1 and DD2 filters with the same $l_c$ indicates that, when $\delta$ is relatively small (say $\delta=1$), the DD2 filter outperforms the DD1 filter, as one might expect. However, if $\delta$ is relatively large (say $\delta =8$), the DD1 filter may instead outperform the DD2 filter. In fact, within the ranges of the parameters we have tested, the DD2 filter and the CDF are always divergent, in the sense that their relative rms errors are always larger than the noise level $e_r^{obv} \approx 0.22$. In contrast, in some regions of Fig.~\ref{fig:ch4_DD1_delta_vs_lc_rms}, e.g., the area surrounded by the contour level curve marked by the value of $0.2$, the axis on the right (corresponding to $\delta=10$) and the axis on the top (corresponding to $l_c=390$, not marked in the figure), the relative rmse of the DD1 filter is actually less than $e_r^{obv}$, and thus non-divergent.

\begin{figure*}[!t]
\centering
\hspace*{-0.5in} \includegraphics[width=1.15\textwidth]{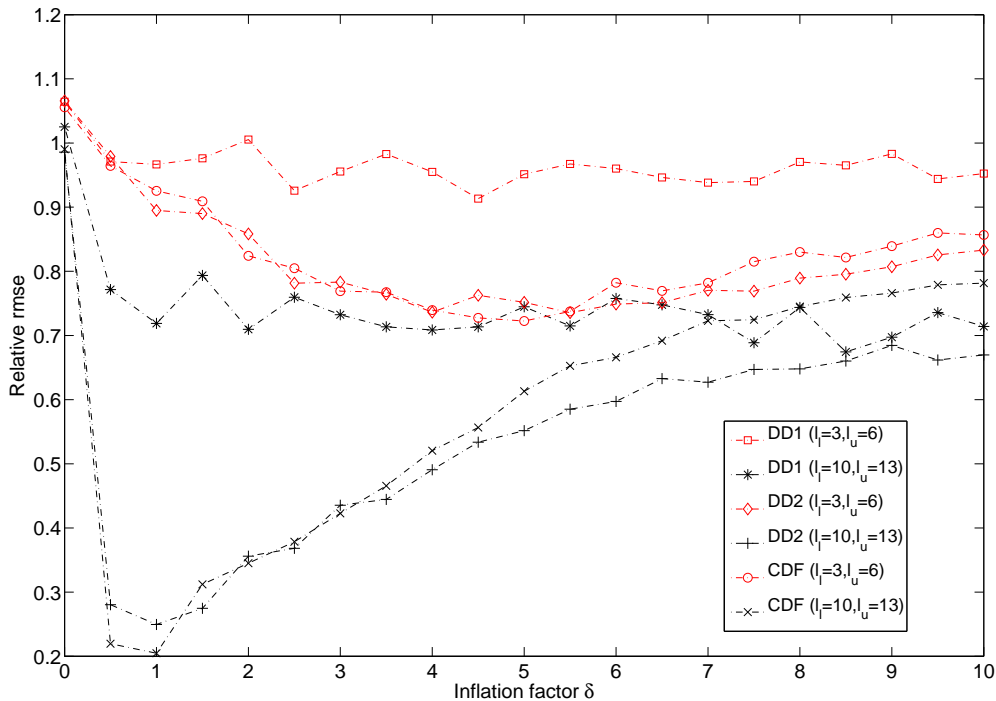} 
\caption{ \label{fig:ch4_DDFs_rmse_vs_delta_no_cov_filtering} The relative rms errors of the DDFs as functions of the inflation factor $\delta$ when there is no covariance filtering. Here we consider two scenarios: lower bound $l_l=3$, upper bound $l_u=6$ and lower bound $l_l=10$, upper bound $l_u=13$, with other parameters following the same setting at the beginning of \S~\ref{ch4:sec_ex_result_delta_vs_lc}.}
\end{figure*} 

Our explanation for the above counter-intuitive phenomenon is that it is the introduction of the covariance filtering technique that makes the DD1 filter outperform its second order counterparts. As shown in Fig.~\ref{fig:ch4_DDFs_rmse_vs_delta_no_cov_filtering}, when there is no covariance filtering, the DD2 filter and the CDF always outperform the DD1 filter, given the same parameter settings (either $l_l=3$ and $l_u=6$, or $l_l=10$ and $l_u=13$). Moreover, by comparing Fig.~\ref{fig:ch4_DDFs_rmse_vs_delta_no_cov_filtering} with Fig.~\ref{fig:EnUKF_rmse_vs_delta_no_inflation}, one can see that, where there is no covariance filtering, the SUKF always outperforms the DDFs within the range of $\delta$ tested. The DD2 filter and the CDF are comparable with the ETKF in general, while the DD1 filter underperforms the ETKF. In this sense, introducing covariance filtering to a filter might significantly influence its performance. Here it substantially improved the performance of the DD1 filter in some parameter regions (cf. Figs.~\ref{fig:ch4_DDFs_rmse_vs_delta_no_cov_filtering} and \ref{fig:ch4_DD1_delta_vs_lc_rms}). However, from the theoretical point of view, in the literature there is still no in-depth understanding of how the covariance filtering technique affects the performance of a filter. 

\begin{figure*}[!t]
\centering
\hspace*{-0.5in} \includegraphics[width=1.15\textwidth]{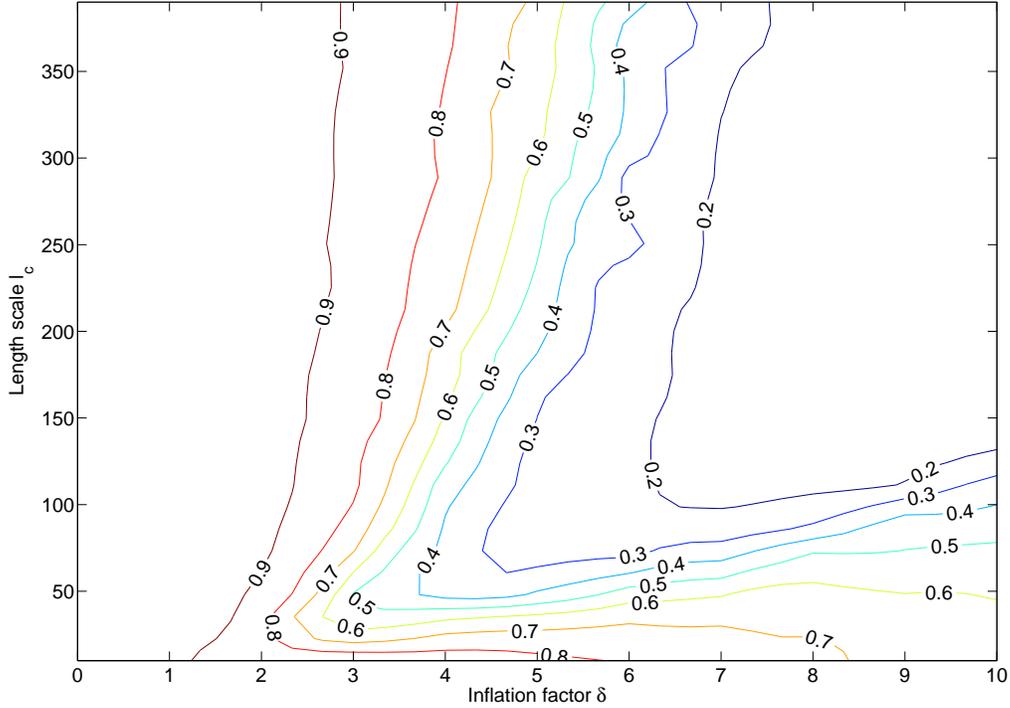} 
\caption{ \label{fig:ch4_DD1_delta_vs_lc_ratio} The rms ratio of the DD1 filter as a function of the inflation factor $\delta$ and the length scale $l_c$. }
\end{figure*} 

\begin{figure*}[!t]
\centering
\hspace*{-0.5in} \includegraphics[width=1.15\textwidth]{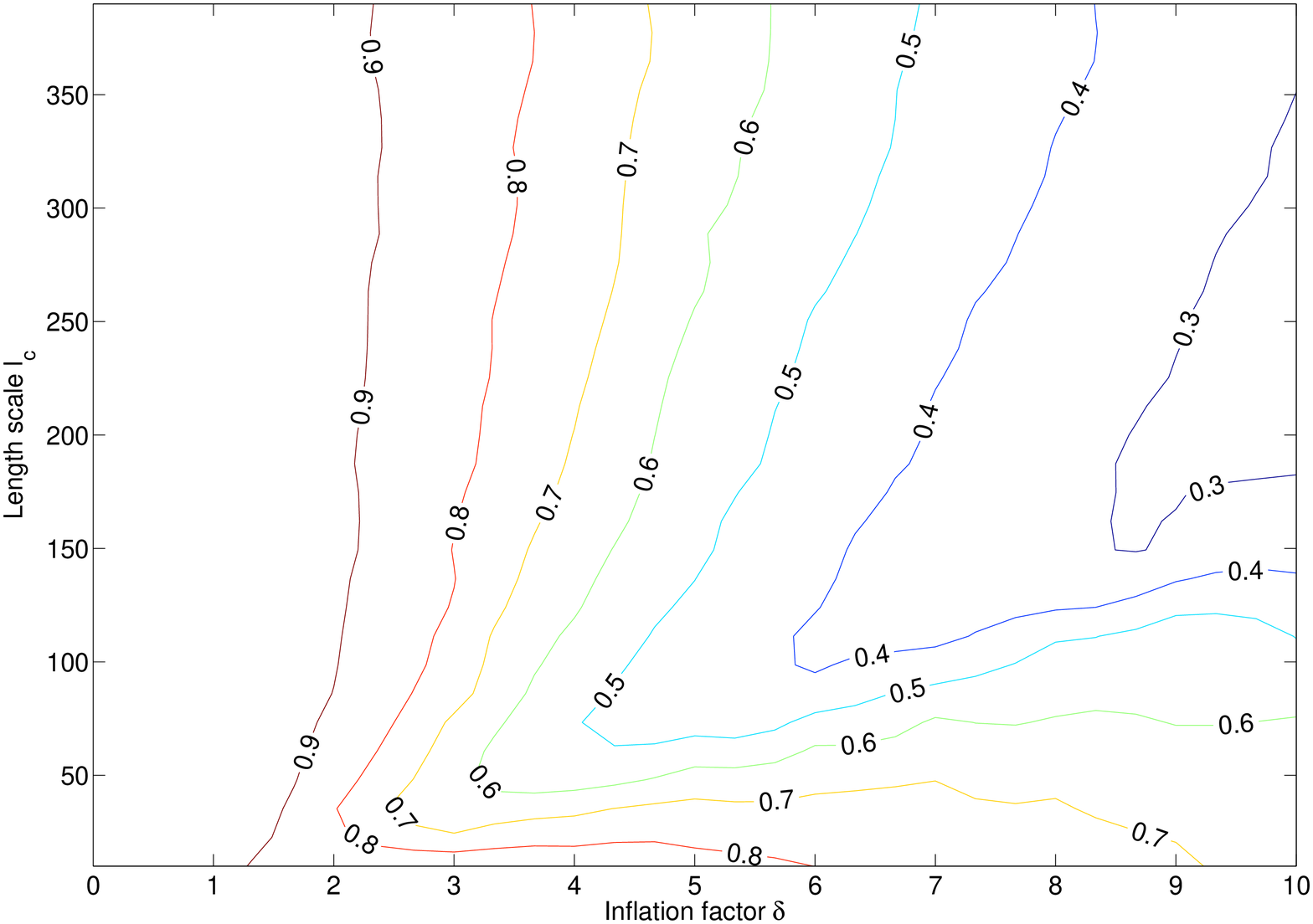} 
\caption{ \label{fig:ch4_DD2_delta_vs_lc_ratio} The rms ratio of the DD2 filter as a function of the inflation factor $\delta$ and the length scale $l_c$. }
\end{figure*} 

\begin{figure*}[!t]
\centering
\hspace*{-0.5in} \includegraphics[width=1.15\textwidth]{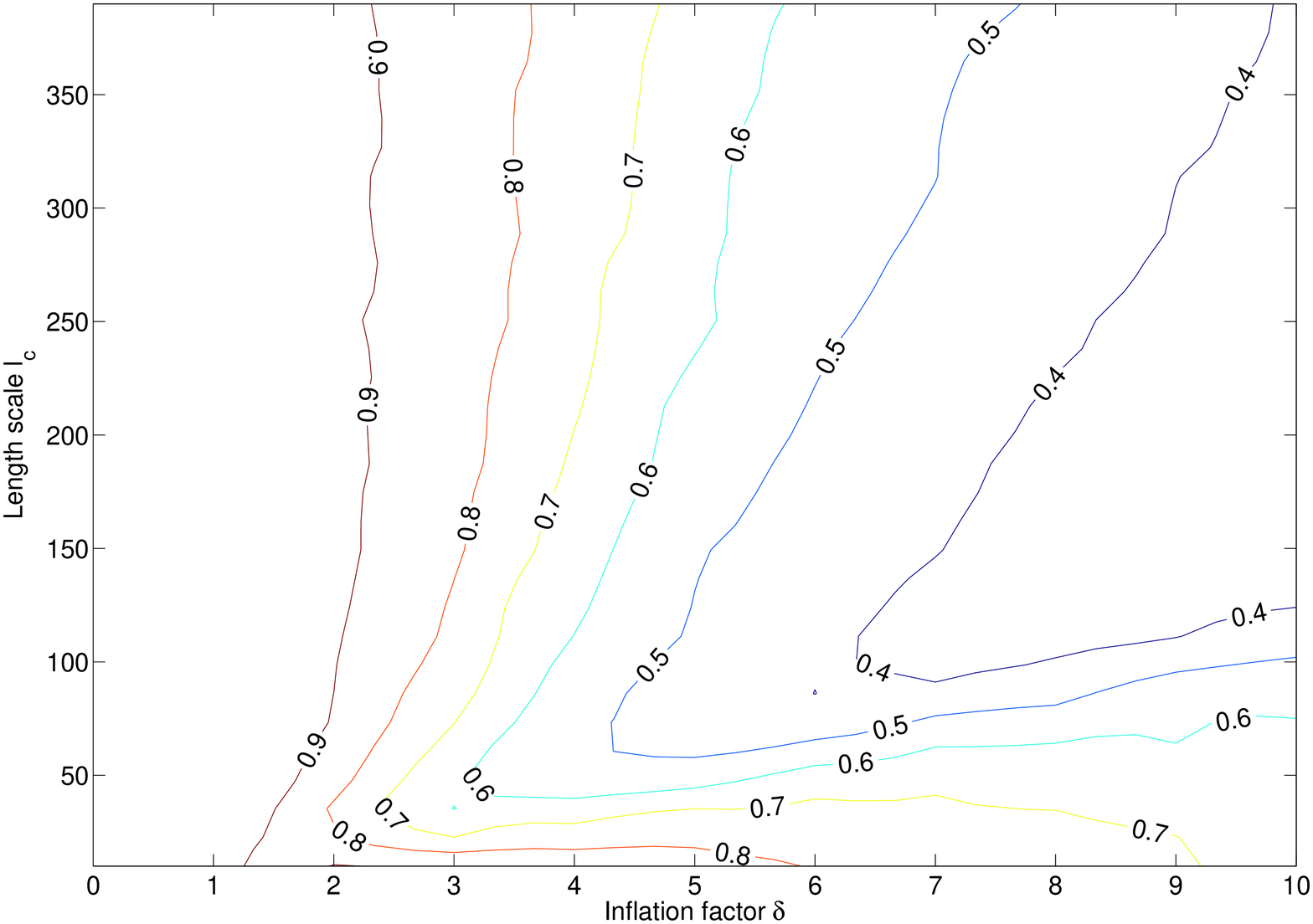} 
\caption{ \label{fig:ch4_CDF_delta_vs_lc_ratio} The rms ratio of the CDF as a function of the inflation factor $\delta$ and the length scale $l_c$. }
\end{figure*} 

Next we examine the rms ratios of the DDFs. We plot the rms ratios of the DD1, DD2 filters and the CDF as functions of $\delta$ and $l_c$ in Figs.~\ref{fig:ch4_DD1_delta_vs_lc_ratio}, \ref{fig:ch4_DD2_delta_vs_lc_ratio} and \ref{fig:ch4_CDF_delta_vs_lc_ratio}, respectively. In all these figures, when fixing $\delta$, for relatively small $\delta$ (say $\delta = 2$ for the DD1 filter), the rms ratios of the DDFs appear to be a monotonically increasing function of $l_c$. If $\delta$ is relatively large (say $\delta =5$ for the DD1 filter), then the rms ratios also exhibit the $U$-turn behaviour as $l_c$ increases. On the other hand, when fixing $l_c$, for relatively large $l_c$ (say $l_c =150$ for the DD1 filter), the rms ratios of the DDFs tend to decrease as $\delta$ increases. If $l_c$ is relatively small (say $l_c = 80$ for the DD1 filter), the rms ratios also exhibit the $U$-turn behaviour as $\delta$ increases.

To make sigma points indistinguishable from the truth (i.e., $R \approx 0.71$), one should take the parameter values of $\delta$ and $l_c$ within the strip between the contour levels of $0.7$ and $0.8$. However, overestimation of the analysis covariance (e.g. $R<0.71$) will improve the performances of all the DDFs in the sense that they can achieve lower relative rms errors. This is consistent with the situations in the EnKF and the reduced rank SUKF, as we have shown in Chapters~\ref{ch2: EnKF} and \ref{ch3:ukf} respectively.

\subsubsection{Effect of the interval length $h$ on the performances of the reduced rank DDFs} \label{ch4:sec_ex_result_h_vs_delta}

For all the DDFs, we set the threshold $\Gamma_1=1000$, the lower bound $l_l=3$, the upper bound $l_u=6$, the length scale of covariance filtering $l_c =240$, the covariance inflation factor $\delta$ and the interval length $h$ take values from the set $0:0.5:10$ and $1:0.5:5$, respectively.    

\begin{figure*}[!t]
\centering
\hspace*{-0.5in} \includegraphics[width=1.15\textwidth]{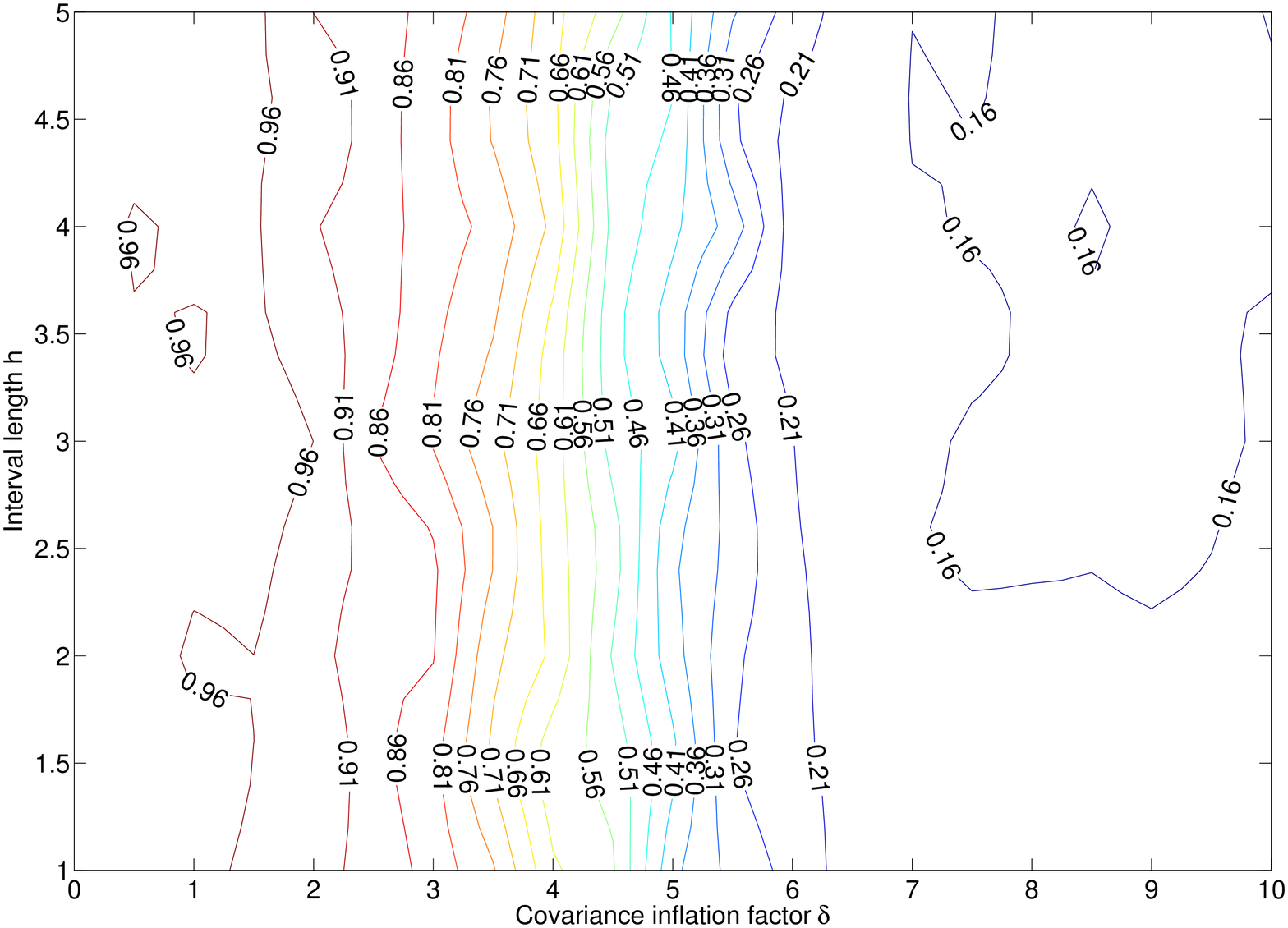} 
\caption{ \label{fig:ch4_DD1_h_vs_delta_rms} The relative rmse of the DD1 filter as a function of the inflation factor $\delta$ and the interval length $h$. }
\end{figure*} 

\begin{figure*}[!t]
\centering
\hspace*{-0.5in} \includegraphics[width=1.15\textwidth]{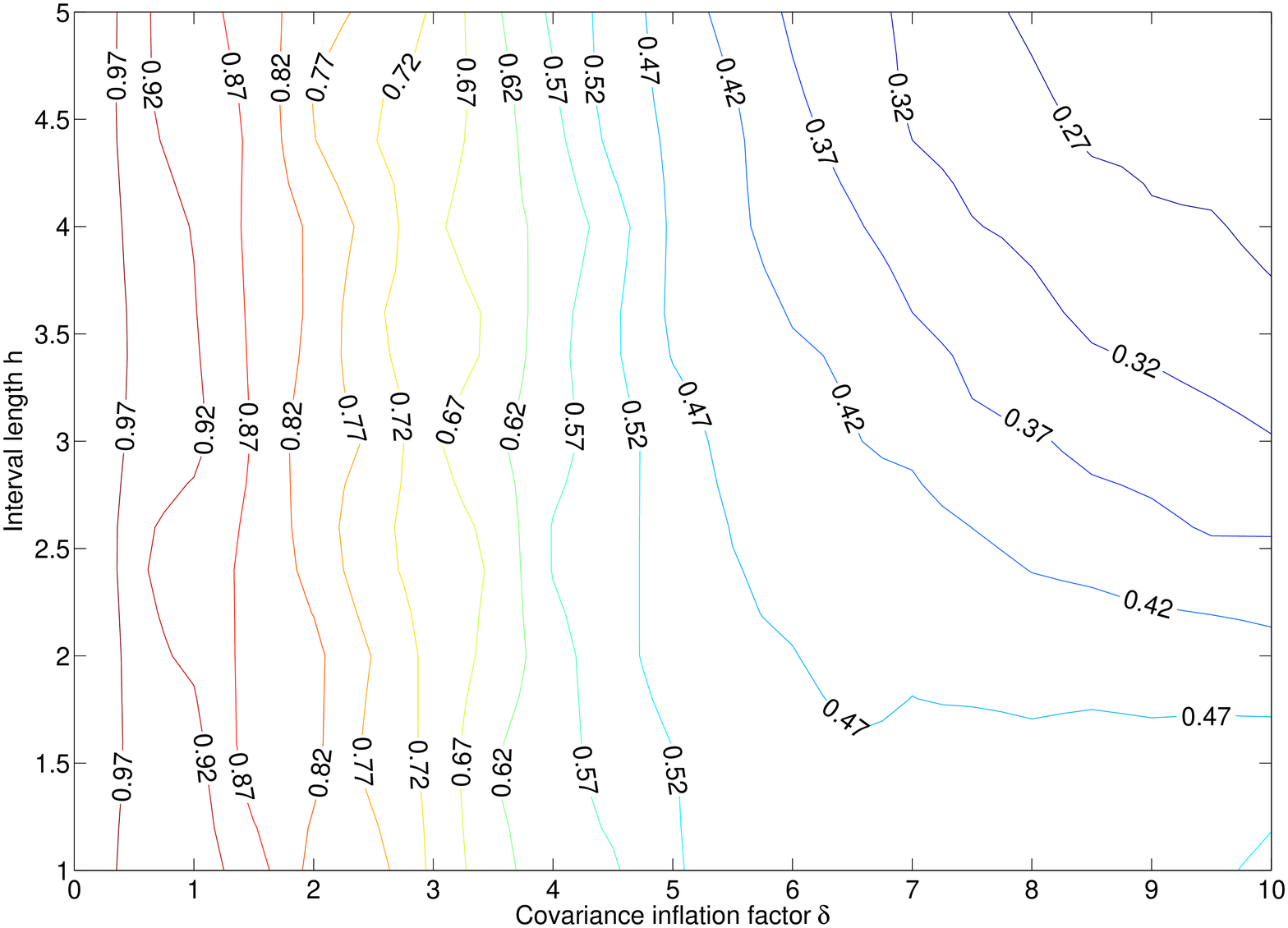} 
\caption{ \label{fig:ch4_DD2_h_vs_delta_rms} The relative rmse of the DD2 filter as a function of the inflation factor $\delta$ and the interval length $h$. }
\end{figure*} 

\begin{figure*}[!t]
\centering
\hspace*{-0.5in} \includegraphics[width=1.15\textwidth]{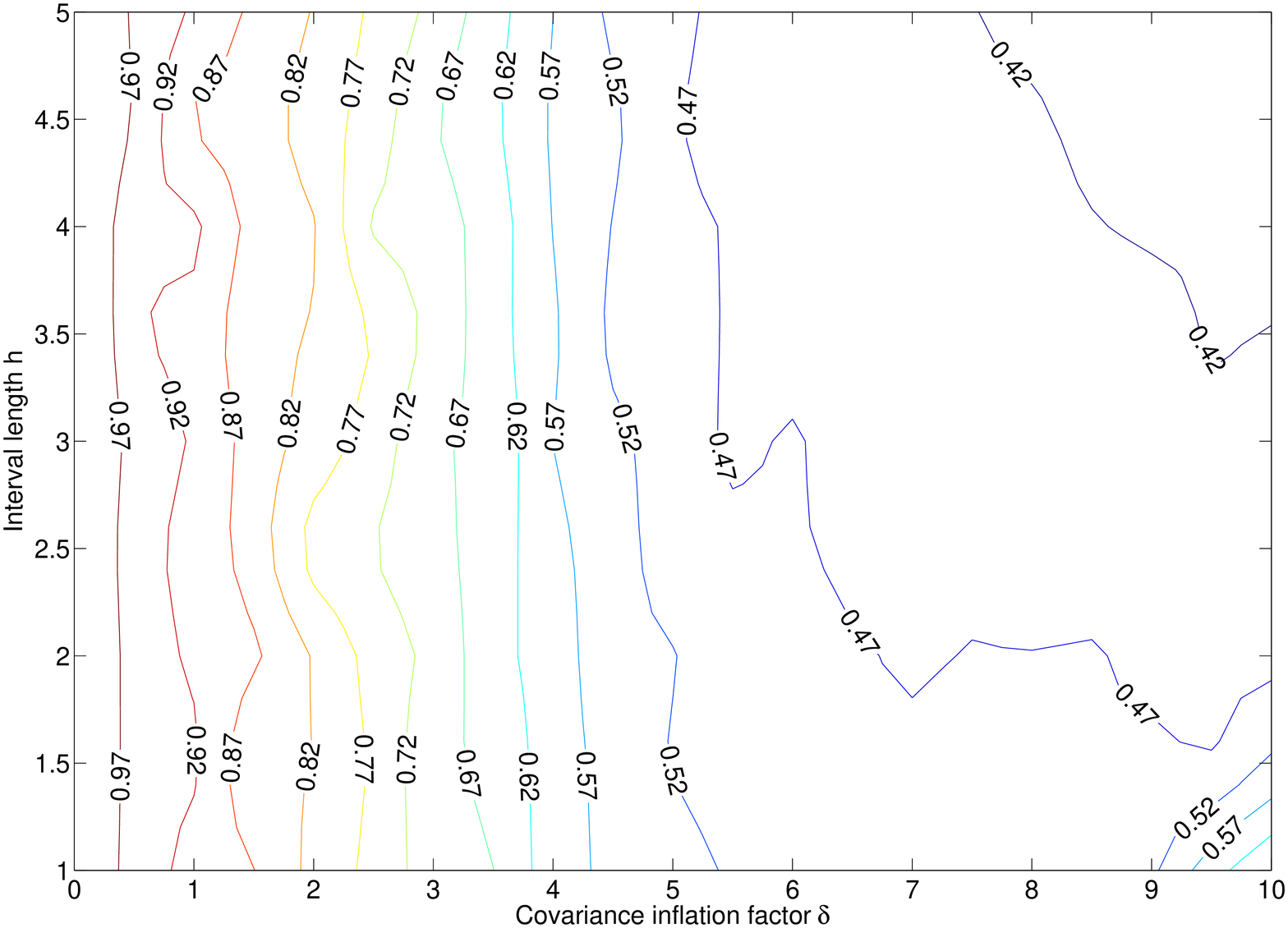} 
\caption{ \label{fig:ch4_CDF_h_vs_delta_rms} The relative rmse of the CDF as a function of the inflation factor $\delta$ and the interval length $h$.}
\end{figure*} 

First we examine the effect of $h$ on the relative rms errors. As shown in Figs.~\ref{fig:ch4_DD1_h_vs_delta_rms}, \ref{fig:ch4_DD2_h_vs_delta_rms} and \ref{fig:ch4_CDF_h_vs_delta_rms}, when fixing $\delta$, for relatively small $\delta$ (say $\delta<6$ for the DD1 filter), the rms errors of the DDFs appear insensitive to the change of $h$. But for relatively large $\delta$ (say $\delta >7$ for the DD2 filter), the rms errors of the DDFs tend to decrease monotonically as $h$ increases. When fixing $h$, the rms errors of the DDFs either decrease monotonically or exhibit the U-turn behaviour as $\delta$ increases, similar to what we have already seen in \S~\ref{ch4:sec_ex_result_delta_vs_lc}.  

\begin{figure*}[!t]
\centering
\hspace*{-0.5in} \includegraphics[width=1.15\textwidth]{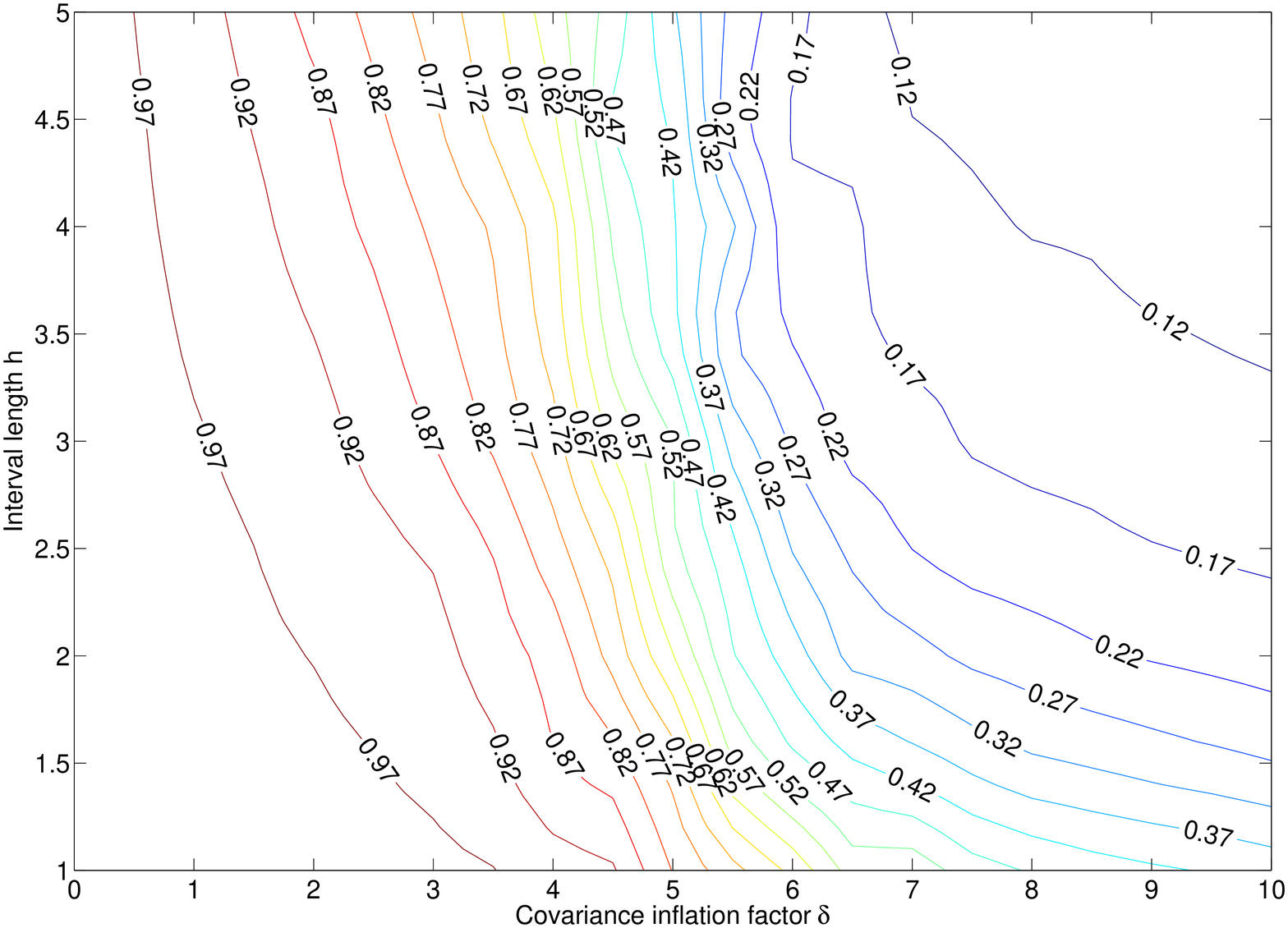} 
\caption{ \label{fig:ch4_DD1_h_vs_delta_ratio} The rms ratio of the DD1 filter as a function of the inflation factor $\delta$ and the interval length $h$. }
\end{figure*} 

\begin{figure*}[!t]
\centering
\hspace*{-0.5in} \includegraphics[width=1.15\textwidth]{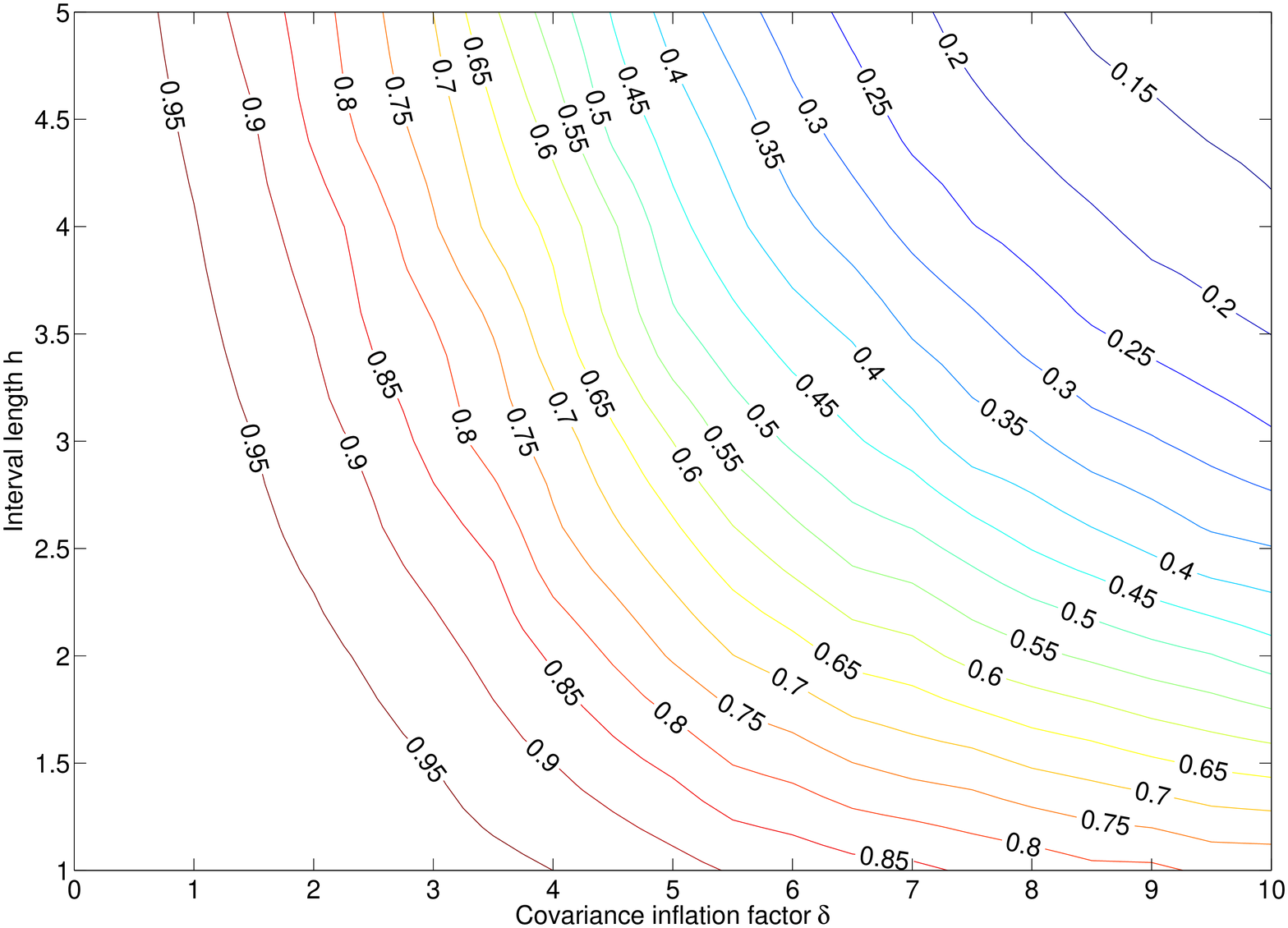} 
\caption{ \label{fig:ch4_DD2_h_vs_delta_ratio} The rms ratio of the DD2 filter as a function of the inflation factor $\delta$ and the interval length $h$. }
\end{figure*} 

\begin{figure*}[!t]
\centering
\hspace*{-0.5in} \includegraphics[width=1.15\textwidth]{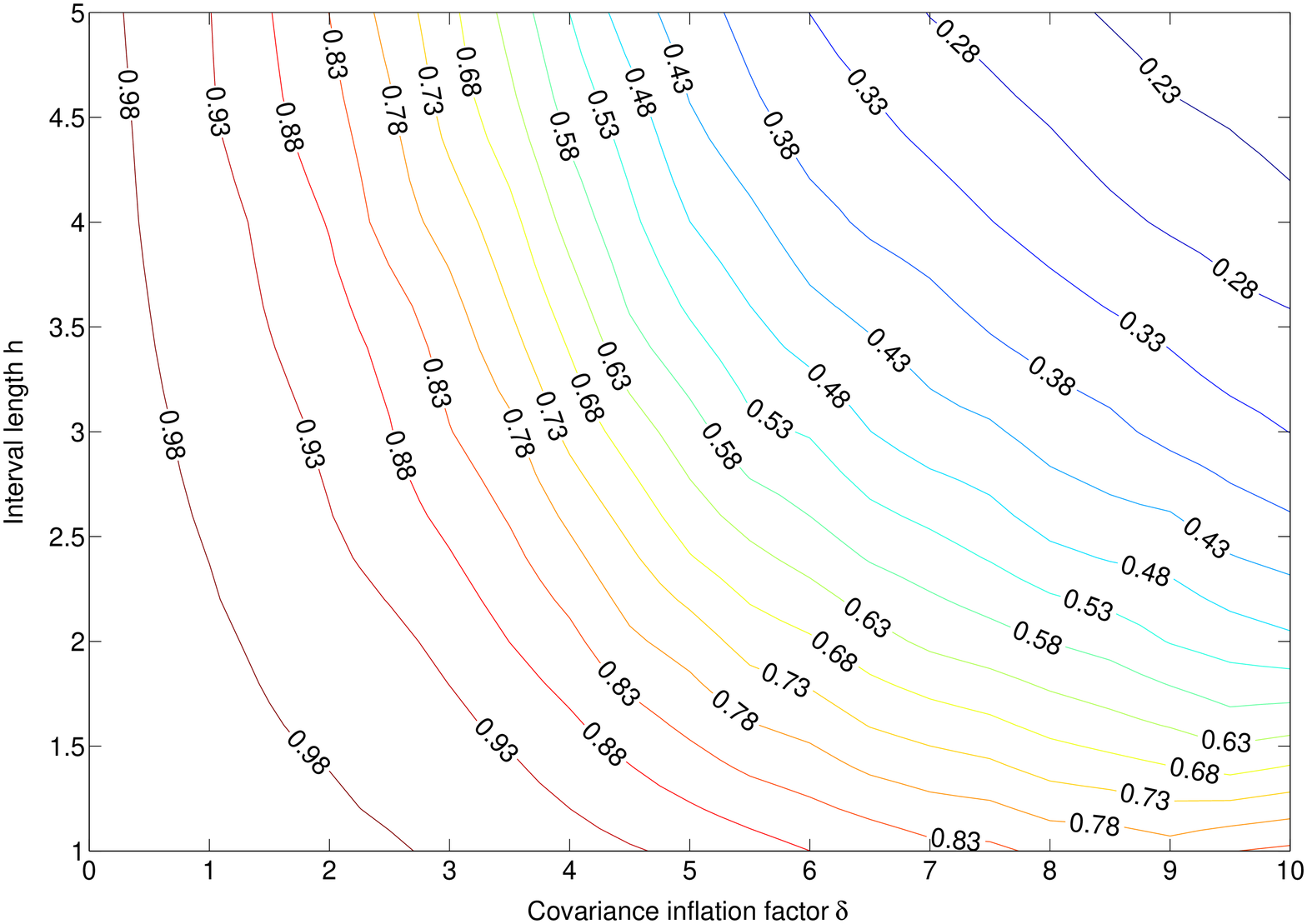} 
\caption{ \label{fig:ch4_CDF_h_vs_delta_ratio} The rms ratio of the CDF as a function of the inflation factor $\delta$ and the length scale $l_c$.}
\end{figure*} 

Next we examine the effect of $h$ on the rms ratios. As shown in Figs.~\ref{fig:ch4_DD1_h_vs_delta_ratio}, \ref{fig:ch4_DD2_h_vs_delta_ratio} and \ref{fig:ch4_CDF_h_vs_delta_ratio}, the rms ratios of the DDFs are all monotonically decreasing functions of $h$. To make sigma points indistinguishable from the truth, one should take values of $h$ and $\delta$ that make the rms ratios close to $0.71$. However, overestimation of the analysis covariance (e.g. $R<0.71$) can also improve the performances of the DDFs in the sense that they can achieve lower relative rms errors.

\subsubsection{Effects of the threshold $\Gamma_1$ and the upper bound $l_u$ on the performances of the reduced rank DDFs}  \label{ch4:sec_ex_result_threshold_bound}

Now we examine the effects of the threshold $\Gamma_1$ and the upper bound $l_u$ on the performances of the DDFs. Like the experiments for the reduced rank SUKF, here we only examine the effects of these parameters on the relative rms errors of the DDFs.

\begin{figure*}[!t]
\centering
\hspace*{-0.5in} \includegraphics[width=1.15\textwidth]{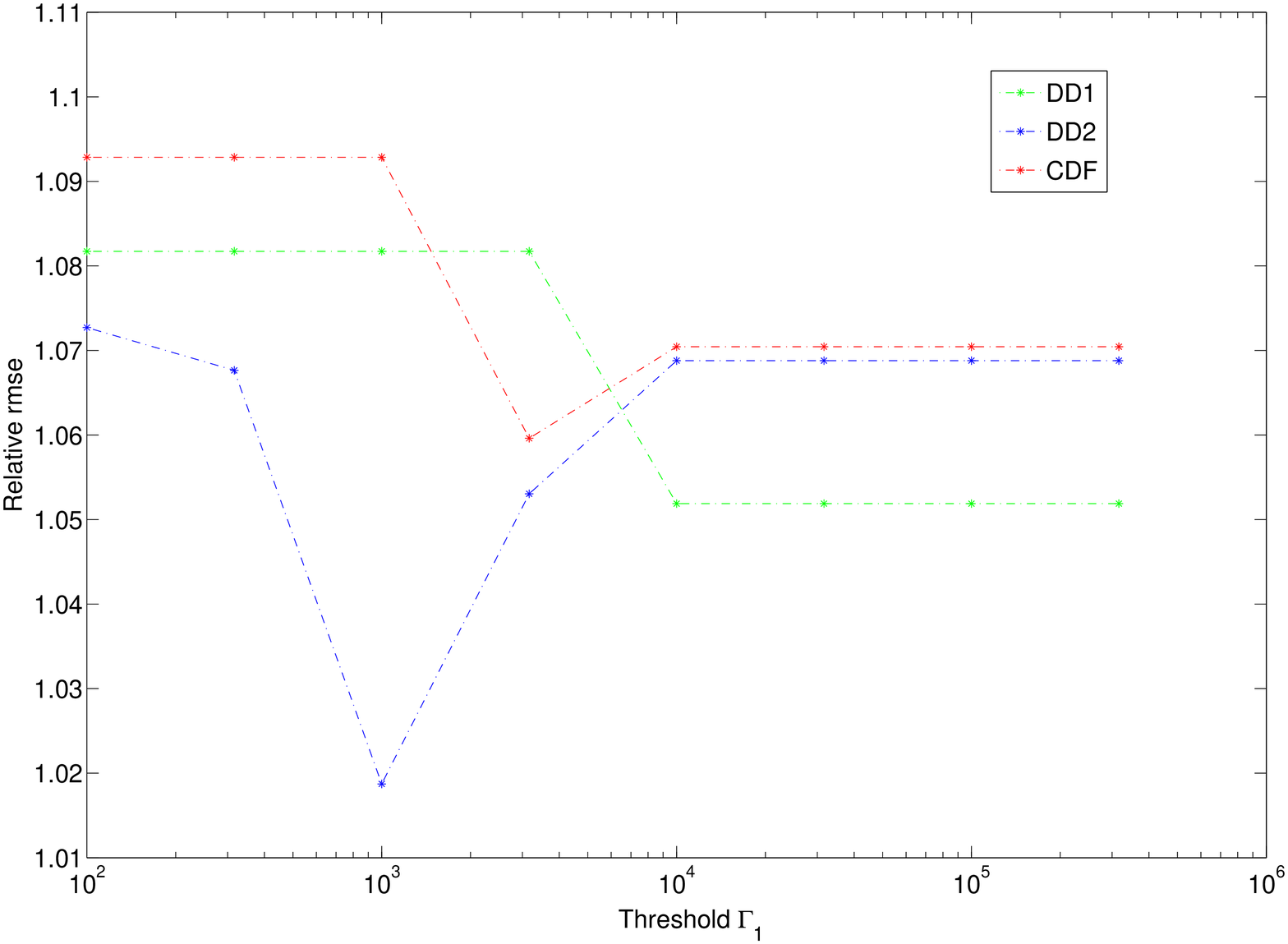} 
\caption{ \label{fig:DDFs_rmse_vs_threshold} Relative rms errors of the DDFs as functions of the threshold $\Gamma_1$.}
\end{figure*} 

In the first experiment, we let the covariance inflation factor $\delta=0$, the length scale $l_c=240$, the initial ensemble size $n=6$, the interval length $h=3$, the lower bound $l_l=3$ and the upper bound $l_u=6$. We vary the threshold $\Gamma_1$ such that $\log_{10}\Gamma_1$ takes values from the set $2:0.5:5.5$. 

We show the numerical results in Fig.~\ref{fig:DDFs_rmse_vs_threshold}. As for the case of the reduced rank SUKF, a larger threshold $\Gamma_1$ in the DDFs does not necessarily lead to lower rms errors. For example, in the DD2 filter, the optimal $\Gamma_1$ is $\Gamma_1=10^3$, rather than $10^5$. A possible explanation for this phenomenon was given in \S~\ref{ch3:sec_experiments_threshold_bounds}. 

\begin{figure*}[!t]
\centering
\hspace*{-0.5in} \includegraphics[width=1.15\textwidth]{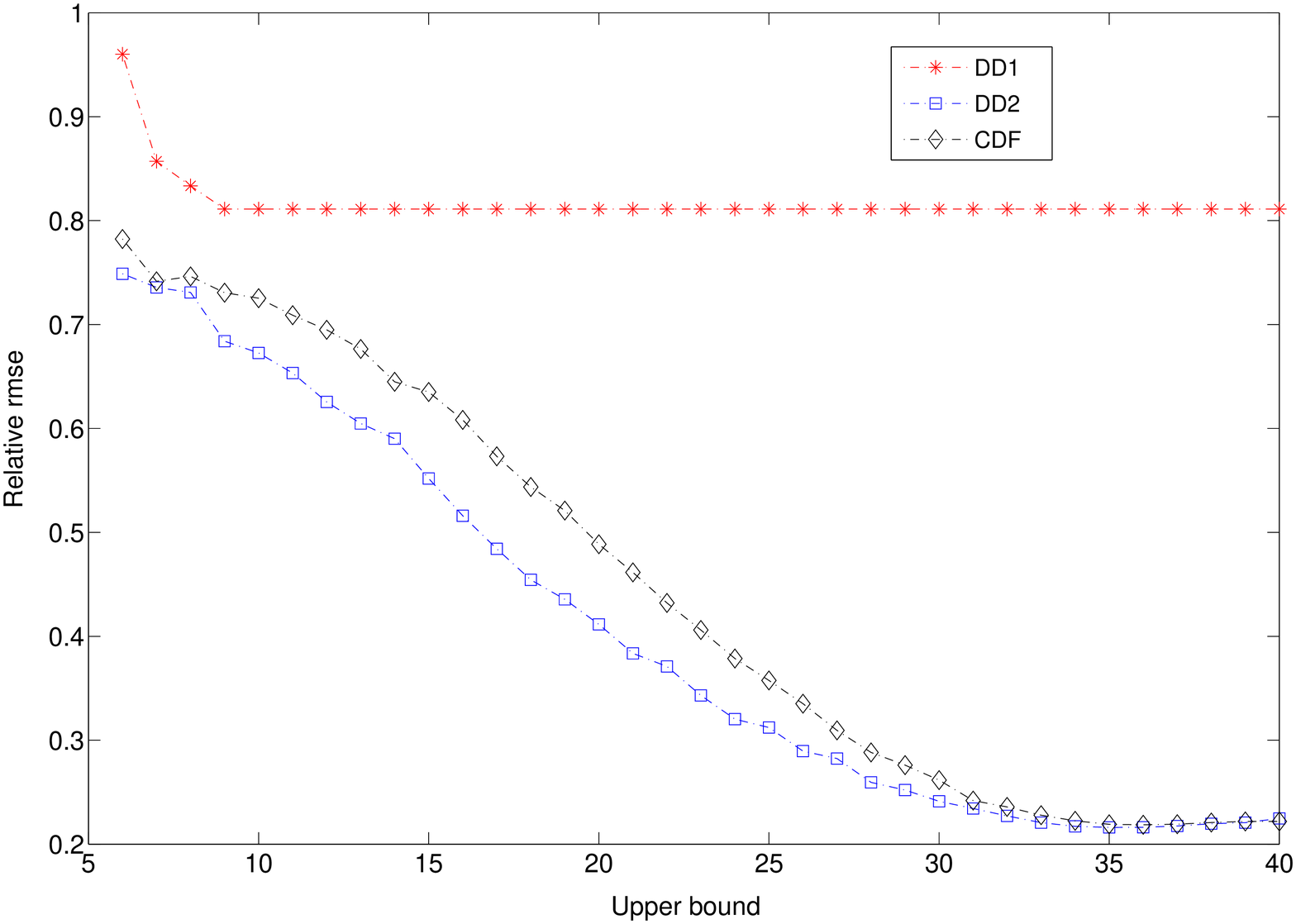} 
\caption{ \label{fig:DDFs_rmse_vs_upper_bound} Relative rms errors of the DDFs as functions of the upper bound $l_u$.}
\end{figure*} 

In the second experiment, we take $\delta=6$, $l_c=\infty$ (i.e., no covariance filtering), $n=10$,$\Gamma_1=1000$, and $h=3$. We fix the lower bound $l_l=3$, but take the values of the upper bound $l_u$ from the set $6:1:40$. 

In Fig.~\ref{fig:DDFs_rmse_vs_upper_bound} we plot the relative rms errors of the DDFs as functions of the upper bound $l_u$. For the DD1 filter, the relative rmse enters a plateau for $l_u \ge 9$, so that choosing $l_u > 9$ does not significantly improve the performance of the DD1. For the DD2 filter and the CDF, however, their relative rms errors appear to be monotonically decreasing until $l_u$ reaches $35$, after which the relative rms errors do not change significantly as $l_u$ further increases. 

\section{Summary of the chapter}\label{ch4:sec_summary}

In this chapter we introduced the idea of using Stirling's interpolation formula to solve the recast problem in Fig.~\ref{ch3:fig_problem_recast}, and presented three divided difference approximation schemes. We conducted accuracy analyses for these three schemes through Taylor series expansions. Analytical results showed that the first order divided difference (DD1) approximation is less accurate than the second order and central (divided) difference approximation (DD2 and CD) schemes, while the DD2 and CD approximation schemes themselves are comparable with the specific scaled unscented transform (SUT) with the scale factor $\alpha=1$ and the compensation parameter $\beta=0$ (cf. Chapter~\ref{ch3:ukf}). 

Incorporating the above approximation schemes into the propagation step of recursive Bayesian estimation (RBE) leads to the corresponding divided difference filters (DDFs). To reduce the computational cost of the DDFs in high dimensional systems, we also introduced the reduced rank DDFs following the idea in \S~\ref{ch3:sec_reduced_sukf}.

For illustration, we used the $40$-dimensional LE 98 system as the testbed to demonstrate the details in implementing the DDFs. We also investigated the effects of filter parameters (e.g., $h$, $\Gamma_1$ etc) on the performances of the DDFs. Numerical results showed that introducing the covariance filtering technique may lead to counter-intuitive results: the DD1 filter outperformed both the DD2 filter and the CDF in some situations. However, when there was no covariance filtering, numerical results were still consistent with the results of accuracy analyses in \S~\ref{sec:SIF_accuracy_analysis}.
\chapter{Gaussian sum filters for data assimilation} \label{ch5:spgsf}

\section{Overview}
In the previous chapters we considered the data assimilation problem in nonlinear/Gaussian systems. In practice, however, the Gaussianity assumption in the previous chapters is often not realistic. Instead, the dynamical and observation noise, and the states of the system under assimilation are non-Gaussian more often than not. For this reason, in this chapter we consider the data assimilation problem in nonlinear/non-Gaussian systems. We show that one may approximately solve such a problem by using the nonlinear Kalman filters established in the previous chapters.

The main idea in this chapter is to approximate the prior and posterior probability density functions (pdfs) of system states, and the pdfs of dynamical and observation noise in recursive Bayesian estimation (RBE) Eq.~(\ref{ch0:BRR}) via some Gaussian distributions. This is known as the Gaussian sum approximation, or the Gaussian mixture model (GMM) in the literature \cite{Alspach-nonlinear,Smith-cluster,Sorenson-recursive}. In the extreme situation, by letting the covariance matrix of a Gaussian distribution tend to zero, the Gaussian distribution approaches a Dirac delta function defined in Eq.~(\ref{ch2:ps_Dirac_delta_function}). For this reason, when all of the covariance matrices of the Gaussian distributions in a GMM tend to zero, the Gaussian sum approximation approaches a Monte Carlo approximation.

By adopting the GMM, one can approximately decompose a nonlinear/non-Gaussian system into a mixture of a set of sub-systems, each of which takes the form of a nonlinear/Gaussian system. Thus, for each sub-system, one can apply the nonlinear Kalman filters introduced in the previous chapters for data assimilation. Incorporating the estimations of the sub-systems into the GMM gives an approximate explicit form for the pdf. This is normally regarded as a ``complete'' solution to the data assimilation problem, since all of the statistical information of interest can be obtained from the explicit form \cite{Arulampalam2002}.

The remainder of this chapter is organized as follows. In \S~\ref{ch5:sec_ps} we state the problem of interest, which differs from those in the previous chapters in that the systems considered in this chapter are possibly nonlinear/non-Gaussian. By incorporating the idea of GMM for pdf approximation into the framework of RBE, in \S~\ref{ch5:GSF_approx_solution} we derive a ``new'' filter (relative to the nonlinear Kalman filters introduced previously), called the Gaussian sum filter (GSF), as the approximate solution to the data assimilation problem in \S~\ref{ch5:sec_ps}. We also propose an auxiliary technique to conduct pdf re-approximations, which aims to reduce the computational cost of the GSF in some situations and increase its numerical stability. For illustration, in \S~\ref{ch5:sec_experiments}, we use the Lorenz-Emanuel 98 system as the testbed and examine the performances of some GSFs. Finally, we draw our conclusions for this chapter in \S~\ref{ch5:sec_summary}.  

\section{Problem statement}\label{ch5:sec_ps}
We consider data assimilation in the following systems:
\begin{subequations} \label{ch5:ps}
\begin{align} 
 \label{ch5:ps_dyanmical_system} & \mathbf{x}_k  = \mathcal{M}_{k,k-1} ( \mathbf{x}_{k-1} ) + \mathbf{u}_{k}  \, ,  \\
  \label{ch5:ps_observation_system} &  \mathbf{y}_k  = \mathcal{H}_{k} ( \mathbf{x}_{k} ) + \mathbf{v}_{k} \, ,
\end{align}
\end{subequations}
where the transition operator $\mathcal{M}_{k,k-1}$ and the observation operator $\mathcal{H}_{k}$ are both possibly nonlinear. The dynamical and observation noise, $\mathbf{u}_{k}$ and $\mathbf{v}_{k}$ respectively, are non-Gaussian, but their approximated pdfs are assumed to be known to us, in terms of the following GMMs:
\begin{subequations}
\begin{align}
\label{ch5:ps_dyn_error} p( \mathbf{u}_k ) \approx & \sum_{i=1}^{n_{k}^u} \alpha_{k,i}^u N ( \mathbf{u}_k: \mathbf{0}, \mathbf{Q}_{k,i} ) \, ,\\
\label{ch5:ps_obv_error} p( \mathbf{v}_k ) \approx & \sum_{i=1}^{n_{k}^v} \alpha_{k,i}^v N ( \mathbf{v}_k: \mathbf{0}, \mathbf{R}_{k,i} ) \, ,
\end{align}
\end{subequations} 
where the notation $N ( \mathbf{x}: \mathbf{\mu}, \mathbf{\Sigma} )$ means that the pdf of a random variable $ \mathbf{x}$ follows a Gaussian distribution with mean $\mathbf{\mu}$ and covariance $\mathbf{\Sigma}$, and $\alpha_{k,i}^u \in [0, 1]$ is the weight associated with $N ( \mathbf{u}_k: \mathbf{0}, \mathbf{Q}_{k,i} )$, which shall satisfy $\alpha_{k,i}^u \in [0, 1]$ and $\sum_{i=1}^{n_{k}^u} \alpha_{k,i}^u =1$. The weights $\alpha_{k,i}^v$ are defined similarly.

\section{Gaussian sum filter as the approximate solution} \label{ch5:GSF_approx_solution}
To solve the data assimilation problem for Eq.~(\ref{ch5:ps}), one can approximate the pdf of the system states through a Gaussian sum approximation, and then substitute the approximated pdf into the framework of RBE Eq.~(\ref{ch0:BRR}). The assimilation algorithm obtained in this way is known as the Gaussian sum filter in the literature \cite{Alspach-nonlinear,Sorenson-recursive}. 
  
Concretely, let the prior pdf of the initial condition $\mathbf{x}_0$ be $p ( \mathbf{x}_0  ) = p ( \mathbf{x}_0|\mathbf{Y}_{-1})$ ($\mathbf{Y}_{-1}$ can be treated as an empty set if no observation is available before the assimilation), which can be approximated by a set of $n_0^{xb}$ Gaussian distributions, so that
\begin{equation}
p ( \mathbf{x}_0) \approx \sum_{i=1}^{n_0^{xb}} \gamma_{0,i} N ( \mathbf{x}_0: \hat{\mathbf{x}}_{0,i}^{b}, \hat{\mathbf{P}}_{0,i}^b ) \, ,
\end{equation}
where $\gamma_{0,i} \in [0,1]$ and $\sum_{i=1}^{n_0^{xb}} \gamma_{0,i}=1$. Then by applying the rules of RBE in Eqs.~(\ref{BRR:update}) and (\ref{BRR:prediction}), one can recursively compute the prior and posterior pdfs of the state $\mathbf{x}_k$, in terms of $p ( \mathbf{x}_{k} | \mathbf{Y}_{k-1} )$ and $p ( \mathbf{x}_{k} | \mathbf{Y}_{k} ) $ respectively. Consequently, we can also divide the procedures of the GSF into the propagation and filtering steps.

\subsection{Propagation step}
Without lost of generality, we assume that at instant $k-1$, we have the posterior pdf $p ( \mathbf{x}_{k-1} | \mathbf{Y}_{k-1} )$ of the system states, which is approximated by $n_{k-1}^{xa}$ Gaussian distributions, so that
\begin{equation}
p ( \mathbf{x}_{k-1} | \mathbf{Y}_{k-1}) \approx \sum_{i=1}^{n_{k-1}^{xa}} \beta_{k-1,i} N ( \mathbf{x}_{k-1}: \hat{\mathbf{x}}_{k-1,i}^{a}, \hat{\mathbf{P}}_{k-1,i}^a ) \, ,
\end{equation}
where $\beta_{k-1,i} \in [0,1]$ and $\sum_{i=1}^{n_{k-1}^{xa}} \beta_{k-1,i} =1$. Moreover, by Eqs.~(\ref{ch5:ps_dyanmical_system}) and (\ref{ch5:ps_dyn_error}),
\begin{equation}
p ( \mathbf{x}_{k} | \mathbf{x}_{k-1} ) \approx \sum_{i=1}^{n_{k}^u} \alpha_{k,i}^u N ( \mathbf{x}_{k}: \mathcal{M}_{k-1,k} ( \mathbf{x}_{k-1}), \mathbf{Q}_{k,i} ) \, .
\end{equation}

Then, according to Eq.~(\ref{BRR:prediction}), the prior pdf $p ( \mathbf{x}_{k} | \mathbf{Y}_{k-1} )$ is given by
\begin{equation}
\begin{split}
p ( \mathbf{x}_{k} | \mathbf{Y}_{k-1} ) & = \int p ( \mathbf{x}_{k} | \mathbf{x}_{k-1} )  p ( \mathbf{x}_{k-1} | \mathbf{Y}_{k-1} ) d\mathbf{x}_{k-1} \, \\
& = \sum_{i=1}^{n_{k-1}^{xa}} \sum_{j=1}^{n_{k}^u} \alpha_{k,j}^u \beta_{k-1,i} \, \mathbf{I}_{i,j} ( \mathbf{x}_{k} )\, ,
\end{split}
\end{equation}
where
\begin{equation}
\mathbf{I}_{i,j} ( \mathbf{x}_{k} ) = \int N ( \mathbf{x}_{k}: \mathcal{M}_{k-1,k} ( \mathbf{x}_{k-1}), \mathbf{Q}_{k-1,j} )  N ( \mathbf{x}_{k-1}: \hat{\mathbf{x}}_{k-1,i}^{a}, \hat{\mathbf{P}}_{k-1,i}^a ) d \mathbf{x}_{k-1}.
\end{equation}
The evaluation of $\mathbf{I}_{i,j} ( \mathbf{x}_{k} )$ can be treated as a nonlinear/Gaussian estimation problem, which has been discussed in the previous chapters. Consequently, the previously introduced ensemble or sigma point Kalman filter can be applied to approximate $\mathbf{I}_{i,j} ( \mathbf{x}_{k} )$ as a Gaussian distribution $N ( \mathbf{x}_{k}: \hat{\mathbf{x}}_{k,(i,j)}^b,  \hat{\mathbf{P}}_{k,(i,j)}^b)$, where $\hat{\mathbf{x}}_{k,(i,j)}^b$ and $\hat{\mathbf{P}}_{k,(i,j)}^b$ are the mean and covariance of the background. These are evaluated by propagating forward the analysis ensemble or sigma points at instant $k-1$, with mean $\hat{\mathbf{x}}_{k-1,i}^a$ and covariance $\hat{\mathbf{P}}_{k-1,i}^a$, through the following nonlinear/Gaussian system:
\begin{equation}\label{ind dynamical system}
\begin{split}
&\mathbf{x}_{k+1}=\mathcal{M}_{k,k+1} ( \mathbf{x}_{k}) + \mathbf{u}_{k,j} \, ,\\
& p ( \mathbf{u}_{k,j} ) = N ( \mathbf{u}_{k,j}: \mathbf{0}, \mathbf{Q}_{k,j}) \, .
\end{split}
\end{equation}

Therefore, as an approximation we can re-write $p ( \mathbf{x}_{k} | \mathbf{Y}_{k-1} )$ as 
\begin{equation} \label{prior}
\begin{split}
p ( \mathbf{x}_{k} | \mathbf{Y}_{k-1} ) & \approx \sum_{i=1}^{n_{k-1}^{xa}} \sum_{j=1}^{n_{k}^u} \alpha_{k,j}^u \beta_{k-1,i} \, N ( \mathbf{x}_{k}: \hat{\mathbf{x}}_{k,(i,j)}^b,  \hat{\mathbf{P}}_{k,(i,j)}^b) \,  \\
& = \sum_{s=1}^{n_{k}^{xb}} \gamma_{k,s} N ( \mathbf{x}_{k}: \hat{\mathbf{x}}_{k,s}^b,  \hat{\mathbf{P}}_{k,s}^b) \, ,  \\
\end{split}
\end{equation}
where $n_{k}^{xb} = n_{k-1}^{xa} n_{k}^u$, $\gamma_{k,s} = \alpha_{k,j}^u \beta_{k-1,i} $ with the integer index $s$ being a one-dimensional representation of the index $(i,j)$, e.g., $s=i+n_{k-1}^{xa} (j-1)$, $1 \le i \le n_{k-1}^{xa}$ and $1 \le j \le n_{k}^u$.

\subsection{Filtering step}
After a new observation $\mathbf{y}_k$ is available, one can update the prior pdf $p ( \mathbf{x}_{k} | \mathbf{Y}_{k-1} )$ to the posterior $p ( \mathbf{x}_{k} | \mathbf{Y}_{k} )$, according to Bayes' rule Eq.~(\ref{BRR:update}). Also note that, by Eqs.~(\ref{ch5:ps_observation_system}) and (\ref{ch5:ps_obv_error}), 
\begin{equation} \label{likelihood}
p ( \mathbf{y}_{k} |  \mathbf{x}_{k} ) \approx \sum_{i=1}^{n_{k}^v} \alpha_{k,i}^v N ( \mathbf{y}_{k}: \mathcal{H}_{k} ( \mathbf{x}_{k}), \mathbf{R}_{k,i} ) \, .
\end{equation}

Substituting Eqs.~({\ref{prior}}) and (\ref{likelihood}) into Eq. (\ref{BRR:update}), one has
\begin{equation} \label{posterior}
\begin{split}
p ( \mathbf{x}_{k} | \mathbf{Y}_{k} ) & \propto p ( \mathbf{y}_{k} |  \mathbf{x}_{k} ) p ( \mathbf{x}_{k} | \mathbf{Y}_{k-1} ) \\
& = \sum_{i=1}^{n_{k}^{xb}}  \sum_{j=1}^{n_{k}^v} \gamma_{k,i} \alpha_{k,j}^v N ( \mathbf{x}_{k}: \hat{\mathbf{x}}_{k,i}^b,  \hat{\mathbf{P}}_{k,i}^b) N ( \mathbf{y}_{k}: \mathcal{H}_{k} ( \mathbf{x}_{k}), \mathbf{R}_{k,j} ) \, \\
& =  \sum_{i=1}^{n_{k}^{xb}}  \sum_{j=1}^{n_{k}^v} \gamma_{k,i} \alpha_{k,j}^v N ( \mathbf{y}_{k}: \mathcal{H}_{k} ( \hat{\mathbf{x}}_{k,i}^b ), \hat{\mathbf{P}}^{pr}_{k,i}+\mathbf{R}_{k,j} )  \mathbf{J}_{i,j}  ( \mathbf{x}_{k} )\, , \\
\end{split}
\end{equation}
where in the first line of Eq.~(\ref{posterior}), ``$\propto$'' means ``proportional to'' (by discarding the constant $\int p ( \mathbf{y}_{k} |  \mathbf{x}_{k} ) p ( \mathbf{x}_{k} | \mathbf{Y}_{k-1} ) d \mathbf{x}_{k}$ in Eq. (\ref{BRR:update})). $\hat{\mathbf{P}}^{pr}_{k,i}$ in the third line is the projection covariance of the Gaussian random variable with mean $\hat{\mathbf{x}}_{k,i}^b$ and covariance $\hat{\mathbf{P}}_{k,i}^b$. This can be computed in the context of either the ensemble Kalman filter or the sigma point Kalman filter as introduced in the previous chapters. Finally, 
\begin{equation} \label{ch5:sec_similar_to_Bayes_rule}
\mathbf{J}_{i,j}  ( \mathbf{x}_{k} ) = \dfrac{N ( \mathbf{x}_{k}: \hat{\mathbf{x}}_{k,i}^b,  \hat{\mathbf{P}}_{k,i}^b) N ( \mathbf{y}_{k}: \mathcal{H}_{k} ( \mathbf{x}_{k}), \mathbf{R}_{k,j} )}{N ( \mathbf{y}_{k}: \mathcal{H}_{k} ( \hat{\mathbf{x}}_{k,i}^b ), \hat{\mathbf{P}}^{pr}_{k,i}+\mathbf{R}_{k,j} )} \, .
\end{equation}
Similar to the situation in \S~\ref{ch2:sec_ps_filtering_step} (in particular, Eq.~(\ref{ch2:ps_analysis_pdf_approx})), Eq.~(\ref{ch5:sec_similar_to_Bayes_rule}) can be interpreted from the following point of view: one has a prior pdf $N ( \mathbf{x}_{k}: \hat{\mathbf{x}}_{k,i}^b,  \hat{\mathbf{P}}_{k,i}^b)$ of $ \mathbf{x}_{k}$, and a new observation $\mathbf{y}_{k}$ is obtained through the following observation system
\begin{equation} \label{ind observer}
\begin{split}
&\mathbf{y}_{k}=\mathcal{H}_{k} ( \mathbf{x}_{k}) + \mathbf{v}_{k,j} \, ,\\
& p ( \mathbf{v}_{k,j} ) = N ( \mathbf{v}_{k,j}: \mathbf{0}, \mathbf{R}_{k,j}) \, .
\end{split}
\end{equation}
According to Bayes' rule, $\mathbf{J}_{i,j}  ( \mathbf{x}_{k} )$ is then the posterior pdf of $ \mathbf{x}_{k}$ with the observation $\mathbf{y}_{k}$ made by the observation system Eq.~(\ref{ind observer}).

Consequently, $\mathbf{J}_{i,j}  ( \mathbf{x}_{k} )$ can be approximated by a Gaussian pdf $N ( \mathbf{x}_{k}: \hat{\mathbf{x}}_{k,(i,j)}^a,  \hat{\mathbf{P}}_{k,(i,j)}^a )$, with mean $\hat{\mathbf{x}}_{k,(i,j)}^a$ and covariance  $\hat{\mathbf{P}}_{k,(i,j)}^a$ computed by
\begin{subequations}
\begin{align}
 & \hat{\mathbf{x}}_{k,(i,j)}^{a} =  \hat{\mathbf{x}}_{k,i}^{b} + \mathbf{K}_{k,(i,j)}  ( \mathbf{y}_k - \mathcal{H}_k ( \hat{\mathbf{x}}_{k,i}^{b} ) ),\\
& \hat{\mathbf{P}}_{k,(i,j)}^a = \hat{\mathbf{P}}_{k,i}^b -  \mathbf{K}_{k,(i,j)} ( \hat{\mathbf{P}}^{cr}_{k,i} )^T.
\end{align}
\end{subequations}
Here
\begin{equation}
\mathbf{K}_{k,(i,j)} = \hat{\mathbf{P}}^{cr}_{k,i}  ( \hat{\mathbf{P}}^{pr}_{k,i} + \mathbf{R}_{k,j})^{-1} \, ,
\end{equation}
while $\hat{\mathbf{P}}^{cr}_{k,i}$ and $\hat{\mathbf{P}}^{pr}_{k,i}$ are the cross and projection covariances of the background ensemble of the Gaussian distribution $N ( \mathbf{x}_{k}: \hat{\mathbf{x}}_{k,i}^b,  \hat{\mathbf{P}}_{k,i}^b)$.

Analogous to Eq.~(\ref{prior}), we let $n_{k}^{xa} = n_{k}^{xb} n_{k}^v$, $s=i+n_{k}^{xb} (j-1)$ ($1 \le i \le n_{k}^{xb}$ and $1 \le j \le n_{k}^v$), and 
\begin{equation} \label{ch5:GSF_GSF_weights_filtering_step}
\beta_{k,s} =\beta_{k,(i,j)}=\dfrac{\gamma_{k,i} \alpha_{k,j}^v N ( \mathbf{y}_{k}: \mathcal{H}_{k} ( \hat{\mathbf{x}}_{k,i}^b ), \hat{\mathbf{P}}^{pr}_{k,i}+\mathbf{R}_{k,j} )}{\sum_{i=1}^{n_{k}^{xb}}  \sum_{j=1}^{n_{k}^v} \gamma_{k,i} \alpha_{k,j}^v N ( \mathbf{y}_{k}: \mathcal{H}_{k} ( \hat{\mathbf{x}}_{k,i}^b ) , \hat{\mathbf{P}}^{pr}_{k,i}+\mathbf{R}_{k,j} )} \, .
\end{equation}
Then we have 
\begin{equation} \label{approx posterior}
p ( \mathbf{x}_{k} | \mathbf{Y}_{k} ) \approx \sum_{s=1}^{n_{k}^{xa}} \beta_{k,s} N ( \mathbf{x}_{k}: \hat{\mathbf{x}}_{k,s}^a,  \hat{\mathbf{P}}_{k,s}^a) \, .
\end{equation}

\subsection{Statistics estimation based on the posterior pdf}
The posterior pdf $p ( \mathbf{x}_{k} | \mathbf{Y}_{k} ) $, given in Eq.~(\ref{approx posterior}), embodies all of the necessary statistical information. In particular, one may be interested in estimating the conditional mean $\hat{\mathbf{x}}_{k}^a = \mathbf{E} ( \mathbf{x}_{k} | \mathbf{Y}_{k} ) $ and the conditional covariance $\hat{\mathbf{P}}_{k}^a = \mathbf{Cov} ( \mathbf{x}_{k} | \mathbf{Y}_{k} )$, which are given by \cite[ch. 8]{Anderson-optimal}
\begin{subequations} \label{ch5:GSF_statistics}
\begin{align}
\label{estimated mean} \hat{\mathbf{x}}_{k}^a &= \sum_{s=1}^{n_{k}^{xa}} \beta_{k,s} \hat{\mathbf{x}}_{k,s}^a \, , \\
\label{estimated cov} \hat{\mathbf{P}}_{k}^a &= \sum_{s=1}^{n_{k}^{xa}} \beta_{k,s} (\hat{\mathbf{P}}_{k,s}^a + (\hat{\mathbf{x}}_{k,s}^a - \hat{\mathbf{x}}_{k}^a  ) ( \hat{\mathbf{x}}_{k,s}^a - \hat{\mathbf{x}}_{k}^a  )^T ) \, .
\end{align}
\end{subequations}

Note that the above computations can be done in parallel. For example, one may use $n_{k}^{xa}$ independent processor units, each of which carries out a nonlinear Kalman filter algorithm to assimilate a sub-system described by Eqs.~(\ref{ind dynamical system}) and (\ref{ind observer}). The final results are simply the weighted averages of the outputs of the individual processors. 

\section{An auxiliary algorithm to reduce potential computational cost} \label{ch5:auxiliary}
For convenience, we call the nonlinear Kalman filter adopted in a GSF the ``base filter'' of the GSF. This can be any filter introduced in the previous chapters, e.g., the ensemble Kalman filter (EnKF) and the reduced rank sigma point Kalman filter (SPKF). 

One potential problem of the GSF is that, the number of Gaussian distributions in the GMM may grow very rapidly in certain circumstances. To see this, let the number of Gaussian distributions used to approximate the distributions of the background, the analysis, the dynamical noise and the observation noise at time $k$ be $n_k^{xb}$, $n_k^{xa}$, $n_k^{u}$ and $n_k^{v}$, respectively. In the previous section we have shown that 
\begin{equation} \label{gs number}
\begin{split}
& n_k^{xb} = n_{k-1}^{xa} n_k^{u} \, , \\
& n_k^{xa} = n_{k}^{xb} n_k^{v} \, .
\end{split}
\end{equation}
Therefore, if $n_k^{u} > 1$ or $n_k^{v} >1$ at all times, $n_k^{xb}$ and $n_k^{xa}$ will grow exponentially with time, which will substantially increase the computational cost of the GSF. 

To reduce the computational cost, the authors in \cite{Alspach-nonlinear,Sorenson-recursive} suggested that ``it is possible to combine many terms into a single term without seriously affecting the approximation''. In addition, some weights in the Gaussian sum approximation, i.e., some $\gamma_{k,s}$'s in Eq.~(\ref{prior}) and some $\beta_{k,s}$'s in Eq.~(\ref{approx posterior}), may be sufficiently small compared to the others so that they can be simply neglected \cite{Alspach-nonlinear,Sorenson-recursive}. 

Another possible strategy is to conduct pdf re-approximations: at each assimilation cycle one uses a new GMM, with the specified number of Gaussian distributions, to approximate the prior or the posterior pdf that itself is expressed in terms of a GMM. For example, see \cite{Smith-cluster}. To estimate the parameters of the new GMM (i.e., the weights, the means and covariances of individual Gaussian distributions), the author in \cite{Smith-cluster} suggested an adoption of the expectation-maximization (EM) algorithm. However, the EM algorithm is an iterative method, which may require many iterations for convergence. Thus, using the EM algorithm in high-dimensional systems might be computationally intensive.
 
In this dissertation, we propose another method for the purpose of reducing the computational cost, which is also based on the idea of pdf re-approximation. Our criterion is that the mean and covariance of a new GMM match those of the original one. The benefit of this re-approximation scheme is that, if one chooses a reduced rank SPKF as the base filter and implements the GSF in parallel, then in principle the computational speed of the GSF can be almost the same as that of the reduced rank SPKF.

For illustration, let $p ( \mathbf{x} )$ be the pdf of a random variable $\mathbf{x}$, which is expressed in terms of a GMM with $n$ Gaussian distributions, i.e.,
\begin{equation} \label{ch5:original_GMM}
p ( \mathbf{x} ) = \sum\limits_{i=1}^{n} a_i N (\mathbf{x}: \mathbf{\mu}_i, \mathbf{\Sigma}_i) \, ,
\end{equation}
where $a_i$ is the weight associated with the Gaussian distribution $N (\mathbf{x}: \mathbf{\mu}_i, \mathbf{\Sigma}_i)$ with mean $\mathbf{\mu}_i$ and covariance $\mathbf{\Sigma}_i$. Our objective is to approximate $p ( \mathbf{x} )$ by another GMM $\tilde{p}( \mathbf{x} )$ with $m$ Gaussian distributions ($m<n$):
\begin{equation} \label{ch5:approx_GMM}
\tilde{p} ( \mathbf{x} ) = \sum\limits_{i=0}^{m-1} b_i N (\mathbf{x}: \mathcal{Z}_i, \mathbf{\Delta}_i) \, ,
\end{equation}
where $b_i$ is the weight associated with the distribution $N (\mathbf{x}: \mathcal{Z}_i, \mathbf{\Delta}_i)$. We want to choose proper values of $b_i$, $\mathcal{Z}_i$ and $\mathbf{\Delta}_i$ so that the mean $\tilde{\mathbf{x}} $ and covariance $\tilde{\mathbf{P}}$ of $\tilde{p}( \mathbf{x} )$ match the mean $\bar{\mathbf{x}}$ and covariance $\bar{\mathbf{P}}$ of $p ( \mathbf{x} )$, respectively. From Eq.~(\ref{ch5:GSF_statistics}), 
\begin{subequations}
\begin{align}
\bar{\mathbf{x}} &= \sum_{i=1}^{n} a_i \mathbf{\mu}_i \, , \\
\bar{\mathbf{P}} &= \sum_{s=1}^{n} a_i (\mathbf{\Sigma}_i + (  \mathbf{\mu}_i - \bar{\mathbf{x}} ) (  \mathbf{\mu}_i - \bar{\mathbf{x}}  )^T ) \, ,  
\end{align}
\end{subequations}
while the mean $\tilde{\mathbf{x}}$ and covariance $\tilde{\mathbf{P}}$ of $\tilde{p} ( \mathbf{x} )$ are given by
 \begin{subequations}
\begin{align}
\tilde{\mathbf{x}} &= \sum_{i=0}^{m-1} b_0 \mathcal{Z}_i \, , \\
\label{ch5:auxiliary_cov_whole_part}\tilde{\mathbf{P}} &= \sum_{s=0}^{m-1} b_i (\mathbf{\Delta}_i + (  \mathcal{Z}_i - \tilde{\mathbf{x}} ) (  \mathcal{Z}_i - \tilde{\mathbf{x}}  )^T ) \, .
\end{align}
\end{subequations}

For our purpose, we need to choose proper $b_i$, $\mathcal{Z}_i$ and $\mathbf{\Delta}_i$ so that $\tilde{\mathbf{x}} = \bar{\mathbf{x}}$ and $ \tilde{\mathbf{P}} = \bar{\mathbf{P}}$. To this end, we employ the idea of the unscented transform to generate a set of sigma points to capture the specified mean and covariance. Concretely, let 
\begin{equation}
\bar{\mathbf{S}} = \left [\mathbf{s}_1, \mathbf{s}_2, \dotsb, \mathbf{s}_p \right]
\end{equation} 
be a square root matrix (with $p$ column vectors $\mathbf{s}_i$, $i=1,\dotsb, p$) of $\bar{\mathbf{P}}$, which can be obtained through some numerical decomposition algorithm (e.g., SVD). We generate a set of $2q+1$ sigma points $\mathcal{Z}_i$ \footnote{This means that the number $m=2q+1$ of sigma points, hence the number of Gaussian distributions, is always odd. If one wants to let $m$ be even, one can use the set of sigma points $\left \{ \mathcal{Z}_i \right \}_{i=1}^{2q}$ (by excluding $\mathcal{Z}_0$) in Eq.~(\ref{ch5:auxiliary_sp_generation}) and the associated weights $\left \{ b_i \right \}_{i=1}^{2q}$ in Eq.~(\ref{ch5:auxiliary_sp_weights}) for the pdf re-approximation. It can be shown that the sample mean and covariance of $\left \{ \mathcal{Z}_i \right \}_{i=1}^{2q}$, with $\left \{  b_i \right \}_{i=1}^{2q}$ being the weights, also capture the mean $\bar{\mathbf{x}}$ and covariance $\bar{\mathbf{P}}$, respectively \cite{Julier2000}.} with respect to the triplet $(\eta, \bar{\mathbf{x}}, \tilde{\mathbf{S}}_1 )$, so that
\begin{equation} \label{ch5:auxiliary_sp_generation}
\begin{split}
& \mathcal{Z}_0 = \bar{\mathbf{x}},\\
& \mathcal{Z}_i = \bar{\mathbf{x}} + c \, \sqrt{q+\eta}  \, \mathbf{s}_i, \, i=1, 2, \dotsb, q ,\\
& \mathcal{Z}_i = \bar{\mathbf{x}} - c \, \sqrt{q+\eta} \, \mathbf{s}_i, \, i=q+1, q+2, \dotsb, 2q ,\\
\end{split}
\end{equation}
where $c \,  \mathbf{s}_i$ is the $i$-th column of the square root matrix 
\begin{equation}
\tilde{\mathbf{S}}_1 = c \left[ \mathbf{s}_1, \mathbf{s}_2, \dotsb, \mathbf{s}_q \right]
\end{equation}
for $0<c<1$ ($q \le p$) \footnote{For high dimensional systems like a weather forecast model, $p$ may be in the order of $10^2$ or even higher. Thus for computational efficiency, $q \le p$ is a reasonable choice. This is the reason that we stick to this setting in this chapter. However, if one wishes to let $q>p$, the re-approximation scheme can be adjusted accordingly. The idea is to produce $l$ sets of sigma points in the spirit of Eq.~(\ref{ch5:auxiliary_sp_generation}), either with the same column vectors $\mathbf{s}_i$, or replacing $\mathbf{s}_i$ therein by their rotations in the vector space. Each of these sets consists of $q_0$ sigma points so that $q_0 \le p$ and $l \times q_0 > p$, but with a different coefficient $c_i$ in Eq.~(\ref{ch5:auxiliary_sp_generation}). In this case, a further constraint on the coefficients $c_i$ is that $\sum\limits_{i=1}^{l} c_i^2 < 1$. Moreover, the weights in Eq.~(\ref{ch5:auxiliary_sp_weights}) also have to change accordingly.}. Accordingly, we let $\left \{b_i \right \}_{i=0}^{2q}$ be the weights associated with the set of sigma points $\left \{ \mathcal{Z}_i \right \}_{i=0}^{2q}$ so that
\begin{equation} \label{ch5:auxiliary_sp_weights}
\begin{split}
&b_0 = \frac{\eta}{q+\eta},\\
&b_i = \frac{1}{2(q+\eta )}, \, i=1,2,\dotsb, 2q, \\
\end{split}
\end{equation}    
where $\eta$ is a free parameter analogous to $\lambda$ in the unscented transform (cf. \S~\ref{ch3:sec_ut}). In particular, $\eta=1/2$ means that $b_0 =b_i$ for $i=1,2,\dotsb, 2q$, so that all Gaussian distributions in $\tilde{p} ( \mathbf{x} )$ are equally weighted. According to Eq.~(\ref{ch3:ut_weighted_stat}) in \S~\ref{ch3:sec_idea_of_ut}, we have
\begin{subequations}  
\begin{align}
& \tilde{\mathbf{x}} = \sum_{i=0}^{2q} b_i \mathcal{Z}_i = \bar{\mathbf{x}}\, , \\
\label{ch5:auxiliary_cov_part1} & \sum\limits_{i=0}^{2q} b_i (\mathcal{Z}_i-\tilde{\mathbf{x}} ) (\mathcal{Z}_i-\tilde{\mathbf{x}} )^T = \tilde{\mathbf{S}}_1 ( \tilde{\mathbf{S}}_1)^T = c^2 \sum\limits_{i=1}^{q} \mathbf{s}_i ( \mathbf{s}_i )^T \, .
\end{align}
\end{subequations} 
For simplicity, we let the covariances $\mathbf{\Delta}_i$ of all the Gaussian distributions $N (\mathbf{x}: \mathcal{Z}_i, \mathbf{\Delta}_i)$ in $\tilde{p} ( \mathbf{x} )$ be the same, say $\mathbf{\Delta}_i =\mathbf{\Delta}$ for $i=0,\dotsb, 2q$. Moreover, we further express $\mathbf{\Delta}$ in terms of $\mathbf{\Delta} = \tilde{\mathbf{S}}_2 \tilde{\mathbf{S}}_2^T$, where $\tilde{\mathbf{S}}_2$ is a square root matrix of $\mathbf{\Delta}$.

Substituting Eq.~(\ref{ch5:auxiliary_cov_part1}) into Eq.~(\ref{ch5:auxiliary_cov_whole_part}) and noting that $\sum\limits_{i=0}^{2q} b_i=1$, $\mathbf{\Delta}_i =\mathbf{\Delta}$, we have
\begin{equation}
\tilde{\mathbf{P}} = \mathbf{\Delta} + c^2 \sum\limits_{i=1}^{q} \mathbf{s}_i ( \mathbf{s}_i )^T \, .
\end{equation}
On the other hand, we note that
\begin{equation}
\bar{\mathbf{P}} = \sum\limits_{i=1}^{p} \mathbf{s}_i ( \mathbf{s}_i )^T \, .
\end{equation}
Thus in order to satisfy $\tilde{\mathbf{P}} =\bar{\mathbf{P}}$, we require 
\begin{equation}
\mathbf{\Delta} = (1-c^2) \, \sum\limits_{i=1}^{q} \mathbf{s}_i ( \mathbf{s}_i )^T + \sum\limits_{i=q+1}^{p} \mathbf{s}_i ( \mathbf{s}_i )^T \,.
\end{equation}
Therefore, we can choose 
\begin{equation}
\tilde{\mathbf{S}}_2 = \left [ d \mathbf{s}_1, \dotsb, d \mathbf{s}_q, \mathbf{s}_{q+1}, \dotsb,  \mathbf{s}_{p} \right] 
\end{equation}
as a square root matrix of $\mathbf{\Delta}$, where $d=(1-c^2)^{1/2}$ is the coefficient complementary to $c$. For convenience, hereafter we call $d$ the complementary coefficient. The role of $d$ in influencing the GMM can be illustrated through the following scenario. Suppose $p=q$, then, when $d \rightarrow 0$, we have $\mathbf{\Delta} \rightarrow 0$ and the Gaussian distributions $N (\mathbf{x}: \mathcal{Z}_i, \mathbf{\Delta})$ ($i=0,\dotsb,2q+1$) approach the delta functions with point masses located at $\mathcal{Z}_i$. Thus the GMM in Eq.~(\ref{ch5:approx_GMM}) will approach a Monte Carlo approximation with $\mathcal{Z}_i$ being the samples. On the other hand, when $d \rightarrow 1$, we have $c \rightarrow 0$. Hence all the Gaussian distributions $N (\mathbf{x}: \mathcal{Z}_i, \mathbf{\Delta})$ ($i=0,\dotsb,2q+1$) approach $N (\mathbf{x}: \bar{\mathbf{x}}, \bar{\mathbf{P}})$. Therefore, the GSF will approach its base filter, e.g., the EnKF or the SPKF. 

In the previous chapters we have seen that, at each assimilation cycle of a reduced rank SPKF, the filter requires an SVD in order to produce sigma points. Therefore, for the GSF with a reduced rank SPKF as its base filter and equipped with the auxiliary algorithm, by letting the covariances of all the Gaussian distributions in the re-approximated GMM be the same, one only needs to perform an SVD once (at the filtering step for the SUKF, or at the propagation step for the DDFs) for both the purpose of generating sigma points for its base filter, and that of conducting a pdf re-approximation. Therefore, if the SPKF-based GSF is implemented in parallel, in principle it may achieve almost the same computational speed as the reduced rank SPKF itself. 

In contrast, if one chooses an EnKF (e.g., the ensemble transform Kalman filter) as its base filter and implements the GSF in parallel, then there are extra costs in conducting matrix factorization (e.g., SVD) if one equips the GSF with the auxiliary algorithm. The computational speed of the EnKF-based GSF will become almost the same as that of the SPKF-based GSF.       

\rems
In a GSF, even if both $n_k^{u}$ and $n_k^{v}$ are equal to $1$ in Eq.~(\ref{gs number}) (and so the number of Gaussian distributions does not grow), we still suggest an implementation of the auxiliary algorithm in the filter. The reasons are twofold.

Firstly, the GSF may suffer from the outlier problem. For some Gaussian distributions, say $N (\mathbf{x}: \mathbf{\mu}_i, \mathbf{\Sigma}_i )$ in the GMM, the observation $\mathbf{y}$ may be too far way from the projection $\mathcal{H} (\mu_i)$ of the mean $\mathbf{\mu}_i$ onto the observation space. Thus the distance between $y$ and $\mathcal{H} (\mu_i)$ is so large that it makes the weight of the Gaussian distribution $N (\mathbf{x}: \mathbf{\mu}_i, \mathbf{\Sigma}_i )$ in the GMM negligible compared with other Gaussian distributions (cf. Eq.~(\ref{ch5:GSF_GSF_weights_filtering_step})). In such circumstances, if the tiny weights are continually carried forward to subsequent assimilation cycles, the weights of the GMM might ``collapse'' as in a particle filter \cite{Bengtsson2008}: the weight of one particular Gaussian distribution in the GMM is very close to 1, while the weights of the others are almost zero. In this case, the GSF is in effect reduced to a nonlinear Kalman filter and may suffer from some numerical problems as very tiny values are involved in computation. In such circumstances, the auxiliary algorithm is similar to the re-sampling technique in the particle filter, with the attempt to adjust the weights of the Gaussian distributions in the GMM by replacing the original Gaussian distributions by new ones.

Secondly, the auxiliary algorithm may also help to decrease the computational cost of the GSF with the reduced rank SPKF as its base filter. To see this, note that if the SPKF-based GSF is not equipped with the auxiliary algorithm, the covariances of all Gaussian distributions may not be the same. Therefore to produce sigma points for all the reduced rank SPKFs, one may have to perform an SVD for each different covariance in the reduced rank SPKFs. In contrast, if the SPKF-based GSF is equipped with the auxiliary algorithm, one only needs to conduct one SVD to generate sigma points for all the reduced rank SPKFs, since through a pdf re-approximation, one can choose to let the covariances of all the Gaussian distributions in the new GMM be the same. 
 
\section{Example: Assimilating the 40-dimensional Lorenz-Emanuel 98 system} \label{ch5:sec_experiments}

\subsection{The testbed and the measures of filter performance} \label{ch4:sec_testbed}
The testbed and the measure of filter performance are also the same as those in \S~\ref{ch3:sec_testbed}. The dynamical system (LE 98) is governed by
\begin{equation} \tagref{ch2:ex_LE98}
\frac{dx_i}{dt} = ( x_{i+1} - x_{i-2} ) x_{i-1} - x_i + 8, \, i=1, \dotsb, 40 \, , 
\end{equation}
while the observation system is
\begin{equation} \tagref{ch2:ex_observer}
\mathbf{y}_k = \mathbf{x}_k + \mathbf{v}_k \, ,
\end{equation} 
where $\mathbf{v}_k$ follows the Gaussian distribution $N ( \mathbf{v}_k: \mathbf{0}, \mathbf{I} )$. Note that there is no dynamical noise (except for some discretization errors) in the dynamical system Eq.~(\ref{ch2:ex_LE98}), but for convenience in using the formulae established in \S~\ref{ch5:GSF_approx_solution}, technically we can model the dynamical noise $\mathbf{u}_k$ at instant $k$ by a Gaussian distribution $N ( \mathbf{u}_k: \mathbf{0}, \mathbf{0} )$ with zero mean and zero covariance.  

We integrate the dynamical system Eq.~(\ref{ch2:ex_LE98}) through a fourth-order Runge-Kutta method \cite[Ch.~16]{Vetterling-numerical}, and choose the integration time window to be from $0$ to $50$ dimensionless time units, with the integration step being $0.05$. This setting is denoted by $0:0.05:50$. Similar notations will also be used later. We make the observations at every integration step.

The measure of filter performance is the time averaged relative rmse introduced in \S~\ref{ch2:two_measures}, which is given by
\begin{equation}\tagref{Eq:reltiave rmse}
e_r=\frac{1}{k_{max}}\sum\limits_{k=1}^{k_{max}} \lVert \hat{\mathbf{x}}_k^{a}-{\mathbf{x}}_k^{tr}\rVert_2/ \lVert{\mathbf{x}}_k^{tr}\rVert_2  \, ,
\end{equation}
where $k_{max}$ is the number of assimilation cycles (here $k_{max}=1001$), ${\mathbf{x}}_k^{tr}$ is the truth at the $k$-th assimilation cycle, and $\hat{\mathbf{x}}_k^{a}$ is the estimation of ${\mathbf{x}}_k^{tr}$ obtained by a GSF.  

\subsection{Numerical results}
For illustration, we implement three GSFs with different nonlinear Kalman filters, namely, the reduced rank scaled unscented Kalman filter (SUKF), the reduced rank first order divided difference (DD1) filter (as the representative of the divided difference filters), and the ensemble transform Kalman filter (ETKF) with the centering matrix given by Eq.~(\ref{ch2:EnSRF_spherical_orthornormal_matrix}) (as the representative of the ensemble Kalman filters). 

For the GSFs with the SPKFs as their base filters, to reduce the computational cost, at each assimilation cycle we conduct pdf re-approximations at the step where the SPKFs produce sigma points. With this choice, we perform an SVD only once for both the purposes of pdf re-approximation and the generation of sigma points. Specifically, for the GSF with the SUKF as its base filter, we conduct a pdf re-approximation at the filtering step, thus it is the posterior pdf, in terms of a GMM, that is re-approximated. However, for the GSF with the DD1 filter as its base filter, we conduct a pdf re-approximation at the propagation step, thus it is the prior pdf that is re-approximated. For the GSF with the ETKF as its base filter, conducting a pdf re-approximation at either the propagation step or the filtering step has the same computational cost. In our experiments we choose to conduct a re-approximation at the propagation step. 

Note that in our experiments, both the dynamical and observation noise are characterized by a single Gaussian distribution. Therefore the number of Gaussian distributions in a GSF does not grow with time. However, for the reasons given in \S~\ref{ch5:auxiliary}, we still choose to perform pdf re-approximations in all subsequent experiments. In those circumstances, a re-approximated GMM will have the same number of Gaussian distributions as does the original GMM. 

Concretely, the parameters with respect to the GSFs are set as follows. We let the scale parameter $\eta = 1/2$ in Eq.~(\ref{ch5:auxiliary_sp_weights}), so that all Gaussian distributions in a GMM are equally weighted\footnote{This implies that we do not prefer any particular Gaussian distribution in the GMM.}. We let the number $q =0:1:5$ in \S~\ref{ch5:auxiliary}, which implies that the number $m=2q+1$ of Gaussian distributions\footnote{A re-approximated GMM and the original one have the same number $m$ each time.} takes values from the set of odd integers $1:2:11$. Finally, we let the complementary coefficient $d$ take values from the set $0.05:0.1:0.95$. 

\subsubsection{The Gaussian sum filter with the reduced rank scaled unscented Kalman filter as the base filter}
For convenience, hereafter we call a GSF with the reduced rank SUKF as its base filter the ``SUKF-based GSF'' (similar terminologies has been used previously and will be adopted for the GSFs with other base filters). Since in the previous chapters, we have already studied the effects of the intrinsic parameters on the performances of the base filters, we do not vary these parameters in subsequent experiments. 

The intrinsic parameters of the reduced rank SUKF are set as follows. The length scale of covariance filtering $l_c=240$, the covariance inflation factor $\delta=7$ (cf.~\S~\ref{ch2:sec_two_techniques} for the meanings of these two parameters), the parameters $\alpha=1$, $\beta=2$ and $\lambda=-2$ (cf.~\S~\ref{ch3:sec_UKF}), the lower bound $l_l = 10$, the upper bound $l_u=10$, and the threshold at the first assimilation cycle $\Gamma_1=1000$ (cf.~\S~\ref{ch3:sec_reduced_sukf} or \S~\ref{ch3:sec_numerical_results}). The ensemble size of the background at the first assimilation cycle is $10$.  

\begin{figure*}[!t]
\centering
\hspace*{-0.5in} \includegraphics[width=1.15\textwidth]{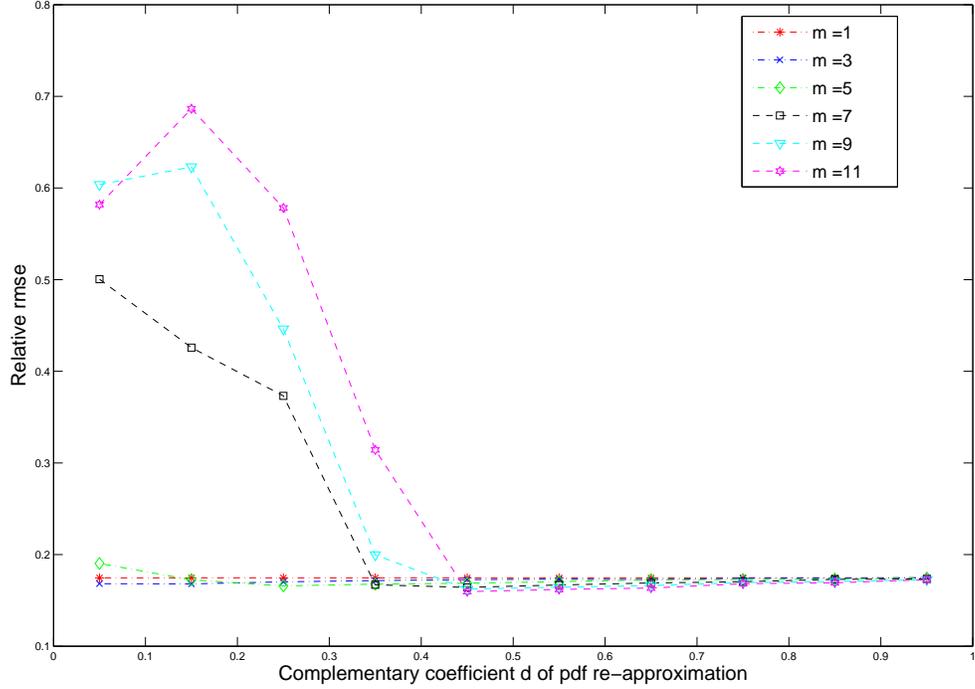} 
\caption{ \label{fig:ch5_sukf_rmse_vs_c} The relative rms errors of the SUKF-based GSFs (with different number of Gaussian distributions) as functions of the complementary coefficient $d$. }
\end{figure*} 

In Fig.~\ref{fig:ch5_sukf_rmse_vs_c} we show the relative rms errors of the SUKF-based GSFs (with different numbers of Gaussian distributions) as functions of the complementary coefficient $d$. As one can see, when the number $m$ of Gaussian distributions is relatively small, say $m=1,3,5$, the relative rmse does not change significantly as $d$ increases from $0.05$ to $0.95$. In particular, when $m=1$, the SUKF-based GSF is equivalent to the SUKF itself, and $d$ does not affect the relative rmse at all, since pdf re-approximations do not take effect in this case. In contrast, when $m$ is relatively large, say $m=7, 9, 11$, the relative rmse exhibits a different behaviour as $d$ increases. The relative rmse with a relatively small value for $d$, say $d=0.05$, is much larger than that with a relatively large value for $d$, say $d=0.95$. Note also that when $d=0.95$ (close to $1$), the relative rms errors of all the GSFs are close to that of the reduced rank SUKF itself. The reason for this was given in \S~\ref{ch5:auxiliary}.   

The under-performance of the GSF with a relatively large $m$ but small $d$ might have a connection with the slow convergence rate of a Monte Carlo approximation. When $d$ is small, the GMM approaches a Monte Carlo approximation, with a convergence rate possible in the order of $\mathcal{O}(m^{-1/2})$. Since the relatively large values for $m$ (say $m=11$) used in our experiments are typically very small for the purpose of convergence, it leads to relatively large estimation errors. On the other hand, the fact that when $d$ is small (say $d=0.05$), the SUKF-based GSF with a small $m$ (say $m=3$) performs better than that with a relatively large $m$ (say $m=11$), is less well understood. A possible explanation might be that, when $m$ is small, the GSF is close to the reduced rank SUKF, which implicitly assumes that the system states follow a Gaussian distribution. Although the Gaussianity assumption might not be realistic, it still works better than the Monte Carlo approximation with a small number of samples.   

\begin{figure*}[!t]
\centering
\hspace*{-0.5in} \includegraphics[width=1.15\textwidth]{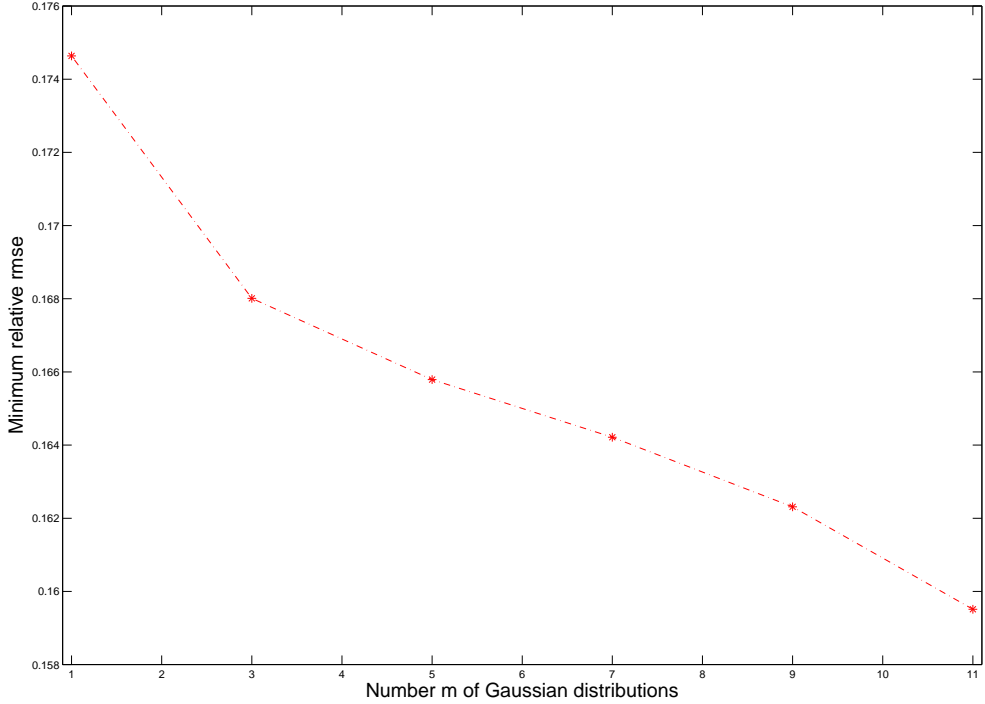} 
\caption{ \label{fig:ch5_sukf_min_rmse_vs_m} The minimum relative rmse $e_r^{min}$ of the SUKF-based GSF as a function of the number $m$ of Gaussian distributions. }
\end{figure*} 

Fig.~\ref{fig:ch5_sukf_rmse_vs_c} indicates that, with other conditions being the same, a larger number $m$ of Gaussian distributions does not necessarily guarantee a better performance. For example, when $d=0.15$, the relative rmse of the GSF with $m=3$ is much lower than that of the GSF with $m=11$. For this reason, in order to compare the performances of the GSFs with different numbers of Gaussian distributions, we need to adopt a new measure. Since in the context of our experiments, the relative rmse $e_r$ in Eq.~(\ref{Eq:reltiave rmse}) is a function of $m$ and $d$, we choose the minimum conditional relative rmse $e_r^{min}(m)$ as the new measure, which reads
\begin{equation} \label{ch5:sec_unnumbered_eq_ermin}
e_r^{min}(m) = \underset{d}{\text{argmin}} \, e_r(m,d) \, .
\end{equation} 
In Eq.~(\ref{ch5:sec_unnumbered_eq_ermin}), $e_r^{min}(m)$ means the minimum value of $e_r(m,d)$ within the range of $d$ tested for a given $m$. For convenience, hereafter we call the minimum conditional relative rmse ``the minimum relative rmse'' for short when it causes no confusion. 

In Fig.~\ref{fig:ch5_sukf_min_rmse_vs_m} we plot $e_r^{min}$ of the SUKF-based GSF as a function of the number $m$ of Gaussian distributions. As one can see, $e_r^{min}$ decreases monotonically as $m$ increases. Thus a larger number of Gaussian distributions can benefit the performance of the SUKF-based GSF in the sense that it can achieve a lower minimum relative rmse.    

\subsubsection{The Gaussian sum filter with the reduced rank first order divided difference filter as the base filter}
For convenience, hereafter we call the GSF with the reduced rank DD1 filter as its base filter the ``DD1-based GSF''. The intrinsic parameters of the reduced rank DD1 filter are set as follows. The length scale of covariance filtering $l_c=240$, the covariance inflation factor $\delta=7$, the interval length of interpolation $h=3$ (cf.~\S~\ref{ch4:sec_stirling}), the lower bound $l_l = 10$, the upper bound $l_u=10$, and the threshold at the first assimilation cycle $\Gamma_1=1000$. The ensemble size of the background at the first assimilation cycle is $10$.  
\begin{figure*}[!t]
\centering
\hspace*{-0.5in} \includegraphics[width=1.15\textwidth]{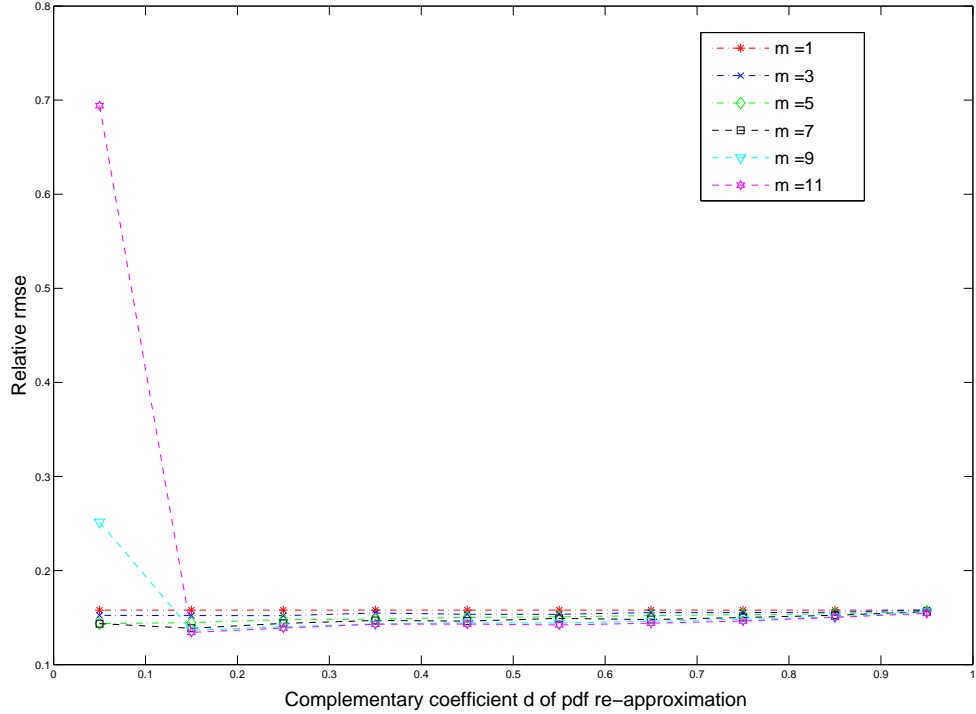} 
\caption{ \label{fig:ch5_dd1_rmse_vs_c} The relative rms errors of the DD1-based GSFs (with different number of Gaussian distributions) as functions of the complementary coefficient $d$ of pdf re-approximation. }
\end{figure*} 

In Fig.~\ref{fig:ch5_dd1_rmse_vs_c} we plot the relative rms errors of the DD1-based GSFs (with different numbers of Gaussian distributions) as functions of $d$. The DD1-based GSF exhibits similar behaviour to the SUKF-based GSF in terms of the relative rmse. Indeed, when the number $m$ of Gaussian distributions is relatively small, say $m=1,3,5$, the relative rms errors do not change significantly as $d$ increases from $0.05$ to $0.95$. Again, when $m=1$, the DD1-based GSF is equivalent to the DD1 filter itself, thus the value of the relative rmse is independent of $d$. For relatively large $m$, say $m=9, 11$, the relative rmse is relatively large when $d=0.05$, but decreases rapidly as $d$ increases. When $d=0.95$, the relative rms errors of all the GSFs again approach that of the DD1 filter.

\begin{figure*}[!t]
\centering
\hspace*{-0.5in} \includegraphics[width=1.15\textwidth]{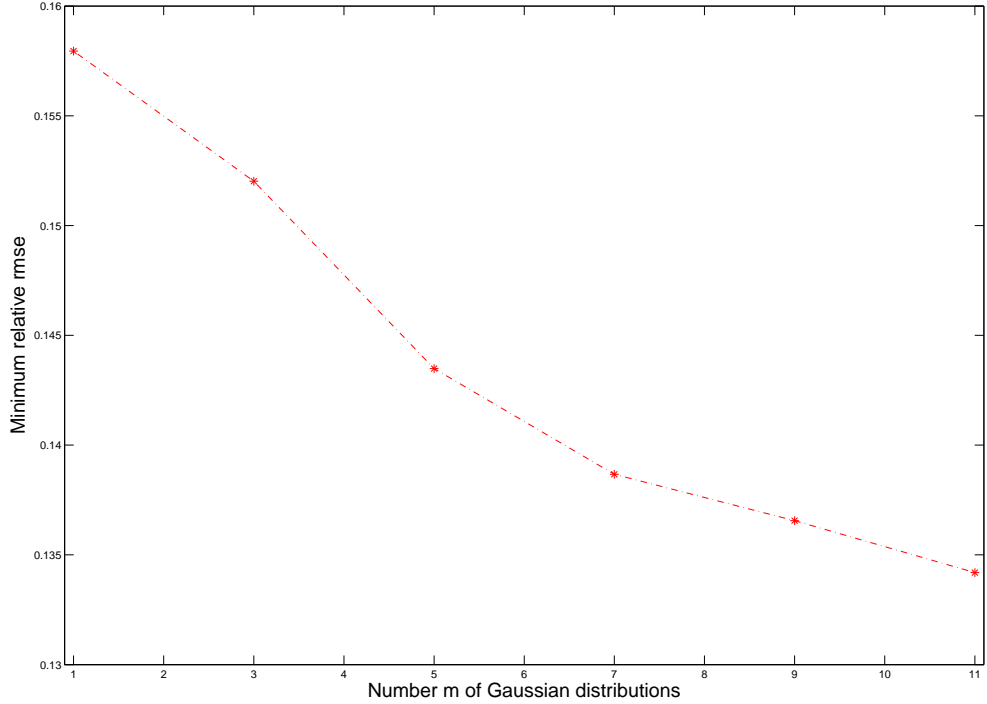} 
\caption{ \label{fig:ch5_dd1_min_rmse_vs_m} The minimum relative rmse of the DD1-based GSF as a function of the number $m$ of Gaussian distributions. }
\end{figure*} 

In Fig.~\ref{fig:ch5_dd1_min_rmse_vs_m} we plot the minimum relative rmse $e_r^{min}$ of the DD1-based GSF as a function of the number $m$ of Gaussian distributions. As one can see, $e_r^{min}$ also decreases monotonically as $m$ increases. Thus in this sense, a larger number of Gaussian distributions benefits the performance of the DD1-based GSF in the context of our experiment settings.  

\subsubsection{The Gaussian sum filter with the ensemble transform Kalman filter as the base filter}
Similarly, we call the GSF with the ETKF as its base filter the ``ETKF-based GSF''. The intrinsic parameters of the ETKF are set as follows. The length scale for covariance filtering $l_c=50$, the covariance inflation factor $\delta=5$. The ensemble size of the background at the first assimilation cycle is $10$.  

\begin{figure*}[!t]
\centering
\hspace*{-0.5in} \includegraphics[width=1.15\textwidth]{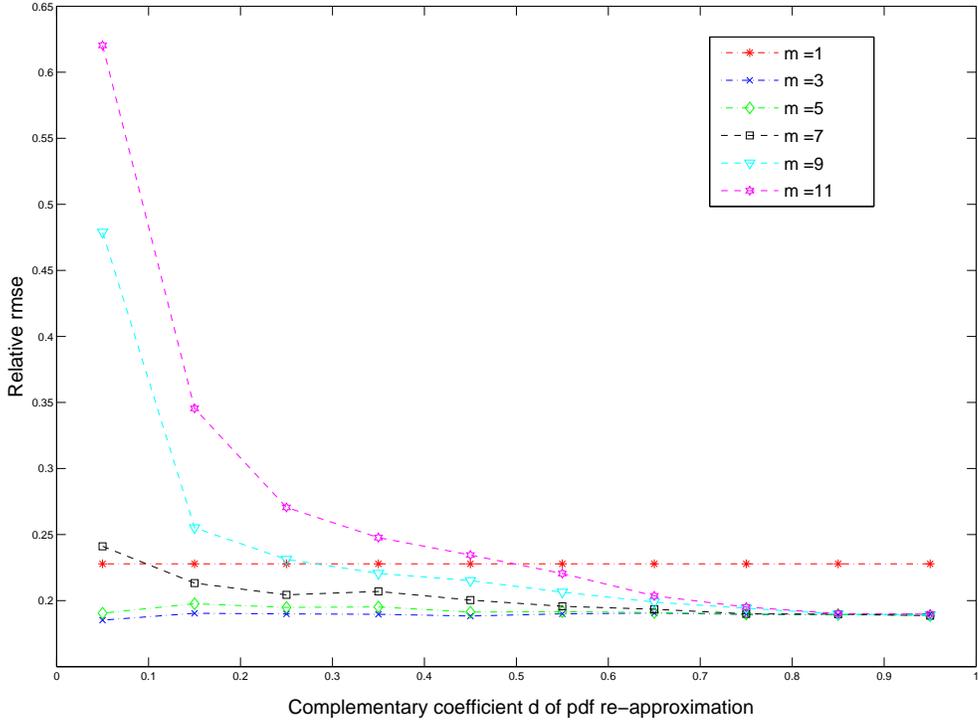} 
\caption{ \label{fig:ch5_etkf_rmse_vs_c} The relative rms errors of the ETKF-based GSFs (with different number of Gaussian distributions) as functions of the complementary coefficient $d$. }
\end{figure*} 
In Fig.~\ref{fig:ch5_etkf_rmse_vs_c} we plot the relative rms errors of the ETKF-based GSFs as functions of $d$. As is evident, the ETKF-based GSF also exhibits similar behaviour to the SUKF and DD1 based GSFs, in terms of the relative rmse. Indeed, when the number $m$ of Gaussian distributions is relatively small, say $m=1,3,5$, the relative rms errors do not change significantly as $d$ increases from $0.05$ to $0.95$. For relatively large $m$, say $m=7, 9, 11$, the relative rmse is relatively large when $d$ is small, say $d=0.05$. The relative rmse also drops as $d$ increases, but not as rapidly as the SUKF and DD1 based GSFs. When $d=0.95$, the relative rms errors of the GSFs with $m>1$ also appear to converge, but not to the relative rmse of its base filter, as in the SUKF and DD1 based GSFs. Our explanation for this difference is that, when $m=1$ we chose to implement the ETKF-based GSF in the same way as that in Chapter~\ref{ch2: EnKF}, where the square root of the background covariance is directly obtained from the background ensemble (without conducting any SVD). But when $m>1$, the square root of the background covariance is obtained through an SVD, as is required for the purpose of pdf re-approximation. 

\begin{figure*}[!t]
\centering
\hspace*{-0.5in} \includegraphics[width=1.15\textwidth]{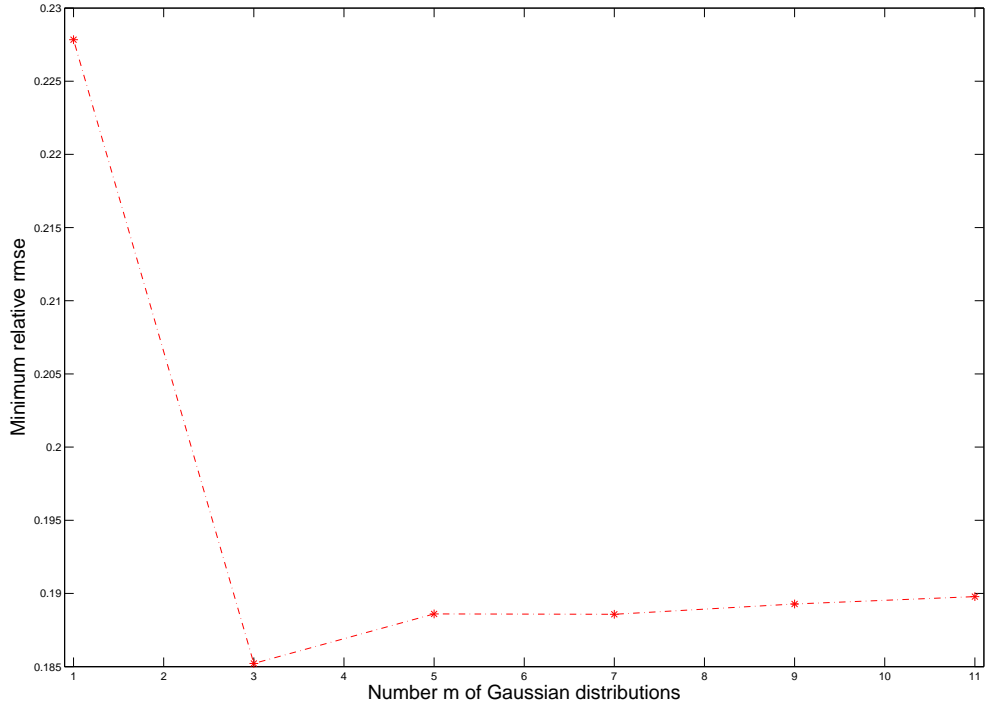} 
\caption{ \label{fig:ch5_etkf_min_rmse_vs_m} The minimum relative rmse of the ETKF-based GSF as a function of the number $m$ of Gaussian distributions. }
\end{figure*} 

In Fig.~\ref{fig:ch5_etkf_min_rmse_vs_m} we show the minimum relative rmse $e_r^{min}$ of the ETKF-based GSF as a function of the number $m$ of Gaussian distributions. Unlike the cases of the SUKF and DD1 based GSFs, the minimum relative rmse of the ETKF-based GSF decreases as $m$ increases from $1$ to $3$. After that, if one further increases $m$, one will instead obtain larger values for $e_r^{min}$. Nevertheless, all these values in the case $m>1$ are lower than that in the case $m=1$. Thus the ETKF-based GSF with $m>1$ also performs better than its base filter, the ETKF. However, a larger number $m$ of Gaussian distributions does not guarantee a lower relative rmse. Instead, $m=3$ is the best choice in the context of our experiment settings. This phenomenon is not understood yet. 

\section{Summary of the chapter}\label{ch5:sec_summary}

In this chapter, we introduced the idea of conducting pdf approximations through Gaussian sum approximations, also known as Gaussian mixture models, to approximate the integrals in recursive Bayesian estimation (RBE). Applying this idea leads to a ``new'' algorithm (relative to the filters discussed in the precious chapters), called the Gaussian sum filter, that can be used to solve the data assimilation problem in nonlinear/non-Gaussian systems approximately. As a feature, a Gaussian sum filter consists of a set of nonlinear Kalman filters, which in principle can be any one that was introduced in the previous chapters. 

A potential problem of the Gaussian sum filter is that, the number of distributions in a mixture model might increase very rapidly with time. This can substantially increase the computational cost. To overcome this problem, we suggested conducting pdf re-approximations by using some Gaussian mixture models to approximate others. To do this, we proposed an auxiliary algorithm such that the re-approximated Gaussian mixture model preserves the mean and covariance of the original one. 

For a SPKF-based Gaussian sum filter, if it is implemented in parallel, then in principle its computational speed can be almost the same as that of the reduced rank SPKF. Note that, even if the number of Gaussian distributions in a Gaussian sum filter does not grow with time, we still suggest conducting pdf re-approximations, as it can avoid some numerical problems.

For illustration, we used the 40-dimensional Lorenz-Emanuel 98 system as the testbed and examined the performances of three Gaussian sum filters, with the reduced rank SUKF, the reduced rank DD1 filter, and the ETKF as their base filters, respectively. Numerical experiments showed that all three Gaussian sum filters outperformed their base filters in terms of their minimum (conditional) relative rms errors.  

\chapter{Conclusions and future work} \label{ch6:conclusions}
\section{Concluding summary}
Since the pioneering work of Evensen \cite{Evensen-sequential}, the ensemble Kalman filter (EnKF) has become a popular method in the data assimilation community. Various versions of the EnKF have been proposed. For examples, see \cite{Anderson-ensemble,Beezley-morphing,Bishop-adaptive,Burgers-analysis,Evensen-sequential,Evensen-ensemble,Evensen2006,Evensen-assimilation,Houtekamer1998,Sakov2007,Wang-which,Whitaker-ensemble,Zupanski-maximum}. 

One of the objectives of this dissertation is to understand the various versions of the EnKF, as well as other recursive filters introduced in the previous chapters, based on the uniform framework of recursive Bayesian estimation (RBE). From the point of view of RBE, all of the filters discussed in the previous chapters can be deemed as different approximation schemes adopted to approximate the integrals of RBE in possibly different scenarios. 

Another objective is to introduce a few nonlinear filters, which include a new family of nonlinear Kalman filter, called the sigma point Kalman filter (SPKF), and the corresponding Gaussian sum filter (GSF) with the SPKF as its base filter, called the sigma point Gaussian sum filter (SPGSF) consequently. The SPKF encompasses two types of nonlinear Kalman filters: the scaled unscented Kalman filter (SUKF) and the divided difference filter (DDF). They provide better estimation accuracy than some of the conventional methods like the extended Kalman filter (EKF). A further advantage of the SPKF is that it does not require evaluations of the derivatives of a nonlinear function, which is convenient in practice.   

Our last objective is to increase the computational efficiency of the SPKF and the SPGSF in high dimensional systems. For each type of the SPKF, we introduced a reduced rank version for it by conducting singular value decompositions (SVDs) on the covariance matrices of the system states. In this way, the number of sigma points in the SPKF can be much less than the dimension of the system under assimilation. Therefore the computational cost of the SPKF can be reduced in high dimensional systems.  

For the GSF, one potential problem is that the number of Gaussian distributions may increase rapidly with time. To address this problem, we suggested conducting a pdf re-approximation at each assimilation cycle, such that one uses a new Gaussian mixture model (with less Gaussian distributions) to approximate the original one. To this end, we proposed an auxiliary algorithm based on the idea of the unscented transform. If a SPGSF equipped with the auxiliary algorithm is implemented in parallel, then in principle its computational speed can be almost the same as that of the SPKF.  

In Chapter~\ref{ch0: introduction} we introduced some basic concepts in data assimilation. We discussed the objectives of this dissertation and introduced two frameworks, least squares estimation (LSE) and RBE, to derive the data assimilation algorithms in this dissertation.

Chapter~\ref{ch1: KF linear} was the starting point of discussing various nonlinear filters in subsequent chapters. We considered the data assimilation problem in linear/Gaussian systems. Applying RBE to solve the problem led to the well-known Kalman filter. We also derived the same result from the point of view of LSE. Based on it, we obtained a useful variant of the conventional Kalman filter, called the Kalman filter with fading memory (KF-FM), which can improve the robustness of the filter. Apart from the KF-FM, we also introduced the square root filter (SRF) as another variant of the conventional Kalman filter in order to increase the numerical accuracy and stability of a filter. The nonlinear filters introduced in subsequent chapters were all implemented in both the forms of the KF-FM and the SRF.

Chapters~\ref{ch2: EnKF} - \ref{ch4:ddfs} studied different extensions of the conventional Kalman filter to nonlinear/Gaussian systems from the point of view of RBE. Chapter~\ref{ch2: EnKF} focused on the ensemble Kalman filter (EnKF). In the literature there are two major types of the EnKF: the stochastic EnKF and its deterministic counterpart, the ensemble square root filter (EnSRF). In principle, all implementations of the EnKF may be deemed as different approximation schemes that are designed to approximate the integrals in RBE numerically. To improve the performance of the EnKF, two auxiliary techniques, namely covariance inflation and filtering, were frequently adopted. Covariance inflation not only compensates for systematic underestimation of the error covariances in the EnKF due to the effect of small ensemble size, but also makes the EnKF behave like the KF-FM. On the other hand, covariance filtering aims to smooth out spuriously large correlations between distant locations. Through some numerical experiments, we examined the performances of the stochastic EnKF and the ensemble transform Kalman filer (ETKF), one of the EnSRFs. Numerical results showed that the ETKF consistently outperformed the stochastic EnKF.          

In Chapter~\ref{ch3:ukf} we introduced one type of nonlinear Kalman filter, called the scaled unscented Kalman filter (SUKF), which is based on the concept of the scaled unscented transform (SUT). One advantage of the SUKF is that it does not require linearizing a nonlinear system, and so is convenient in implementation. We performed an accuracy analysis of the SUKF through Taylor series expansions and compared the accuracy of the unscented Kalman filter (UKF), a special case of the SUKF, with that of the EnKF in Appendix~\ref{appendix:accuracy analysis}. We showed that the UKF can achieve better accuracy than the EnKF under the assumption that the system under assimilation is nonlinear/Gaussian. To reduce the computational cost of the SUKF in high dimensional systems, we proposed a reduced rank version of the filter. Using the Lorenz-Emanuel 98 model, we conducted numerical experiments to examine the effects of the intrinsic parameters of the reduced rank SUKF on the performance of the filter. We also compared the performance of the reduced rank SUKF with that of the ETKF. Numerical results showed that, given the same amount of information, the reduced rank SUKF may consistently outperform the ETKF if there is no covariance filtering conducted on both filters.   

In Chapter~\ref{ch4:ddfs} we presented another type of nonlinear Kalman filter, called the divided difference filter (DDF), which is based on Stirling's Interpolation Formula. Like the SUKF, the DDFs do not require linearizing a nonlinear system under assimilation. Instead, they also generate sigma points for the purpose of approximation. For this reason, in the literature the SUKF and the DDFs are uniformly called the sigma point Kalman filters (or derivative-free filters). We conducted accuracy analyses on the DDFs through Taylor series expansions. For data assimilation in high dimensional systems, we also proposed reduced rank versions of the DDFs. We examined the effects of the intrinsic parameters on the performances of the reduced rank DDFs. We also made a comparison between the DDFs, the SUKF and the ETKF. Numerical results showed that, when there is no covariance filtering, the order of the filter, with performance ranked from best to worst, was the SUKF, the DD2 filter, the CDF, the ETKF, and the DD1 filter. However, when covariance filtering is introduced, it may significantly affect their performances and lead to counter-intuitive results. For example, we showed that the performance of the DD1 filter was much better than that of the DD2 filter for certain parameter values.

In Chapter~\ref{ch5:spgsf} we introduced the Gaussian sum filter (GSF) for data assimilation in nonlinear/non-Gaussian systems. A GSF essentially consists of a set of parallel nonlinear Kalman filters. All the nonlinear Kalman filters introduced in the previous chapters, e.g., the EnKF, the SUKF and the DDFs, can be adopted as the base filters of the GSF. A potential problem of the GSF is that, in some situations, the number of Gaussian distributions in the GSF may increase very rapidly with time. To tackle this problem, we suggested conducting pdf re-approximations. To this end, we proposed an auxiliary algorithm based on the concept of the unscented transform. If a GSF adopts the SUKF or the DDFs as its base filter and is implemented in parallel, then in principle the GSF can achieve almost the same computational speed as its base filter, the SPKF. But if an EnKF is chosen as the base filter, then there is an extra cost at each assimilation cycle in conducting an SVD for the purpose of pdf re-approximation, and the computational speed of the EnKF-based GSF becomes roughly the same as that of the SPGSF. We also conducted numerical experiments to examine the performances of the GSFs with different base filters. Numerical results showed that the GSFs consistently outperformed their base filters in terms of the minimum conditional relative rmse. In this sense, whenever feasible, it would be beneficial to implement the GSF rather than use its base filter.

\section{Main results of this dissertation}
The main results of this dissertation are the following. 

Firstly, we studied the SPKF, including the SUKF and the DDFs. For data assimilation in high dimensional systems, we proposed the reduced rank SPKF to reduce the computational cost. The relevant materials presented in this dissertation were partially drawn from the research works \cite{Luo2008,Luo-ensemble,Luo-sgkf}. 

Secondly, we combined the reduced rank SPKF and the GSF to assimilate nonlinear and non-Gaussian systems. To increase the computational efficiency, we proposed an auxiliary algorithm to conduct pdf re-approximations. The relevant materials presented in this dissertation were partially drawn from the research works \cite{Luo2008-spgsf1,Luo2008-spgsf2}.
 
\section{Outstanding problems and future works}
\subsection{Intrinsic parameters of various nonlinear filters}
One important aspect in implementing the nonlinear filters in the previous chapters is to specify their parameters, which include the covariance inflation factor, the length scale of covariance filtering, and filter-specific parameters, such as parameters $\alpha$, $\beta$ and $\lambda$ in the SUKF. We explained, in a qualitative way, the effects of some parameters on the performances of these filters. For example, the U-turn behaviour of the relative rmse as a function of the covariance inflation factor. However, we are far from being able to fully understand these effects in a quantitative way. Thus in practice, we may have to resort to intensive searches to find the proper values for those filter parameters. 

\subsection{Performance analysis}
Here performance analysis of a nonlinear filter refers to convergence or error bound analysis of the filter. Convergence analysis aims to analyze if the estimated pdf of the system states asymptotically converges to the true pdf. If the answer is yes, one may continue to study the convergence rate. If the answer is no, then one may instead study if there exists any bound of the estimation errors. For examples, see \cite[Ch.~2]{Doucet2001} and \cite{Crisan2002,Xiong2006a}.  

The authors in \cite{Xiong2006a} derived an error bound for the full rank UKF under some assumptions. Similar arguments may also be applied to analyze the filters presented in this dissertation. However, we expect that analyzing the filters equipped with the covariance filtering technique would be more complicated. For this purpose, understanding how the covariance filtering technique affects the behaviour of a filter will be a preliminary step.     

Another interesting topic is the convergence analysis of the GSF. In a recent work \cite{Ali-Loytty2008}, the author reported a proof of weak convergence \footnote{Let $\mathbf{x}$ be a random variable, and $\left\{ \mathbf{x}_n \right\}$ be a sequence of random variables. Then we say the sequence $\left\{ \mathbf{x}_n \right\}$ converges weakly to $\mathbf{x}$, if the cumulative density function $F_{\mathbf{x}_n} (\mathbf{\eta})$ of $\mathbf{x}_n$ tends to the cumulative density function $F_{\mathbf{x}} (\mathbf{\eta})$ of $\mathbf{x}$ for all $\mathbf{\eta}$ when $n \rightarrow \infty$ \cite{Ali-Loytty2008}.} of the GSF with the extended Kalman filter as its base filter. This highlights a possible way to prove the weak convergence of the GSFs with other base filters.     

\subsection{Parameter estimation and smoothing problems}
We may extend the methods presented in this dissertation to parameter estimation and smoothing problems. 

A convenient strategy to address a parameter estimation problem is to treat the parameters to be estimated as (unobserved) system states, so that the problem is recast as a state estimation problem \cite{Wan2002}. From this point of view, in principle, all the filters presented in this dissertation can be applied to (approximately) solve the parameter estimation problem.

Solving a smoothing problem in general is more complicated as it may require a dynamical system to run backwards. Currently, a popular smoothing algorithm in the data assimilation community is the four-dimensional variational data assimilation (4D-Var) \cite[ch. 5]{Kalnay-atmospheric} \footnote{4D-Var is essentially a smoothing algorithm, although it is customary in numerical weather prediction (NWP) to use it for prediction. In doing this, one assumes that an improved initial condition obtained by 4D-Var yields an improved forecast \cite{Zupanski2002}.}. A shortcoming of 4D-Var is that it only generates an improved initial condition, but without the associated error covariance to indicate how good the estimation might be. To overcome this problem, one may adopt a Kalman smoother \cite[ch. 9]{Simon2006}, which can provide both an estimation of a system state and its associated error covariance\footnote{In fact, it can be shown that in linear/Gaussian systems, 4D-VAR is equivalent to the fixed-lag Kalman smoother (but without the calculations of the associated error covariances) \cite{Li2001-Optimality}.}. Again, the three problems in data assimilation, i.e., nonlinearity, non-Gaussianity and high dimensionality, might also arise in a smoothing problem. Thus one may apply the methods presented in this dissertation to tackle these problems. 

\subsection{Particle filter}
The particle filter (PF) \cite{Arulampalam2002,Gordon1993} is another type of filter designed for nonlinear/non-Gaussian scenarios, which can also be interpreted from the point of view of RBE \cite{Arulampalam2002}. The PF also conducts pdf approximations to approximate the integrals in RBE. But unlike the GSF, the PF adopts a Monte Carlo method, rather than the Gaussian mixture model (GMM), for the purpose of approximation.

With sufficient samples, it can be shown that the PF can approach the optimal solution to a nonlinear/non-Gaussian estimation problem \cite{Doucet2001,Gordon1993}. The major difficulty preventing the use of the PF in high dimensional systems is that the number of samples from the pdf of the system states needs to scale exponentially with the dimension of the system under assimilation (often known as the curse-of-dimensionality) \cite{Bengtsson2008, Snyder2008}. Otherwise the weights associated with the samples may often collapse: the weights will concentrate onto a single sample so that its weight is very close to one, while the other samples only have negligible (almost zero) weights \cite{Bengtsson2008}. For example, the authors in \cite{Snyder2008} showed that for a 200-dimensional system, at least $10^{11}$ samples are required with the PF approach. Thus it may need substantial computational cost in dealing with the large samples, for example, updating the weight of each sample at the filtering step.

As particle filtering is a relatively new field, there shall be improvements to make the PF perform better. For example, one may design new sampling algorithms to make the approximation error of the PF decrease more rapidly. As a result, one may generate, on average, less samples for the PF equipped with new sampling algorithms to achieve the same performance as the PF equipped with conventional sampling algorithms. One example in this aspect can be dated back to an early research paper \cite{Rlchtmyer1958}, where the author conjectured that, in certain circumstances, ``systematic sampling'' \footnote{Systematic sampling involves taking a sample from the available data in a set pattern, rather than at random. See, for example, \cite[ch. 7]{Chaudhuri2005} for more details.} can generate samples that converge faster than those generated by random sampling. In a recent work \cite{Gentil2008}, the authors proved that the above conjecture holds under certain conditions. Numerical results in \cite{Gentil2008} also showed that, although with fewer samples, the PF equipped with systematic sampling can achieve better accuracy than the PF equipped with some of the random sampling methods. Note that, however, even with systematic sampling, the problem of the curse-of-dimensionality remains in the PF. Thus further efforts are needed to tackle this problem.

\appendix
\chapter{An alternative derivation of analysis update formula of the conventional Kalman filter} \label{appendix:weighted LSE}

Here we follow \cite{Bouttier-data} to show that one can also derive analysis update formula Eq.~(\ref{ch1:LSE_analysis_update}) at the filtering step of the conventional Kalman filter by minimizing the following weighted quadratic cost function
\begin{equation} 
J( \mathbf{x}_k ) = \dfrac{1}{2} (\mathbf{x}_k - \mathbf{x}^b_k )^T
(\mathbf{P}_k^b )^{-1} (\mathbf{x}_k - \mathbf{x}^b_k ) + \dfrac{1}{2}
(\mathbf{y}_k - \mathcal{H}_{k}\mathbf{x}_k )^T \mathbf{R}_k^{-1}
(\mathbf{y}_k - \mathcal{H}_{k} \mathbf{x}_k ) \, .
\end{equation}
To see this, we solve
\begin{equation}
\dfrac{d J( \mathbf{x}_k )}{d \mathbf{x}_k} \vert_{\mathbf{x}_k=\mathbf{x}_k^a}=0.
\end{equation}
This leads to
\begin{equation}
\begin{split}
0 & =(\mathbf{P}_k^b )^{-1} (\mathbf{x}_k^a - \mathbf{x}^b_k ) - \mathcal{H}_{k}^T \mathbf{R}_k^{-1}
(\mathbf{y}_k - \mathcal{H}_{k} \mathbf{x}_k^a ) \, \\
& = (\mathbf{P}_k^b )^{-1} (\mathbf{x}_k^a - \mathbf{x}^b_k ) - \mathcal{H}_{k}^T \mathbf{R}_k^{-1}
(\mathbf{y}_k - \mathcal{H}_{k} \mathbf{x}_k^b ) + \mathcal{H}_{k}^T \mathbf{R}_k^{-1}
\mathcal{H}_{k} (\mathbf{x}_k^a - \mathbf{x}_k^b )  \, . \\
\end{split}
\end{equation}
With some algebra, we have
\begin{equation} \label{apendix_kalman_gain}
\mathbf{x}_k^a = \mathbf{x}_k^b + \left[ (\mathbf{P}_k^b )^{-1} + \mathcal{H}_{k}^T \mathbf{R}_k^{-1}
\mathcal{H}_{k} \right]^{-1} \mathcal{H}_{k}^T \mathbf{R}_k^{-1}
(\mathbf{y}_k - \mathcal{H}_{k} \mathbf{x}_k^b ) \, . 
\end{equation}
Comparing Eq.~(\ref{apendix_kalman_gain}) with Eqs.~(\ref{ch1:LSE_analysis_update}) and (\ref{ch1:LSE_optimal_weight}), it is clear that we need to prove that the Kalman gain $\mathbf{K}_k$ in Eq.~(\ref{ch1:LSE_analysis_update}) satisfies
\begin{equation}
\begin{split}
\mathbf{K}_k & = \mathbf{P}_k \mathcal{H}_{k}^T
\left( \mathcal{H}_{k} \mathbf{P}_k^b \mathcal{H}_{k}^T +  \mathbf{R}_k \right)^{-1} \\
& = \left[ (\mathbf{P}_k^b )^{-1} + \mathcal{H}_{k}^T \mathbf{R}_k^{-1}
\mathcal{H}_{k} \right]^{-1} \mathcal{H}_{k}^T \mathbf{R}_k^{-1} \, . 
\end{split}
\end{equation}

To this end, we need to employ the matrix inversion lemma \cite[ch.~1]{Simon2006}, which reads
\begin{equation}\label{appendix:matrix inversion}
\left( \mathbf{A} + \mathbf{B}\mathbf{D}^{-1}\mathbf{C} \right)^{-1} = \mathbf{A}^{-1}
- \mathbf{A}^{-1}  \mathbf{B}\left( \mathbf{D}+ \mathbf{C}\mathbf{A}^{-1}\mathbf{B}\right)^{-1} \mathbf{C}\mathbf{A}^{-1} \, ,
\end{equation}
where $\mathbf{A}$, $\mathbf{B}$, $\mathbf{C}$ and $\mathbf{D}$ are matrices with suitable dimensions.

Applying the matrix inversion lemma, we have
\begin{equation}
\begin{split}
 & \left(\left(\mathbf{P}_k^b \right)^{-1} + \mathcal{H}_{k}^T \mathbf{R}_k^{-1} \mathcal{H}_{k} \right)^{-1} \mathcal{H}_{k}^T \mathbf{R}_k^{-1}  \\
&=  \left[ \mathbf{P}_k^b - \mathbf{P}_k^b \mathcal{H}_{k}^T \left( \mathcal{H}_{k} \mathbf{P}_k^b \mathcal{H}_{k}^T + \mathbf{R}_k \right)^{-1}  \mathcal{H}_{k} \mathbf{P}_k^b \right] \mathcal{H}_{k}^T \mathbf{R}_k^{-1} \\
&= \mathbf{P}_k^b \mathcal{H}_{k}^T \left[ \mathbf{R}_k^{-1} -\left( \mathcal{H}_{k} \mathbf{P}_k^b \mathcal{H}_{k}^T + \mathbf{R}_k \right)^{-1} \mathcal{H}_{k} \mathbf{P}_k^b \mathcal{H}_{k}^T  \mathbf{R}_k^{-1} \right] \\
&= \mathbf{P}_k^b \mathcal{H}_{k}^T \left[ \mathbf{R}_k^{-1} -\left( \mathcal{H}_{k} \mathbf{P}_k^b \mathcal{H}_{k}^T + \mathbf{R}_k \right)^{-1} \left( \mathbf{R}_k \left( \mathcal{H}_{k} \mathbf{P}_k^b \mathcal{H}_{k}^T \right)^{-1} \right)^{-1}  \right] \\
&= \mathbf{P}_k^b \mathcal{H}_{k}^T \left[ \mathbf{R}_k^{-1} -\left( \mathbf{R}_k \left( \mathcal{H}_{k} \mathbf{P}_k^b \mathcal{H}_{k}^T \right)^{-1} \mathbf{R}_k+ \mathbf{R}_k \right)^{-1} \right] \\
&=  \mathbf{P}_k^b \mathcal{H}_{k}^T \left[ \mathbf{R}_k^{-1} - \mathbf{R}_k^{-1} \left( \left( \mathcal{H}_{k} \mathbf{P}_k^b \mathcal{H}_{k}^T \right)^{-1} + \mathbf{R}_k^{-1} \right)^{-1} \mathbf{R}_k^{-1} \right] \\
& = \mathbf{P}_k \mathcal{H}_{k}^T
\left( \mathcal{H}_{k} \mathbf{P}_k^b \mathcal{H}_{k}^T +  \mathbf{R}_k \right)^{-1} \, .
\end{split}
\end{equation}
The last step is obtained by letting $\mathbf{A} = \mathbf{R}_k$, $\mathbf{B} =\mathbf{C} =\mathbf{I}$, and $\mathbf{D} = \left( \mathcal{H}_{k} \mathbf{P}_k^b \mathcal{H}_{k}^T \right)^{-1}$ in Eq.~(\ref{appendix:matrix inversion}).

\chapter{Proof of Eq.~(\ref{ch1:RBE_propagation_3})} \label{appendix:KF-deduction}

The proof provided here mainly follows the procedures given in \cite[\S~2.2]{Simandl2006}. We suppose that $\mathbf{x}_k$ is an $m$-dimensional state vector defined on the $m$-dimensional real space $\mathbb{R}^{m}$ for any $k$. For notational convenience, we will drop the domain of definition of $\mathbf{x}_k$ in subsequent deductions. 

For our purpose, given
\begin{equation} \tag{$\ref{ch1:RBE_propagation_2}$}
p \left( \mathbf{x}_{k} | \mathbf{Y}_{k-1} \right) = \int N \left( \mathbf{x}_{k}: \mathcal{M}_{k,k-1} \, \mathbf{x}_{k-1}, \mathbf{Q}_{k} \right)   N \left( \mathbf{x}_{k-1}: \mathbf{x}_{k-1}^a, \mathbf{P}_{k-1}^a \right) d\mathbf{x}_{k-1} \, ,
\end{equation}
we need to show that
\begin{equation}
p \left( \mathbf{x}_{k} | \mathbf{Y}_{k-1} \right) = N \left( \mathbf{x}_{k}: \mathbf{x}_{k}^b, \mathbf{P}_{k}^b \right) \, ,
\end{equation}
where $\mathbf{x}_{k}^b$ and $\mathbf{P}_{k}^b$ are given by Eq.~(\ref{ch1:RBE_propagation_3}).

Note that
\begin{subequations}
\begin{align}
& N \left( \mathbf{x}_{k}: \mathcal{M}_{k,k-1} \, \mathbf{x}_{k-1}, \mathbf{Q}_{k} \right) \\
\nonumber &=\left( 2 \pi \right)^{-m/2}  \left( \text{det} \, \mathbf{Q}_{k} \right)^{-1/2} \text{exp} \left \{ -\dfrac{1}{2} \left( \mathbf{x}_k - \mathcal{M}_{k,k-1} \, \mathbf{x}_{k-1}\right)^T \mathbf{Q}_{k}^{-1} \left( \mathbf{x}_k - \mathcal{M}_{k,k-1} \, \mathbf{x}_{k-1}\right)  \right \} , 
\end{align}
\end{subequations}
and
\begin{subequations}
\begin{align}
& N \left( \mathbf{x}_{k-1}: \mathbf{x}_{k-1}^a, \mathbf{P}_{k-1}^a \right) \\
\nonumber &=\left( 2 \pi \right)^{-m/2}  \left( \text{det} \, \mathbf{P}_{k-1}^a \right)^{-1/2} \text{exp} \left \{ -\dfrac{1}{2} \left(  \mathbf{x}_{k-1} - \mathbf{x}_{k-1}^a\right)^T \left( \mathbf{P}_{k-1}^a \right)^{-1} \left(  \mathbf{x}_{k-1} - \mathbf{x}_{k-1}^a \right)  \right \} ,
\end{align}
\end{subequations}
where $\text{det} \bullet$ denotes the determinant of a matrix, and $\text{exp} \left( \bullet \right)$ means the exponential function.

Thus 
\begin{equation}\label{appendix:Gaussian pdf product}
\begin{split}
& N \left( \mathbf{x}_{k}: \mathcal{M}_{k,k-1} \, \mathbf{x}_{k-1}, \mathbf{Q}_{k} \right)   N \left( \mathbf{x}_{k-1}: \mathbf{x}_{k-1}^a, \mathbf{P}_{k-1}^a \right) \\
& = \left( 2 \pi \right)^{-m}  \left( \text{det} \, \mathbf{P}_{k-1}^a \right)^{-1/2}  \left( \text{det} \, \mathbf{Q}_{k} \right)^{-1/2} \times \\
& ~ \quad \text{exp} \left \{ -\dfrac{1}{2} \left( \mathbf{x}_k - \mathcal{M}_{k,k-1} \, \mathbf{x}_{k-1}\right)^T \mathbf{Q}_{k}^{-1} \left( \mathbf{x}_k - \mathcal{M}_{k,k-1} \, \mathbf{x}_{k-1}\right) \right. \\
& \left. ~ \quad \qquad - \dfrac{1}{2} \left(  \mathbf{x}_{k-1} - \mathbf{x}_{k-1}^a\right)^T \left( \mathbf{P}_{k-1}^a \right)^{-1} \left(  \mathbf{x}_{k-1} - \mathbf{x}_{k-1}^a \right) \right \} \\
& = \left( 2 \pi \right)^{-m}  \left( \text{det} \, \mathbf{P}_{k-1}^a \right)^{-1/2}  \left( \text{det} \, \mathbf{Q}_{k} \right)^{-1/2} \times \\ & ~ \quad \text{exp} \left \{ \mathbf{x}_{k-1}^T \mathbf{A}_{k-1}^{-1} \mathbf{x}_{k-1} - \mathbf{x}_{k-1}^T \mathbf{b}_{k-1} -  \mathbf{b}_{k-1}^T \mathbf{x}_{k-1} + \mathbf{c}_{k-1} \right \} \, ,
\end{split}
\end{equation}
where 
\begin{subequations}\label{appendix:RBE_misc_notations}
\begin{align}
& \mathbf{A}_{k-1}^{-1} =  \left( \mathbf{P}_{k-1}^a \right)^{-1} + \mathcal{M}_{k,k-1}^T \mathbf{Q}_{k}^{-1} \mathcal{M}_{k,k-1} \, ,\\
& \mathbf{b}_{k-1} =  \left( \mathbf{P}_{k-1}^a \right)^{-1} \mathbf{x}_{k-1}^a + \mathcal{M}_{k,k-1}^T \mathbf{Q}_{k}^{-1} \mathbf{x}_{k} \, ,\\
& \mathbf{c}_{k-1} = \left( \mathbf{x}_{k-1}^a \right)^T \left( \mathbf{P}_{k-1}^a \right)^{-1} \mathbf{x}_{k-1}^a + \mathbf{x}_{k}^T \mathbf{Q}_{k}^{-1} \mathbf{x}_{k} \, .
\end{align}
\end{subequations}
After integration, one has
\begin{equation} \label{appendix:RBE_misc_int}
\begin{split}
& \int \text{exp} \left \{ \mathbf{x}_{k-1}^T \mathbf{A}_{k-1}^{-1} \mathbf{x}_{k-1} - \mathbf{x}_{k-1}^T \mathbf{b}_{k-1} -  \mathbf{b}_{k-1}^T \mathbf{x}_{k-1} + \mathbf{c}_{k-1} \right \} \, d\mathbf{x}_{k-1}  \\
& = \left( 2 \pi \right)^{m/2}  \left( \text{det} \, \mathbf{A}_{k-1} \right)^{1/2} \text{exp} \left \{ \dfrac{1}{2} \left( \mathbf{b}_{k-1}^T  \mathbf{A}_{k-1} \mathbf{b}_{k-1} - \mathbf{c}_{k-1} \right ) \right \} .
\end{split}
\end{equation}
Substituting Eq.~(\ref{appendix:RBE_misc_notations}) into Eq.~(\ref{appendix:RBE_misc_int}), with some algebra (also cf. \cite[\S~2.2]{Simandl2006}), it can be shown that
\begin{equation}
\mathbf{b}_{k-1}^T  \mathbf{A}_{k-1} \mathbf{b}_{k-1} - \mathbf{c}_{k-1} = - \left( \mathbf{x}_k - \mathbf{x}_k^b \right)^T \left( \mathbf{P}_k^b \right)^{-1} \left( \mathbf{x}_k - \mathbf{x}_k^b \right) \, ,
\end{equation}
where 
\begin{subequations}\label{appendix:RBE_propagation_3}
\begin{align}
\label{appendix:RBE_background}& \mathbf{x}_{k}^b = \mathcal{M}_{k,k-1} \, \mathbf{x}_{k-1}^a \, , \\
\label{appendix:RBE_background_covariance}& \mathbf{P}_{k}^b = \mathcal{M}_{k,k-1} \, \mathbf{P}_{k-1}^a \, \mathcal{M}_{k,k-1}^T + \mathbf{Q}_{k}\, . 
\end{align}
\end{subequations}

With the above results, one has
\begin{equation} \label{appendix:RBE_misc_1}
\begin{split}
p \left( \mathbf{x}_{k} | \mathbf{Y}_{k-1} \right) &= \int N \left( \mathbf{x}_{k}: \mathcal{M}_{k,k-1} \, \mathbf{x}_{k-1}, \mathbf{Q}_{k} \right)   N \left( \mathbf{x}_{k-1}: \mathbf{x}_{k-1}^a, \mathbf{P}_{k-1}^a \right) d\mathbf{x}_{k-1} \\
&= \left( 2 \pi \right)^{-m/2}  \left( \text{det} \, \mathbf{P}_{k-1}^a \right)^{-1/2}  \left( \text{det} \, \mathbf{Q}_{k} \right)^{-1/2} \left( \text{det} \, \mathbf{A}_{k-1} \right)^{1/2} \times \\ 
& ~ \quad \text{exp} \left \{ - \dfrac{1}{2} \left( \mathbf{x}_k - \mathbf{x}_k^b \right)^T \left( \mathbf{P}_k^b \right)^{-1} \left( \mathbf{x}_k - \mathbf{x}_k^b \right) \right \} \, .
\end{split}
\end{equation}

Now we examine the determinants before the exponential function in Eq.~(\ref{appendix:RBE_misc_1}). To this end, the following identities, which hold for arbitrary matrices $\mathbf{U}$ and $\mathbf{V}$ with suitable dimensions, will be used:
\begin{equation}
\begin{split}
& \text{det} \left( \mathbf{U}^{-1} \right) = \left(  \text{det} \mathbf{U} \right)^{-1} \, ,\\
&  \text{det} \left( \mathbf{U} \mathbf{V} \right) = \left(  \text{det} \mathbf{U} \right) \left(  \text{det} \mathbf{V} \right) \, ,\\
&  \text{det} \left( \mathbf{I} + \mathbf{U} \mathbf{V} \right) =  \text{det} \left( \mathbf{I} + \mathbf{V} \mathbf{U} \right) \, .
\end{split}
\end{equation}
Thus, 
\begin{equation}
\begin{split}
&\left( \text{det} \, \mathbf{P}_{k-1}^a \right)^{-1/2}  \left( \text{det} \, \mathbf{Q}_{k} \right)^{-1/2} \left( \text{det} \, \mathbf{A}_{k-1} \right)^{1/2} \\
&= \left( \text{det} \, \mathbf{P}_{k-1}^a \right)^{-1/2}  \left( \text{det} \, \mathbf{Q}_{k} \right)^{-1/2}   \left( \text{det} \left( \left( \mathbf{P}_{k-1}^a \right)^{-1} + \mathcal{M}_{k,k-1}^T \mathbf{Q}_{k}^{-1} \mathcal{M}_{k,k-1} \right) \right)^{-1/2} \\
& = \left( \text{det} \, \mathbf{Q}_{k} \right)^{-1/2} \left( \text{det} \left( \mathbf{I} + \mathbf{P}_{k-1}^a \mathcal{M}_{k,k-1}^T \mathbf{Q}_{k}^{-1} \mathcal{M}_{k,k-1} \right) \right)^{-1/2} \\
& = \left( \text{det} \, \mathbf{Q}_{k} \right)^{-1/2} \left( \text{det} \left( \mathbf{I} + \mathcal{M}_{k,k-1} \mathbf{P}_{k-1}^a \mathcal{M}_{k,k-1}^T \mathbf{Q}_{k}^{-1} \right) \right)^{-1/2} \\
& = \left( \text{det} \left( \mathbf{Q}_{k} + \mathcal{M}_{k,k-1} \mathbf{P}_{k-1}^a \mathcal{M}_{k,k-1}^T \right) \right) ^{-1/2} \\
& = \left( \text{det}  \mathbf{P}_{k}^b \right) ^{-1/2} \, .
\end{split}
\end{equation}
Therefore 
\begin{equation} \label{appendix:RBE_misc_2}
\begin{split}
p \left( \mathbf{x}_{k} | \mathbf{Y}_{k-1} \right) &= \int N \left( \mathbf{x}_{k}: \mathcal{M}_{k,k-1} \, \mathbf{x}_{k-1}, \mathbf{Q}_{k} \right)   N \left( \mathbf{x}_{k-1}: \mathbf{x}_{k-1}^a, \mathbf{P}_{k-1}^a \right) d\mathbf{x}_{k-1} \\
&= \left( 2 \pi \right)^{-m/2} \left(  \text{det}  \mathbf{P}_{k}^b \right) ^{-1/2}  \text{exp} \left \{ - \dfrac{1}{2} \left( \mathbf{x}_k - \mathbf{x}_k^b \right)^T \left( \mathbf{P}_k^b \right)^{-1} \left( \mathbf{x}_k - \mathbf{x}_k^b \right) \right \} \\
& = N \left( \mathbf{x}_{k}: \mathbf{x}_{k}^b, \mathbf{P}_{k}^b \right) \, .
\end{split}
\end{equation}
\chapter{Accuracy comparison between the ensemble Kalman filter and the unscented Kalman filter}\label{appendix:accuracy analysis}

Here we make an accuracy comparison between the ensemble Kalman filter (EnKF) and the unscented Kalman filter (UKF) in solving the recast problem in Fig.~\ref{ch3:fig_problem_recast}. The analysis of the UKF follows the works \cite{Julier2000,Julier2004}, while the analysis of the EnKF follows our original work \cite{Luo-ensemble}.

Given a random variable $\mathbf{x}$ with the known mean $\bar{\mathbf{x}}$ and covariance $\mathbf{P}_{x}$, we are interested in estimating the mean and covariance of the transformed random variable $\mathbf{\eta} =
\mathcal{F} \left(\mathbf{x} \right)$ . In analysis, we assume that the nonlinear function $\mathcal{F} \left(\mathbf{x} \right)$ can be expanded as a Taylor series, which converges to the true value of the variable $\mathbf{\eta}$ \cite{Julier2000,Julier2004}.

Note that the unscented transform (UT) can be applied to more general
situations \cite{Julier2000,Julier2004}. For example, the transformed random variable may be described by $\mathbf{\eta} =
\mathcal{F} \left(\mathbf{z}, \mathbf{u} \right)$, where $\mathbf{z}$ is a Gaussian random variable with mean $\bar{\mathbf{z}}$ and covariance $\mathbf{P}_{z}$, and the dynamical
noise $\mathbf{u}$ is independent of $\mathbf{z}$, following a Gaussian distribution 
with mean zero and covariance $\mathbf{Q}$. To apply the UT, we
adopt the joint state $\mathbf{x} = [\mathbf{z}^T, \mathbf{u}^T ]^T$,
so that the transformation becomes $\mathbf{\eta} = \mathcal{F}
\left(\mathbf{x} \right)$. In this case, the sigma points of the
joint state $\mathbf{x}$ can be generated in the spirit of Eq.~(\ref{ch3:ut_sigma_points}), i.e.,
\begin{equation} \label{joint state ut}
\begin{split}
& \mathcal{X}_0 = \left[\bar{\mathbf{z}}^T, \mathbf{0}^T \right]^T \\
& \mathcal{X}_i = \mathcal{X}_0 + \sqrt{L+\lambda} \, \left( \sqrt{\mathbf{P}_{x}} \right)_i, \, i=1, 2, \dotsb, L ,\\
& \mathcal{X}_i = \mathcal{X}_0 - \sqrt{L+\lambda} \, \left( \sqrt{\mathbf{P}_{x}} \right)_{i-L}, \, i=L+1, L+2, \dotsb, 2L ,\\
\end{split}
\end{equation}
where $\mathbf{0}$ means the zero vector,
\begin{equation}
\mathbf{P}_{x} = \begin{pmatrix} \mathbf{P}_{z} & \mathbf{0} \\ \mathbf{0} & \mathbf{Q} \end{pmatrix} 
\end{equation}
is the covariance matrix of the joint state $\mathbf{x}$, and $\left( \sqrt{\mathbf{P}_{x}} \right)_i$ means the $i$-th column vector of the square root matrix $\sqrt{\mathbf{P}_{x}} \, $. Therefore we can just use $\mathbf{\eta} =\mathcal{F} \left(\mathbf{x} \right)$ as the general form in our discussion. 

\section*{The actual mean and covariance of the transformed variable in terms of Taylor series}
Given a vector $\bar{\mathbf{x}}$ and a Gaussian perturbation  $\delta \mathbf{x}$ with zero mean and covariance $\mathbf{P}_{x}$, we first expand the transform $\mathbf{\eta} = \mathcal{F} (\bar{\mathbf{x}}+\delta \mathbf{x})$ in a Taylor series around the point $\bar{\mathbf{x}}$. The mean $\bar{\mathbf{\eta}}$ is then given by (cf. Eq.~(\ref{ch3:sut_symmetric_y_mean_in_expansion}))
\begin{equation}\label{series expectation}
\begin{split}
\bar{\mathbf{\eta}} &= \mathbb{E} \left( \mathcal{F} (\bar{\mathbf{x}}+\delta \mathbf{x}) \right) \\
\quad &=\mathcal{F} (\bar{\mathbf{x}}) +  \frac{1}{2!}\left( \nabla^T \mathbf{P}_{x}  \nabla  \right) \mathcal{F} + \frac{1}{4!} \, \mathbb{E} \left( \mathbf{D}_{\delta \mathbf{x}}^4 \mathcal{F} \right) + \dotsb \, ,
\end{split}
\end{equation}
where the operator $\mathbf{D}_{\delta \mathbf{x}}^i$ is defined as
\begin{equation} \label{D def}
\mathbf{D}_{\delta \mathbf{x}}^i \equiv \left( \delta \mathbf{x}^T \nabla \right)^i 
\end{equation}
for a positive integer number $i$.

Similarly, the covariance matrix $\mathbf{P}_{\eta}$ is given by (cf. Eq.~(\ref{ch3:sut_y_cov_in_expansion}))
\begin{equation} \label{expanded cov of y}
\begin{split}
\mathbf{P}_{\eta} =& \mathbb{E} \left( \left( \mathbf{\eta}- \bar{\mathbf{\eta}}\right)   \left( \mathbf{\eta}- \bar{\mathbf{\eta}}\right)^T \right) \\
= &   (\nabla \mathcal{F})^T \mathbf{P}_{x}  (\nabla \mathcal{F}) + \mathbb{E} \left[ \frac{ \mathbf{D}_{\delta \mathbf{x}} \mathcal{F} \left( \mathbf{D}_{\delta \mathbf{x}}^3 \mathcal{F} \right)^T}{3!}   \right. \\
&  +  \frac{ \mathbf{D}_{\delta \mathbf{x}}^2 \mathcal{F} \left( \mathbf{D}_{\delta \mathbf{x}}^2 \mathcal{F} \right)^T}{2! \times 2!} \left.  +  \frac{ \mathbf{D}_{\delta \mathbf{x}}^3 \mathcal{F} \left( \mathbf{D}_{\delta \mathbf{x}} \mathcal{F} \right)^T}{3!} \right]    \\
& - \left[  \left( \frac{ \nabla^T \mathbf{P}_{x}  \nabla }{2!}
\right) \mathcal{F} \right] \left[  \left( \frac{ \nabla^T
\mathbf{P}_{x}  \nabla }{2!}  \right) \mathcal{F} \right]^T + \dotsb \, .
\end{split}
\end{equation}

\section*{Accuracies of the EnKF}

Given an ensemble $\{ \mathbf{x}_i\}_{i=1}^n$, the sample mean $\hat{\mathbf{\eta}}$ of the transformed variable $\mathbf{\eta}$ in the EnKF is given by
\begin{equation} \label{enkf series mean}
\begin{split}
\hat{\mathbf{\eta}} &= \dfrac{1}{n} \sum\limits_{i=1}^{n} \mathcal{F} \left( \mathbf{x}_i \right)\\
& =  \mathcal{F} \left( \bar{\mathbf{x}} \right) +
\frac{\sum_{i=1}^{n} \mathbf{D}_{\delta \mathbf{x}_i}
\mathcal{F}}{n} + \frac{\sum_{i=1}^{n}  \mathbf{D}_{\delta \mathbf{x}_i}^2
\mathcal{F}}{n \times 2!} + \frac{\sum_{i=1}^{n}  \mathbf{D}_{\delta
\mathbf{x}_i}^3 \mathcal{F}}{n \times 3!} + \dotsb ,
\end{split}
\end{equation}
where $\delta \mathbf{x}_i = \mathbf{x}_i -  \bar{\mathbf{x}}$.

On the rhs of Eq. (\ref{enkf series mean}), 
\[
\frac{\sum_{i=1}^{n}  \mathbf{D}_{\delta \mathbf{x}_i}^2 \mathcal{F}}{n \times 2!} = \frac{1}{2!}  \nabla^T \left( \frac{1}{n} \sum\limits_{i=1}^{n}  \delta \mathbf{x}_i  \delta \mathbf{x}_i^T \right) \nabla  \mathcal{F},
\]
which is a biased estimation of the second term on the rhs of Eq. (\ref{series expectation}). Moreover, due to the effect of finite ensemble size, the odd-order derivative terms like 
\begin{equation}
\frac{1}{n} \sum_{i=1}^{n}  \mathbf{D}_{\delta \mathbf{x}_i} \mathcal{F}= \left( \frac{1}{n} \sum\limits_{i=1}^{n} \delta \mathbf{x}_i \right)^T \nabla \mathcal{F}
\end{equation}
and
\begin{equation} \label{sample D3}
\frac{1}{n} \sum_{i=1}^{n}  \mathbf{D}_{\delta \mathbf{x}_i}^3 \mathcal{F}= \nabla^T \left( \frac{1}{n} \sum\limits_{i=1}^{n} \delta \mathbf{x}_i  \delta \mathbf{x}_i^T \nabla \delta \mathbf{x}_i^T \right) \nabla \mathcal{F}
\end{equation}
may not vanish.

Similarly, we have the sample covariance $\hat{\mathbf{P}}_{\eta}$ given by
\begin{equation} \label{series enkf cov}
\begin{split}
\hat{\mathbf{P}}_{\eta} &= \dfrac{1}{n-1} \sum\limits_{i=1}^{n} \left( \mathcal{F} \left( \mathbf{x}_i \right) - \hat{\mathbf{\eta}} \right) \left( \mathcal{F} \left( \mathbf{x}_i \right) - \hat{\mathbf{\eta}} \right)^T \\
& = (\nabla \mathcal{F})^T \hat{\mathbf{P}}_{x}  (\nabla \mathcal{F}) + \frac{1}{(n-1) \times 2! } \sum\limits_{i=1}^{n} \left[ \mathbf{D}_{\delta \mathbf{x}_i} \mathcal{F}  \left( \mathbf{D}_{\delta \mathbf{x}_i}^2 \mathcal{F}\right) ^T + \mathbf{D}_{\delta \mathbf{x}_i}^2  \mathcal{F}  \left( \mathbf{D}_{\delta \mathbf{x}_i} \mathcal{F}\right) ^T \right]\\
&+ \frac{1}{n-1} \left( \frac{ \sum_{i=1}^{n}  \mathbf{D}_{\delta \mathbf{x}_i} \mathcal{F}  \left( \mathbf{D}_{\delta \mathbf{x}_i}^3 \mathcal{F}\right) ^T }{3!} + \frac{ \sum_{i=1}^{n}  \mathbf{D}_{\delta \mathbf{x}_i} \mathcal{F}^2  \left( \mathbf{D}_{\delta \mathbf{x}_i}^2 \mathcal{F}\right) ^T }{ 2! \times 2!}   \right. \\
& \left. + \frac{ \sum_{i=1}^{n}  \mathbf{D}_{\delta \mathbf{x}_i}^3 \mathcal{F}  \left( \mathbf{D}_{\delta \mathbf{x}_i} \mathcal{F}\right) ^T }{3!} \right) - \frac{n-1}{n}  \left[  \left( \frac{ \nabla^T \hat{\mathbf{P}}_{x}  \nabla }{2!}  \right) \mathcal{F} \right] \left[  \left( \frac{ \nabla^T \hat{\mathbf{P}}_{x}  \nabla }{2!}  \right) \mathcal{F} \right]^T  + \dotsb \, .
\end{split}
\end{equation}
Note that here 
\[
\hat{\mathbf{P}}_{x} = \frac{1}{n-1} \sum\limits_{i=1}^{n} \delta \mathbf{x}_i \delta \mathbf{x}_i^T  
\]
is an unbiased estimation of $\mathbf{P}_{x}$.

Comparing Eq. (\ref{series enkf cov}) with Eq. (\ref{expanded cov of y}), we note that
\begin{itemize}
\item there are also some spurious modes, for example, terms like $\sum\limits_{i=1}^{n} \left[ \mathbf{D}_{\delta \mathbf{x}_i} \mathcal{F}  \left( \mathbf{D}_{\delta \mathbf{x}_i}^2 \mathcal{F}\right) ^T \right]$ and $\sum\limits_{i=1}^{n} \left[ \mathbf{D}_{\delta \mathbf{x}_i}^2 \mathcal{F}  \left( \mathbf{D}_{\delta \mathbf{x}_i} \mathcal{F}\right) ^T \right]$, arising in Eq. (\ref{series enkf cov});
\item the term $\dfrac{n-1}{n} \left[  \left( \nabla^T \hat{\mathbf{P}}_{x}  \nabla   \right) \mathcal{F} \right] \left[  \left( \nabla^T \hat{\mathbf{P}}_{x}  \nabla   \right) \mathcal{F} \right]^T$ in Eq. (\ref{series enkf cov}) is biased.
\end{itemize}

\section*{Accuracies of the unscented transform}
For the UT, given a set of sigma points $\left \{ \mathcal{X}_i\right \}_{i=0}^{2L}$ with mean $\bar{\mathbf{x}}$ and covariance $\mathbf{P}_{x}$, the sample mean is given by
\begin{equation} \label{series ut mean}
\begin{split}
\hat{\mathbf{\eta}} &= \sum\limits_{i=0}^{2L} W_i \mathcal{F} \left( \mathcal{X}_i \right) \\
& = \mathcal{F} \left(  \bar{\mathbf{x}} \right) + \frac{1}{2!} \left( \nabla^T \mathbf{P}_{x}  \nabla \right) \mathcal{F} + \frac{1}{2(L+\lambda)} \sum\limits_{i=1}^{2L} \left( \mathbf{D}_{\delta \mathbf{x}_i}^4 \mathcal{F} + \dotsb \right) \, ,
\end{split}
\end{equation}
where $\delta \mathbf{x}_i = \mathcal{X}_i - \bar{\mathbf{x}}$. Note that in Eq. (\ref{series ut mean}), the first and third order derivative terms vanish because of the symmetry in sigma points.

Comparing Eq. (\ref{series ut mean}) with Eq. (\ref{series expectation}), we note that, second and third order derivative terms in Eq. (\ref{series ut mean}) match those in Eq. (\ref{series expectation}) exactly, while the difference starts from fourth order terms.

Similarly, the sample covariance is given by
\begin{equation} \label{series ut cov}
\begin{split}
\hat{\mathbf{P}}_{\eta} =& \sum\limits_{i=0}^{2L} W_i \left(  \mathcal{F} \left( \mathcal{X}_i \right) - \hat{\mathbf{\eta}} \right) \left(  \mathcal{F} \left( \mathcal{X}_i \right) - \hat{\mathbf{\eta}} \right) ^T \\
= &  (\nabla \mathcal{F})^T \mathbf{P}_{x}  (\nabla \mathcal{F}) + \frac{1}{2(L+\lambda)} \left( \frac{ \sum_{i=1}^{2L}  \mathbf{D}_{\delta \mathbf{x}_i} \mathcal{F}  \left( \mathbf{D}_{\delta \mathbf{x}_i}^3 \mathcal{F}\right) ^T }{3!} + \right. \\
& \left. \frac{ \sum_{i=1}^{2L}  \mathbf{D}_{\delta \mathbf{x}_i} \mathcal{F}^2  \left( \mathbf{D}_{\delta \mathbf{x}_i}^2 \mathcal{F}\right) ^T }{ 2! \times 2!}   + \frac{ \sum_{i=1}^{2L}  \mathbf{D}_{\delta \mathbf{x}_i}^3 \mathcal{F}  \left( \mathbf{D}_{\delta \mathbf{x}_i} \mathcal{F}\right) ^T }{3!} \right) \\
& - \left[  \left( \frac{ \nabla^T \mathbf{P}_{x}  \nabla }{2!}  \right) \mathcal{F} \right] \left[  \left( \frac{ \nabla^T \mathbf{P}_{x}  \nabla }{2!}  \right) \mathcal{F} \right]^T  + \dotsb \, .
\end{split}
\end{equation}

Clearly, unlike Eq. (\ref{series enkf cov}), there are no terms like $\sum\limits_{i=1}^{2L} \left[ \mathbf{D}_{\delta \mathbf{x}_i} \mathcal{F}  \left( \mathbf{D}_{\delta \mathbf{x}_i}^2 \mathcal{F}\right) ^T \right]$ and $\sum\limits_{i=1}^{2L} \left[ \mathbf{D}_{\delta \mathbf{x}_i}^2 \mathcal{F}  \left( \mathbf{D}_{\delta \mathbf{x}_i} \mathcal{F}\right) ^T \right]$ arising in Eq.~(\ref{series ut cov}) because of the symmetry in sigma points. Moreover, there is also no bias in the term $\left[  \left( \nabla^T \mathbf{P}_{x}  \nabla   \right) \mathcal{F} \right] \left[  \left( \nabla^T \mathbf{P}_{x}  \nabla   \right) \mathcal{F} \right]^T$.

\addcontentsline{toc}{chapter}{Bibliography}

\bibliographystyle{elsart-num-sort}
\bibliography{./thesis}

\end{document}